\documentclass[
		twoside,openright,titlepage,numbers=noenddot,headinclude,
                footinclude=true,cleardoublepage=empty,dottedtoc,
                BCOR=5mm,paper=a4,fontsize=11pt, 
                ngerman,american, 
                ]{scrreprt} 
                
\PassOptionsToPackage{eulerchapternumbers,listings, pdfspacing, subfig,beramono,parts}{classicthesis}

\newcommand{\myTitle}{Wandering in cities}
\newcommand{\mySubtitle}{A statistical physics approach to urban theory}

\newcommand{\myName}{R\'emi Louf\xspace}

\newcommand{\myFaculty}{Put data here\xspace}

\newcommand{\myUni}{Université Pierre et Marie Curie\xspace}
\newcommand{\myTime}{August 2015\xspace}
\newcommand{\myVersion}{version 1.0\xspace}

\newcounter{dummy} 
\providecommand{\mLyX}{L\kern-.1667em\lower.25em\hbox{Y}\kern-.125emX\@}

\usepackage{lipsum} 

\PassOptionsToPackage{latin9}{inputenc} 
\usepackage{inputenc}
 
\usepackage{babel}

\PassOptionsToPackage{square,numbers}{natbib}
 \usepackage{natbib}
 
\usepackage{bibtopic}

\PassOptionsToPackage{fleqn}{amsmath} 
 \usepackage{amsmath}
 
\PassOptionsToPackage{T1}{fontenc} 
\usepackage{fontenc}

\usepackage{xspace} 

\usepackage{mparhack} 

\usepackage{fixltx2e} 

\PassOptionsToPackage{smaller}{acronym} 
\usepackage{acronym} 

\PassOptionsToPackage{pdftex}{graphicx}
\usepackage{graphicx} 

\usepackage{tabularx} 
\usepackage{caption}
\usepackage{subfig}  

\usepackage{listings} 
\PassOptionsToPackage{pdftex,hyperfootnotes=false,pdfpagelabels}{hyperref}
\usepackage{hyperref}  
\hypersetup{
colorlinks=true, linktocpage=true, pdfstartpage=3, pdfstartview=FitV,
breaklinks=true, pdfpagemode=UseNone, pageanchor=true, pdfpagemode=UseOutlines,
plainpages=false, bookmarksnumbered, bookmarksopen=true, bookmarksopenlevel=1,
hypertexnames=true, pdfhighlight=/O, urlcolor=webbrown, linkcolor=RoyalBlue, citecolor=webgreen,
pdftitle={\myTitle},
pdfauthor={\textcopyright\ \myName, \myUni, \myFaculty},
pdfsubject={},
pdfkeywords={},
pdfcreator={pdfLaTeX},
pdfproducer={LaTeX with hyperref and classicthesis}
}   

\usepackage{ifthen} 
\newboolean{enable-backrefs} 
\setboolean{enable-backrefs}{false} 

\newcommand{\backrefnotcitedstring}{\relax} 
\newcommand{\backrefcitedsinglestring}[1]{(Cited on page~#1.)}
\newcommand{\backrefcitedmultistring}[1]{(Cited on pages~#1.)}
\ifthenelse{\boolean{enable-backrefs}} 
{
\PassOptionsToPackage{hyperpageref}{backref}
\usepackage{backref} 
\renewcommand*{\backref}[1]{}  
\renewcommand*{\backrefalt}[4]{
\ifcase #1 
\backrefnotcitedstring
\or
\backrefcitedsinglestring{#2}
\else
\backrefcitedmultistring{#2}
\fi}
}{\relax} 

\makeatletter
\@ifpackageloaded{babel}
{
\addto\extrasamerican{

}
\addto\extrasngerman{

}
}{\relax}
\makeatother

\usepackage{classicthesis} 

\usepackage{hfoldsty} 

\begin{document}

\frenchspacing 

\raggedbottom 

\selectlanguage{american} 


\pagenumbering{roman} 

\pagestyle{plain} 



\begin{titlepage}

\begin{addmargin}[-1cm]{-3cm}
\begin{center}
\large

\hfill
\vfill

\begingroup
\color{Maroon}\spacedallcaps{\myTitle} \\ \medskip 
\emph{\mySubtitle} \\ \bigskip 
\endgroup

\spacedlowsmallcaps{\myName} 

\vfill



\vfill

\end{center}
\end{addmargin}

\end{titlepage}

\thispagestyle{empty}

\hfill

\vfill

\noindent\myName: \textit{\myTitle,} \mySubtitle, 
\textcopyright\ \myTime







\cleardoublepage

\thispagestyle{empty}
\refstepcounter{dummy}

\pdfbookmark[1]{Opening}{Opening} 

\vspace*{3cm}

\begin{center}{\slshape 
    Jazz is not dead,\\ 
    It just smells funny.} \\ \medskip
    ---\\
    Frank Zappa,\,\it{Be-Bop Tango}\,(1973)
\end{center}

\bigskip
\cleardoublepage

\thispagestyle{empty}
\refstepcounter{dummy}

\pdfbookmark[1]{Dedication}{Dedication} 

\vspace*{3cm}

\bigskip

\begin{center}
    \`A mes parents, \\ \smallskip
    Qui ont toujours plac\'e l'\'education avant tout.

    \bigskip

    \`A tous mes enseignants : \\
    Votre main n'\'etait pas tendue en vain.
\end{center}
\cleardoublepage

\pdfbookmark[1]{Abstract}{Abstract} 

\begingroup
\let\clearpage\relax
\let\cleardoublepage\relax
\let\cleardoublepage\relax

\chapter*{Abstract} 

The amount of data that is being gathered about cities is increasing in size and
specificity. However, despite this wealth of information, we still have little
understanding of what really drives the processes behind urbanisation. In this
thesis we apply some ideas from statistical physics to the study of cities.\\

We first present a stochastic, out-of-equilibrium model of city growth that describes
the structure of the mobility pattern of individuals. The model explains the
appearance of secondary subcenters as an effect of traffic congestion. We are
also able to predict the sublinear increase of the number of centers with
population size, a prediction that is verified on American and Spanish data. 

Within the framework of this model, we are further able to give a prediction for the
scaling exponent of the total distance commuted daily, the total length of the
road network, the total delay due to congestion, the quantity of
CO\textsubscript{2} emitted, and the surface area with the population size of
cities. Predictions that agree with data gathered for U.S. cities.

In the third part, we focus on the quantitative description of the
patterns of residential segregation. We propose a unifying theoretical framework
in which segregation can be empirically characterised. We propose a measure of
interaction between the different categories. Building on the information about
the attraction and repulsion between categories, we are able to define classes
in a quantitative, unambiguous way. The framework also allows us to identify the
neighbourhoods where the different classes concentrate, and characterise their
properties and spatial arrangement. Finally, we revisit the traditional
dichotomy between poor city centers and rich suburbs; we provide a measure that
is adapted to anisotropic, polycentric cities.

In the fourth and last part, we present the most important results of our
studies on spatial networks. We first present an empirical
study of $131$ street patterns across the world, and propose a method to
classify the patterns based on the geometrical shape of the blocks. We then present a cost-benefit
analysis framework to understand the properties and growth of spatial networks.
We introduce an iterative model that can explain the emergence of a hierarchical
structure (`hubs and spokes') in growing spatial networks. Starting from the
cost-benefit framework of this model, we finally show that the length, number of
stations and ridership of subways and rail networks can be estimated knowing the
area, population and wealth of the underlying region.\\

Throughout this thesis, we try to convey the idea that the complexity of cities is --
almost paradoxically -- better comprehended through simple approaches. 
Looking for structure in data, trying to isolate the most important processes,
building simple models and only keeping those which agree with data, constitute
a universal method that is also relevant to the study of urban systems.

\endgroup			

\vfill
\cleardoublepage

\pdfbookmark[1]{Publications}{Publications} 

\chapter*{Publications} 
\label{p:publications}

Most of the ideas and figures presented in this thesis have appeared previously in the following publications:

\smallskip

\bibliographystyle{unsrt}
\begin{btSect}{biblio/thesis-articles}
\btPrintAll
\end{btSect}

\bigskip

This thesis was the occasion to study other topics, that I chose
not to address in the present dissertation. Some of this work appeared in the following
publication:

\smallskip

\bibliographystyle{unsrt}
\begin{btSect}{biblio/thesis-others}
\btPrintAll
\end{btSect}

\bigskip

Parts of the work presented here will also be the subject of a book that
is being written concurrently with this dissertation

\smallskip

\bibliographystyle{unsrt}
\begin{btSect}{biblio/thesis-book}
\btPrintAll
\end{btSect}
\cleardoublepage

\pdfbookmark[1]{Acknowledgements}{Acknowledgements} 

\begingroup

\let\clearpage\relax
\let\cleardoublepage\relax
\let\cleardoublepage\relax

\chapter*{Acknowledgements} 

\begin{flushright}{\slshape    
    I had the idea that the world's so full of pain \\
    it must sometimes make a kind of singing.} \\\medskip
    --- Robert Hass,\, \emph{Faint Music}~\cite{Hass:1998}
\end{flushright}

\bigskip

I would first and foremost like to express my gratitude to Alain Barrat, Michael
Batty, Pablo Jensen, Renaud Lambiotte, Jean-Pierre Nadal and Lena Sanders for accepting to be in
my thesis committee. I can only imagine how daunting reading a PhD thesis must
be, and I am really grateful for their time and for helping me to jump over this
last hurdle. I sincerely hope this was not as awful as reading a 200 pages
document (written by a student) sounds.\\

Thank you, also, to the Institut de Physique Th\'eorique. By that, I mean its
director Michel Bauer, the PhD coordinators Olivier Golinelli and St\'ephane
Nonnenmacher, and the incredibly efficient (and kind!) administrative and IT
staff.\\

This thesis would not have been the same without the supervision of Marc
Barthelemy. I still remember the sunny, winter morning of 2012 when the bus
droppped me in the middle of nowhere, not far from the Institut de Physique
Th\'eorique. The fields were frozen, and vaguely white. I remember thinking, as
I stepped out of the bus and felt the cold air biting my face, that this was
going to be temporary. Needless to say, I did not sign up for three more years
because of my love of long commutes, or the (too) familiar smell of the RER B. Neither
was it for the office with a view, not on the fields---or the odd rabbit coming
out of the nearby forest---, but on a brick wall. And faded flowers. 

Now that this thesis is coming to end, I do know why I stayed, what I am
grateful for. I am grateful to Marc for his trust, and for giving me the
appreciated freedom to choose and explore the topics I liked. I am thankful, not
only for his scientific supervision, our insightful discussions, but also for
teaching me the things that cannot be found in textbooks: how to navigate the
research world. Marc also knew when to push me and how to push me, and this is
hopefully reflected in the work presented here. I am greatly indebted to him.\\

Je tiens \`{a} remercier mes parents, sans qui, plus encore que toutes les
personnes sus-cit\'ees, cette th\`ese n'existerait pas.  Souvent (!), on me
demande: ``Comment est-ce qu'un fils d'agriculteur fait pour devenir
physicien?''.  Je ne sais pas exactement, parce que c'est juste arriv\'e, et
parce que cela a pris tant d'ann\'ees. Je sais en revanche---c'est ce que je
r\'eponds souvent---que c'est que \c{c}a n'aurait jamais \'et\'e possible sans
vous, et l'\'education que vous m'avez donn\'ee. Vous m'avez offert le plus
beau et le plus utile des cadeaux. Merci. Je vous en suis infiniment
reconnaissant.

I finally wanted to thank Nikki, Cl\'ement, Pu, Rosa-L\`y, K\'evin (and my
parents, again) for being there when it really mattered. And all my other
friends and colleagues, for being there.\\
\endgroup

\pagestyle{scrheadings} 
\cleardoublepage

\refstepcounter{dummy}

\pdfbookmark[1]{\contentsname}{tableofcontents} 

\setcounter{tocdepth}{1} 

\setcounter{secnumdepth}{3} 

\manualmark
\markboth{\spacedlowsmallcaps{\contentsname}}{\spacedlowsmallcaps{\contentsname}}
\tableofcontents 
\automark[section]{chapter}
\renewcommand{\chaptermark}[1]{\markboth{\spacedlowsmallcaps{#1}}{\spacedlowsmallcaps{#1}}}
\renewcommand{\sectionmark}[1]{\markright{\thesection\enspace\spacedlowsmallcaps{#1}}}

\cleardoublepage

\pagenumbering{arabic} 

\cleardoublepage 


\ctparttext{We begin this part with a general introduction that stresses the
    ever growing importance of cities in the world, and highlights the
    difficulties encountered when trying to reach a scientific understanding of
    these systems. We briefly outline the history of the quantitative tradition
    in the study of urban systems, and argue that we may be witnessing a second
quantitative revolution. We then  succintly present the methodology that we
followed during the past 3 years, and end this part with an outline of the
content presented in this thesis.} 

\part{Introduction} 
\label{part:introduction}

\chapter{Studying cities}
\label{chap:studying_cities}

\begin{flushright}{\slshape    
Chaos was the law of nature;\\
Order was the dream of man.} \\ \medskip
--- Henry Adams~\cite{Adams:1990}
\end{flushright}

\bigskip

Cities appeared some $10,000$ years ago~\cite{Bairoch:1985, Mumford:1961}
concomitantly with the agriculture revolution, and really started to
thrive after the industrial revolution~\cite{Bairoch:1985}.  In England first,
where the revolution was born; London was the first city in the modern world
to reach $1,000,000$ inhabitants at the beginning of the $19$th century. The
urban growth then slowly spread through the end of the $19$th and the $20$th to
the rest of the Western world. Now, while western countries are already
mostly urban (as of $2014$, the United States' population was $82\%$ urban,
Japan's cities hosted $93\%$ of the population, and most countries in the European Union were around the $80\%$
mark), most of what has been dubbed the 'urban \graffito{Source:\\ UN Population
Division (2011)} revolution' is happening in developing countries. A symbolic
barrier was reached in $2005$, when it was estimated by the U.N. that more than
$50\%$ of the world total population was living in cities. It is not difficult
to convince oneself that urbanisation is not an accident in human history, and
that cities' influence and impact are not going to stop growing any time soon.

In fact, the impact of cities is already tremendous. First, they have a
disproportionately large importance in the world's economy. A 2012 report by
McKinsey noted that while cities represented respectively $79\%$ and $19\%$ of the Unites
States' and India's population, their share in the countries' GDP was
respectively $85\%$ and $39\%$. 
Data from the NASA indicate that urban areas cover a total of $5\%$ of the total
land surface area in the world, roughly the equivalent of the superficy of the
European Union. Yet, despite their little spatial fooprint, cities have a great
impact on the environment. The United Nations indeed estimated in $2011$ that cities were
responsible for $70\%$ percent of the world's CO\textsubscript{2} emissions.

We could multiply the statistics, but the few examples given above should
convince the reader of the importance to understand cities if we want to
improve the world we built for ourselves. The dramatic growth of urban areas in
developing countries brings unprecedented challenges. The cause, and the
solution of some of the world's most pressing challenges certainly find their origin in
cities. By improving the way cities work, we can hopefully make dramatic changes
to the way people live. To be able to do so however, we first need to understand
how they work.

\section{We need data}
\label{sub:we_need_data}

Walk a few steps in your favourite city, feel the streets bustling all around
you. The sound of the cars, of people chatting, the pavement lined with
 homogeneously diverse buildings. The sense of familiarity we feel when stepping
back in a city that was once our home, years later. And that smell you had
forgotten you knew. Maybe the hardest thing, when studying cities, is the
impression that we know them closely. The belief that our impression of what
they are, the way we experience them, gives a true picture of what they really
are, the purpose they serve. This familiarity is what makes the study of
macroscopic, human-made systems so difficult compared to the study of natural
systems. 

There are indeed only so many ways one can get acquainted with, say, electrons, and
therefore just so many things one can say about them. This, in a sense, makes the
study of electrons easy. Think about cities now. All the memories, habits,
knowledge you have gathered over the years. As individuals, we know too many and
too little things about them at the same time. We can have a very detailed
recollection of the city we have experienced. But this information is not
organised, and it is too local, too provincial. Therefore, we cannot infer what
cities are solely from our own experience.  We are a single piece of a puzzle
that counts hundreds of thousands, millions of them, all with a different
opinion of what their environment is like. 

No, to understand cities, how they work as a system, we need to be told these
thousands of stories, we need to analyse them and see how similar, or dissimilar
they really are. To understand cities, we need data.\\

\section{Cities as complex systems}
\label{sec:cities_as_complex_systems}

\subsection{A paradigmatic example}
\label{sub:a_paradigmatic_example}

Cities are paradigmatic examples of complex systems~\cite{Ladyman:2013}.  First,
they comprise thousands, millions of individuals that are moving and interacting
constantly. Cities are indeed more than the mere agglomeration of residences,
factories and shops in the same region; they exist and thrive through the
resulting facilitated interaction between individuals~\cite{Bettencourt:2013,
Sim:2015}. Cities are built so that many people can live together and interact. 

Second, cities are incredibly resilient systems. There are multiple examples in
History of cities that were completely destroyed -- Dresden and Hiroshima, for
instance, completely burnt to ashes during WWII -- but were later rebuilt and
thrived again.

Finally, cities exhibit very particular shapes and behaviours. Because of these
identifiable properties, they are patterns that stand out in their
environment~\cite{Dennett:1991}. We can recognise cities because of their
particular structure, even though the details of the structure differ from one
city, country to another. The road network, for instance, is such that cities
can be readily identified when looking at a map (even though the layout of
 say American cities is different from that of most European cities). The high density of
population, hence nightlights, also make urban environments identifiable on
satellite pictures. These are two obvious, visual particularities of cities, but
some of their regularities are more subtle. In this thesis, we will be
interested in some of these particular behaviours.\\

\subsection{An organised complexity}
\label{sub:an_organised_complexity}

The systems studied in Physics can be roughly divided in two
categories~\cite{Parisi:1999}

\begin{itemize}
    \item Simple systems with only a few variables. Their dynamics is described
        by \emph{deterministic} equations. For instance, the motion of planets
        can be described with high accuracy by General Relativity.
    \item Weakly, locally interacting systems, with a very large number of
        particules. Their properties are described using \emph{probabilitistic}
        language. For instance, monoatomic gases in usual conditions of pressure
        and temperature are well described by Statistical Mechanics.
\end{itemize}

Cities, however, do not fit in any of the above categories. They are clearly not
simple, deterministic systems, and cannot be described in their entirety with
only a few variables. On the other hand, the traditional approach of Statistical
Mechanics is also bound to fail. Although they can contain several million of
individuals, cities are not maximally disordered systems, and thus cannot be
described in the same way we describe gases. Cities, while being disorganised,
have structure. Our goal is to identify and quantify this structure.

At the individual level, interactions are weak: one individual is very unlikely
to radically change the system's dynamics. But the multiplication of individual
interactions can create robust and influent structures (the activity centers
discussed in Chapter~\ref{part:polycentricity}, for instance). Interactions can
occur locally -- during face-to-face meetings -- but also non-locally
 -- through the phone, or the use of information systems. 
Individuals are not aimless particles, but usually have a purpose whenever they
move. But at the same time, the sheer number of individuals leaves room for unexpected situations
and encounters. As a result, cities are neither completely organised systems,
nor are they completely disorganised. They are thus very different to the
kind of systems natural sciences have traditionally studied. 


\section{Layers and scales}
\label{sub:layers_and_scales}

A first step in the identification of order consists in identifying the different
spatial and temporal scales involved in the dynamics within and of cities. The
goal of any theory of how cities work would be to understand the phenomena
occuring at each scale, to understand how scales interact with one another,
and to establish a hierarchy of
mechanisms, as in natural sciences~\cite{Simon:1962}.

\subsection{Layers}
\label{ssub:layers}

At the smallest scale, we have the individuals who live in urban
systems. They make decisions about where they live, where they work, etc. and
interact constantly with one another. Individuals are, in a way, the building
blocks of cities, and it is therefore crucial to understand the way they
interact with their environment to understand the structure and behaviour of
cities.

At a larger scale, cities can be considered as systems characterised by specific
behaviours~\cite{Bettencourt:2007}. Besides, they do not evolve in isolation and
belong to larger scale structures. To quote the geographer B.J.L. Berry, `cities
[are] systems within systems of cities'~\cite{Berry:1964}, and their
interactions---migrations, commodity and capital flows---ought to constrain
their evolution~\cite{Pumain:2010}. 

Finally, there is a great amount of evidence to show that systems of cities also
exhibit very particular behaviours: the rank-size plot of the population of
cities that belong to these systems is indeed strikingly regular (a regularity
known as `Zipf's law'), and breaks down for other geographical units or when the
chosen set of cities is not geographically and economically
coherent~\cite{Cristelli:2012}.\\

\begin{figure}[!h]
    \centering
    \includegraphics[width=\textwidth]{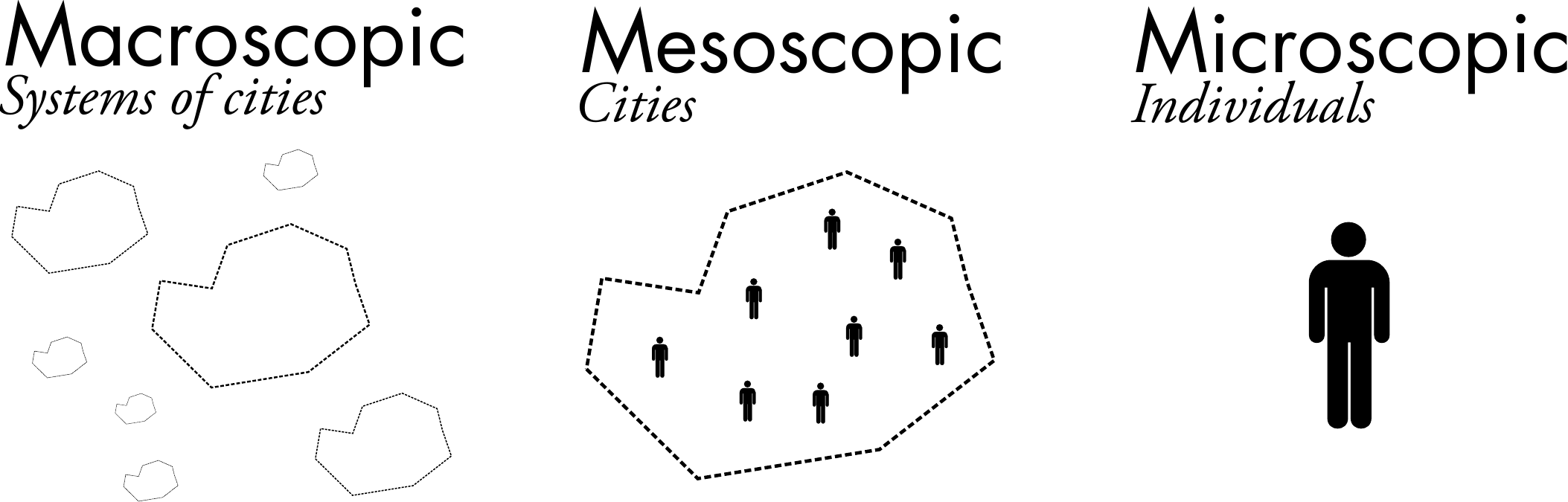}
    \caption{{\bf Interactions at different spatial scales.} Cities are the
    result of interactions occuring at different spatial scales. The movement
and interactions of individuals result in the properties of the city as a whole.
But cities are not closed systems, and interact with other cities in a system of
cities.\label{fig:spatialscale}}
\end{figure}

Cities are therefore the result of interactions occuring at different spatial
scales. Furthermore, they are not static: they evolve in time, through various
processes taking place at different time scales.

\subsection{Time scales}
\label{ssub:time_scales}

First we have time scales of the order of a day, which span the daily commuting
of inhabitants. This incessant movement of people has been traditionally
explored through surveys, but new data now allow more thorough studies. The
digital traces that are left by people at all times (through their mobile phone,
metro pass or GPS device) indeed allow us to explore the structure of flows and
the pace of life in cities at unprecedently fine spatial and time resolutions.

Then, at the order of a year one can see the variation in terms of wealth,
population, etc. of cities, as recorded by statistical agencies. Data about
demographic, social and economic aspects of urban systems allow us to
characterise more specifically the structure and behaviour of these systems.

Finally, at time scales of the order of a decade, we can see the city's
infrastructure as well as its spatial footprint evolve. The study of the
underlying processes is made possible by various projects lead by the GIS
community, historians and geographers which aim at digitizing historical maps of
the road and rail networks in different regions of the world. Also, since the
$1970s$, many satellites have been taking pictures of the Earth's surface, and
the remote sensing community has been treating these data to get information
about the spatial extension of cities. These data should give us some insight
about the processes responsible for the long-term evolution of cities'
structure.\\

\begin{figure}[!h]
    \centering
    \includegraphics[width=0.9\textwidth]{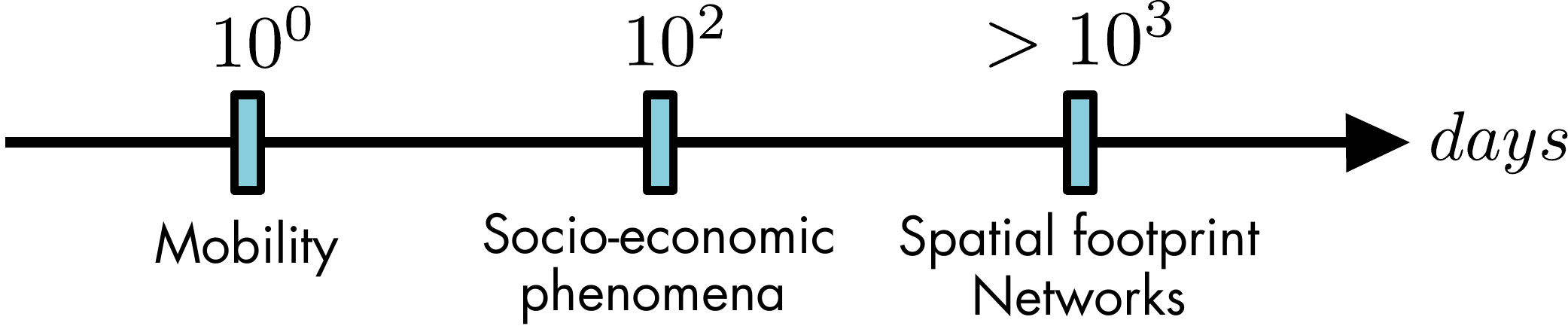}
    \caption{{\bf Different time scales.} The various data available about
    cities are associated with different time scales.\label{fig:timescale}}
\end{figure}

These time scales are summarised on Fig.~\ref{fig:timescale}. The long-term goal
of our studies is to understand exactly how cities and systems of cities behave,
and how interactions between these three layers lead to the behaviours we
observe. 


\pdfbookmark[1]{Quantitative revolutions}{Introduction}

\chapter{Quantitative revolution(s) in urban science}
\label{chap:quantitative_revolutions}

\begin{flushright}{\slshape    
And the first one now\\
Will later be last\\
For the times, they are a-changin'} \\ \medskip
--- Bob Dylan 
\end{flushright}

\bigskip

It is difficult to make a concise summary of what is known and not known about
urban systems. The vast amount of knowledge that has been gathered so far seems
very little in comparison to the bewildering complexity of the object being
studied~\cite{Batty:2008}. Every map, every satellite view, every statistic, every step
in cities elicits a question yet to be answered. What do we have to answer them?
A surprisingly small array of empirical tools and models. A surprisingly small
amount of solid, undisputed empirical facts.

Having said that, previous contributions are by no mean negligible. The body of
quantitative knowledge about cities has dramatically grown since the
quantitative revolution that took place in Geography after the $1950$s.

People have recently suggested that we may be witnessing the dawn of a second
quantitative revolution~\cite{Batty:2013}. In the following Chapter, we will try to get some perspective on this claim, and
see to what extent it is justified. We will start with a (very) brief account of the
first quantitative revolution and the main themes around which it articulated
knowledge (a more comprehensive account can be found in~\cite{Sanders:2011}). We
will then critically review the factors usually invoked to justify the
use of the expression 'second quantitative revolution'.

\section{The first quantitative revolution}
\label{sec:the_first_quantitative_revolution}

Quantitative efforts in the study of human activities find their origin in Von
Th\"unen's model of agricultural land in $1826$. More than a century later,
in $1933$, the German geographer Walter Christaller published his Central Place
Theory~\cite{Christaller:1933}, which aimed at explaining the size and location
of settlements in a system of cities. Needless to say, these early efforts are
theoretical in nature, and the empirical aspect -- studying things as they are
-- is left out. Likely because of the lack of available data.\\

The quantitative effort really starts to spread in the US in the
$1950$-$1960$~\cite{Berry:1993}. From the very beginning, the objective to make
geography a science is clearly stated, starting with the introduction of Bunge's
seminal \emph{Theoretical Geography}, published in $1962$~\cite{Bunge:1962}.
According to the author, geographers can and should go beyond the mere
accumulation of facts, and try to discover the laws that rule the human and
physical phenomena occuring on the Earth's surface.   

Bunge proposed geometry as a tool to understand the observed patterns and
describe objectively the geographical space. The range of tools used quickly
expanded~\cite{Haggett:1966,Chorley:1968}, spanning stastistical
models~\cite{King:1969, Brunsdon:1998} -- whose importance is demonstrated by
the publication in $1969$ of Leslie King's \emph{Statistical Analysis in
geography} -- and graph theory  -- as early as $1963$ with the publication of
Kansky's PhD thesis~\cite{Kansky:1963}. An early review of the use of graph
theory in geography can be found in Hagget and Chorley's
book~\cite{Haggett:1969}.\\

The research undertaken in the quantitative tradition can be -- tentatively --
divided in three different categories. First, the study of spatial
differentiation aims at characterising the spatial patterns that result from
human activities. For instance, the study of population or employment densities
(see Part~\ref{part:polycentricity}), the local concentration of population
categories (see Part~\ref{part:segregation}), or the repartition of cities
inside a territory. 

Second, the study of spatial interactions. The progressive realisation that
distance is a critical factor to understand the arrangement of different spatial
phenomena led Tobler to state the First Law of Geography~\cite{Tobler:1970}. 

\begin{quote}
    Everything is related to everything else. But near things
    are more related than distant things.
\end{quote}

Linked to the study of spatial interactions is the (in)famous gravity model,
which states that the flow $F_{ij}$ between two locations $i$ and $j$ is given
by a function of the form

\begin{equation}
    F_{ij} = C\, P_i^\alpha\,P_j^\beta\, f(d_{ij})
\end{equation}

where $f$ is a decreasing function of distance. Although the analogy with
Newton's gravitation law was used by Reilly in $1931$ to find the retail market
boundaries between cities~\cite{Reilly:1931}, the above formulation in terms of
flows was formulated by Stewart in~\cite{Stewart:1948}. Note the competing
existence of Stouffer's theory of intervening
opportunities~\cite{Stouffer:1940}, according to which the flow between $i$ and
$j$ is proportional to the number of opportunities at $j$ and inversely
proportional to the number of opportunities between $i$ and $j$. It was
mathematically formulated much later by Simini et al.~\cite{Simini:2012}.

Finally, the study of infrastructure, which started with Kansky in
$1963$~\cite{Kansky:1963}. The study of the shape and growth of road networks,
railway networks and other infrastructure has recently witnessed a renewed interest
thanks to the study of spatial networks~\cite{Barthelemy:2011}.

\section{A second quantitative revolution?}
\label{sec:a_second_quantitative_revolution_}

People can be forgiven for believing that the present time bears any sort of
special character. But when we look closely enough, the change is perpetual, and
what is new now will be outdated tomorrow. During the past $3$ years, I have
at many times overheard discussions about the fact that we were currently witnessing a 'second
quantitative revolution' in the study of geographical systems. But is it really
the case? What differences with past tools or methods could justify such a
claim? In the following, we explore the three following hypotheses

\begin{itemize}
    \item The new quantitative revolution is due to the use of new methods coming
        from interdisciplinary studies;
    \item The new quantitative revolution is due to the availability of `new data';
    \item The new quantitative revolution is due to a technological convergence.      
\end{itemize}

\subsection{New methods}
\label{sub:new_methods}

The recent years have seen the application of new methods, mainly coming from
Physics or Computer Science, to the study of cities~\cite{Dupuy:2015}. Either by geographers, or
outsiders who imported well-established methods from another
field~\cite{Batty:1995}. These collaborations, or incursions, are however not
new. For instance, John Stewart, an american astrophysicist is famous for the first use of allometric
scaling in the study of cities~\cite{Stewart:1947}, or for his work 
on the gravitation model~\cite{Stewart:1948}. Another interesting example is
given by the collaboration in $1971$ between Waldo Tobler -- a geographer -- and
Leon Glass -- a chemist -- who plot the radial distribution function of Spanish
cities, a method that is traditionally used to study the property of
liquids~\cite{Glass:1971}.

So, the application of well-established methods from other fields to cities is
not new, and neither are the contributions made by outsiders. Yet, we can
identify two qualitative changes: the number, and nature of these contributions.
If some authors have continued to import directly methods and models from other
disciplines (for instance, the use of diffusion-limited aggregation models,
traditionally studied in physics, to explain the growth of
cities~\cite{Makse:1995}), this type of theoretical contribution is becoming
marginal. Contributions are more and more empirical; and if theoretical, are not
direct applications of another domain's theories. For
instance, Rozenfeld and co-authors used percolation on census tracts to define
cities~\cite{Rozenfeld:2008} in an original way. Masucci et al. use percolation
on the road network for the same purpose~\cite{Masucci:2015}, while Li et al.
use percolation to study the properties of congestion~\cite{Li:2015}. New
approaches to spatial network~\cite{Barthelemy:2011} have yielded new insights
into the structure and evolution of road, railway and subway networks~\cite{Strano:2012,
Barthelemy:2013,Louf:2013_emergence,Louf:2014_scaling,Louf:2014}.
Original out-of-equilibrium models that are inspired by the studied system allow
a better understanding: Simini's radiation model~\cite{Simini:2012,Simini:2013}
 -- which is nothing else that the mathematical transposition of Stouffer's
 intervening opportunities theory -- or our model to explain the polycentric
 transition of cities~\cite{Louf:2013_polycentric} are examples of such models.
 Not to forget the important literature on scaling
 relationships~\cite{Bettencourt:2007, Bettencourt:2013, Louf:2014_mobility,
 Arcaute:2014, Louf:2014_smog}, and other empirical analyses -- such as the
 study of residential segregation we present in Part~\ref{part:segregation}.

At the same time, the number of contributions to the field from authors who do
not have a geography (or economics, urbanism, etc. for that matter) affiliation
seens to have increased over the past years. After all, I am a theoretical physicist
by training, and this thesis is officially a Theoretical Physics thesis. So, if
the contributions of outsiders are not new, they are changing in number and
nature. To the point where we can wonder whether some of these `outsiders'
should still be considered as such.

\subsection{New data?}
\label{sub:new_data}

Besides the import of methods from other disciplines, it is often argued that
the influx of new data, thanks to the digitization of our lives, is a revolution
in itself. 

\begin{figure}
    \centering
    \includegraphics[width=\textwidth]{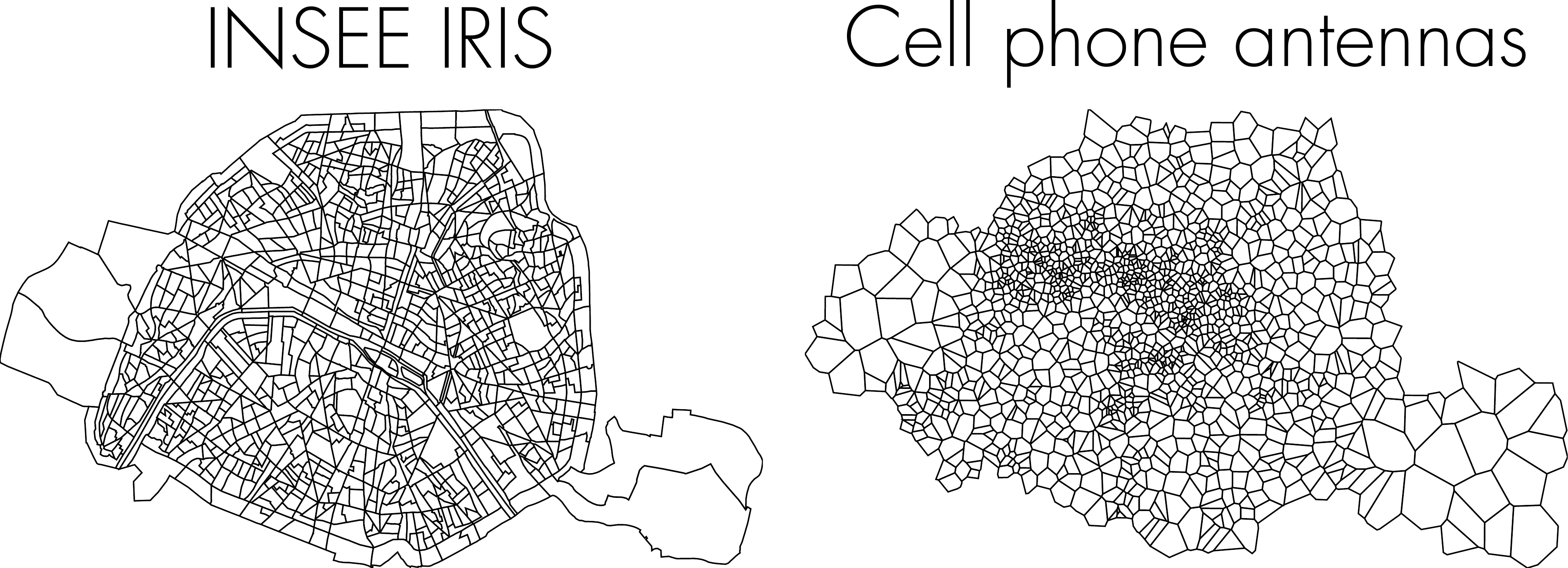}
    \caption{(Left) IRIS zones in Paris, the smallest statistical units defined
    by the national statistics institute, INSEE. (Right) Voronoi tessellation
    built from the position of antennas of a popular french mobile phone carrier.
    There are $40\%$ more antennas than there are IRIS, and they tend to be more
    concentrated in zones of high daily activity (8th and 9th
    arrondissements).\label{fig:IRIS_phone}}
\end{figure}

The most important new source of data come from the wide use of mobile phones
across the world~\cite{Gonzalez:2008,Fen-Chong:2012}. They consist, for each individual, of a
list of antenna locations to which the individual was the closest at a given time
(either when she used the phone, or when she switched from an antenna to another).
Naively, one could think that mobile phone data are better than census-based
data: they give a \emph{continuous} information about the flow of individuals
within the city (and are not limited to commuting), they cover a larger part of
the population (which is critical in developing countries: censuses are not
widely used due to the costs involved, but mobile phones have a high
penetration rate), and are more spatially precise than released census data in
urban areas (see Figure~\ref{fig:IRIS_phone} for a comparison between the smallest
INSEE areal units, and mobile phone antennas in Paris). But one needs to be
careful. If mobile phone data are fine to monitor aggregate quantities (such as
origin-destination commuting matrices~\cite{Lenormand:2014}, to map
population changes during the day~\cite{Louail:2014}, or year~\cite{Deville:2014}), one should be careful with the study of
individual trajectories (such as in the
seminal~\cite{Gonzalez:2008,Song:2010_modelling,Song:2010_limits}). Indeed, the
fact that positions are recorded every time a call is made by the user -- 
events with a powerlaw inter-event time~\cite{Song:2010_modelling} and 
probably correlated with locations -- is likely to introduce an important biais in
the obtained trajectories. Not mentioning the spatial sampling introduced by the
fact that positions are attached to a finite number of antennas. Unfortunately,
no study has looked at the impact of these two types of sampling on the
properties of the observed trajectories yet. In the meantime, one should refrain
from using such data to study individual trajectories.

Mobile phone data are not the only `new' source of data. Because mobile phones carry
GPS chips that are used by applications such as
FourSquare~\cite{Noulas:2012} or Twitter~\cite{Lenormand:2014_tweets}. Last, but
not least, credit card companies have recently started to release datasets
regarding the spending of individuals~\cite{Lenormand:2015}.\\

So, new data (mainly mobile phone data) are now available and allow to give a
picture of the city that was not accessible before. The contribution of these new
data is particularly useful for the mobility of people besides commuting
pattern~\cite{Louail:2014}, or for developing country where there are little
census data available~\cite{Blondel:2012}. Are they so overwhelmingly different
from previously available data to deserve the title of `revolution'? Nothing is
less certain: in this thesis, for instance, I have only used traditional data
sources, and we are still waiting for important results that `new data' could teach
us (and that we could not access with more traditional data). Only time will tell,
and the term `revolution' is not warranted yet.

\subsection{A technological convergence}
\label{sub:a_technological_convergence} 

Interdisciplinary collaborations already existed, data were already there. So
what is the qualitative difference between the state of the field say $20$ years
ago, and the state of the field as it is now, if any? A factor that is often
overlooked is the recent technological leap in the treatment of information,
including spatial information. Thanks to the development of GIS software as well
as spatial databases and libraries, the treatment of geographical data has never
been simpler. Added to this is the emergence of powerful scripting languages, R
and Python, which allow to quickly implement complex data analysis workflow or
simulations, and reduce dramatically the time spent writing code. 

Internet is also progressively changing the way research is done. Census data
are more and more easily accessible available online. Open data repositories,
although far from perfect, are emerging. Online platforms such as
\url{www.github.com} allow to share and collaborate on code. All in all,
the access and processing of information is getting easier and easier.\\

Taken individually, the introduction of methods from other disciplines, the
increasing amount and specificity of available data and the technological
progress in the treatment of information are probably not enough to justify
the term `revolution'. Taken together, however, they could mark the beginning of
a qualitative rupture in the way we understand cities.

It is too premature to conclude that the convergence of the aforementioned will
necessarily deeply change our understanding of cities. Only the future can tell
us whether new regularities, new laws are about to be discovered and more
phenomena to be understood. But where there is data, there is hope.
As long as the correct methodology is followed. 

In the following Chapter, we will introduce the broad methodological principles
that we adopted during this thesis.

\pdfbookmark[1]{Introduction}{Introduction}

\chapter{Methodology}
\label{chap:methodology}

\begin{flushright}{\slshape    
If it disagrees with experiment, it's wrong.\\ 
In that simple statement is the key to Science.} \\ \medskip
--- Richard Feynman~\cite{Feynman:1965}
\end{flushright}

\bigskip

The success of natural sciences lies in their great emphasis on the role of
quantifiable data and their interplay with models. Data and models are both
necessary for the progress of our understanding: data generate stylized facts
and put constraints on models. Models on the other hand are essential to
comprehend the processes at play and how the system works. If either is missing,
our understanding and explanation of a phenomenon are questionable. This issue
is very general, and affects all scientific domains, including the study of
cities.\\

Until recently, the field of urban economics essentially consisted in untested
laws and theories, unjustified concepts that supersede empirical
evidence~\cite{Bouchaud:2008}. Without empirical validation, it is not clear
what these models teach us about cities. The tide has turned in recent years,
however: the availability of data is increasing in size and specificity, which
has led to the discovery of new stylized facts and opened the door to a new
science of cities~\cite{Batty:2013}. Yet, the situation is not perfect: while
the recent deluge of data have triggered the apparition of many empirical
analyses, in the absence of convincing models to explain these regularities, it
is not always clear what we learn about cities.

In this chapter, we will try to specify what we mean by model, and explain with
a concrete example why data analysis is not enough understand the behaviour of
systems.

\section{Of models and theories}

\subsection{For what purpose?}
\label{sub:why_bother_}

As scientific sceptics often like to remind us, all models, all theories are
wrong. But surely, there must be some interest in models to make them deserve
the months, sometimes years of work that scientist devote to them. 

Models' two main functions are, broadly speaking, to understand, and to predict.
The benefits linked with the ability to predict the behaviour of a system need
not be recounted. Understanding is a more complicated notion, and a
philosophical discussion of the concept lies far beyond the scope of this
thesis. Roughly, to understand is to untangle the mechanisms involved so as to have a
simplified, barebone description of the processes that shape the system.\\

\subsection{Theory, not analogy}
\label{sub:theory_not_analogy}

Unfortunately, expressive words and metaphors are too often used as a substitute
for a real understanding of the system. But, however intellectually appealing
they are, metaphors are not a theory. For instance, what do we understand from
the comparison of cities with biological systems? What new knowledge do we gain?
Metaphors do not provide interesting ideas that are ready to be applied to a
specific field. Rather, they trigger very different ideas into different people,
which explains their recurrent success. Yet, what we need to highlight are
regularities, not similarities.\\

We also need to avoid models that are only loosely connected to reality, analogy
or metaphor. There is a lot of confusion, and little understanding to be gained
that way. In the words of Einstein, Podolsky and Rosen

\begin{quote}
    In a complete theory, there is an element corresponding to each element of
    reality.~\cite{Einstein:1935}
\end{quote}

In this thesis, we tried to make sure that most -- if not all -- elements
(variables) of our models are related to a quantity that is measurable. We also
paid a special attention to the rigour in the language used. We qualify
suggestions, by presenting them as such.  This kind of work may be less
suggestive, the vocabulary used less expressive, but it is a necessary step
towards a science of cities. We need to clear the language of unfruitful
metaphors and fill the gap with mechanisms.

\section{Quantitative stands for 'data'}
\label{sec:quantitative_stands_for_data_}

Richard Feynman's statement used as an epigraph in this chapter might be an
oversimplified, narrow view of what Science is and how it proceeds. It
nevertheless hits the nail right in the head, by isolating the core component of
what Science is: a tight relation with empirical analysis. Data are needed, at
first, to give us ideas about how the system works: stylized facts. We then
usually try to build a simplified version of the system, a model, that is able
to reproduce the stylized facts. Because of the simplification entailed, the
model highlights the most important features of the phenomenon and allows us to
understand the behaviour of the system. Finally, we use data again to test the
predictions of the model and assess its validity and/or limitations.

In this thesis, we adopt a quantitative approach to the the study of cities. In
other words, we extract information about urban systems using measured quantities: data. As we will argue in the next section,
however, data are not enough.

\subsection{Against data}
\label{sec:against_data}

In `Againt Method', the philosopher of science Paul Feyerabend argued against
the idea that Science proceeds through the application of a single, monolithic
method; what people usually call `The Scientific Method'~\cite{Feyerabend:1975}.
The reference is not innocent, and I will argue here that, although empirical
analysis constitutes the alpha and the omega of our enquiry for knowledge, data
are not enough.
There is common confusion, often innocent, that because data are at the core of
scientific enquiry, one only needs data analysis to understand how a system
works and predict its behaviour -- especially so when we have a lot of data. A
very extreme view of this statement has recently been put forth by Big Data
supporters. An article in the magazine `Wired'~\cite{Anderson:2008} recently
argued that the current deluge of data marked the end of Science as we know it.
That models were not necessary anymore, that they were to be replaced with the
extensive correlation analysis that a vast amount of data allow. This view is
completely misguided.\\

For one, pure data analysis is, at best, a myth: as Pierre Duhem argued in
$1906$~\cite{Duhem:1997}, all empirical observations are theory-laden. That is,
they are necessarily affected by the theoretical presuppositions held by whoever
is making the observation. Measuring the population of a city, for instance,
presupposes that there are such objects as cities, and that we can delineate
them. A deluge of data does not relieve the investigator from defining the
objects she is studying, from implicitely thinking about the relation between
the different elements in the system.

Then, correlations are science, indeed. But they are rudimentary science, and
there is nothing new about them. Arguably, the reason why we are able to
function at all as individuals is because our brain is capable of computing
correlations all the time. Take chairs. Chairs are fairly simple objects. Yet,
they come in all kind of colors, material and shapes. And despite this
potentially infinite diversity, we are able to recognise a chair when we see
one. We also have a notion of what a chair is to be used for. Although we do not
ackowledge it often, we are capable of surprisingly high levels of abstraction
and generalisation. Because our brains correlate, all the time. 

Science starts with the observation of these regularities. For instance, that
the sun always appears at the same place and disappears in the opposite
directions. That seasons come and go regularly. That after the night always
comes the day. Are pure correlations useful? Yes, for limited applications. Do
they constitute science? No. Science is when one goes beyond the simple
observation of correlations, and tries to understand the mechanisms responsible
for the correlations we observe.\\

In short, data is not enough: we must build models, theories.

\subsection{An example: The law of metropolises}
\label{sec:an_example_the_law_of_metropolises}

\subsubsection{Statement}
\label{sub:statement}

The above discourse may seem a bit abstract, so let us observe the shortcomings
of pure data analysis on a simple example, related to cities.

\begin{figure}
    \centering
    \includegraphics[width=\textwidth]{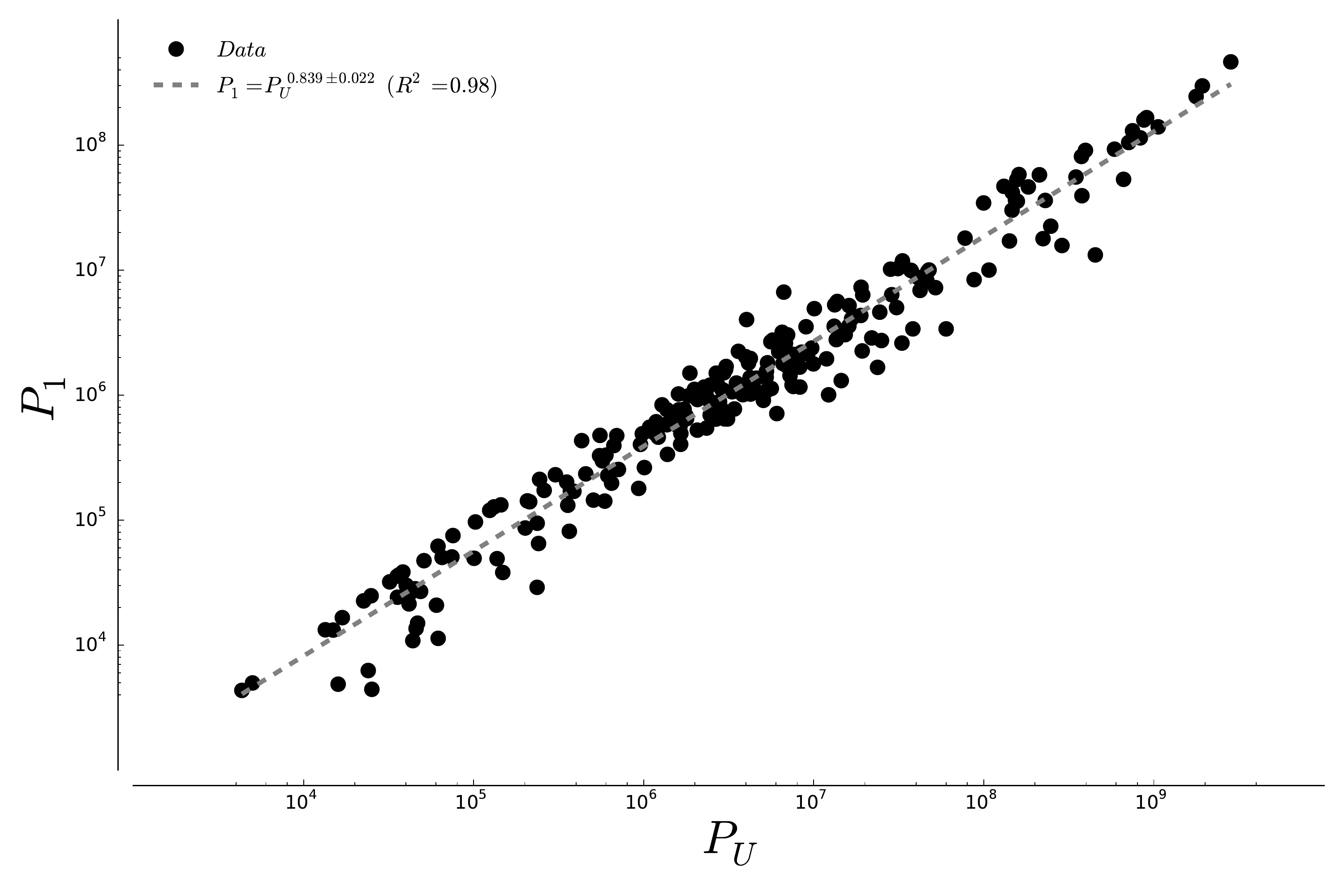}
    \caption{{\bf The law of metropolises.} Population of the largest city of
    systems of cities $P_1$ versus the total urban population $P_u$ in that
system. The dashed line shows the result of a powerlaw fit, whose exponent
agrees well with the one found in~\cite{Pumain:1997}. Data for the total urban
population and the population of the largest city of countries in the year
$2000$ were obtained from
the World Bank.\label{fig:metropolises}}
\end{figure}

Using the GEOPOLIS database, Moriconi-Ebrard and Pumain derived a general transversal rule about system
of cities, that they called \emph{law of metropolises}~\cite{Pumain:1997}. If we
note $P_U$ the urban population of systems of cities (here countries), and $P_1$ the size of
their largest city~\graffito{The original regularity was observed for what the
author calls 'metropolises', which are roughly equivalent to the largest city in
terms of population.}, we can
plot $P_1$ versus $P_U$ for all systems of cities and obtain the plot on
Figure~\ref{fig:metropolises}.

Assuming a powerlaw relationship between the two quantities, one finds

\begin{equation}
    P_1 \sim P_U^{\,0.84}\:(r^2=0.98)
    \label{eq:metropolis}
\end{equation}

which agrees very well with the empirical data (for all years where data are
available). It is tempting, at first, to consider this as yet-another emprical
regularity exhibited by urban systems, and try to find a coherent interpretation
in geographical terms. However, as we will show, if we assume that the Auerbach-Zipf
law~\cite{Auerbach:1913,Zipf:1949} holds for each system of cities
individually

\begin{enumerate}
    \item We can derive a relation that fits the data as well as
        Eq.~\ref{eq:metropolis};
    \item The relation is not a powerlaw.
\end{enumerate}

\subsubsection{Deriving the `law of metropolises'}
\label{sub:deriving_the_law_of_metropolises_}

Let us consider a system of cities comprised of $N$ cities, with total
population $P_U$. The size of the largest city is noted $P_1$. We assume that
the distribution of city sizes follows the Auerbach-Zipf law, so that the city
of rank $r$ (the $r$th largest city) has a population

\begin{equation*}
    P_r = P_1\,r^{-\mu}
\end{equation*}

So the total population in the system of cities can be written

\begin{equation}
    P_U = \sum_{r=1}^N P_r = P_1\,\sum_{r=1}^{N} \frac{1}{r^\mu}
\end{equation}

If we assume that $\mu=1$, $P_U$ is given by the harmonic series, and thus

\begin{equation}
    P_U = P_1 \left[ \ln(N) + \gamma + O\left(\frac{1}{N}\right)\right]
\end{equation}

where $\gamma \approx 2.58$ is Euler's constant. This gives us a first relation
between $P_1$, $P_U$ and $N$.\\

Still using the assumption that the distribution of city size follows the
Auerbach-Zipf law with $\mu=1$, we can show (using extremal value
theory)~\cite{Clauset:2009} that on average\graffito{'Average' as in {\bf
ensemble average}} the size of the largest city is proportional to the total
number of cities

\begin{equation*}
    P_1 \propto N
\end{equation*}

Thus, when the number of cities in the system is large, $N \gg 1$ the following
relation holds 

\begin{equation}
    \boxed{P_1\,\ln(P_1) = P_U}
    \label{eq:metropolises_debunked}
\end{equation}

\begin{figure}
    \centering
    \includegraphics[width=\textwidth]{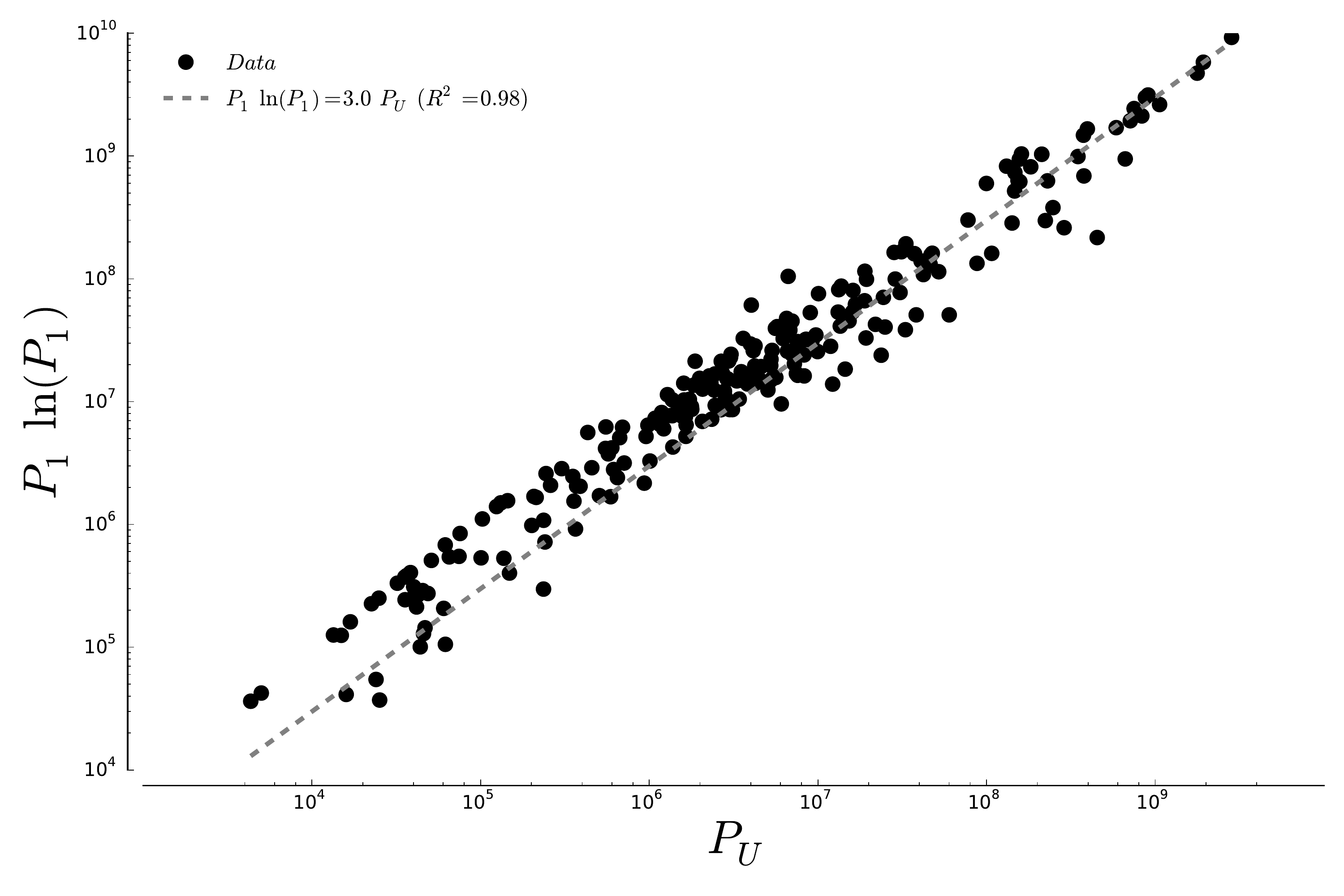}
    \caption{{\bf The law of metropolises revisited.} $P_1 \ln(P_1)$ versus the total urban population $P_u$ in that
system. The dashed line shows the result of a linear fit, which agrees as well
with the data as does the powerlaw relation assumed in~\cite{Pumain:1997}. Data for the total urban
population and the population of the largest city of countries in the year
$2000$ were obtained from
the World Bank.\label{fig:metropolises_debunked}}
\end{figure}

As one can see on Figure\ref{fig:metropolises_debunked}, the formula given by
Eq.~\ref{eq:metropolises_debunked} fit the data as well as the previous one.

It is therefore impossible to determine which of
Eq.~\ref{eq:metropolis} or Eq.~\ref{eq:metropolises_debunked} describes the
`true' relation between $P_1$ and $P_U$ based on data analysis alone.
Nevertheless, the later finds a very simple explanation in the fact that cities
in systems of cities follow the Zipf-Auerbach law up to a good
approximation. In the absence of any theoretical explanation for the powerlaw
relationship and given the empirical equivalence of both forms, it
least-assuming to consider $P_1 \ln P_1 \sim P_u$.

\subsubsection{Lessons learned}
\label{sub:lessons_learned}

So, the \emph{law of metropolises} is not a fundamental relation.
This teaches us that, given the range of variation of the measured
quantities, it is very difficult to distinguish empirically a powerlaw
relationship from something qualitatively different such as $Y \ln Y \sim P$, as
recently argued by Shalizi in~\cite{Shalizi:2011}.
One should therefore be wary of interpreting empirical relationships,
like the one originally found in~\cite{Pumain:1997}, unless a mechanistic
explanation of the fitted relationship is provided. As shown above, what
was thought as a fundamental law might end up being trivial and without great
interest.

We will further discuss the limitations of data analysis in
Chapter~\ref{chap:scaling_implications}, after having studied scaling relationships.


\pdfbookmark[1]{Outline}{Outline}

\chapter{About this thesis}
\label{chap:methodology}


\begin{flushright}{\slshape    
Anybody can plan weird, that's easy.} \\ \medskip
--- Charles Mingus
\end{flushright}

\bigskip

The following thesis might surprise the reader used to the monographs usually
produced by PhD students in Social Sciences, articulated around a single,
general question. The outline of this thesis reflects more the
line of thoughts and of research that has been undertaken than the answer to a
single question that would have been asked a priori and answered during the
last three years. For that reason, the four Parts of this thesis are mostly
independent. There is not single thread holding them together. But rather multiple
wires; common themes and similar ideas. 

\section{Outline}

Part~\ref{part:polycentricity} tackles the problem of measuring and
understanding urban form, an issue that has been running through the $3$ years
of my PhD. In this Part, we first (Chapter~\ref{chap:monocentric_introduction})
present a brief historical overview of the monocentric and polycentric
representations of the city, before enumerating the methods that are used in the
literature to count the number of activity centers. We end with the observation
that the number of activity centers increases in a regular way with population
size. The following chapter (Chapter~\ref{chap:monocentric_model}) is devoted to
an out-of-equilibrium model that we built in order to explain the previous
empirical regularity. The model is able to predict the sublinear increase of the
number of centers that we observe on American and Spanish data. In the last
chapter (Chapter~\ref{chap:monocentric_discussion}), we question the assumptions
of the model and the current empirical methods to quantify urban form.\\

Part~\ref{part:scaling} is concerned with scaling relationships. We first propose
(Chapter~\ref{chap:scaling_introduction}) a non-exhaustive overview of the dawn
and surge of allometric scalings, from Stewart's $1949$ to the recent wealth of
studies. Then, using the model developped in the preceding part, we show in
Chapter~\ref{chap:scaling_model} how the structure of mobility patterns allow us
to understand the qualitative and quantitative values of the exponents related
to urban form and mobility. We conclude this part with a discussion on the
interpretation of these scaling laws, and their important shortcomings
(Chapter~\ref{chap:scaling_implications}).\\

Part~\ref{part:segregation} departs from the preceding chapters and turns to the
study of residential segregation. Driven by the desire to extend the model
presented in Chapter~\ref{chap:monocentric_model}, we soon realised there was a
lack of robust empirical description of patterns of segregation that could be
reproduced by a model. In Chapter~\ref{chap:segregation_introduction} we tackle
the problem of defining what segregation is; we propose a brief review of the
existing literature, and subsequently define a null model -- the segregated
city. In the next chapter (Chapter~\ref{chap:patterns_segregation}), we build on
this null model to propose a set of measures to quantify patterns of residential
segregation.\\

Part~\ref{part:networks} concerns the original topic of this thesis: spatial
networks. Because my interests have shifted towards the study of
socio-economical phenomena over the years, we only briefly present the most
important results in the present thesis. The three chapters are, for the most
part, reprints of articles that have been previously published in peer-reviewed
journals. We first (Chapter~\ref{chap:typology}) present an empirical study of
$131$ street patterns across the world where we propose a method to classify the
patterns based on the geometrical shape of the blocks. In the following chapter
(Chapter~\ref{chap:cost-benefit}), we present a cost-benefit analysis framework
to understand the properties and growth of spatial networks.  We introduce an
iterative model that can explain the emergence of a hierarchical structure
(`hubs and spokes') in growing spatial networks. Starting from the cost-benefit
framework of this model, we show that the length, number of stations and
ridership of subways and rail networks can be estimated knowing the area,
population and wealth of the underlying region.\\

Finally, Part~\ref{part:conclusion} ties everything together, highlights the lessons
learned and concludes this thesis with some potentially interesting research avenues for the
years to come.

\section{Miscellaneous notes}

\subsection{Style}
\label{sub:style}

I will be using the pronoun 'we' for most of the manuscript, to reflect the fact
that the work presented here was, for the most part, done in the context of
collaboration with others. For the sake of clarity, the technical details of
calculations have been omitted in this manuscript. Most of these calculations are
relatively simple anyway, and the interested reader can find them in the
publications mentioned on page~\ref{p:publications} of this thesis.

\subsection{Tools}
\label{sub:tools}

Unless otherwise specified, all figures in this manuscript have been prepared
using Python $2.7$~\footnote{Available at \url{http://www.python.org}} and the
Matplotlib library~\cite{Hunter:2007}. Inkscape~\footnote{Available at
\url{https://inkscape.org/en/}} was used to prepare most diagrams. This document
was typeset using Vim and \LaTeX. The template used is the typographical look-and-feel
\texttt{classicthesis} developed by Andr\'e
Miede.\footnote{Available at 
\url{http://code.google.com/p/classicthesis/}.}

\begin{center}
\end{center}

\cleardoublepage 


\ctparttext{

The monocentric model of cities -- where all activities are organised around a
single activity center -- has pervaded the literature on urban systems for more
than $4$ decades. However, as it was repeatedly demonstrated, the model is empirically
inadequate.

The contribution of this part is threefold. First, we recount the history of
ideas about urban form, from the monocentric hypothesis and its origins, to the
various methods proposed to identify and count subcenters. We then demonstrate
empirically the existence of a polycentric transition for cities, and that the
number of centers increases as a sublinear function of population size. Finally,
we propose an out-of-equilibrium model that explains the emergence of new
subcenters as cities expand, and predicts the sublinear increase of the number
of centers with population size.

}

\part{Polycentri-city}
\label{part:polycentricity}

%
\chapter{The (end of the) monocentric city}
\label{chap:monocentric_introduction}

\begin{flushright}{\slshape    
It may be a small irony that  
just as\\
the phenomenon of polycentricity\\ is getting considerable attention,\\
The world is moving beyond it.} \\ \medskip
--- Peter Gordon \& Harry Richardson~\cite{Gordon:1996}
\end{flushright}

\bigskip

The hypothesis that cities organise themselves around a single center of
activities -- often called Central Business District (CBD) in the US -- may well
be one of the strongest hypotheses in urban studies. Although no one seriously
believes in its validity anymore, its influence is still noticeable in many empirical and theoretical works.  In order to deconstruct the
monocentric model, we first need to understand where it came from in the first
place, why it was introduced, and what evidence it was based on. 

In this chapter, we present a historical perspective on the monocentric
hypothesis. First, the context in which it was introduced, how it was gradually
realised that cities had a decentralised structure, and the emergence of the
notion of center. We then present a brief review of the methods and tools
developed to count their number. Finally, using American and Spanish data, we
show that larger cities are more polycentric. This suggests the existence of a
transition from a monocentric to a polycentric structure when the population of
cities increases.

\section{From monocentric to polycentric cities}
\label{sec:introduction}

Maybe the least assuming way to represent the density profiles in cities is
through either choropleth maps, or 3-dimensional representations. On choropleth
maps, the $x$ and $y$ coordinates correspond to the original coordinates
projected on the plane. In the former case, the different values of density are
expressed by the use of different colours. This approach can be traced back as
far as $1898$ in Meuriot's \emph{Des agglom\'erations urbaines dans l'Europe
contemporaine}~\cite{Meuriot:1898} who drew a large number of density maps of
large Europen cities. He was later followed by Jefferson in
$1909$~\cite{Jefferson:1909} who did the same for several cities in the US,
Europe and Australia.\\

\begin{figure}
    \centering
    \includegraphics[width=\textwidth]{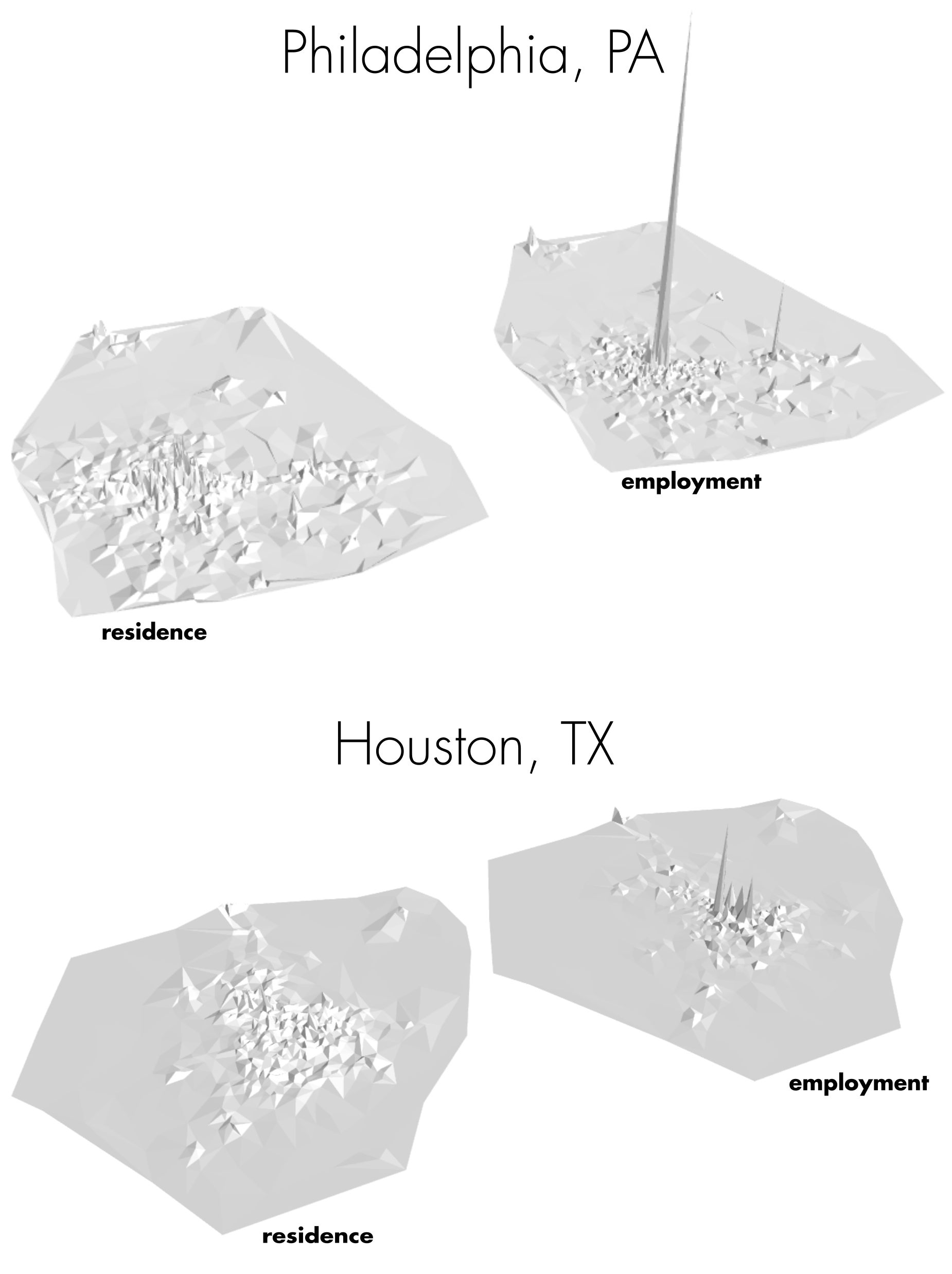}
    \caption{{\bf 3D representations of densities.} Residential and
        employment densities in (Top) the Metropolitan Statistical Area (MSA) of
        Philadelphia, PA and (Bottom) the MSA of Houston, TX. Employment and
        residential densities are represented at the same scale.  Employment
        densities are sensibly more peaked than residential densities,
        suggesting that the notion of `center' is more relevant in the context of
        activies. Data were obtained from the 2000 US Census.  \label{fig:density_3d}}
\end{figure}

3-dimensional representations, on the other hand, use the $z$ coordinate to the
represent the density values. On Figure~\ref{fig:density_3d} we represent the
density profiles of two metropolitan areas in the US: Minneapolis-St.Paul, MN
and Houston, TX. These two cities are enough to illustrate the difficulties associated with
studying density profiles. 

\emph{What densities we are talking about?} People are constantly moving
throughout the city during the day, and density profiles can only be
(approximate) snapshots of the city at different instants.  Traditionally,
scholars have only considered residence densities (nightime city) and employment
densities (daytime city). The recent availability of mobile phone data may
however give us a more precise, continuous picture of the densities during the
day~\cite{Louail:2014}. In this part, we will be focusing on employment
densities.

\emph{How can we makes sense of these density patterns?} The densities represented
on Figure~\ref{fig:density_3d} are indeed very complex, and we would like to
isolate some particular structure. Arguably, the notion of center stems from
this desire to find some structure in the complex, messy empirical reality. \\

Realising that districts of large population tend to be central, and
districts of small population in the periphery, Clark proposes in
$1951$~\cite{Clark:1951} to write the density $\rho$ as a function of the
distance $d$ from the center

\begin{equation}
    \rho = a\,e^{-d/b} 
\end{equation}

Where $a$ is the density at the center, and $b$ the typical distance over which
the density decreases. To justify his assumption, Clark plots the population
density of various cities as a function of the distance to the
center~\cite{Clark:1951}. Some structure was found. The monocentric hypothesis
was born.\\

\begin{figure}
    \centering
    \includegraphics[width=\textwidth]{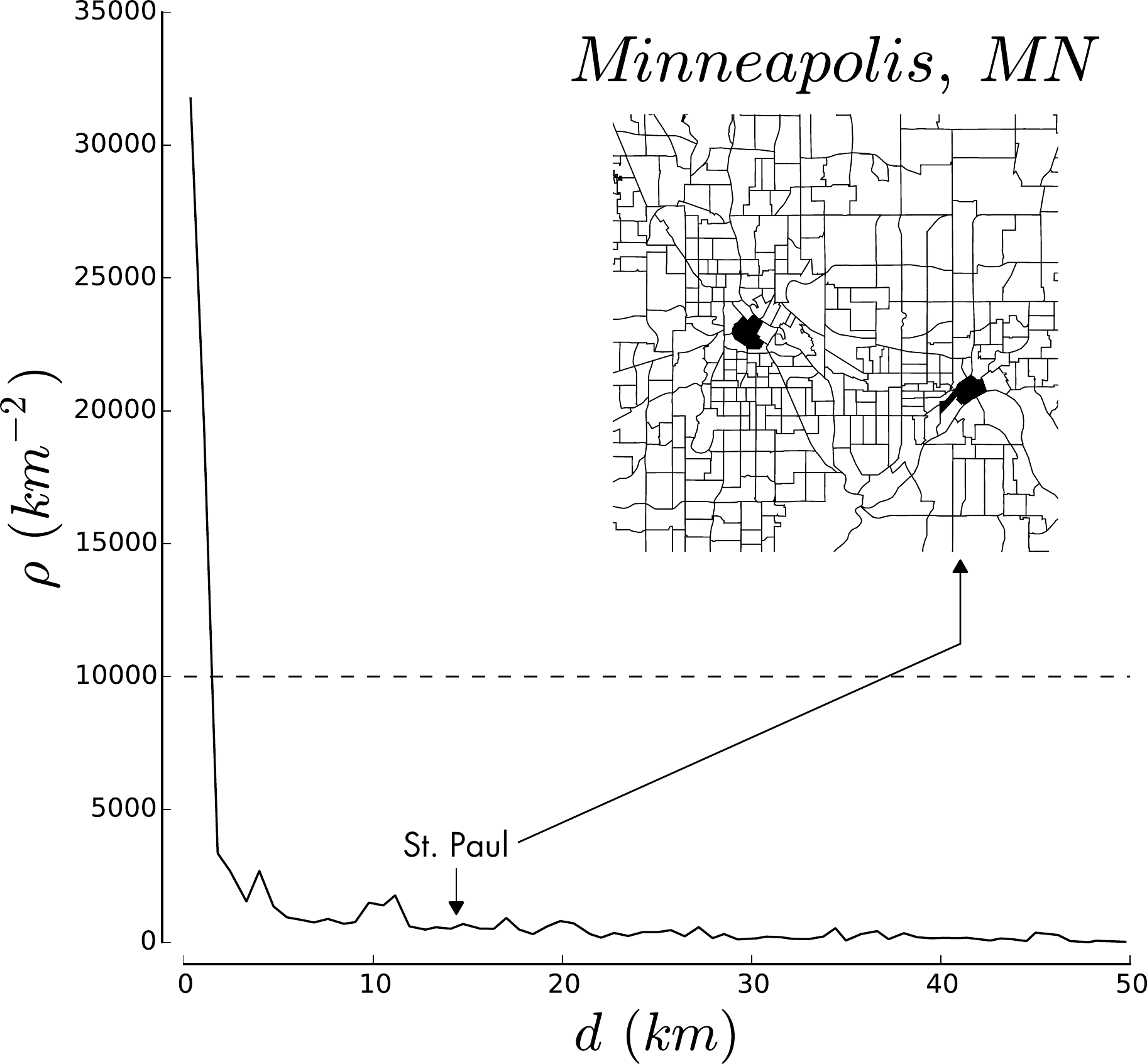}
    \caption{{\bf The limitations of density profiles.} Employment density as a function of distance to the center for the
    Minneapolis-St. Paul MSA in $2000$. The center is defined here as the
    tract with the highest employment density, and corresponds to the historical
    Central Business District of Minneapolis. The curve exhibits a very sharp
    decay, giving the illusion of a monocentric structure. (Inset) The census tracts
    of Minneapolis-St. Paul in 2000. In black, the census tracts where the
    employment density reaches values above $10,000\,\text{km}^{-2}$. The two tracts
    coincide with the historical centers of the Twin Cities, and are distant from
    $14\,\text{km}$. This fragmented structure cannot be infered from the density
    profile (arrow on the curve).
    \label{fig:distance_center_minneapolis}}
\end{figure}

Looking at the density profiles plotted by Clark in 1951~\cite{Clark:1951} for
many cities across the world, or on Figure~\ref{fig:distance_center_minneapolis}
for the Minneapolis-St. Paul MSA, one can be
forgiven for thinking that cities have a monocentric structure. Such profiles
indeed almost always exhibit a sharp decrease as we go farther from the city
center -- defined here as the areal unit with the highest density. 

However, density profiles are not enough to prove the existence of a monocentric
structure. Unless one other hypothesis is verified: namely that the pattern of
employment densities is symmetric under rotations around the center. This is
however never the case: cities are nowhere isotropic but in the imagination of
modelers. To make this point clearer, we show on
Figure~\ref{fig:distance_center_minneapolis} both the density profile of the
Minneapolis-St. Paul MSA and a map where we highlight in black the tracts with
an employment density greater than $10000\,\text{km}^{-2}$. As one can see, two
tracts (respectively the historical centers of Minneapolis, and of St. Paul) are
highlighted. However, the peak in density corresponding to St. Paul is not
distinguishable on the density profile.  Indeed, it is averaged out with smaller
densities that are located at equidistance from Minneapolis. The decreasing
exponential model, however appealing, is thus mispecified.\\

So why did Clark's methods and plots did not become a simple curiosity, but were
instead so widely adopted? Although it is sometimes difficult to trace back the
reasons for the adoption of ideas, there is little doubt that the echo this idea
had in urban economics had something to do with it (besides the simplicity of
the hypothesis). Indeed, beginning as an implied assumption in Clark's empirical
analyis, the monocentric hypothesis first became clearly stated in the
theoretical work of economists.

The Alonso-Muth-Mills model (inspired by Von Th\"unen's land rent model)  might
well be the reason for the long-lasting influence of the monocentric
model\footnote{A concise exposition of the AMM model can be found
in~\cite{Brueckner:1987,Fujita:1989}}. In 1964, Alonso introduced the bid-rent
curve as a function of the distance to the city center~\cite{Alonso:1964}. The
assumption that all firms in a city are concentrated in a single, fixed-size
part of the city naturally followed. Later, in $1967$ and $1969$,
Mills~\cite{Mills:1967} and Muth~\cite{Muth:1969} show how we can can obtain an
exponentially decreasing function for the density as a function of the distance
from the center, using the monocentric hypothesis. The monocentric
Alonso-Muth-Mills (AMM) model was born, and was seemingly backed by empirical
evidence.

One should not underestimate how the monocentric model influenced people's
perception of what a city is. In the US, the name of Central Business District
is casually used as a way to designate the principle activity center in a city.
Many, if not most, measures of the spatial variation of quantities inside cities
actually use the notion of `distance to the city center'. Many authors are
relying on the monocentric hypothesis for their empirical analysis -- sometimes
without being aware of it. This biais can still be found in the recent
literature. For instance, in a recent study by Glaeser, Kahn and Rappaport on the
repartition of income classes in cities~\cite{Glaeser:2008}, the authors comment
on plots of the average income as a function of the distance to the center. This
only makes sense, however, under the assumption of monocentricity.\\

This persistence of the monocentric hypothesis is all the more surprising that
authors repeatedly suggested and showed that the hypothesis was not adequate. In
$1974$, Kemper and Schmenner~\cite{Kemper:1974} explore industry and
employment density data, trying to fit a negative exponential function. Their
conclusion is clear: ``A declining exponential function fails to explain much of
the spatial variation of manufacturing density''. A few years later,
Odland~\cite{Odland:1978} explores the possibility of polycentric cities on a
theoretical basis. As explained in~\cite{Griffith:1981}, scholars subsequently started to
explore the density patterns of cities by fitting multi-center exponential
functions of the form

\begin{equation}
    \rho_i = \sum_{j=1}^{q} A_j\,e^{-d_{ij}/b_j}
    \label{eq:multi-exponential}
\end{equation}

where $\rho_i$ is the density at location $i$, $q$ the number of centers, $A_j$
the local maximum of density at $j$, $b_j$ the characteristic size of the center
$j$, and $d_{ij}$ the distance between locations $i$ and $j$. The idea of
polycentricity, originally as the generalisation of the monocentric hypothesis,
is progressively gaining ground. 

Trying to fit equations like Eq.~\ref{eq:multi-exponential} is cumbersome, and
requires some a-priori knowledge of the density patterns. It requires to
determine in advance which parts of the cities are going to be subcenters
\graffito{{\bf sub}center because they are subsidiary to the traditional CBD},
before attempting to fit the density profile. As noted in~\cite{Giuliano:1991},
authors used arbitrary definitions of subcenters, either designating them based
on their own intuition, or refering to the centers defined by planning agencies.
The centers were thus determined \emph{exogenously}.\\

In this context, the first definitions of employment centers independent from
the exponential model start to emerge, and subcenters start an existence of
their own. By the $90$s, the idea that cities can be polycentric is
well-established, and more and more empirical analyses confirm the existence of
several employment centers. For instance, McDonald~\cite{McDonald:1987} identifies the
employment subcenters in the region of Chicago, IL; Giuliano and Small~\cite{Giuliano:1991} in the
region of Los Angeles, CA; Dokmeci et al.~\cite{Dokmeci:1994} show that Istanbul's employment
is spread across several centers, etc. 

The concept of subcenters is further expanded in $1991$~\cite{Garreau:1991},
when Garreau shows that secondary centers are not necessarily `subcenters'.
Indeed, activities do not always accumulate in the traditional downtown. He
introduces the concept of `Edge cities': the concentration of business, shopping
and entertainment at the outskirts of cities, in regions that were previously
rural, or purely residential.\\

\section{How to count centers}
\label{sec:how_to_measure_polycentrity}

The methods designed to identify employment subcenters can be divided in three
categories. The clustering methods, which appeared first, were progressively
abandonned for regression-based methods due to their reliance on arbitrary
cut-offs. Distribution-based methods have emerged recently, and leave
aside the spatial aspect of the density distribution.

\subsubsection{Clustering methods}
\label{ssub:clustering_methods}

In $1987$, McDonald~\cite{McDonald:1987} remarks that despite being mentioned
in the empirical and theoretical literature, the features that an employment
subcenter should have are nowhere discussed. For the first time, he proposes a method to
determine the number of subcenters empirically. Given
a number $T$ of areal units, we will say that $i$ with employment $E_i$,
population $P_i$ and surface area $A_i$ is an employment subcenter if:
\graffito{He also proposes a definition based on the employment-to-population
ratio}

\begin{quote}
    The gross employment density $\rho_i = E_i/A_i$ is greater than that of the
contiguous units; 
\end{quote}

Giuliano and Small~\cite{Giuliano:1991} acknowledge the necessity to
consider employment densities to define subcenters put forward by
McDonald~\cite{McDonald:1987}. However, they deplore that the method does not
allow for adjacent units with a high employment density to be centers -- as only
the larger one would be selected. Thus, they propose an alternate definition.
Namely that a contiguous set of units $\mathcal{S}$ is a subcenter if 

\begin{itemize}
    \item The employment density $\rho$ of every areal units in the set $\mathcal{S}$ is greater than
        a threshold value  $\overline{D}$;
    \item And the total employment $E$ contained in $\mathcal{S}$ is greater than a threshold
        $\overline{E}$.
\end{itemize}

where the thresholds $\overline{D}$ and $\overline{E}$ are imposed arbitrarily.
Using this definition, all areal units with a high employment densities are part
of a subcenter, unless they are small (contain less than $\overline{E}$
employees) or isolated (i.e. they do not belong to a cluster containing at
least $\overline{E}$ employees).\\

As mentioned by Anas et al. in~\cite{Anas:1998}, because density landscapes are highly
irregular at a small scale (see Figure~\ref{fig:density_3d} for instance), the
subcenter boundaries are very sensitive to the threshold values. Because there
is no a priori reason to choose a threshold rather than another, the obtained
subcenter boundaries are arbitrary and may vary from one author, one situation
to another. Instead, it would be preferable to have a method based on first
principles, that adapts to the local specificities.  In McMillen's words,
threshold methods lack a proper consideration of how large is `large' supposed
to mean for the threshold values~\cite{McMillen:2003}. 

Another problem highlighted in~\cite{Anas:1998} is that the number of centers
depends on the size of the areal unit, an issue that is tied to scale problem
discussed in the Modifiable Areal Unit Problem (MAUP)~\cite{Openshaw:1984}
literature. On the one hand, small areal units will lead to several low
employment density units in otherwise very high density areas. On the other
hand, large areal units are likely to smooth over local employment peaks. This
begs the question of whether we should use contiguity of units, or rather
distance, as a measure of proximity.

\subsubsection{Regression-based methods}
\label{ssub:regression_based_methods}

In an attempt to address these concerns, McMillen~\cite{McMillen:2001} proposes
a two stage procedure. In the first allegedly non-parametric stage, he uses a
geographically weighted regression (GRW, see~\cite{Brunsdon:1998} for more
details on the topic) to `smooth' the employment density, using distance rather
than contiguity as a measure of proximity, thus partially solving the issue
linked with the size of areal units.  The units that have unusually high
employment densities compared to the broad spatial trends obtained with the GWR are designated as
candidate subcenters. If we note $\rho_i$ the employment density at site $i$,
$\hat{\rho}_i$ the density estimated with GWR and $\hat{\sigma}_i$ the standard
deviation around this estimate, $i$ is said to be a \emph{candidate} subcenter
if 

\begin{equation*}
    \rho_i - \hat{\rho}_i > 1.96\,\hat{\sigma}_i
\end{equation*}

Candidate, because the GWR only identifies fluctuations in the density profile
with no consideration of whether these local fluctuations have a sensible impact
on the employment density. Identifying which of these candidates are actually
centers is the goal of the second, semi-parametric procedure. This second
procedure uses somewhat arbitrary criteria (the first and second largest
candidates are omitted in the regression, candidates at less than $1$ mile from
the CBD are omitted) to produce a second reference global trend, to which real values are
compared to identify the `real' centers among the candidates.\\

Redfearn critizes the first procedure~\cite{Redfearn:2007}, on the ground that candidate
subcenters are defined as outliers with respect to an average that uses half of the total
number of points (in the GWR), thus losing the local information about employment
density. The author proposes another non-parametric method that aims at correcting the issues
with McMillen's\cite{Redfearn:2007}. The estimation of the employment density is
done locally in order to keep intact the local structure of the density profile.
However, arbitrariness still lies in the choice of the span (the amount of data
that are considered to estimate the slopes at a given point) for the GRW. In
other words, regression-based methods are \emph{not} truly non-parametric.\\

\subsubsection{Distribution-based methods}
\label{ssub:distribution_based_methods}

The approach that we originally took in this thesis is radically different from
that of regression-based methods~\cite{Louf:2013_polycentric}. We start with the
remark that one does not need to know the spatial arrangement of areal units
with different densities in order to know which ones are most important. Indeed,
the local fluctuations that are registered as centers in the regression-based
methods are very likely to have a negligible contribution to the total
employment. They can thus be left out in a first approximation. A good estimate
of the number of centers should thus be given by the shape of the employment
density distribution alone. Because it does not require any spatial knowledge,
it makes the extraction of centers fairly easy and quick to compute compared to
the previous methods.\\

\begin{figure}
    \centering
    \includegraphics[width=1\textwidth]{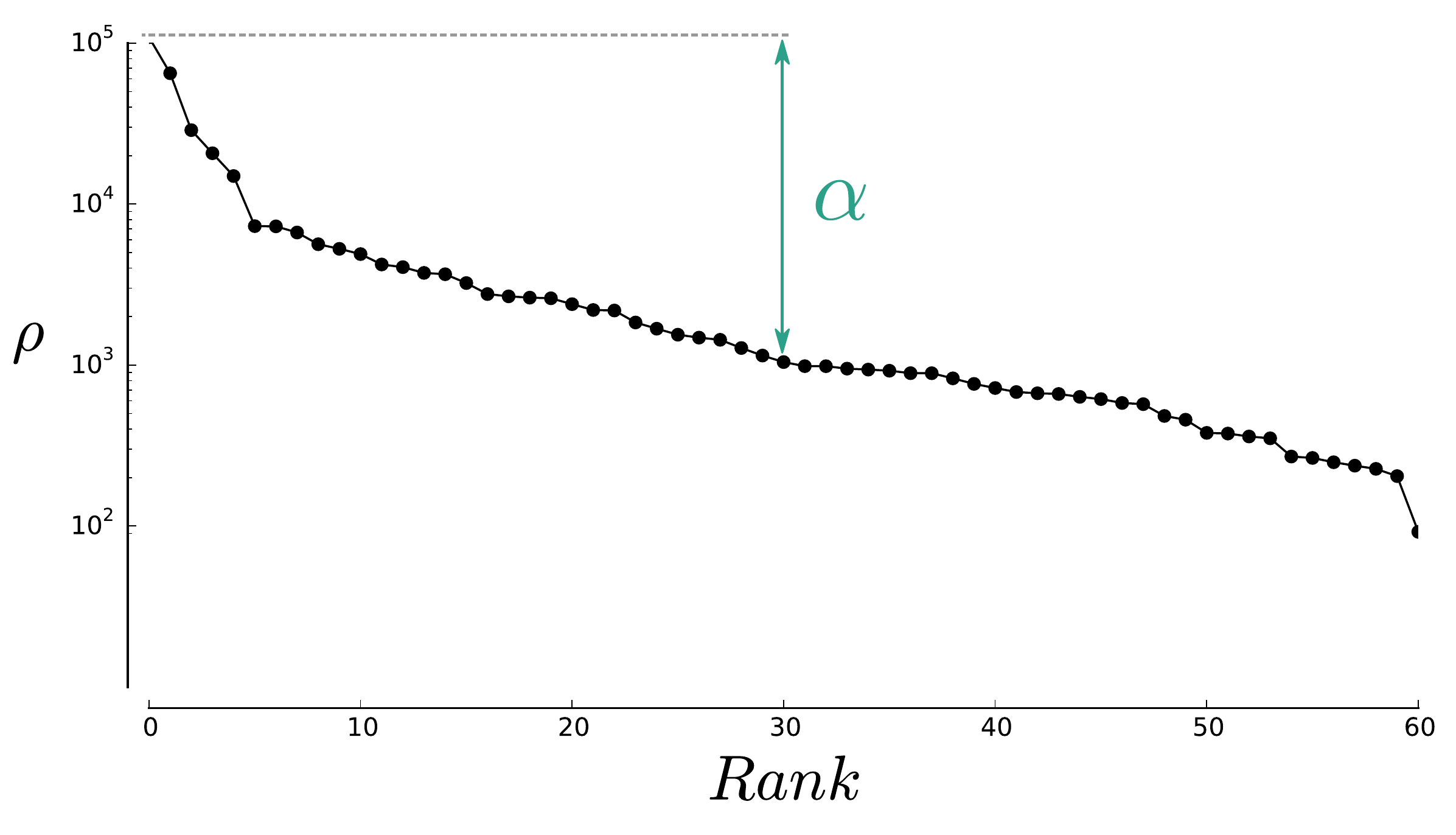}
    \caption{{\bf The rank-plot method.} Rank plot of the employment density in
        the Zip Code Tabulation Areas of Los Angeles, CA.
    \label{fig:rank_plot}}
\end{figure}

We start by building the rank plots of employment density $\rho$ inside the
areal units (see Figure~\ref{fig:rank_plot}). These plots display a decay at least as fast as that of an
exponential. If they were an exact exponential, they could be modeled by a
function of the form

\begin{equation}
    \rho(r) = \rho_0\,e^{-r/r_c}
    \label{eq:}
\end{equation}

where $\rho(r)$ is the $r$th highest value of the density inside the city,
$\rho_0$ the maximum density value. This exponential decrease implies that there
exists a natural scale for the rank, $r_c$, that we interpret here as the number
of centers. In order to get the number of centers, one would either need to
compute the slope on a lin-log plot, of find the value of $r^*$ for which

\begin{equation}
    \rho(r^*) = \frac{\rho_0}{e} 
\end{equation}

in which case $r^*=r_c$. However, empirical rank plots are not strictly exponential,
and we define the number of centers using a threshold value $\alpha$. We define
$\rho_m$ as 

\begin{equation}
    \rho_m = \frac{\rho_0}{\alpha}
\end{equation}

and the number of subcenters $k$ is equal to the number of values $\rho_c$ of
the density such that $\rho_c \in \left[ \rho_m, \rho_0 \right]$. 

In the case where the rank plot would be strictly exponential, we would have

\begin{equation}
    k = \rho_0\,\ln \alpha
\end{equation}

so that the number of centers is mainly determined by $\rho_0$. Small variations
in $\alpha$ should not sensibly change the number $k$ of centers obtained.

The method however suffers from two flaws. First, the use of an arbitrary
parameter, the threshold $\alpha$ to extract the number of centers. All the criticisms listed earlier also apply: we are not sure to extract the
`true' number of centers. Moreover, the method assumes a particular form for
the density distribution, which is likely to biais the estimation.\\

\begin{figure}
    \centering
    \includegraphics[width=0.6\textwidth]{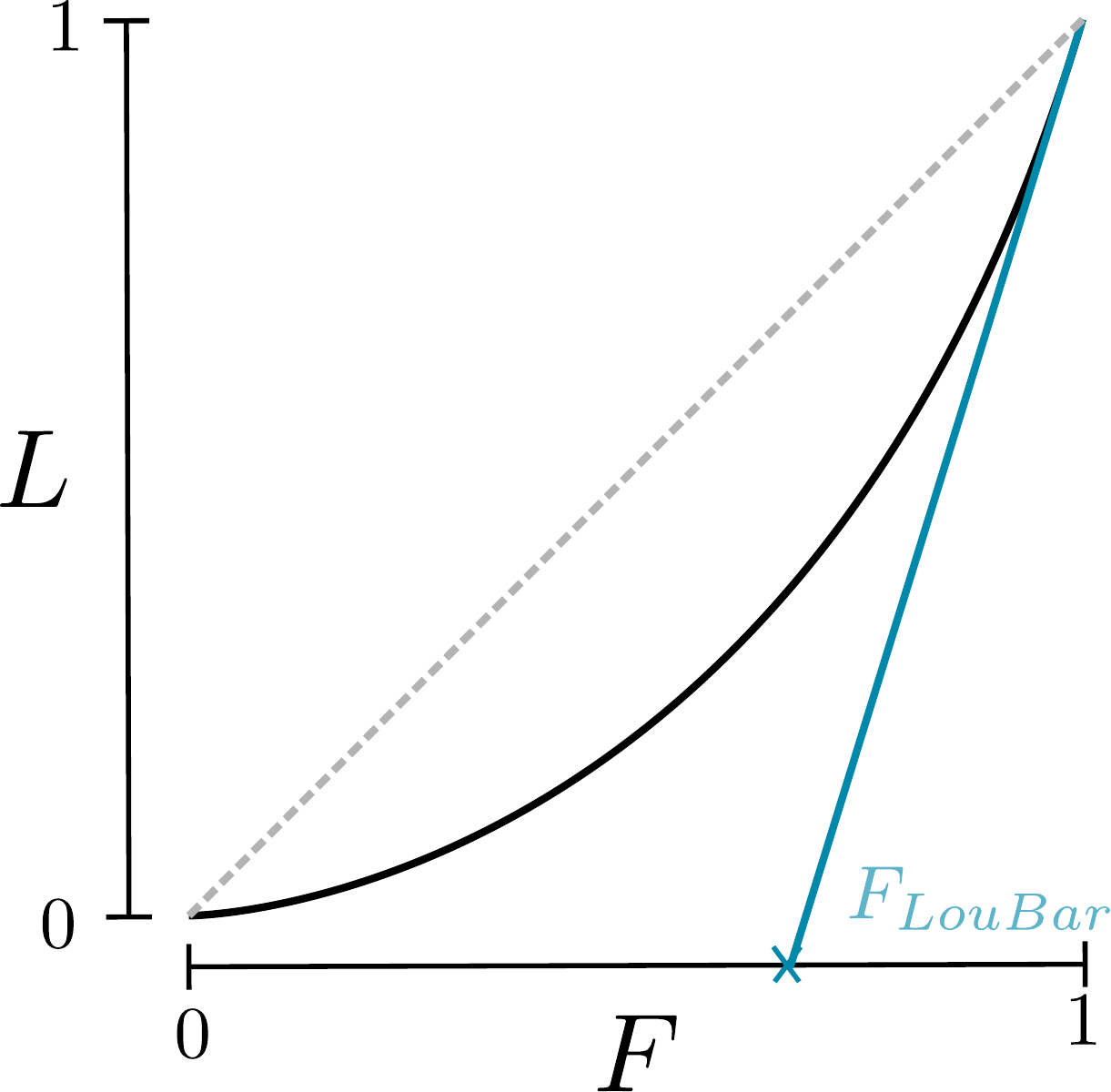}
    \caption{{\bf Lorentz curve and Loubar method.} An example of realistic Lorentz curve (solid black line), the curve
    that would be obtained in a city with uniformly distributed density (dashed
    grey line), and the tangent at the point $L(F) = 1$ (blue line) used to determine the number
of centers in the LouBar method.\label{fig:loubar}}
\end{figure}

Louail and Barthelemy~\cite{Louail:2014} propose a generalisation of the
previous method based on the Lorenz curve. \graffito{The Lorentz curve is often
used in Economics to quantify income inequality.} Given the ordered set of
densities $\rho_1 < \rho_2 < \dots < \rho_T$ in the $T$ units, we plot the
proportion of cells $F_i=i/T$ as a function of the corresponding proportion of
employment density

\begin{equation}
    L_i = \frac{\sum_{n=1}^i \rho_n}{\sum_{n=1}^T \rho_n}
\end{equation}

so that both $F_i$ and $L_i$ take their values between $0$ and $1$ (see 
Figure~\ref{fig:loubar}). It is easy
to see that, in the case of a city with a uniform employment density, the
Lorentz curve is a straight line. In the general case, however, the curve has a
convex shape, with a more or less pronounced curvature.  The higher the
curvature of the Lorentz curve, the higher the inequality in terms of employment
density, and thus the smaller the number of potential centers. 

Following this observation, the authors define a new criterion to determine the number of
centers. They consider the intersection $F^*$ between the tangent of the
Lorentz curve at the point $L(F) = 1$ and the axis $F=0$ (see
Figure~\ref{fig:loubar}. The units that correspond to the values of $F$ between
$F^*$ and $1$ are defined as centers. This definition has the merit to only
depend on the distribution of density inside the areal units; it is \emph{genuinely
non-parametric}, while being easily tractable and understandable.\\

Of course, all the methods presented here have issues (that we discuss 
in Chapter~\ref{chap:monocentric_discussion}), and there is currently no
consensus on what method should be used to find the employment centers.
More work is needed before we arrive at a satisfactory description of
urban form. Nevertheless, the results given by these methods -- although slightly
different -- provide together a compelling evidence for the polycentric
structure of cities.

\section{The polycentric transition}
\label{sec:the_polycentric_transition}

Occasionnally mentioned in the empirical literature~\cite{McMillen:2003,
Redfearn:2007}, and hinted at in urban economics models~\cite{Fujita:1982},
the greater polycentricity of larger cities was not firmly established before
this thesis. Almost all cities (apart from the notable exception of twin cities)
start growing around a single center of activity. Yet, as we will see, no large
city adopts a strict monocentric structure. Therefore, it seems that, as
they grow and expand, urban systems develop a more and more polycentric form. We
call this phenomenon the `polycentric transition' of cities. 

\subsection{Empirical evidence}
\label{sub:empirical_evidence}

\subsubsection{American cities (Census data)}
\label{ssub:american_cities_census_data_}

Historical data over long periods of time, on a consistent set of areal units,
are very difficult -- if not impossible to find. However, we do, for one point in
time, have many cities with very different population values. We can thus compute
and plot the number of centers as a function of population. Of course, as we
will discuss in more details in Part~\ref{part:scaling}, there is a gap between
time series and transversal studies that is not completely obvious to
bridge. Some cities can be, for historical reasons, locked into a monocentric
state when the average city would not. For different reasons, another city
might as well have developed a polycentric structure more pronounced than
other cities of the same size have. The idea here is to look at a large number
of cities and measure the average behaviour of this ensemble of cities, hoping
that marginal cases are indeed marginal.


\subsubsection{American cities (census data)}
\label{ssub:american_cities_census_data_}

During this thesis~\cite{Louf:2013_polycentric}, we used data on the employment in the Zip Codes of US cities
every year between 1994 and 2010. We first extracted the number of
centers for every city, for every year between 1994 and 2010. Using the
rank-plot method described earlier. We then applied the following treatment to
the data:

\begin{itemize}
    \item If there is only one Zip Code in the given city, $k=1$;
    \item We perform a Kolmogorov-Smirnov test~\cite{Massey:1951} between the
        distributions of a given city for consecutive years. If there is a
        significant difference (above a threshold $p_{KS}$) between the
        distribution at $t$ and $t+1$, we keep the point at $t+1$. If there is
        no sensible difference, we discard it.
\end{itemize}

\begin{figure}
    \centering
    \includegraphics[width=0.9\textwidth]{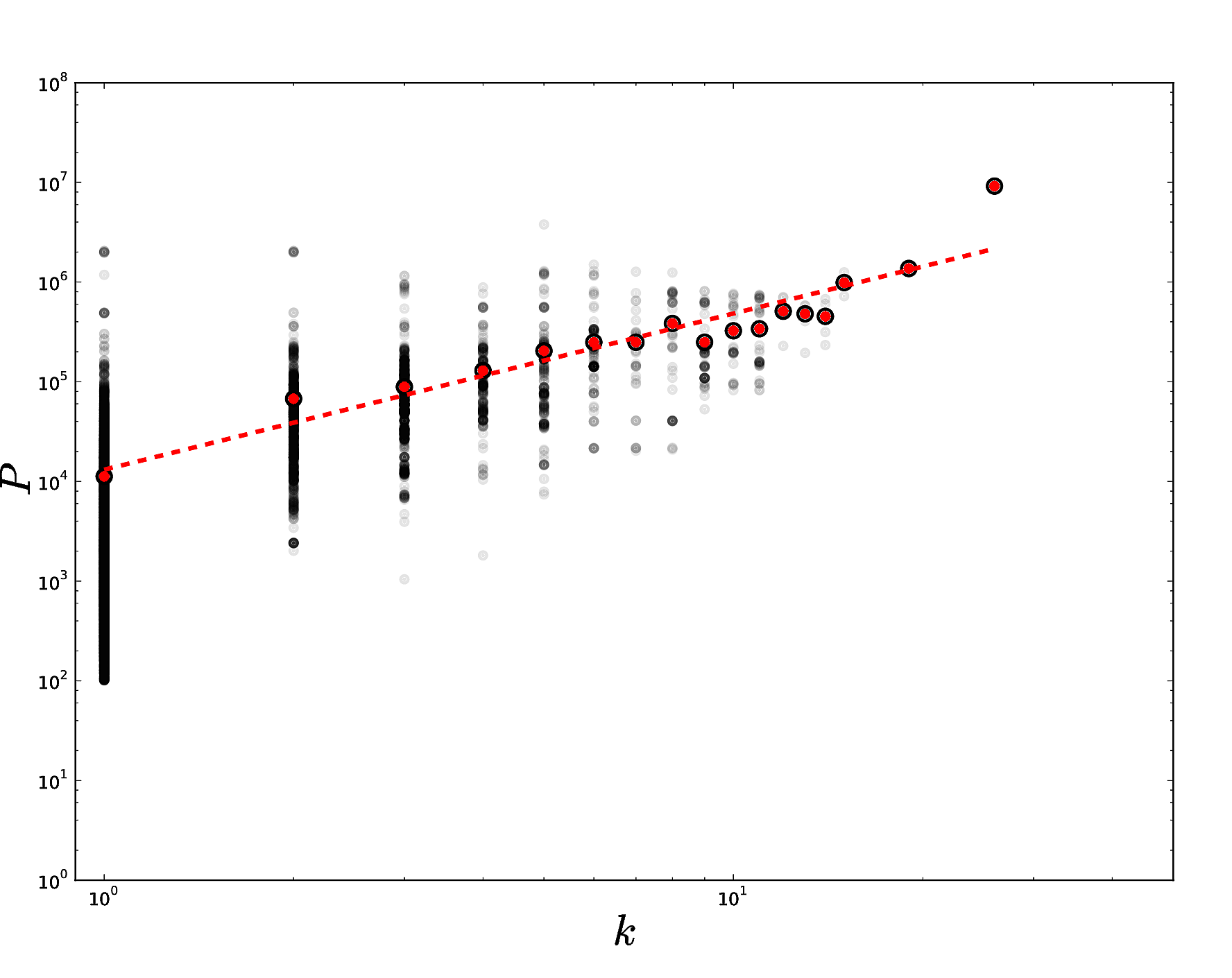}
    \caption{{\bf Centers in American cities.} Scatter plot for the estimated number of centers versus the
    population for about 9000 cities (different realisations) in the US. The red
dots represent the average population for a given number of subcenters. We fit
this average assuming a power-law dependence giving an exponent $\delta = 1.56
    \pm 0.15\,(R^2=0.87)$. Data were obtained from the US Census Bureau's Zip
Code Business Patterns for every year between $1994$ and $2010$. \label{fig:us_centers}}
\end{figure}

At the end of this process, we obtain points that can be understood as coming
from different realisations of a city. We then plot the number of centers
computed for all these realisations as a function of the total population and
obtain the curve obtained on Figure~\ref{fig:us_centers}.

A power-law fit on the average per population bin gives an exponent $\delta =
1.56 \pm 0.15\,(95\%\,C.I.)$. Thus, we find that on average, the number of
centers in US cities scales with population size as

\begin{equation}
    k_{\,US} \sim P^{\,0.64}
\end{equation}

\subsubsection{Spanish cities (mobile phone data)}
\label{sub:spanish_cities_mobile_phone_data_}

\begin{figure}
    \centering
    \includegraphics[width=\textwidth]{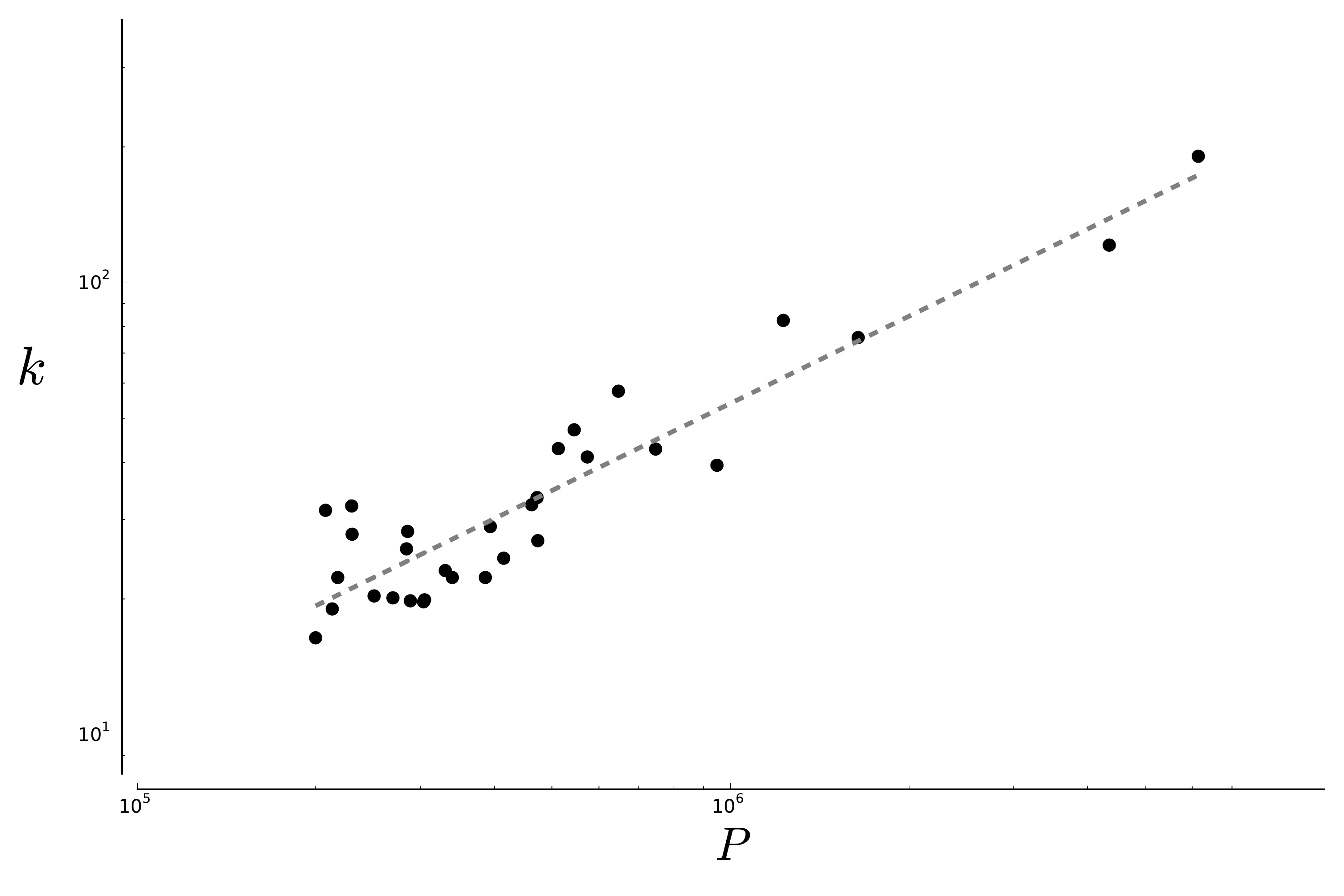}
    \caption{{\bf Centers in Spanish cities.} Scaling of the number of centers with population for Spanish
        metropolitan areas. Assuming a powerlaw relationship, the authors
    of~\cite{Louail:2014} find an exponent $\beta = 0.64$ ($r^2=0.93$). The data
were kindly provided by Thomas Louail.\label{fig:centers_spain}}
\end{figure}

Using mobile phone data and the LouBar method to determine the number of
centers, Louail et al.~\cite{Louail:2014} also computed the number of centers
versus population for Spanish cities. 

\begin{equation}
    k_{\,Spain} \sim P^{\,0.64}
\end{equation}

Strikingly, the exponent they found is very close (equal) to the one we found on a
different system of city, using a different method to count centers, and a
radically different data collection method.\\

Taken together, the previous empirical analyses teach us that
\begin{itemize}
    \item The larger cities are, the
more polycentric they tend to be; 
    \item The average behaviour is well-approximated
by a power-law relationship between the number of centers and population;
    \item The increase of the number of centers with population is \emph{sublinear}.  
\end{itemize}

These facts for a theoretical explanation. We will present a model to that
effect in the next chapter. But before concluding, let us review quickly the
reasons that are traditionally invoked for the polycentric transition.

\subsection{Reasons invoked for the polycentric transition}
\label{sec:reasons_invoked_for_the_polycentric_transition}

There are numerous examples where polycentrism finds its origin in the fusion of
two Metropolises, or the incorporation of satellite
municipalities~\cite{LeNechet:2015}. The Twin Cities in the US, for instance:
the cities of Minneapolis and St. Paul have grown to such an extent that they now
form a single metropolitan area. The region of the Ruhr in Germany, or the
region of Tokyo in Japan are other examples. However, in this thesis, we are only interested in
an endogeneous polycentrism, caused by the growth of a single city.\\

Already in $1972$, Mills~\cite{Mills:1972} suggests that congestion might be the
cause of decentralisation and suburbanisation in large metropolitan areas.
However, we have to wait until $2003$ for McMillen to propose a thorough
empirical investigation~\cite{McMillen:2003}. Commuting cost is estimated using the peak travel time index index which is
defined as the ratio between the average travel time at peak congestion time
over the average travel time at any other time of the day. Effectively, the
commuting cost is thus a measure of the level of congestion in the city. 

Studying US cities, The author finds a positive correlation between the number
of centers, population, and commuting cost.  In other words, congestion might be
the key factor to understand the polycentric transition of cities.

\section{Summary}
\label{sec:summary}

In this chapter, we have presented a historical perspective on the monocentric
hypothesis, trying to show why it appeared, disappeared, and how it is still hiding in 
some of the empirical literature. We then discussed the polycentric hypothesis,
how it was introduced, and the different methods that have been proposed to
identify and count the subcenters.

We then showed on US and Spanish data that the average number of activity
centers increases sublinearly with population size. This proves, we believe, the
existence of a polycentric transition of urban areas as their population
increases. A transition, we saw, that might be due to increased levels of
congestion in larger cities. In the next chapter, we will present a model to
understand this polycentric transition. 

\chapter{How congestion shapes cities}
\label{chap:monocentric_model}

\begin{flushright}{\slshape    
What is here required is a new kind of statistical mechanics,\\
in which we renounce exact knowledge not\\
of the state of the system but\\
of the nature of the system itself.}  \\ \medskip
--- Freeman J. Dyson~\cite{Dyson:1962}
\end{flushright}

\bigskip

We saw in chapter~\ref{chap:monocentric_introduction} that as cities grow and
expand, they evolve from a monocentric organisation where all the activities are
concentrated in the same geographical area -- usually the central business
district -- to a more distributed, polycentric organisation. In this chapter, we
will try to uncover the mechanisms at play behind this transition. We begin with
a brief introduction of the model of Fujita and Ogawa in urban economics. We
will highlight its shortcomings, and present a stochastic, out-of-equilibrium
model. This model relies on the assumption that the polycentric structure of
large cities might find its origin in congestion, irrespective of the particular
local economic details. We are able to reproduce many stylized facts, and --
most importantly -- to derive a general relation between the number of activity
centers of a city and its population.

\section{Fujita and Ogawa}
\label{sec:fujita_and_ogawa}

In line with the tradition of economic geography~\cite{Fujita:2001}, the model
of Fujita and Ogawa~\cite{Fujita:1982} is based on the concept of agglomeration
economies---to explain why economical activities tend to group---and the spatial
distribution of wages and rents across the urban space. They consider that
cities are constituted of two kinds of actors: the firms, who tend to
concentrate to maximise their production, and the households, who try to
minimise their rent and commuting cost.\\ 

The model is \emph{static}, in the sense that the numbers of firms and
individuals are fixed. It is an \emph{equilibrium} model, and considers that the
city is the realisation of a general optimum. The original model is also
\emph{one-dimensional}, although the hypothesis of one-dimensionality is not
fundamental, and only necessary to make the calculations easier. Because we 
do not try to solve the model, we write equations in the more general
two-dimensional case.

\subsection{Households} 
\label{sub:households}

Fujita and Ogawa assume that there is a fixed number $N$ of households in the
city. The households are considered identical, in the sense that they all have
the same utility function and the same budget constraint. The utility function
of each household is given by the function $U = U(Z)$ where $Z$ is the surplus
of money that is left after budgetary constraints (expressed in monetary units);
basically, the money one has left at the end of the month, once the rent, bills
and petrol (or transportation card) have been paid. 

The utility is assumed to be an increasing function of $Z$ so

\begin{equation}
    \frac{\partial U}{\partial Z} > 0
\end{equation}

The budget constraint on an household living at $i$ (of coordinates $\vec{x}$)
and working at the firm located at $j$ (of coordinates $\vec{y}$) is given by the
equation

\begin{equation}
    Z = W\left(j\right)
      - C_R\left(i\right)
      - C_T\left(i,j\right)
\end{equation} 

where $W\left(j\right)$ is the wage earned at $j$, $C_R\left(i\right)$ the total
rent paid at $i$ and $C_T\left(i,j\right)$ the cost of commuting between
home and work. This equation is very general, and will be our starting point for
the model presented in the next section. The authors of~\cite{Fujita:1982}
further specify the commuting cost

\begin{equation}
    C_T\left(i,j\right) = t\,d_E(i,j) = t\,\left|\vec{y}-\vec{x}\right|
\end{equation}

where $t$ represents the commuting cost per unit distance, and $d_E(i,j) = \left| \vec{y} -
\vec{x} \right|$ the euclidean distance between home and work. The total rent
cost is further written as

\begin{equation}
    C_R\left(i\right) = R(i)\,S_h
\end{equation}

where $R(i)$ is the rent per unit surface at $i$, and $S_h$ the surface area
used by households, which becomes a parameter of the model.  The surplus $Z$
thus finally reads

\begin{equation}
    Z = W\left(j\right)
      - R\left(i\right)\,S_h
      - t\,d_E\left(i,j\right)
\end{equation}

\subsection{Firms}
\label{sub:firms}

The second type of agents taken into consideration in the model are the firms.
It is assumed that all firms employ the same number of individuals, which
amounts to having a fixed number $M$ of firms (once the number of households is
fixed). The profit earned by a firm  located at $j$ reads, in a general
form

\begin{equation}
    \Pi = G\left(j\right) 
        - C_R\left(j\right) 
        -  W\left(j\right)\,L_f
\end{equation}

where $G(j)$ is the total gain realised by the firm selling its production,
$C_R(j)$ the rent paid by the firm, and $L_f$ the total number of employees per
firm---a parameter of the model.\\

To take agglomeration economies into account, Fujita and Ogawa define the
locational potential $F$ defined by

\begin{equation}
    F\left(j\right) = \int_{\mathcal{C}} b(\vec{x})\,e^{-\alpha\,\left|\vec{y}-\vec{x}\right|}\:\mathrm{d}\vec{x}
\end{equation}

where $b(\vec{x})$ is the density of firms at $\vec{x}$. The integral runs over
the entire city's spatial extent $\mathcal{C}$. One can easily see that the
higher the density of firms in a radius of $1/\alpha$ around a firm, the higher
the locational potential is going to be. Balanced by the constraint imposed by
the rent, which prevents too many firms from agglomerating at the same location,
the locational potential likely is the term responsible for the existence of
polycentric solutions in the model. Indeed, the authors further write the total
gain $G$ as a multiple of $F$:

\begin{equation}
    G(j) = \beta\,F(j)
\end{equation}

where $\beta$ integrates both the productivity of the employees and the effect
of the locational potential. The rent, as
in the case of households, is written $C_R(j) = R(j)\,S_f$ where $S_f$, the
surface needed by firms, is a parameter of the model. The profit
of companies therefore reads

\begin{equation}
    \Pi = \beta\, \int_{\mathcal{C}} b(\vec{x})\,e^{-\alpha\,\left|\vec{y}-\vec{x}\right|}\:\mathrm{d}\vec{x}
        - R\left(j\right)\,S_f
        - W\left(j\right)
\end{equation}

\subsection{Equilibrium conditions and results}
\label{sub:equilibrium_conditions}

Once the budget constraints have been explicited, one needs to define
the equilibrium conditions to be able to solve the model. First, the goal of
each household is to maximise their utility under the budget constraint. That
is, to choose $Z$, $S$, $\vec{x}$ and $\vec{y}$ so that $U(S,Z)$ is maximum.

Here, the maximisation of utility under budget constraints is
equivalent to chosing the residential location $i$ and the job location $j$ so
as to maximise $Z$. In other words, the maximisation of utility in this
particular situation is equivalent to performing a cost-benefit analysis.\\ 

The firms have no utility function, and choose to be a the location $j$ that
maximises their profit.\\

A further constraint is given by the bid-rent curve, and determines the spatial
interaction between households and firms. The authors define two intermediate
functions, $\Psi(\vec{x})$ and $\Phi(\vec{x})$ which are respectively the bid
rent function of households and of firms, defined as

\begin{align}
    \Psi\left(\vec{x}\right) &= \max_{\vec{x}} \left\{ \frac{1}{S_h} \left[W(\vec{x} ) - Z -
    t\,d_E\left(\vec{x}-\vec{y}\right)\right] | U(Z) = U\right\}\\
    \Phi\left(\vec{x}\right) &= \frac{1}{S_f} \left[\beta\,F(\vec{y}) - \Pi -
W(\vec{y})\right]
\end{align}

$\Psi(\vec{x})$ represents the maximum rent that the households could pay to be
located at $\vec{x}$ while still having a utility value $U$. $\Phi(\vec{y})$ is
the maximum rent that firms could pay to be located at $\vec{y}$. At
equilibrium, it is assumed that whoever's bid rent function has the highest
value at $\vec{x}$ will be located at $\vec{x}$.

Taken together, the equilibrium conditions determine the spatial distribution of
households and firms, of the wages and land prices.\\

The results of this model, given its intricacy, are somewhat disappointing.
Unsurprisingly, the authors are not able to derive an analytical solution for
their model. What they do, however, is deriving the conditions on the parameters
for the existence of monocentric and polycentric organisations of activities,
using numerical methods.

\section{Problems with the Fujita and Ogawa model}
\label{sec:problems_with_the_fujita_and_ogawa_model}

The approach of Fujita \& Ogawa fails at giving a satisfactory quantitative account  of the
polycentric transition of cities. A lot can be said about the details of the
model and its assumptions. But we choose to only discuss the issues that we
feel are the most important, and that we will try to address in our model. 

\paragraph{It is an equilibrium model.} In line with the rest of Urban
Economics~\cite{Fujita:2001, Fujita:2013}, the authors describe a city as being in
an equilibrium characterised by static spatial distributions of households and
business firms. However, the equilibrium assumption is unsupported as cities are
out-of-equilibrium systems and their dynamics is of particular interest for
practical applications~\cite{Batty:2008}.

\paragraph{It is too complex.} The model integrates so many interactions and
variables that it is difficult to understand the hierarchy of processes
governing the evolution of cities: which ones are fundamental and which ones are
irrelevant. A model is however only interesting when it provides a simple
structure to understand empirical results, whether it reproduces them, or
provides well-understood limiting cases (`null models').

\paragraph{It does not make any prediction.} Worse, due to its complexity, the
model is unsolvable, and does not make any prediction. At best it shows that
polycentric configurations are \emph{possible}. Yet, there are possibly
different models that would admit polycentric activity profiles as a solution.
The constraint is not strong enough, the model is unsupported by data.\\ 

We also note that the model does not take congestion into account in the
commuting cost (which is only a function of the distance). However, as we saw in
Chapter~\ref{chap:monocentric_introduction}, it is
mentioned in the economics literature as being a possible cause of the
polycentric transition of cities~\cite{McMillen:2003}.

\section{Modeling mobility patterns}
\label{sec:an_out_of_equilibrium_model_}

In this section, we start from the model by Fujita \& Ogawa to propose a
dynamical model of city growth. Following recent interdisciplinary efforts to
construct a quantitative description of cities and their
evolution~\cite{Makse:1995,Zanette:1997,Marsili:1998,Bettencourt:2007,Batty:2008},
we deliberately omit certain details and focus instead on basic processes. We
thereby aim at building a minimal model which captures the complexity of the
system and is able to account for -- qualitative as well as quantitive -- stylized
facts.  

The model we propose is by essence dynamical and describes the evolution
of cities' organisation as their population increases. We focus on car
congestion -- mainly due to journey-to-work commutes -- and its effect on the
job location choice for individuals.\\

\subsection{Decoupling the choice of household location and job}
\label{sub:decoupling_the_choice_of_household_and_job}

The time scales involved in the evolution of cities are usually such that the
employment turnover rate is larger than the relocation rate of households. On a
short time scale, we can thus focus on the process of job-seeking alone, leaving
aside the problem of the choice of residence. In other words, we assume the
coupling between both processes to be negligible: we assume that each inhabitant
newly added to the city has a random residence location and we concentrate on
understanding how such an inhabitant chooses its job location.\\

As a result of this assumption, a worker living at $i$
will choose to work at the center $j$ such that the quantity
 
\begin{equation}
    Z_{ij} = W(j) - C_T(i,j)
    \label{eq:Z_workers_general}
\end{equation}

is maximum. Doing so, we give up any hope to describe the spatial structure of the rent
distribution, or the alledged scaling between rent prices and population size in
cities~\cite{Bettencourt:2013}.

\subsection{Decoupling the behaviour of firms and individuals}
\label{sub:decoupling_the_dynamics_of_}

Another difficulty with the Fujita-Ogawa model is the strong coupling between
the behaviour of firms and individuals. The empirical literature on the
behaviour of firms points to a tendency of similar industries to cluster
geographically~\cite{Duranton:2005, Marcon:2009}, and a higher
profit of industries located in Urban environments~\cite{Melo:2009}. Although
theoretical attempts at explaining these behaviours have been
proposed~\cite{Duranton:2004}, the models are yet to be developed in an
out-of-equilibrium framework.\\

Here, we decide to simplify the problem by assuming that firms indeed cluster
into specific locations, that we call activity centers. Each worker can then
choose among a pool of $N_c$ potential activity centers (whose locations are
randomly distributed across the city). The active subcenters are then
defined as the subset of potential centers which have a non zero incoming number
of individuals. We thus assume that the existence of activity centers is defined
by the willingness of workers to work in the possible locations.\\

Let us now discuss the form of the wage $W(j)$ and the commuting cost $C_T(i,j)$ that are
present in equation~\ref{eq:Z_workers_general}. 

\subsection{Determining the wage}
\label{sub:determining_the_wage}

The problem of determining the (spatial) variations of the average wage $W(j)$
at location $j$ is very reminiscent of some problems encountered in fundamental
physics. Indeed, the wage depends on many different factors, ranging from the
type of company, the education level of the inhabitant, the level of
aglomeration, etc., and in this respect is not too different from quantities
that can be measured in a large atom made of a large number of interacting
particles. In this situation, physicists figured that although it is possible
to write down the corresponding equations, not only is it impossible to solve
them, but also not really useful. In fact they found out that
a statistical description of these systems, relying on random matrices could
lead to predictions which agree with experimental results~\cite{Dyson:1962}.\\

We wish to import in spatial economics this idea of replacing a complex quantity
such as wages -- which depends on so many factors and interactions -- by a random
one. The problem is not so much that we cannot write down the equations that
determine the wage that an individual could get in a given company. Even if we
could (and we can't), the sheer number of people living in an urban area would prevent us
from solving these equations. And even if we could solve them, the resulting
information would be too overwhelming to really allow us to \emph{understand} the
behaviour of the system as a whole. We thus need an \emph{effective}
description of the phenomemon.

We account for the interaction between activity centers and
people by taking the wage in location $j$ as proportional to a random variable
$\eta_j \in \left[ 0,1\right]$ such that $W(j) = s\, \eta_j$ where $s$ defines
the maximum attainable average wage in the considered city.\\

We are aware that wages are not determined endogenously but are instead the
result of thousands, millions of interactions between firms and individuals. In
the same way that Dyson did not mean that the interactions between electrons in
large atoms \emph{are} random, our assumptions does not mean that wages are
\emph{really} randomly determined. What we mean, however, is that in the case of
systems containing a large number of individuals, one may do \emph{as
if they were randomly determined}. Although we thereby abandon the possibility
to describe the dynamics of the wages and their spatial distribution, the
resulting model is analytically solvable and makes quantitative predictions.

\subsection{Commuting cost and congestion}
\label{sub:the_commuting_cost}

We choose the transportation cost $C_T(i,j)$ proportional to the commuting time
between $i$ and $j$. In a typical situation where passenger transportation is
dominated by personal vehicles, this commuting time not only depends on the
distance between $i$ and $j$, but also on the traffic between the two places,
the vehicle capacity of the underlying network and its resilience to congestion.
The Bureau of Public Road formula~\cite{Branston:1976} proposes a simple form
taking all these ingredients into account. In our framework, it leads to the
following expression for the commuting costs

\begin{equation}
    C_T(i,j) =  t\, d_{ij} \left[ 1 + \left( \frac{T_{ij}}{c} \right)^{\mu} \right]
    \label{eq:commuting_cost}
\end{equation}

where $T_{ij}$ the trafic per unit of time between $i$ and $j$ and $c$
is the typical capacity of a road (taken constant here). The quantity
$\mu$ is a parameter quantifying the resilience of the transportation
network to congestion. We further simplify the problem by assuming
than the traffic $T_{ij}$ is only a function of the subcenter $j$ and
therefore write $T_{ij}=T(j)$ the total traffic incoming in subcenter
$j$.

\subsection{Summary}
\label{sub:summary}

In summary, our model is defined as follows. At each time step, we add
a new individual $i$ located at random in the city, who will
choose to work in the activity area $j$ (among $N_c$ possibilities
located at random) such that the following quantity

\begin{equation}
    Z_{ij} = \eta_j - \frac{d_{ij}}{\ell} \left[ 1 + \left( \frac{T(j)}{c} \right)^{\mu} \right]
    \label{eq:cost_function}
\end{equation}

is maximum (we omitted irrelevant multiplicative factors). The quantity $\ell =
s/t$ is interpreted as the maximum effective commuting distance that people can
financially withstand. Interestingly, the presence of commuting costs entails the
existence of a second length scale $\ell$ in the system (the first one being the
typical size $L$ of the city).

\section{Monocentric to polycentric transition}
\label{sub:monocentric_to_polycentric_transition}

\begin{figure}
    \centering
    \includegraphics[width=0.49\textwidth]{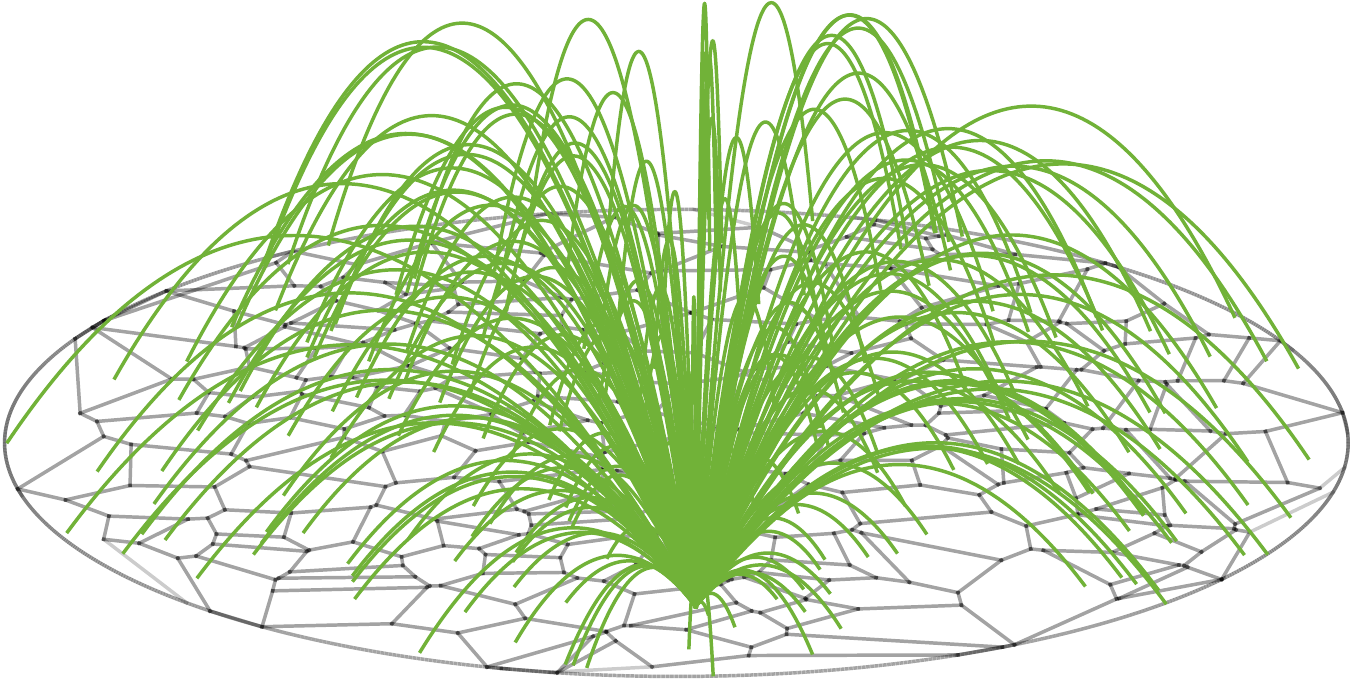}
    \includegraphics[width=0.49\textwidth]{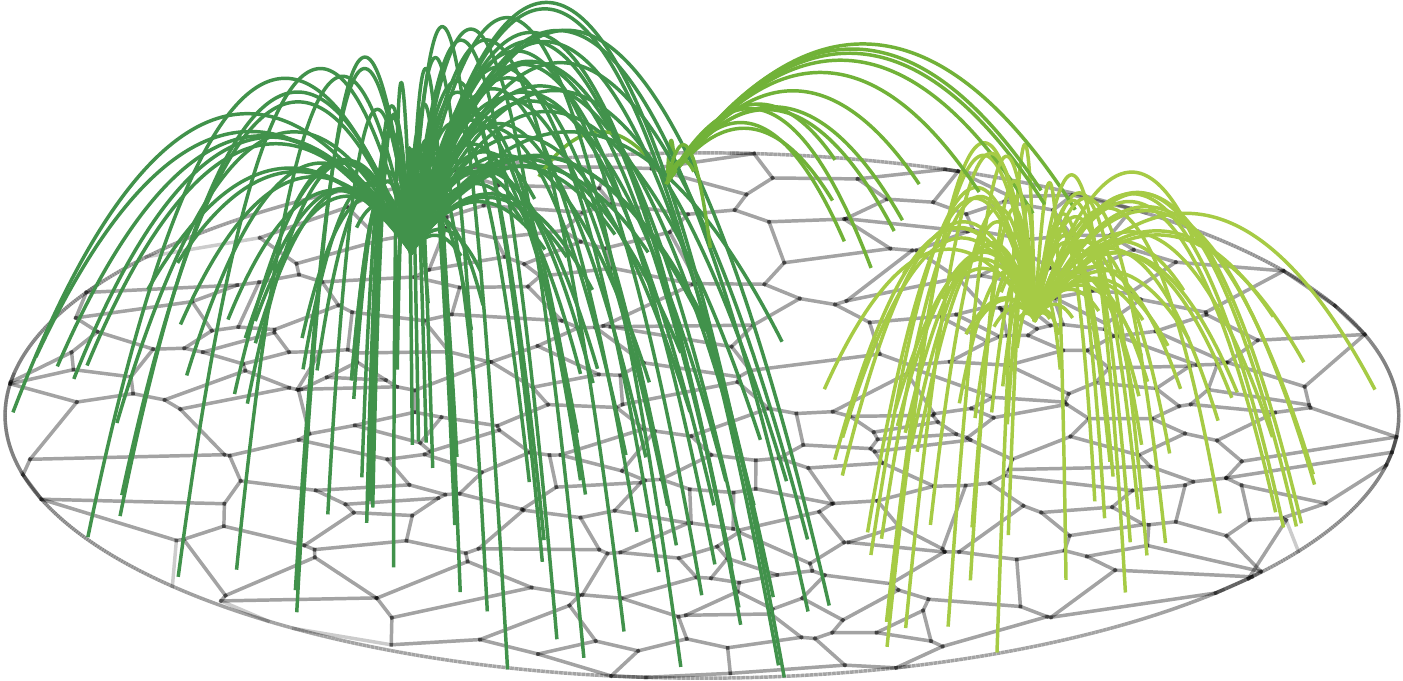}
    \includegraphics[width=0.49\textwidth]{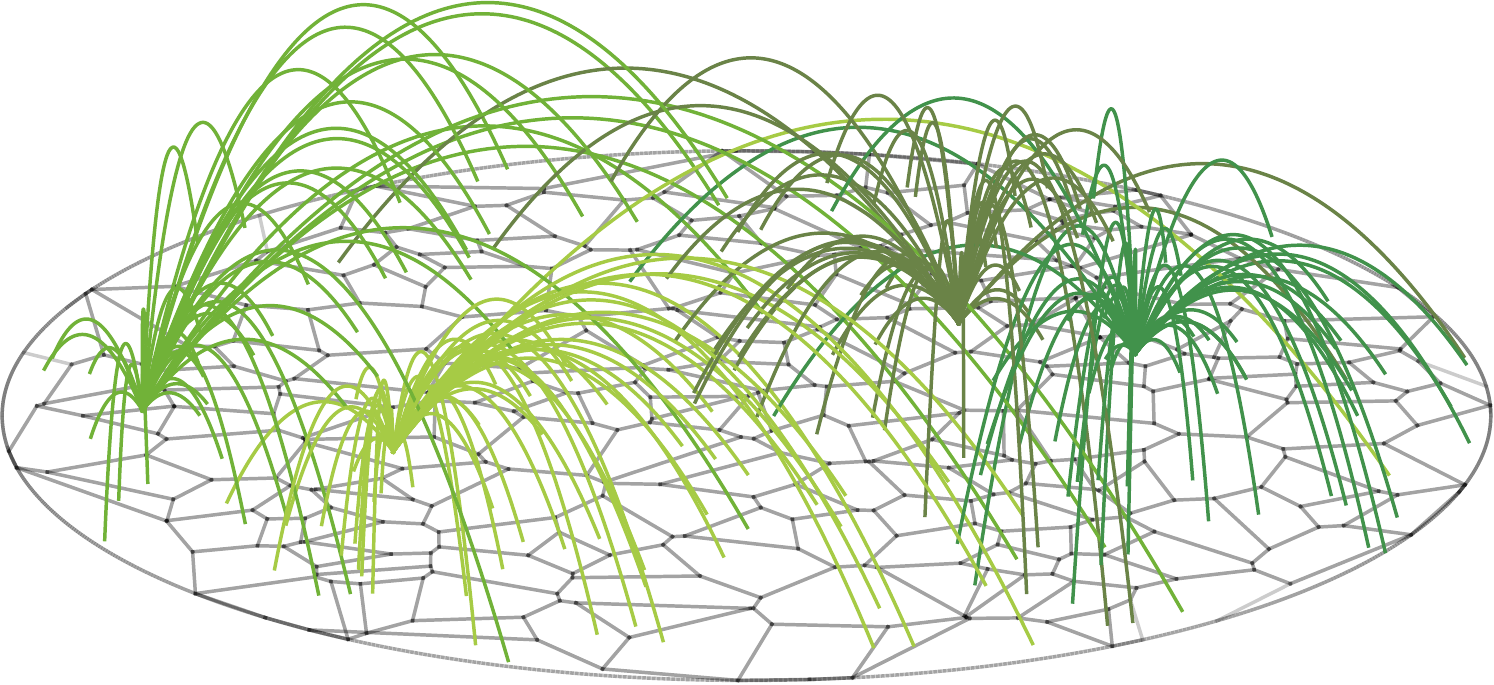}
    \caption{{\bf The different regimes.} The monocentric (top left), distance-driven polycentric (top right)
      and attractivity-driven polycentric (bottom) regimes as produced by
      our model. Each link represents a commuting journey to an activity center. \label{fig:model_results}}
\end{figure}

Depending on the relative importance of wages, distance and congestion, the
model predicts the existence of three different regimes: the monocentric regime
(Top left Figure~\ref{fig:model_results}), the distance-driven polycentric (Top
right Figure~\ref{fig:model_results}) regime and the attractivity-driven
polycentric (Bottom Figure~\ref{fig:model_results}) regime.\\

The existence of a monocentric regime depends on how $\ell$ -- the maximum
commuting distance that people can afford -- compares to the size of the city
$L$. Indeed, people located at a distance $d > \ell$ from the most
attractive center will not be able to afford commuting to this center, and will,
according to our model, choose to commute to a closer center.  As a result, a
monocentric regime is only sustainable as long as people's residence is drawn
close to the most attractive center. Thus, in the limit where $\ell \ll L$, the
attractiveness of a center becomes irrelevant, and a monocentric regime cannot
exist. In this case, we end up in the situation shown on the top-right of
Figure~\ref{fig:model_results}.\\

From now on, we will assume that $\ell$ is large enough so that a
monocentric state exists for small values of the population. In this
regime, the value of $\eta$ prevails and the monocentric state evolves
to an attractivity-driven polycentric structure as the population
increases. 
Starting from a small city with a monocentric
organisation, the traffic is negligible and 

$$Z_{ij}\approx \eta_j$$

which implies that all individuals are going to choose the most attractive
center, with the largest value of $\eta_j$, say $\eta_1$.
When the number $P$ of individuals increases, the traffic will also increase and
some initially less attractive centers (with a smaller values of $\eta$) might
become more attractive, leading to the appearance of a new subcenter. More specifically, a new
subcenter $j$ will appear when for an individual $i$, we have 

$$Z_{ij}>Z_{i1}$$

Because we assumed we originally were in a monocentric state, the traffic at
this point is such that $T(1)=P$ and $T(j)=0$ which leads to the equation

\begin{equation}
    \eta_j-\frac{d_{ij}}{\ell}>\eta_1-\frac{d_{i1}}{\ell}\left[1+\left(\frac{P}{c}\right)^\mu\right]
\end{equation}

We assume that there are no spatial correlations in the subcenter distribution,
so that we can make the approximation $d_{ij}\sim d_{i1}\sim L$. The new
subcenter will thus be such that $\eta_1-\eta_j$ is minimum. It will thus be the
potential subcenter with the second largest value denoted by $\eta_j=\eta_2$.\\

According to order statistics, we have on average for a uniform distribution

$$\overline{\eta_1-\eta_2}\simeq 1/N_c$$

hence a critical value for the population

\begin{equation}
    \boxed{P^*= c \left( \frac{\ell}{L N_c} \right)^{1/\mu}}
    \label{eq:critical_population}
\end{equation}

Whatever the system considered, there will \emph{always} be a critical
value of the population above which the city becomes polycentric. The
monocentric regime is therefore fundamentally unstable with regards to
population increase, which is in agreement with the fact that no major city in
the world exhibits a monocentric structure. We note that the smaller the value
of $\mu$ (or larger the value of the capacity $c$), the larger the critical
population value $P^*$ which means that cities with a good road system capable
of absorbing large traffic should display a monocentric structure for a longer
period of time.

\section{Number of centers}
\label{sec:number_of_centers}

We have so far established that, because of increased levels of congestion as
the population grows, all cities will eventually adopt a polycentric
structure. Although appealing and in agreement with common observations, the
prediction given by Eq.~\ref{eq:critical_population} is impossible to test with
the currently available data. Therefore, we would like to obtain a prediction
for the variation of the number of subcenters with population.\\

We compute the value of the population at which 
the $k^{th}$ center appears. Still in the attractivity-driven regime, we assume
that so far $k-1$
centers have emerged with 

$$\eta_{1} \geq \eta_{2} \geq \ldots \geq \eta_{k-1}$$

with a number of commuters $T(1), T(2), \ldots,
T(k-1)$, respectively. The next worker $i$ will choose the center $k$ if

\begin{equation}
    Z_{ik} > \max_{j \in \left[1,k-1\right]} Z_{ij}
\end{equation}

which reads

\begin{equation}
    \eta_k - \frac{d_{ik}}{\ell} > \max_{j \in \left[1,k-1\right]} \left\{
    \eta_j - \frac{d_{ij}}{\ell} \left[ 1 + \left(
      \frac{T(j)}{c}\right)^\mu\right] \right\}
\end{equation}

According to simulations of the model, we know that the distribution of traffic $T(j)$ is
narrow~\cite{Louf:2013_polycentric}, and we can assume that all the centers have roughly the same number of
commuters $T(j) \sim P/(k-1)$. As above we also assume that there are no spatial
correlations in the position of employment centers so that $d_{ij} \sim d_{ik} \sim L$. 
We can now write the previous expression as

\begin{equation}
\frac{L}{\ell} \left( \frac{P}{(k-1)\,c} \right)^{\mu} > \max_{j \in
  \left[1,k-1\right]} \left( \eta_j \right) - \eta_k
\end{equation}

Following our definitions, $\max_{j \in \left[1,k-1\right]} \left( \eta_j
\right) = \eta_1$. According to order statistics, if the $\eta_j$ are uniformly
distributed, we have on average $$\overline{\eta_1 - \eta_k} = (k-1)/(N_c+1)$$ 

It follows from these assumptions that (1) the $k^{th}$ center to appear is the
$k^{th}$ most attractive one (2) the average value of the population
$\overline{P}_k$ at which the $k^{th}$ center appears is given by:

\begin{equation} 
    \overline{P}_k = P^* \left( k-1 \right)^{\frac{\mu+1}{\mu}}
\end{equation}

Conversely, the number $k$ of subcenters scales sublinearly with population size as

\begin{equation} 
    \boxed{k \sim \left( \frac{P}{P^*} \right)^{\frac{\mu}{\mu
    + 1}}}
    \label{eq:centers_prediction}
\end{equation} 

For positive values of $\mu$, we have $\frac{\mu}{\mu+1}<1$. we can thus
conclude that the number of activity subcenters in urban areas scales sublinearly
with their population where the prefactor and the exponent depend on the
properties of the transportation network of the city under consideration. This
prediction is in agreement with the scalings obtained for Spanish and American
cities in Chapter~\ref{chap:monocentric_introduction}.

\section{Conclusion}
\label{sec:conclusion}

\subsection{A predictive model}
\label{sub:a_predictive_mode}

The model we just presented, although not perfect, exhibits many of the
desirable features of a model we listed in the introduction. First, it goes
beyond the standard models in urban economics by going beyond the explanation of
simple, qualitative, stylized facts. As we saw earlier, one major problem with the model of Fujita
and Ogawa is the absence of quantitative prediction. Instead of providing a
prediction that can be further confirmed or refuted by empirical observation,
the authors merely test the existence of polycentric solutions in the
framework of their model. The link with reality is however very loose, in the
sense that there is a big intellectual leap between the actual prediction of the
model and reality. Even though the model proposed here is very simple, it is not
difficult to link it to reality. Once the notion of activity centers is defined
empirically, it is not difficult to count the number of centers and look at the
dependence of this number on the population size of cities. The
model can then be confirmed, or refuted. Furthermore, as we will see in the
following section, the model serves as a basis to the understanding of some of
the scaling relationships in cities, linking the model even more strongly to 
empirical reality.

\subsection{Understanding the polycentric transition}
\label{sub:understanding_the_polycentric_tranistion}

Second, the model allows us to \emph{understand} why the polycentric transition
occurs. Taking a step back on the assumptions that lead to the prediction of
Eq.~\ref{eq:centers_prediction}, one can see that the transition
in our model is triggered by the congestion term in Eq.~\ref{eq:cost_function}. The positions
of households and firms are indeed taken as random, the wages are also taken at
random. Therefore, we can conclude that our model explains the polycentric
transition of cities through the increasing congestion around employment centers
as the population increases. More mechanisms are probably involved, but the model
shows that congestion alone is enough to lead to a polycentric situation.\\

If we assume that agglomeration economies can explain the
existence of centers in the first place, the model provides evidence that this
centripetal force is balanced by the centrifugal effect of congestion that
tears cities apart. Arguably, the non trivial spatial patterns observed in large cities can
 be understood as a result of the interplay between these competing
processes.\\

The model we propose trades off exhaustivity and complexity for simplicity and
explanatory power. Although some of the hypotheses we made are debatable, it is
striking that we manage to make a prediction on the scaling of the number of
centers with population size. On the other hand, unlike simplistic model, our
model's ontology is hard-wired into the reality we experience. For this reason,
its assumptions can be discussed, possibly changed. The model can be improved
upon in many different ways.

\chapter{Discussion}
\label{chap:monocentric_discussion}

\begin{flushright}{\slshape    
Our progress is narrow;\\
it takes a vast world unchallenged and for granted.}  \\ \medskip
--- J. Robert Oppenheimer~\cite{Oppenheimer:1954}
\end{flushright}

\bigskip

As we stated in the introduction, all models are fundamentally wrong -- at least
incomplete. Although is it able to reproduce key empirical regularities, the
model presented in Chapter~\ref{chap:monocentric_model} is no exception to this
rule. In the following chapter, we will enumerate some of its weaknesses, and
propose possible ways in which it could be extended.

Besides, because they are trying to make sense of a complex reality with a limited number
of tools, empirical analyses are not exempt of limitations either. Before
closing this chapter, we question the validity of the distribution-based methods used to
identify subcenters, and challenge the notion of polycentricity itself.

\section{Questioning and extending the model}
\label{sec:model}

\subsection{What the model does not say}
\label{sec:what_the_model_does_not_say}

The model makes many simplifying assumptions that make it
analytically tractable, but hide some interesting aspects of intra-urban
dynamics. We do not pretend to explain the complexity of
urban dynamics in its entirety, but rather some of its aspects.\\ 

A first feature, hidden in the assumptions of the model, is that we do not
explain the concentration of activities in particular areas of the cities.
Rather, we take the existence of centers for granted, and do not bother with the
behaviour of firms. Of course, this is a topic worthy of investigation, and
should be studied in more depth in order to have a comprehensive understanding
of the mechanisms that shape cities.

A second limitation lies in the fact that we ignore the process of residence
choice, and attribute households' location at random in the city. We
therefore set aside the problem of competition for space between households, and
a theoretical description of the spatial distribution of housing prices
(see~\cite{Gauvin:2013} for a model that explores this aspect).

Another limitation lies in the description of congestion. In a worry to
simplify the problem, we chose to adopt a macro-scale description of
traffic congestion, given by Eq.~\ref{eq:commuting_cost}. The sensitivity of the
road network to congestion is taken into account through the exponent $\mu$ and
the capacity $C$, which are assumed to be the same across the entire city. In
order to derive and compute these parameters, one would need to understand how local
patterns of congestion lead to macroscropic behaviours at the city scale. This
is, of course, a difficult entreprise:  local particularities of the layout may
have dramatic consequences on the fluidity of traffic, and congestions do
propagate through the network so that access to a given center can have an
effect on the travel to another center~\cite{Li:2015}. 

\subsection{Possible avenues}
\label{sub:possible_avenues}

\begin{figure}[!h]
    \centering
    \includegraphics[width=0.8\textwidth]{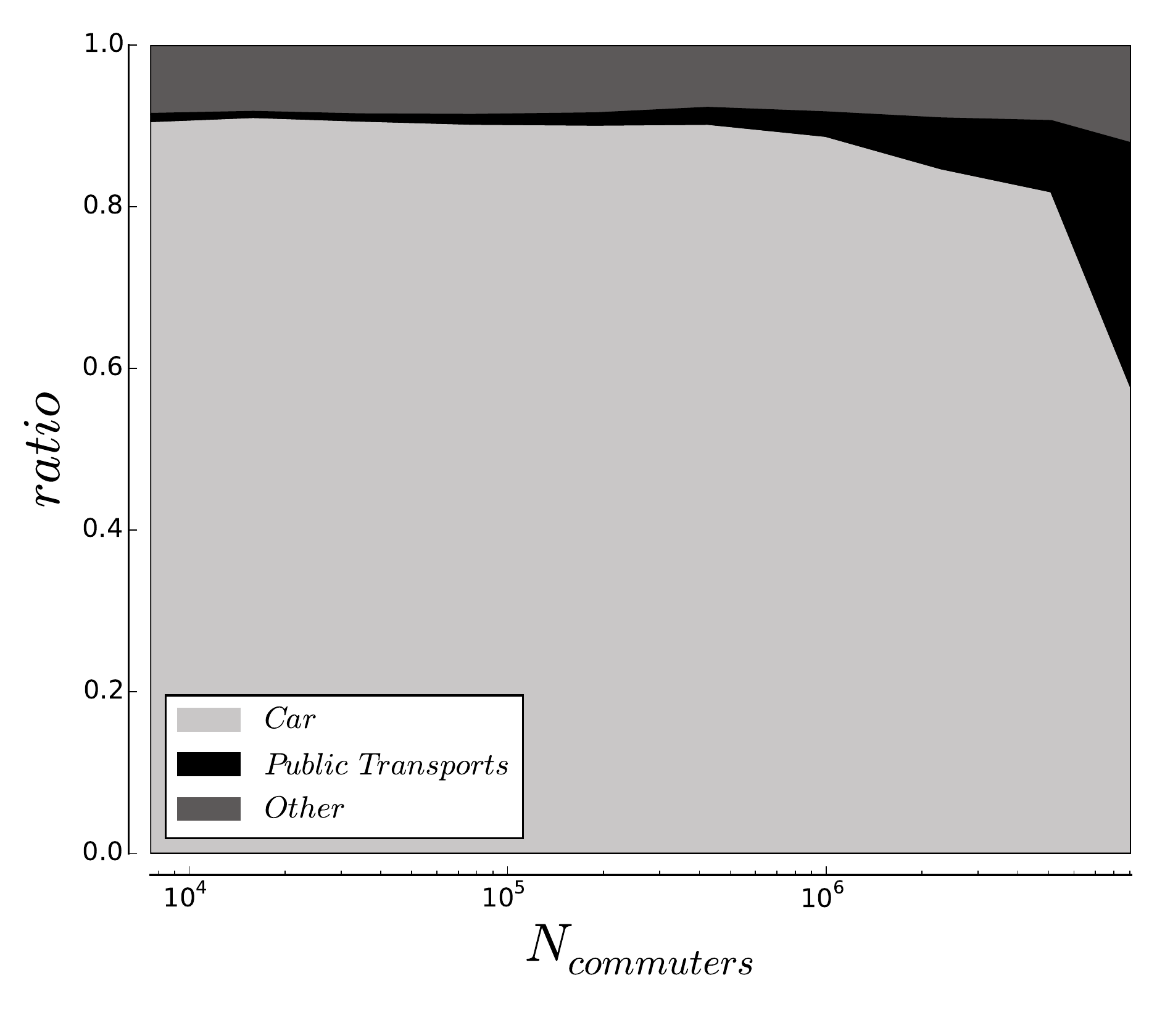}
    \caption{{\bf Mode share in the U.S..} Importance of different transportation modes in U.S. Metropolitan
    Statistical Areas, as a function of the number of commuters. Although the
proportion of individuals using public transportation or other modes (walking,
cycling, working at home) increases with population size, cars stay the dominant
mode of transportation everywhere. Data are from the 2013 American Community
Survey.\label{fig:transportation_mode}}
\end{figure}

Even without considering the difficult problem of modeling the behaviour of the
firms, and the way it is coupled to that of individuals, the model could be
improved in several ways. 
One first possible extension is to take the presence of public transportation
into account. Indeed, the model only considers individual vehicles, prone to
congestion, as a transportation mean.  However, the largest cities in the world
are all served by metro systems~\cite{Roth:2012}, and the share of transports
other than personal vehicles can attain $42\%$\graffito{Number from the 2013
American Community Survey} in cities like New-York. It is therefore far from
being negligible, and should be taken into account in the model. In its defense 
however, cars remain the dominant mode of transportation in the U.S., as shown of
Figure~\ref{fig:transportation_mode}. The use of alternative modes of
transportation is only notable in New-York, which is already a polycentric
city.\\

Another possible (but non-trivial) extension to the model is linked to the
second limitation stated above. Adding an income structure into the model (and
rules concerning the interaction of individuals) could allow us to explore the
spatial patterns of segregation, and see whether they can be understood from
basic economical choices alone~\cite{Gauvin:2013}. We considered this avenue
during this thesis, and realised there was very little of the empirical
knowledge on segregation could be used to test a model. This led us to working
on the material presented in Part~\ref{part:segregation} of this thesis.\\

\medskip

\section{Shadows in the empirical picture}
\label{sec:empirical}

\subsection{Identifying and counting centers}
\label{sub:measuring_the_number_of_centers}

Although non-parametric methods are an improvement over the previous parametric
methods, we are yet to understand the exact meaning of the obtained centers.

In particular, a problem that remains with non-parametric methods is that, no
matter the distribution of employment, population, etc. into the areal units,
the method will output a number. For instance, let us consider the extreme case
of a city where employment is uniformly distributed in space, so that the
employment density is uniform. In this situation, the LouBar method would tell
us that the number of centers is equal to the number of areal units. Yet, can we
really talk about centers in this case? Most would (rightfully) object. But on
what ground?

The difficulty resides in that we do not know what we mean exactly when we talk
about centers: do they reflect an objective reality, or are they a mere artifact
of the way our brains process information? Can they be quantitatively defined,
based on their desired properties or are they merely `unusual'
fluctuations in the distribution of activities? In the latter case, parametric
methods will do just fine. In the former case means we need to understand
what we talk about when we talk about centers. It is somewhat ironic that, more than $15$ years
after the publication of McDonald's seminal paper~\cite{McDonald:1987}, we are
still pondering over the question he originally asked.\\

A further shortcoming of the most recent (distribution-based) methods is that
they do not consider the spatial arrangement of the areal units involved. This
can be problematic, especially when the method identifies as centers areal units
that are contiguous. 

\begin{figure}[!h]
    \centering
    \includegraphics[width=0.75\textwidth]{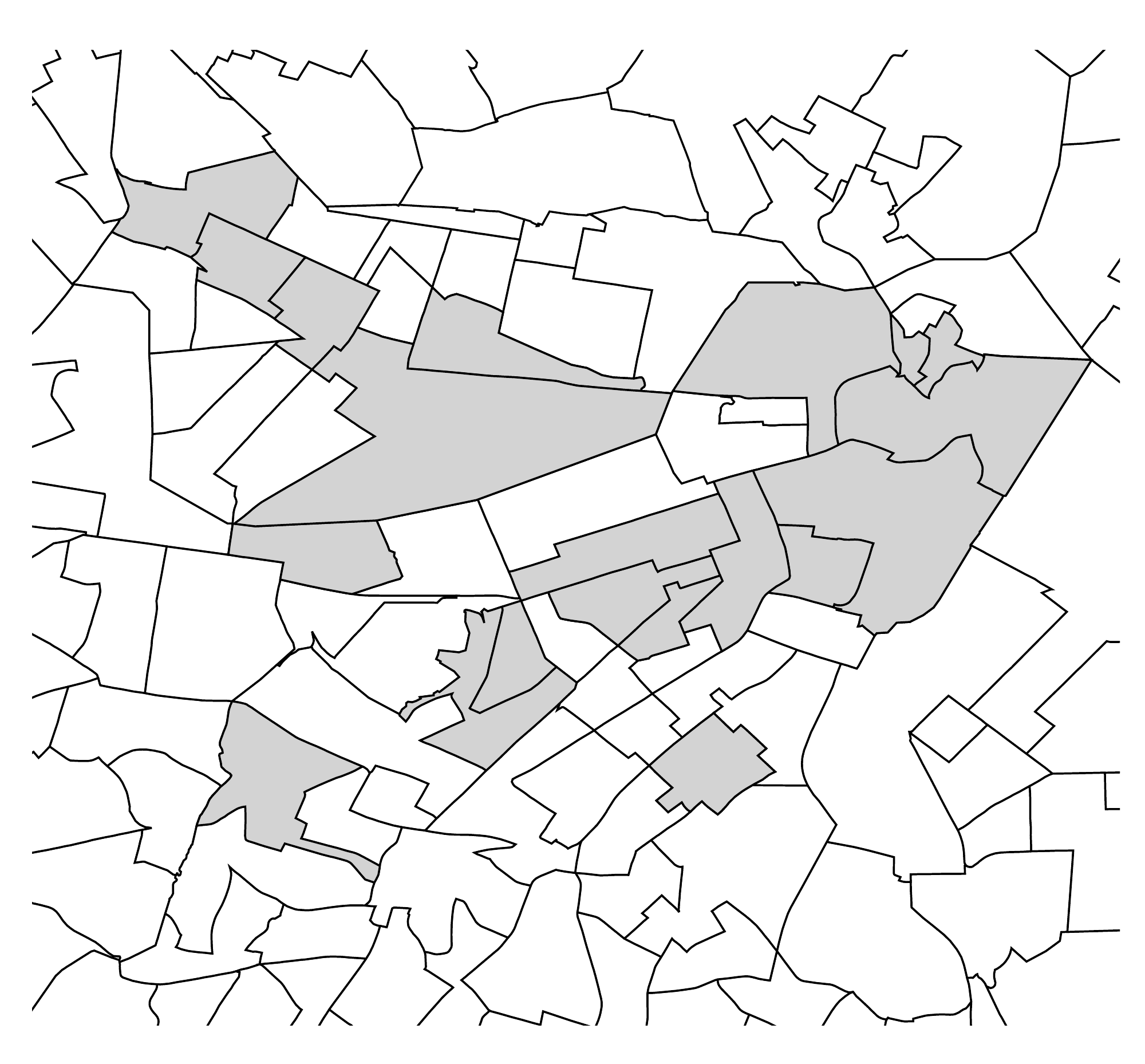}
    \caption{{\bf Subcenters and contiguity.} The census tracts of downtown Boston, MA in the U.S.. In light grey,
    the census tracts that are identified as employment hotspots by the LouBar
method. Although the method designates all light grey tracts as different hotspots,
many of them are contiguous. We can wonder whether such contiguous
hotspots are, in fact, part of a larger hotspot that would include all of them.
This plot was generated using the 2000 Census tract-to-tract
commuting flows and the 2000 Census tracts geometry.\label{fig:hotspots_boston}}
\end{figure}

We show an example of such a situation on Figure~\ref{fig:hotspots_boston}. We use
the LouBar method~\cite{Louail:2014} to extract the employment hotspots in the
Boston, MA MSA using data from the 2000 Census. As one can see, several of the
identified hotspots are contiguous. Should we still count them as separate
hotspots? Or should we consider that all contiguous hotspots are part of a
larger hotspot that encompasses them all?\\

The results of the methods provided in the introduction should not be thrown
away altogether, though. The number of centers they provide probably does not
reflect the `real' number of centers (if there is such a thing) in a particular
city. But, assuming that different cities exhibit similar structures, they
should still provide values that are coherent across different urban areas, and
are thus useful for \emph{comparison} purposes.\\

\subsection{Beyond polycentricity?}
\label{sec:beyond_polycentricity}

\subsubsection{The dispersed city}
\label{sub:the_dispersed_city}

As we saw in Chapter~\ref{chap:monocentric_introduction}, the concept of the
monocentric city was progressively replaced with the more elaborate polycentric hypothesis. It
is, however, not the end of the story. Gordon and Richardson, in a
provocative article~\cite{Gordon:1996}, argue that cities are dispersed more
than they are polycentric. Indeed, studying the employment density in Los Angeles,
they found that the centers they identified only contained $17\%$
of the total employment. Hardly a polycentric situation!  

\begin{figure}
    \centering
    \includegraphics[width=1\textwidth]{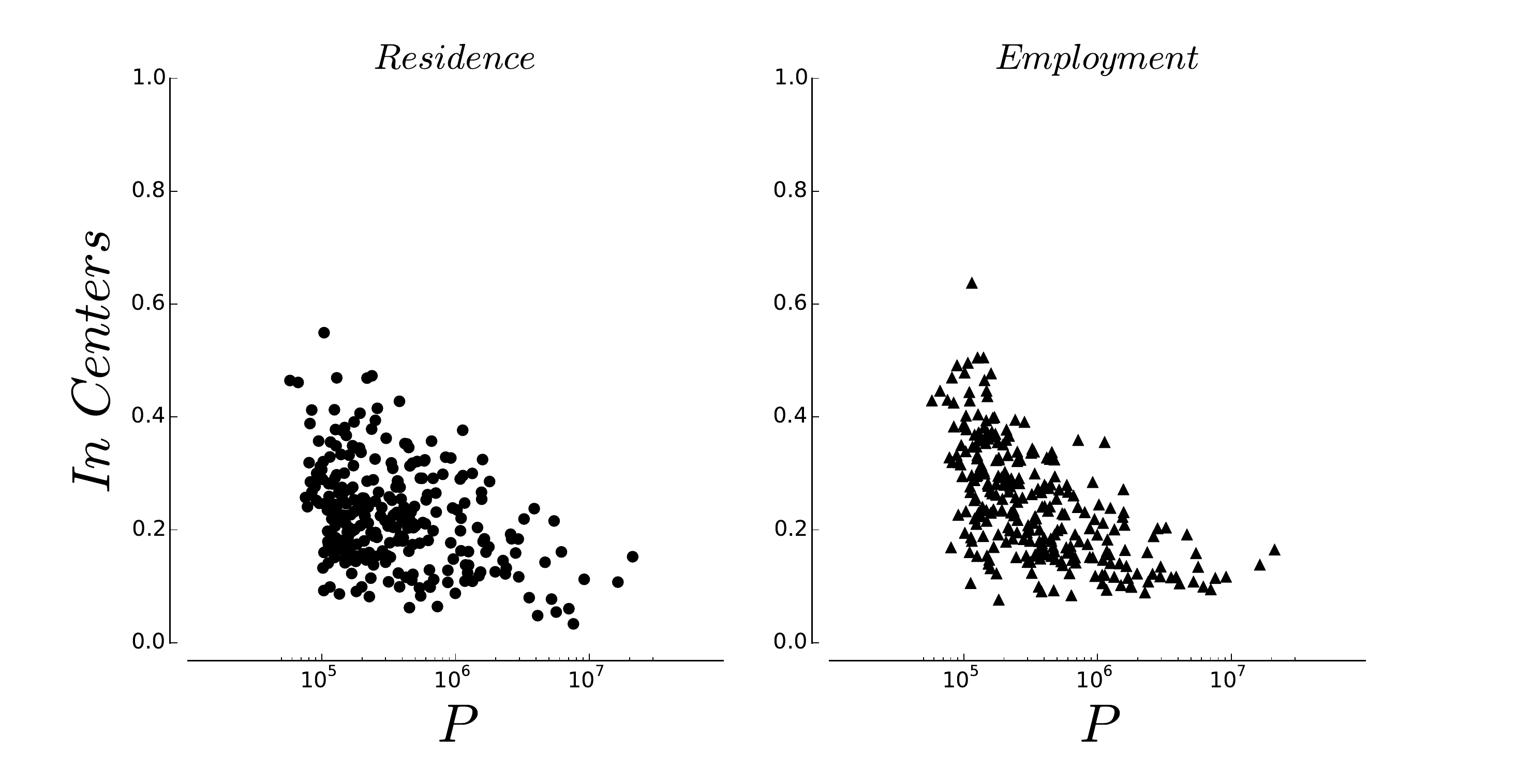}
    \caption{{\bf Concentration in subcenters.} (Left) Ratio of the total
    residential population in U.S. MSAs that lives in the centers identified by the LouBar
method. (Right) Ratio of the total number of employees in U.S. MSAs that work in the centers
identified by the LouBar method. Overall, cities are very dispersed, with only a
few cities having more than $50\%$ of their workforce or residential population
living in centers, confirming the results of Gordon and
Richardson~\cite{Gordon:1996}. Data are from the 2000 U.S.
Census.\label{fig:concentration_loubar}}
\end{figure}

Of course, we can (and should) wonder whether Gordon and Richardson's results
are an artefact of the choice of their case study --Los Angeles, famous for its
sprawl-- or the particular method they used to compute the number of centers. We
thus plot on Figure~\ref{fig:concentration_loubar} the ratio of the total number
of individuals that is contained in the centers defined by the LouBar method.
The results are striking: only a few, small metropolitan area reach the mark
where $50\%$ of individuals (employees or residents belong) to a designed
center. Worse, cities seem to be on average more dispersed as they are bigger.\\

The lesson that should be learned from the article by Gordon and Richardson is
that the notion of polycentricity is \emph{also an hypothesis} on the spatial
structure of densities. While it is arguably more involved than the monocentric
hypothesis, it does indeed implicitly impose some structure onto the data. The
process itself of counting centers implies that these centers exist, that there
is an element of reality attached to what we call centers. A quick look on the
3D plot shown on Figure~\ref{fig:density_3d} should convince the reader that the world
is not as simple as the way we picture it. For intance, while employment
densities indeed exhibit strong peaks that are easily distinguishable (although
that is arguable for Houston), the same cannot be said for population densities.

The point is not that the monocentric or the polycentric model are wrong
altogether. The problem lies in the lack of appropriate tools to describe a
density spatial profile, in the fact that there is no `one size fits all',
unbiaised method
of analysis. Indeed, the exploratory tools presented above try to fit a certain
model of the city to the actual data, be it monocentric or polycentric. The
methods developed to identify centers count the centers \emph{provided} there
are centers. We definitely need more elaborate methods that are also able to
tell us \emph{whether} there are centers. Or that go beyond the notion of
center.

\subsubsection{Quantifying Urban form}
\label{sub:urban_form}

This problem is in fact very general, and pertains to the field of spatial
analysis (including spatial statistics). Finding centers indeed amounts to
finding the proper way to describe a density profile at a meso-scale level and
to devising proper methods to detect the salient feature of this spatial
pattern. The collection of tools and methods to describe the structure
of density patterns in cities consitutes the sub-field of urban
form~\cite{Tsai:2005,Schwarz:2010,LeNechet:2010_these,LeNechet:2015} and reaches far beyond the
determination of subcenters.

Finally, we have focused in this part on the \emph{morphological} aspect of urban form,
as most of the preceding studies. We ackowledge however the existence of a
\emph{functional} aspect (see~\cite{Berroir:2008}), which takes the attraction
range of employment subcenters into account, in addition to the raw number of
employees. Mixing employment densities and the property of the flows to the
center may indeed lead to a better understanding of what a center really is.

\section{Summary}
\label{sec:summary}

In this part, we have presented an historical overview of the monocentric
hypothesis for the structure of cities, and how the view has progressively
shifted towards the picture of a more distributed, polycentric organisation.
Starting with indirect evidence for a polycentric picture, several methods were
then naturally proposed to directly measure the number of centers, from the
first parametric methods to the more recent non-parametric methods. Observing
evidence for an increased polycentricity with population size, we then wondered
what were the possible explanations for this phenomenon. We proposed an
out-of-equilibrium model of city growth that predicts the necessary emergence of
secondary centers as populations grows, and a sublinear increase of the number
of subcenters with population---both verified on empirical data, across
different countries, for several city definitions.

In the next part, we will continue our journey with another, seemingly unrelated
topic: scaling relationships. We will start with a historical perspective on
scaling, showing that scaling relationships did in fact precede Quantitative
Geography, and we will provide a non-exhaustive review of the empirical results.
We will then be ready to show how, using the model exposed in the previous chapter, we
can understand the value of the scaling exponents related to individual mobility. We will
then conclude on a reflection of what scaling relationships can and do tell us
about cities, and highlight their shortcomings.


\ctparttext{The past decade has witnessed a renewed interest for the scaling of some of cities' characteristics with population size
    -- first discovered more than $60$ years ago.

The contribution of this part is threefold. First, we review the exisiting
literature on allometric scalings, sorting the measured
exponents by theme. We then propose a model to explain the scaling exponent of several
indicators related to mobility in cities, and discuss the theoretical and
practical consequences of these exponents.  Finally, we present some of the
challenges posed by scaling relationships: their interpretation,
and the issues they reveal about the definition of cities.}

\part{Scaling} 
\label{part:scaling}

\chapter{Introduction}
\label{chap:scaling_introduction}

\begin{flushright}{\slshape    
The allometric law promises to become\\
an integral part of geography theory.} \\ \medskip
--- David Harvey (1969)~\cite{Harvey:1969} 
\end{flushright}

\section{Probing cities with scaling laws}
\label{sec:probing_cities_with_scaling_laws}

\subsection{Scaling laws}
\label{sub:scaling_laws}

As discussed in the introduction of this thesis
(Chapter~\ref{chap:studying_cities}), cities are paradigmatic examples of
complex systems. As systems, they can be of thought of as `black boxes' with
inputs (people, goods, money, information, etc.), a structure (roads, buildings,
electric cables, etc.) and outputs (Patents, $CO_2$ emissions, etc.).  A simple
way to explore the behaviour of such a system is to look at the way it behaves
when we change its size. That is, how its structure and its outputs change when
the inputs are altered. Formally speaking, we try to find the function $f$ such
that the quantity $Y$ -- a measure of the output or the structure -- varies as

\begin{equation}
    Y = f(S)
    \label{eq:functional_form}
\end{equation}

where $S$ is the size of the system.\\

What is to be considered as the size of the city?  The spatial footprint, the
total volume occupied by its building? The answer adopted by many before this
thesis~\cite{Stewart:1947, Bettencourt:2007}, is the total number of
inhabitants. The real reason is probably pragmatic: ``it
works''. Although, in retrospect, the choice of population makes complete sense. 

Cities are indeed more than roads and buildings: cities are the people who
inhabit them. People are responsible for the changes in wealth, employment,
number of patents. People need new roads, and it is people who build them.
People need electricity, and again it is people who run electric cables between
buildings. Inhabitants of a city, through their actions and interactions, are reponsible for the collective
mechanisms that act on the city as a whole.  In a sense, behind the use of the
population $P$ to measure the size of a city as a system hides the idea 
that cities are, first and foremost, the people that inhabit them.\\

As a matter of fact, when we try to plot quantities as a function of the
population size $P$ of cities, we obtain \emph{allometric scaling
relationships}. That is, a power-law relationship between various quantities $Y$
and the population size $P$ of cities \emph{in a given system of cities} 

\begin{equation}
    Y = Y_0\,P^{\,\beta}
    \label{eq:scaling_definition}
\end{equation}

where the exponent $\beta$ can be different from $1$. This type of scaling
relation, used extensively in Biology~\cite{Thompson:1942} and in
Physics~\cite{Barenblatt:1996}, is a signature of the various processes
governing the phenomenon under study, especially when the exponent $\beta$ is
different from what would be naively expected. Three qualitatively different
regimes are usually distinguished for the exponent
$\beta$~\cite{Bettencourt:2007}

\begin{description}
    \item[Superlinear] when $\beta>1$. In this situation, the $Y$ per capita
        increases with population size. This is associated with the notion of
        increasing returns with scale in economics.
    \item[Linear] when $\beta=1$. In this situation, the $Y$ per capita is
        constant. This behaviour is characteristic of an extensive system, when
        the whole is equal to the sum of its parts.
    \item[Sublinear] when $\beta<1$. In this situation, the $Y$ per capita
        decreases with population size. When $Y$ is the cost in infrastructure,
        this is characteristic of economies of scale.
\end{description}    

\begin{figure}[!h]
    \centering
    \includegraphics[width=\textwidth]{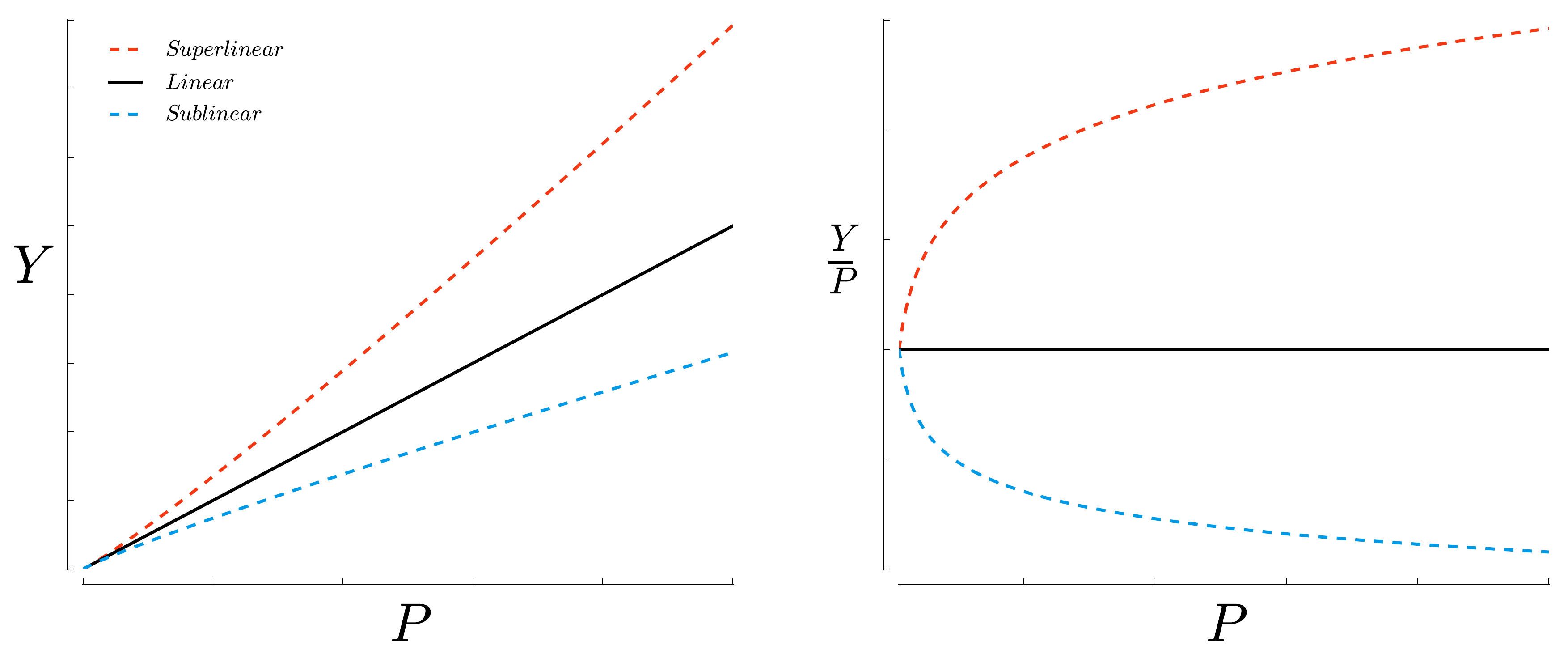}
    \caption{{\bf Sublinear, Linear and Superlinear scaling.} (Left) Example of a linear (black), sublinear (blue) and superlinear (red)
    behaviour. (Right) Evolution of the correspondant per-capita quantities with
population. A superlinear behaviour means that per-capita quantities increase with
population size, while a sublinear behaviour means per-capita quantities
decrease with city size.\label{fig:scaling_scheme}}
\end{figure}

We note that the scaling exponent $\beta$ is also directly related to the \emph{elasticity}
defined in Economics. Indeed, the cities' population elasticity of the quantity
$Y$ is defined as\\

\begin{equation}
    \beta = \frac{dY/Y}{dP/P}
\end{equation}

\subsection{Underlying assumptions}
\label{sub:underlying_asumptions}

Several assumptions, although rarely mentionned, hide behind every exhibited
scaling law. The first one, is that we are able to unambigusouly
delineate cities as systems.  While this is trivial in the case of animals (it
is fairly easy for us to isolate an elephant, or a cat from its environment
before 
measuring its mass and its metabolic rate), it is a much more difficult task in
the case of cities. Indeed, cities do not have fixed boundaries, and their 
geographical limits evolve with time. They are also open system:
people are born and die, change residence and companies do the same. 

Traditionally, people have relied on the definition given by statistical
agencies of the respective countries they were studying -- and we will do the
same in the next chapter. We will however see, in the chapter concluding this
part, that the problem of delineating cities is a sensible issue and affects
greatly scaling analyses. \\

A second issue, rarely -- if ever -- mentioned in the literature, is the
necessity to define the set of cities to study. Scaling laws are essentially
cross-sectional relationships, where we measure the quantity $Y$ on a set of
cities with different populations. But how is the set determined? For instance,
would it make sense to mix French cities, Ukrainian, Canadian and Korean, etc
cities and plot, say, their total GDP as a function of the population? Would we
then observe a neat scaling relationship? 

Intuitively, this is very unlikely to happen, as different countries have
overall different levels of wealth, and this should be reflected in the wealth
of their cities. Therefore, plotting cities from different countries together is
likely to introduce important deviations to the pure scaling relations which are
not due to the fact that cities in different countries do not follow the same
processes, but rather because of systemic differences at the country level. As a
matter of fact, most studies limit themselves to a single country. But one should
bear in mind that this choice is arbitrary. And the problem of choosing
the appropriate set from which to pick the cities is linked to the more general
problem of defining systems of cites.

\subsection{An increasing importance}
\label{sub:an_increasing_place_in_the_landscape}

This chapter's epigraph, from Harvey's 1969
\emph{Explanation in Geography}, is somewhat prophetic.  Allometric scaling
relationships only concern $1$ page out of the $500$ pages that the book
contains, a reflection of the very few empirical results that were available at
the time.  Looking at the extent of the literature on scaling relationships
almost $50$ years after Harvey wrote this sentence, it is difficult to deny the
accuracy of this prophecy. Thanks to the wider availability of data through
statistical agencies, but also the availability of 'new data' (such as mobile
phone data), empirical measurements of scaling laws have multiplied, and now
concern quantities as diverse as the total surface area, the number of new patents,
the quantity of $CO_2$ emitted, the number of phone contacts of individuals, etc. 
The discovery of allometric scaling in cities is not recent~\cite{Stewart:1947},
but it has undoubtedly caused a stir in the literature about urban systems over
the last
decade~\cite{Bettencourt:2007,Pumain:2004,Bettencourt:2013,Louf:2014_scaling,Louf:2014_smog,Arcaute:2014}.\\

In the next section, we will present a non exhaustive historical review of the
empirical results on scaling relationships. This will lay the ground for our contribution to the
debate: a theoretical interpretation of the scalings related to the mobility of
people, and an estimate for the scaling exponent of the surface area.

\section{A brief history of allometric scaling and cities}
\label{sec:a_brief_history_of_allometric_scaling_and_cities} 

Rather than an exposition that is linear in time, we deliberately choose to
classify the proposed studies according to the type of quantity. That way, we  
emphasize the variety of variables that have been studied. Incidentally, this
order also reveals the different waves of interest scaling relationships have
sparked off in the past $6$ decades, and hints at some issues related to scaling laws.

\subsection{Surface area}
\label{sub:surface_area}

The spatial footprint of cities, as can be observed on satellite picture or on
maps, is one of the properties that is easiest to measure. It is therefore not
surprising that the first occurence of scaling relationships in cities was
the scaling of the surface area of cities with their population. In $1947$,
using data about administrative cities obtained from the $1940$ US Census, John
Stewart shows
\graffito{Incidentally, the author of the study, John Stewart, was a physicist.}

\begin{equation}
    A = \frac{P^{\,3/4}}{350}
\end{equation}

The next occurence of this scaling can be found $9$ years later in a study by
the same author~\cite{Stewart:1958}, using UK census data. It isn't
long until the result percolates in Geography with Boyce in $1963$~\cite{Boyce:1963}.
In $1965$, Nordbeck's paper~\cite{Nordbeck:1965} also studies the scaling of surface area
with population, and, for the first time, explicitly refers to allometry in
biology. Later, Tobler~\cite{Tobler:1969} uses some of the first available
satellite images to provide the first confirmation using satellite pictures.
Satellite pictures were also used more recently by Gu\'erois
in~\cite{Guerois:2003} (Table~\ref{tab:area}).\\

When applied to morphological definitions of cities, all studies
(see~\cite{Batty:2011}) give an exponent that varies in the range $[0.70,
0.90]$. However, different results are obtained for functional definitions of
cities~\cite{Batty:2011}, or when the set of studied cities span several systems of
cities~\cite{Fuller:2009}. Thus, despite being the oldest and most trusted scaling relationship in the
literature, the relation between the surface area and population size of cities
exhibits some of the issues we will discuss in
Chapter~\ref{chap:scaling_implications}.

\begin{table*}[!h]
    \centering
\begin{tabular}{|cccc|}
\hline
Exponent & City definition & Year & Study\\
\hline
0.75 & Administrative (US) & 1940 & Stewart~\cite{Stewart:1947}\\
0.75 & Administrative (UK) & 1951 & Stewart \& Warntz~\cite{Stewart:1958}\\
0.86 & Morphological (US) & 1950 & Boyce~\cite{Boyce:1963}\\
0.88 & Administrative (US) & 1950 & Nordbeck~\cite{Nordbeck:1965}\\
0.88 & Built-up (US) & 1969 & Tobler~\cite{Tobler:1969}\\
0.86 & Built-up (Europe) & 1990 & Gu\'erois~\cite{Guerois:2003}\\
0.73 & Administrative (Europe) & 1990 & Gu\'erois~\cite{Guerois:2003}\\
0.78 & Morphological (US) & 2010 & Louf \& Barthelemy~\cite{Louf:2014_scaling}\\
\hline
\emph{1.48} & \emph{Functional} (US) & 2005 & Batty \& Ferguson~\cite{Batty:2011}\\
\hline
\end{tabular}
\caption{{\bf Scaling of the surface area.} Scaling exponents for the surface area of
cities found in the literature. The scaling for administrative cities, built-up
areas or cities defined according to a morphological criterion are consistent
with one another -- at least qualitatively. The exponent for cities with a
functional definition is however qualitatively different.}
\label{tab:area}
\end{table*}

\subsection{Economic diversity and employment}
\label{sub:economic_diversity}

\subsubsection{Employment diversity}
\label{ssub:employment_diversity}

The economic diversity has been of interest to researchers very early on. In
1949, Zipf in \emph{Human behavior and the principle of least
effort}~\cite{Zipf:1949} plots the number of service-business establishments, manufactures and
retail stores per city as a function of population (in log-log scale) using data
from the $1940$ US Census. He finds a linear relationship with population for
the three types of establishments, which agreed at the time with his model. He
also plots the scaling of the diversity, defined as the number of different kinds of
entreprises present in the city being studied. 

In his $1967$ \emph{Geography of market centers and retail
distribution}~\cite{Berry:1967} Berry, hoping to demonstrate the hierarchical
organisation of central places, plots this time the population of cities as a
function of the number of retail and service businesses observed.
Strangely enough, the data imply

\begin{equation}
    D \propto P^{\, \beta}
\end{equation}

with $\beta > 1$, in contradiction with later results. Indeed, Bettencourt et al.\cite{Bettencourt:2014} showed that the professional
diversity $D$, measured as the number of professions of different kind in the
city considered, could be fitted by the following function

\begin{equation}
    D(N_e) = d_0\, \frac{\left(\frac{N_e}{N_0}\right)^\gamma}{1+\left(\frac{N_e}{N_0}\right)^\gamma}
\end{equation}

where $d_0$ is the size of the classification used in the data, $N_0$ is the
typical saturation size, and $\gamma < 1$ is an exponent expressing the extent to
which new activities `appear' as the total employment increases. Far from the
saturation regime, when $N_e \ll N_0$ (the classification is sufficiently
fine-grained), we have 

\begin{equation}
    D(N_e) \sim A\,N_e^{\,\gamma}
\end{equation}

\subsubsection{Employment in different activities}
\label{ssub:employment_in_different_activities}

More recently, Pumain and coauthors~\cite{Pumain:2006}, extending the work done
by Paulus in his PhD thesis~\cite{Paulus:2004}, showed that the
employment $E_a$ in different activities $a$ scaled as

\begin{equation}
    E_a \propto P^{\,\beta}
\end{equation}

with different exponents $\beta$ for the different activities
(Table~\ref{tab:employment}). They observed,
for the year $1999$ in France, that the exponents could be classified in three
categories

\begin{itemize}
    \item $\beta > 1$ for innovative sectors: research and developement,
        consultancy.
    \item $\beta = 1$ for common sectors: hotels, health and social services,
        education.
    \item $\beta < 1$ for `mature' sectors such as the food industry
\end{itemize}

This result was confirmed recently by Youn et al.~\cite{Youn:2014} -- although
they do not refer to this previous work -- who showed that the same behaviour was
observed for the number of business of a given type. 

A particularly interesting result by Pumain et al.~\cite{Pumain:2006} is
the evolution of the different exponents with time, where we can see a clear
increase of the exponents for research and developpement, and a clear decrease
of the exponents related to manufactures of different kinds. We will come back
to the interpretation of this phenomenon in
Chapter~\ref{chap:scaling_implications}.\\

\begin{table*}[!h]
    \centering
\begin{tabular}{|ccc|}
\hline
Exponent & City Definition & Economic sector\\
\hline
1.67 & Functional (France) & Research and development\\
1 & Functional (France) & Hotels and restaurants \\
0.85 & Functional (France) & Manufacture of food products\\
\hline
\end{tabular}
\caption{{\bf Scaling of employment in different economic sectors.} The scaling
behaviour of the number of employees in a given economic sector depends on the
nature of the 
economic sector. We give an example for each of the `innovative'
(superlinear), `common' (linear) and `mature' (sublinear) categories defined by
Pumain et al.~\cite{Pumain:2006}. The exponents were obtained from~\cite{Pumain:2006} and concern
French $1999$ `Aires urbaines'.} 
\label{tab:employment} 
\end{table*}

\subsection{Wealth}
\label{sub:wealth}

The notion of increasing returns with the size of the agglomeration is
often discussed in economics, although emprical proofs are hard to find. The
superlinear scaling of the GDP of american cities as a function of their
population may be the most striking example of such increasing
returns~\cite{Bettencourt:2007}. In the same article, Bettencourt et al. showed
that the number of patents (used as a proxy for creativity), and wages also scaled
superlinearly with population size in the US (see Table~\ref{tab:wealth}).

Because larger cities create proportionally more wealth than smaller cities, we
can wonder whether this supplement of wealth allows to sustain proportionally
more jobs. The answer, as shown in~\cite{Bettencourt:2014} for american cities,
is negative: the total employment of a city is on average proportional to its population.

\begin{table*}[!h]
    \centering
\begin{tabular}{|cccc|}
\hline
Quantity & Exponent & City Definition & Study\\
\hline
GDP & 1.13 & Functional (US) & Bettencourt~\cite{Bettencourt:2013}\\
New patents & 1.27 & Functional (US) & Bettencourt et al.~\cite{Bettencourt:2007}\\ 
Total wages & 1.12 & Functional (US) & Bettencourt et al.~\cite{Bettencourt:2007}\\
\hline
Employment & 1.01 & Functional (US) & Bettencourt et
al.~\cite{Bettencourt:2007}\\
\hline
\end{tabular}
\caption{{\bf Economic vitality.} The scaling of quantities linked to cities'
economic vitality and creativity scale superlinearly with population size.
This does not translate however in larger employment rates, as the number of
employees scales linearly with population size.} 
\label{tab:wealth} 
\end{table*}

\subsection{Human interactions}
\label{sub:human_interactions}

At the heart of Bettencourt's model~\cite{Bettencourt:2013} to explain the
superlinear scaling of quantities associated with wealth and creativity is the
behaviour of the total number of interactions between individuals with the size
of the city. In an attempt to test this hypothesis, Schl\"apfer et
al.~\cite{Schlapfer:2014} looked at the scaling of the cumulative number of
contacts $K$ that people had over the phone, using mobile phone data in
Portugal, and landlines in the UK. They also looked at the cumulative call
volume (total number of minutes called) and the cumulative number of calls, and
found that the three quantities scale superlinearly with population size (see
Table~\ref{tab:interactions}). 

They further found that the number of non-returned calls showed a larger
exponents than the number of calls, meaning that the number of solicitations an
individual gets is greater in large cities.\\

\begin{table*}[!h]
    \centering
\begin{tabular}{|ccc|}
\hline
Quantity & Exponent & City Definition\\
\hline
Cumulative phone contacts & 1.12 & Morphological (Portugal)\\
Cumulative phone contacts & 1.13 & Administrative (Portugal)\\
\hline
Cumulative call volume & 1.11 & Morphological (Portugal)\\
Cumulative call volume & 1.15 & Administrative (Portugal)\\
\hline
Cumulative number of calls & 1.10 & Morphological (Portugal)\\
Cumulative number of calls & 1.13 & Administrative (Portugal)\\
\hline
\end{tabular}
\caption{{\bf Interactions over the phone.} Scaling of the cumulative number of
    phone contacts, phone calls and the cumulative call volume over $409$ days
    in Portugal. As for the scaling of the surface area, administrative and
    morphologically defined cities exhibit similar exponents. The scaling for
    LUZ (european functional definition) shows a behaviour compatible with a
    linear scaling, although the number of points (9) is not large enough to
    conclude. The data were obtained from a mobile phone provider, and all
    quantities are rescaled to take into
    account the variation of the operator's coverage between
cities.\label{tab:interactions}}
\end{table*}

\subsection{Mobility of individuals, and environmental impact}
\label{sub:mobility}

Because cars are widely used (at least in the US), and because peak travel
demand on the roads corresponds to journey-to-work trips, most of the
information available on the mobility of individuals concerns the commuting to
work, often by car. 

Samaniego and Moses~\cite{Samaniego:2008} showed that the total number of miles
driven in US Urban Areas (morphological definition) rescaled by the total
surface area scales sublinearly with population size, with a non-trivial
exponent (that is, different from $1/2$. More details in the next chapter). We
showed in a later study~\cite{Louf:2014_scaling} that the total distance driven
scales linearly with population size in Urban Areas.
Also related to commuting, and the use of personal vehicles, is the
evolution of the total comsumption of gasoline with city size.
Bettencourt et al. showed that gasoline sales in Metropolitan Statistical Areas
scaled sublinearly with population size~\cite{Bettencourt:2007} (see
Table~\ref{tab:mobility} for values). 

Hopefully, new data such as mobile phone data should be
able to inform us about other trips, which represent no less than 80\% of all
trips undertaken in the United States!~\cite{FHWA-PL-11-022}.\\

A diseconomy associated with the mobility of individuals is the
quantity of $CO_2$ emitted due to transportation (and polluting substances).
Using different city definitions, different authors find very different
behaviours. The authors of \cite{Fragkias:2013} find that transport-related
$CO_2$ emissions in Metropolitan Statistical Areas in the US scale sublinearly
with population size, while the authors of
~\cite{Louf:2014_mobility,Oliveira:2014} find that they scale superlinearly with
population size for US Urban Areas (morphological definition). We will come back
to this in the next Chapter.

\begin{table*}[!h]
    \centering
\begin{tabular}{|cccc|}
\hline
Quantity & Exponent & City Definition & Study\\
\hline
Distance driven & 1 & Morphological (US) & Louf \& Barthelemy~\cite{Louf:2014_scaling} \\
Gasoline sales & 0.79 & Functional (US) & Bettencourt et al.~\cite{Bettencourt:2007}\\
\hline
$CO_2$ emissions & 1.42 & Morphological (US) & Oliveira et al.~\cite{Oliveira:2014}\\
$CO_2$ emissions & 1.37 & Morphological (US) & Louf \&
Barthelemy~\cite{Louf:2014_smog}\\
$CO_2$ emissions & \emph{0.93} & \emph{Functional} (US) & Fragkias et al.~\cite{Fragkias:2013}\\
\hline
\end{tabular}
\caption{{\bf Mobility. } Scaling relationships linked to the individual
    mobility in cities. The three scaling exponents regarding the $CO_2$
    emissions \emph{due to transportation} were obtained using the
    Vulcan data (\url{http://vulcan.project.asu.edu/}) which provide
    measurements of the $CO_2$ emissions on a $10\text{ km}\,x\,10\text{ km}$
    grid. The difference between the three studies is in the method used to
    delineate cities: Fragkias et al.~\cite{Fragkias:2013} rely on the Metropolitan Statistical Areas
    defined by the Census Bureau, Oliveira et al.~\cite{Oliveira:2014} rely on
    the City Clustering Algorithm~\cite{Rozenfeld:2008} (morphological
    criterion) while we rely on the Urbans Areas defined by the Census
    Bureau.\label{tab:mobility}
}
\end{table*}

\subsection{Basic commodities}
\label{sub:basic_commodities}

We can also wonder how the consumption of basic commodities (housing, water,
electricity) per capita changes with population size. By far the most expected
result, Bettencourt et al. showed~\cite{Bettencourt:2007} that the total water
consumption (in China), the total electrical consumption (in China),  and the
total housing (in the US) are proportional to the population (see
Table~\ref{tab:commodities}).\\

\begin{table*}[!h]
    \centering
\begin{tabular}{|cccc|}
\hline
Quantity & Exponent & City definition & Study\\
\hline
Total housing & 1.00 & Functional (US) & Bettencourt et al.~\cite{Bettencourt:2007}\\
Total electrical consumption & 1.05 & Administrative (China) & Bettencourt et al.~\cite{Bettencourt:2007}\\
Total water consumption & 1.01 & Administrative (China) & Bettencourt et al.~\cite{Bettencourt:2007}\\
\hline
\end{tabular}
\caption{{\bf Basic commodities. } Scaling of the total housing, electrical
    consumption and water consumption with population size. All exponents are
    compatible with a linear behaviour (within the $95\%$ confidence interval
    error bars).\label{tab:commodities}
}
\end{table*}

\subsection{Infrastructure}
\label{sub:infrastructure}

What about infrastructure, and the alledged economies of scale? Do we need to build less
roads, lay less cables for every individual in larger cities? This question can
be answered by looking at the scaling of the length of roads, cables, etc. in
cities: if the exponent is smaller than one, larger cities need less
infrastructure per capita.

Veregin and Tobler, using the 1980 US Census DIME files (a lot less convenient
to use than shapefiles!) showed that the number of street segments--the portion
of road between two intersections--scaled sublinearly with the size of urban
areas~\cite{Veregin:1997} (see Table~\ref{tab:infrastructure}). 

Arguably, the total length of the street network is more relevant to measure
costs in terms of infrastructure. In~\cite{Louf:2014_scaling}, we provide
evidence for the sublinear scaling of total street length with the population
size of urban areas (Table~\ref{tab:infrastructure}). 

Finally, Bettencourt et al. showed that the length of electric cables in German
cities scaled sublinearly with population size~\cite{Bettencourt:2007}. So far,
studies thus indicate that cities indeed realise some economies of scale.

\begin{table*}[!h]
    \centering
\begin{tabular}{|cccc|}
\hline
Quantity & Exponent & City definition & Study\\
\hline
Street segments & 0.83 & Morphological (US) & Veregin \& Tobler~\cite{Veregin:1997}\\
Street length & 0.86 & Morphological (US) & Louf \& Barthelemy~\cite{Louf:2014_scaling}\\
Electric cables length & 0.87 & Administrative (Germany) & Bettencourt et al.~\cite{Bettencourt:2007}\\
\hline
\end{tabular}
\caption{{\bf Infrastructure. } Scaling of the total number of street segments,
    the total length of roads and the total length of electrical cables of
    cities as a function of population. The three quantities exhibit a sublinear
    scaling behaviour, implying that larger cities need less infrastructure per
    capita, thereby realising economies of scale.\label{tab:infrastructure}
}
\end{table*}

\section{Summary}
\label{sec:summary}

The above review of the literature beggs several questions. 

First, most of the scaling exponents that are found in the literature (all but
linear scalings) are highly non-trivial, in the sense that their values seem 
somewhat arbitrary. We argued at the beginning of this Chapter that these
exponents where the signature of the processes happening within cities. But it
is not clear what mechanisms can lead to these values. In the following Chapter,
we will provide a model that reproduces the exponents observed on quantities
that are relatd to the mobility of individuals.

A second issue has to do with the fact that studies find different exponent for
the exact same quantities. The problem does not lie so much with the numerical
differences, but in the qualitative difference: some quantities are found to
scale sublinearly in a context, and superlinearly in another. For instance, the
$CO_2$ emissions scale differently with population size in different studies.
While studies focusing on Urban Areas or equivalent (in the US) find that
emissions scale superlinearly with population size~\cite{Louf:2014_scaling,
Oliveira:2014}, studies interested in Metropolitan Statistical Areas report a
sublinear scaling~\cite{Fragkias:2013}. This calls for an explanation that we
will sketch in Chapters~\ref{chap:scaling_model}
and~\ref{chap:scaling_implications}.
%
\chapter{From mobility patterns to scaling}
\label{chap:scaling_model}

\begin{flushright}{\slshape    
I remember my friend Johnny von Neumann used to say,\\
`with four parameters I can fit an elephant\\
and with five I can make him wiggle his trunk'} \\ \medskip
--- Enrico Fermi (quoted in~\cite{Dyson:2004})
\end{flushright}

\bigskip

A common trait shared by all complex systems -- including cities -- is the
existence of a large variety of processes occuring over a wide range of time and
spatial scales. The main obstacle to the understanding of these systems
therefore resides in uncovering the hierarchy of processes and in singling out
the few ones which govern their dynamics. Albeit difficult, the hierarchisation
of processes is of prime importance. A failure to do so leads to models which
are  either too complex to give any real insight into the phenomenon, or too
simple and abstract to have any resemblance with reality. As a matter of fact,
despite numerous
attempts~\cite{Fujita:1982,Makse:1995,Batty:2008,Frasco:2014,Bettencourt:2010,Bettencourt:2013},
a theoretical understanding of many observed empirical regularities in cities is
still missing.\\

Here we show that the spatial structure of the mobility pattern
controls the scaling behaviour of many quantities in urban systems. Indeed, cities are
not only defined by the spatial organisation of places fulfilling different
functions -- shops, places of residence, workplaces, etc. -- but also by the way
indivduals move among them. Understanding where people live, where and how they
travel within the city thus appears as a necessary step towards a scientific
theory of cities.\\

\section{A naive approach}
\label{sec:elementary_understanding_of_the_scaling_relationships}

We start by presenting some naive arguments to estimate the scaling exponents
for the area $A$, the total daily distance driven $L_{tot}$ and the total lane
miles $L_N$. Although these predictions turn out to be wrong, naive scalings are
useful as a first approach to the problem as they allow us understand how the
different quantities relate to one another.

\subsection{Surface area}

We first would like to estimate the dependence of the area
$A$ of a city on its population $P$ -- a long standing problem in the
field~\cite{Stewart:1947, Batty:2011}.

\paragraph{Naive argument.}  A first crude approach is to assume
that cities evolve in such a way that their population density $\rho = P/A$
remains constant. This assumption immediately implies that the area should
scale linearly with population
\begin{equation} 
    A \sim \lambda^2\, P 
    \label{eq:area_naive} 
\end{equation}

where $\lambda^2$ is the average surface occupied by each individual (the
assumption of a constant density is then equivalent to the one of a constant
average surface per capita).

\begin{figure}[!h]
    \centering
    \includegraphics[width=\textwidth]{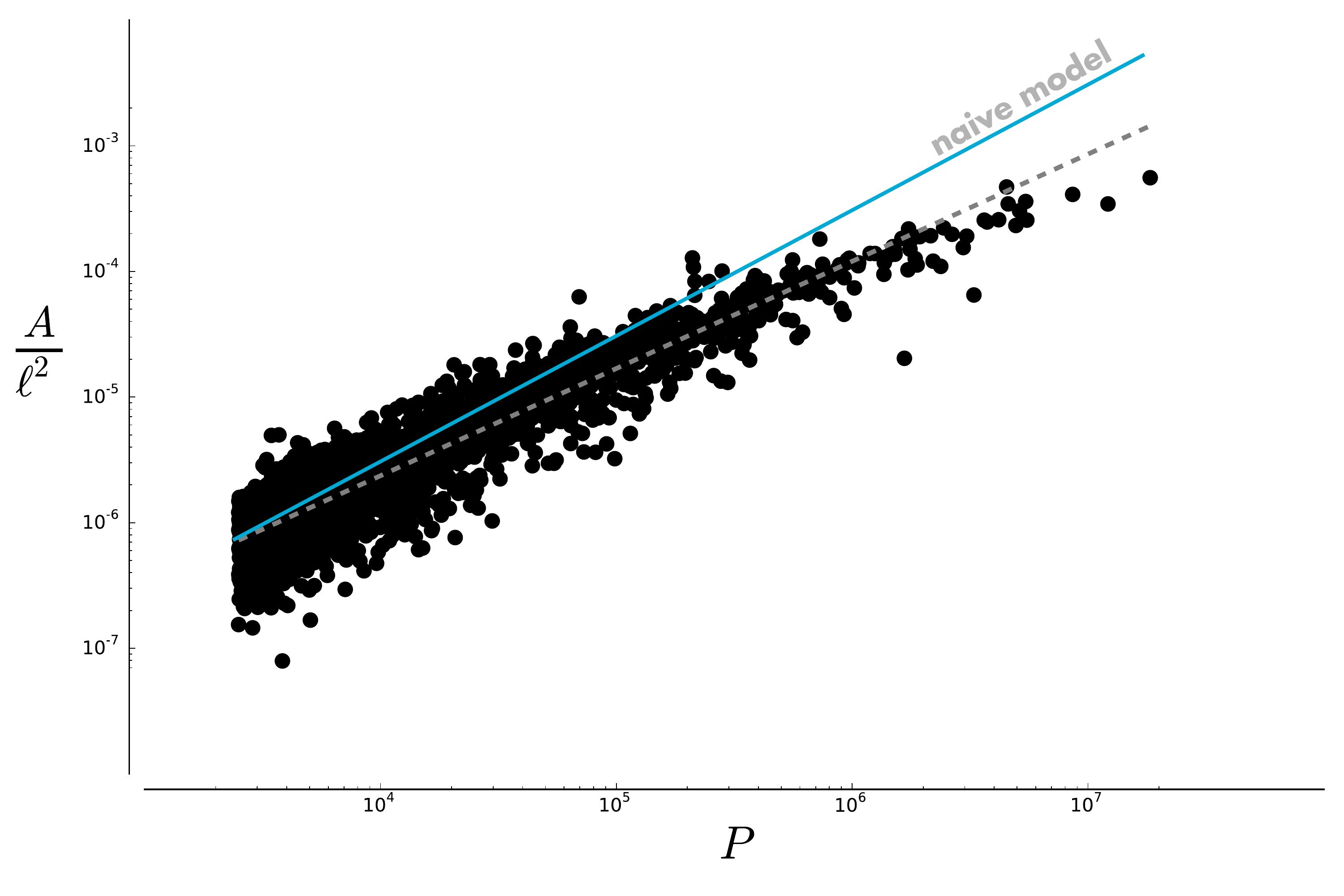}
    \caption{{\bf Spatial footprint.} Scaling of the surface area of US urban areas with population size,
        and what would be expected with a naive model (blue solid line).
    A fit assuming a powerlaw dependence (dashed line) gives an exponent
    $\beta_A = 0.85 \pm 0.01\,(r^2 = 0.93)$.\label{fig:scaling_area}}
\end{figure}

\paragraph{Reality.} The naive argument does not compare well with reality. We
plot the scaling of the surface area versus population for US Urban Areas on
Figure~\ref{fig:scaling_area}. A fit assuming a power-law dependence gives an exponent

\graffito{All $\pm$ intervals are $95\%$ confidence intervals.}
\begin{equation}
    \boxed{\beta_A = 0.85 \pm 0.01\;(r^2 = 0.93)}
\end{equation}

A result which agrees with previous measurements made on morphologically defined
cities (see~\cite{Batty:2011} or Chapter~\ref{chap:scaling_introduction}). This means that the average surface occupied by each individual decreases with
city size. Or equivalently, that the population density increases with city size.
The prediction given by the naive model is therefore quantitatively -- and
worse, qualitatively -- different from the behaviour observed empirically.

\subsection{Total length of road}
\label{sub:total_length_of_road}

\paragraph{Naive model.} We would now like to estimate the total length $L_N$ of all the roads within a
city. If we consider that the network formed by streets is such that all the
nodes (intersections) are connected to their closest neighbour, the typical
length of a road segment is given by

\begin{equation}
    \ell_R \sim \sqrt{\frac{A}{N}}
\end{equation}

where $N$ is the number of intersections~\cite{Barthelemy:2008}. Previous studies of road networks in
different regions, and over extended time
periods~\cite{Strano:2012,Barthelemy:2013}, have shown that the number of
intersections is proportional to the population size. Therefore, the typical
length of a road segment (between two intersections) varies with the population
size $P$ as

\begin{equation} 
    \ell_R \sim \sqrt{\frac{A}{P}} 
    \label{eq:length_nodes}
\end{equation}

and the total length of the network $L_N \sim P\ell_R$ should then scale as

\begin{equation} 
    \frac{L_N}{\sqrt{A}} \sim \sqrt{P} 
\end{equation}

Using the naive scaling for the dependence of $A$ on population size given
previously in Eq.~\ref{eq:area_naive} we finally get 

\begin{equation} 
    L_N \sim P
\end{equation}

\begin{figure}
    \centering
    \includegraphics[width=\textwidth]{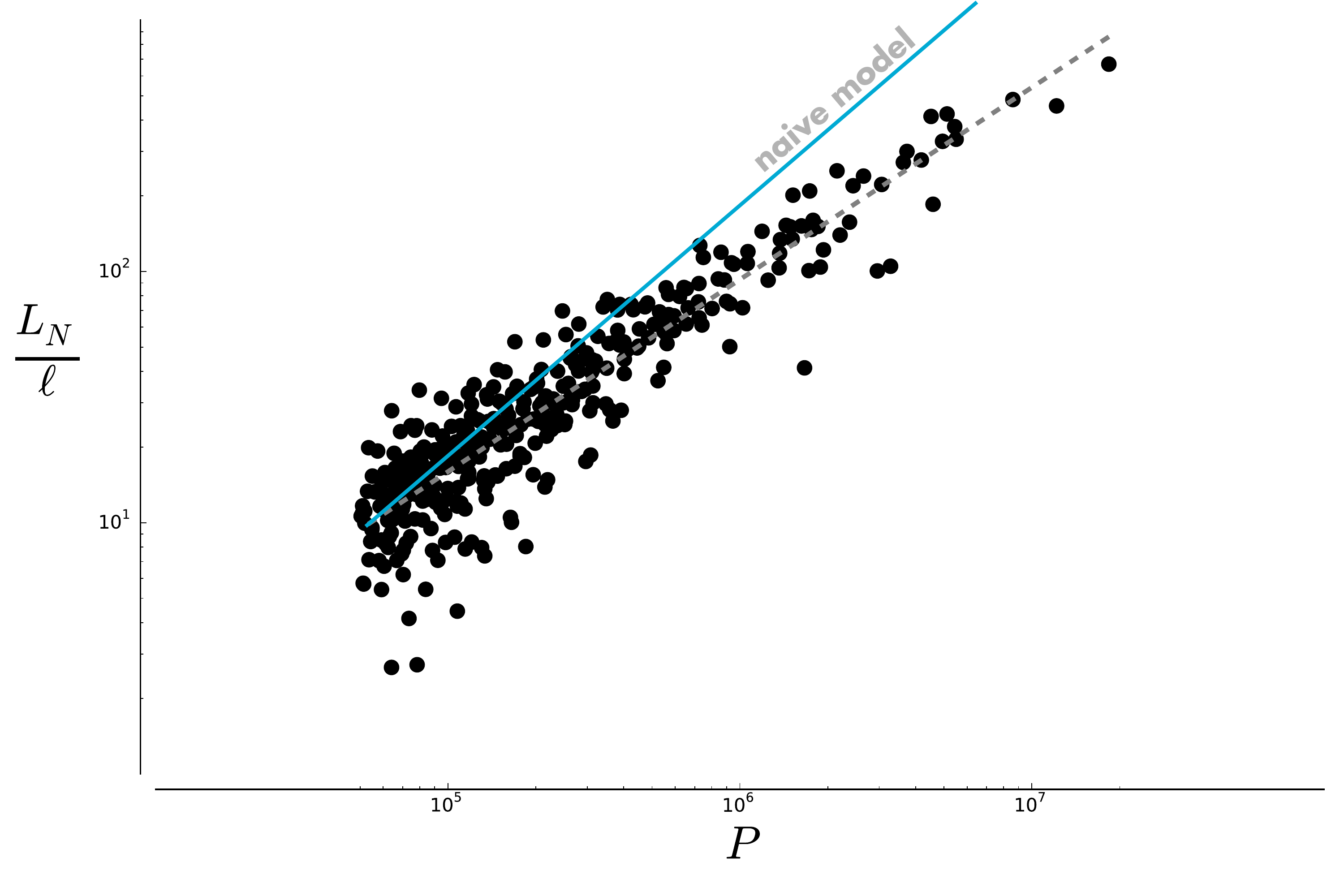}
    \caption{{\bf Length of roads.} Scaling of the total length of roads in US Urban Areas versu the
    total population. A fit assuming a powerlaw dependence (dashed grey line)
gives an exponent $\beta_R \sim 0.765 \pm 0.033$ ($r^2=0.92$). The behaviour is
qualitatively different from what would be expected with a naive model (solid
blue line).\label{fig:scaling_lanemiles}}
\end{figure}

\paragraph{Reality.} Again, the naive argument does not compare well with
reality. We fit the data for US Urban Areas (see
Figure~\ref{fig:scaling_lanemiles}) assuming a powerlaw dependence and find an
exponent

\begin{equation}
    \boxed{\beta_R \sim 0.765 \pm 0.033\;(r^2 = 0.92)}
\end{equation}

Note that the relation between the length and the number of nodes given by
Eq.~\ref{eq:length_nodes}, as well as the relation between number of
intersections and population, have been verified independently in the
literature. The observed discrepancy on the exponent of $L_N$ is therefore
certainly due to the scaling of the surface area.

\subsection{Total commuting distance}
\label{sec:total_length_driven}

The total commuting distance $L_{tot}$ is determined by two different
constraints. First the individual constraint: individuals make the decision
about where they are going to live and work; they have their own behaviour and
limitations. However, the individuals' choices are also limited by the city
structure itself, that is by the respective distributions of jobs and residences across the
city.

\subsubsection{Influence of the individual constraint} 

The first constraint on the commuting distance comes from individuals' limitations and
behaviour. We make here the simple assumption that individuals choose their
residence and work place such that their total commuting distance is fixed (or
at least, is smaller than a certain value) and equal on average to $\ell_C$. In that
case, we would simply have

\begin{equation} 
    \frac{L_{tot}}{P} \sim \text{constant} = \ell_c 
    \label{eq:assum}
\end{equation}

\begin{figure}[!h]
    \includegraphics[width=1.0\textwidth]{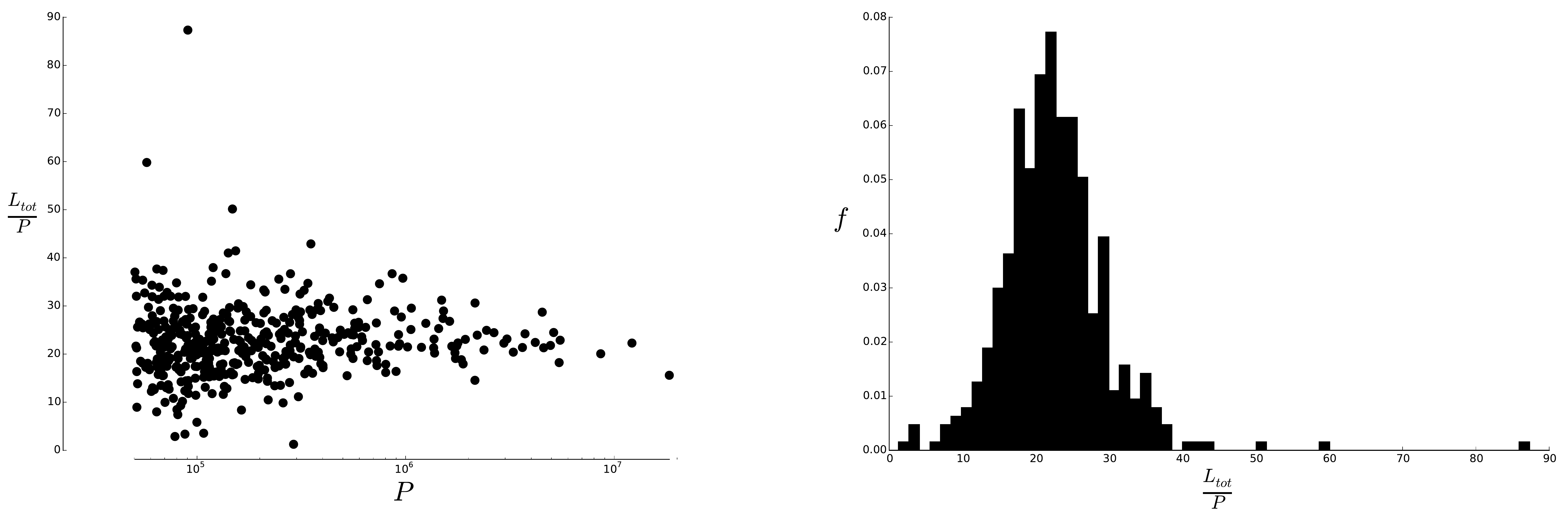}
    \caption{{\bf Commuting distance \& individual choice.} Constant daily driven distance per capita. (a) daily total driven
        distance per capita as a function of population for 441 urbanised area
        in the US in 2010. The data shown in the plot are compatible with a
    population-independent behaviour. (b) Histogram of the daily total driven
distance per capita for the same cities. The average daily driven distance is $23$ miles, and the standard deviation $7$ miles.}
\label{fig:LtotoverP} 
\end{figure}

(by constant, we mean independent from the population size of the city). As
surprising as it may seem, the data show that $L_{tot}/P$ can indeed be considered
 independent from $P$ (with a value of approximately $23$ miles for
the US, see Figure~\ref{fig:LtotoverP}), in agreement with the individual
constraint assumption (Eq. \ref{eq:assum}). This finding is also in agreement
with the results drawn from census data in Germany by~\cite{Wilkerson:2014}.
This does not mean, of course, that the distance driven is the same for every
city. As one can see on Figure~\ref{fig:LtotoverP}, the fluctuations are quite
important between cities.  

\subsubsection{Influence of the city structure}
\label{sub:influence_of_the_city_structure}

\begin{figure}[!h]
    \centering
    \includegraphics[width=1\textwidth]{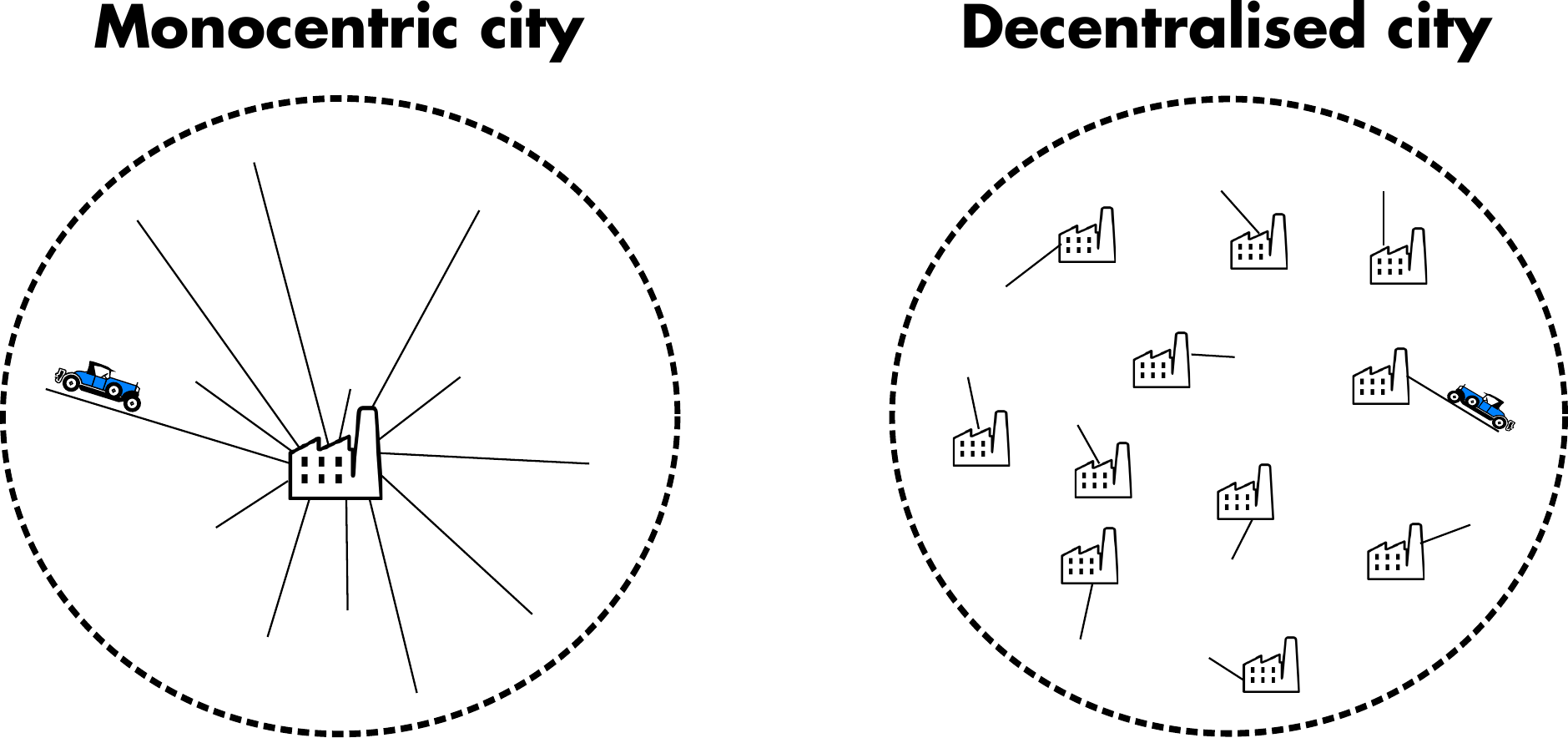}
    \caption{{\bf Limiting cases.} Representation of the monocentric city (left) and the totally
    decentralised city (right), two extreme models for the shape of mobility
patterns.\label{fig:monocentric_decentralised}}
\end{figure}

The easiest way to understand the influence of the city constraints is to
consider two limiting cases: the totally centralised
(monocentric) city where everyone goes to work to a single center, and the
totally decentralised city where everyone goes to work to the nearest
location (see Figure~\ref{fig:monocentric_decentralised})~\cite{Samaniego:2008}.\\

\paragraph{Monocentric.} If we first assume that the city is monocentric, individuals are all commuting
to the same center and the typical commuting distance $\ell^m_c$ is controlled
by the typical size of the city of order $\sqrt{A}$, so that

\begin{equation} 
    \frac{L_{tot}^{m}}{\sqrt{A}} \sim P 
\end{equation}

\paragraph{Decentralised.} On the other hand, if we assume that the city is completely decentralised, the
typical commuting distance is of order the nearest neighbour distance
$\sqrt{A}/\sqrt{P}$, and we obtain

\begin{equation} 
    \frac{L_{tot}^{d}}{\sqrt{A}} \sim \sqrt{P} 
\end{equation}

\begin{figure}
    \centering
    \includegraphics[width=0.9\textwidth]{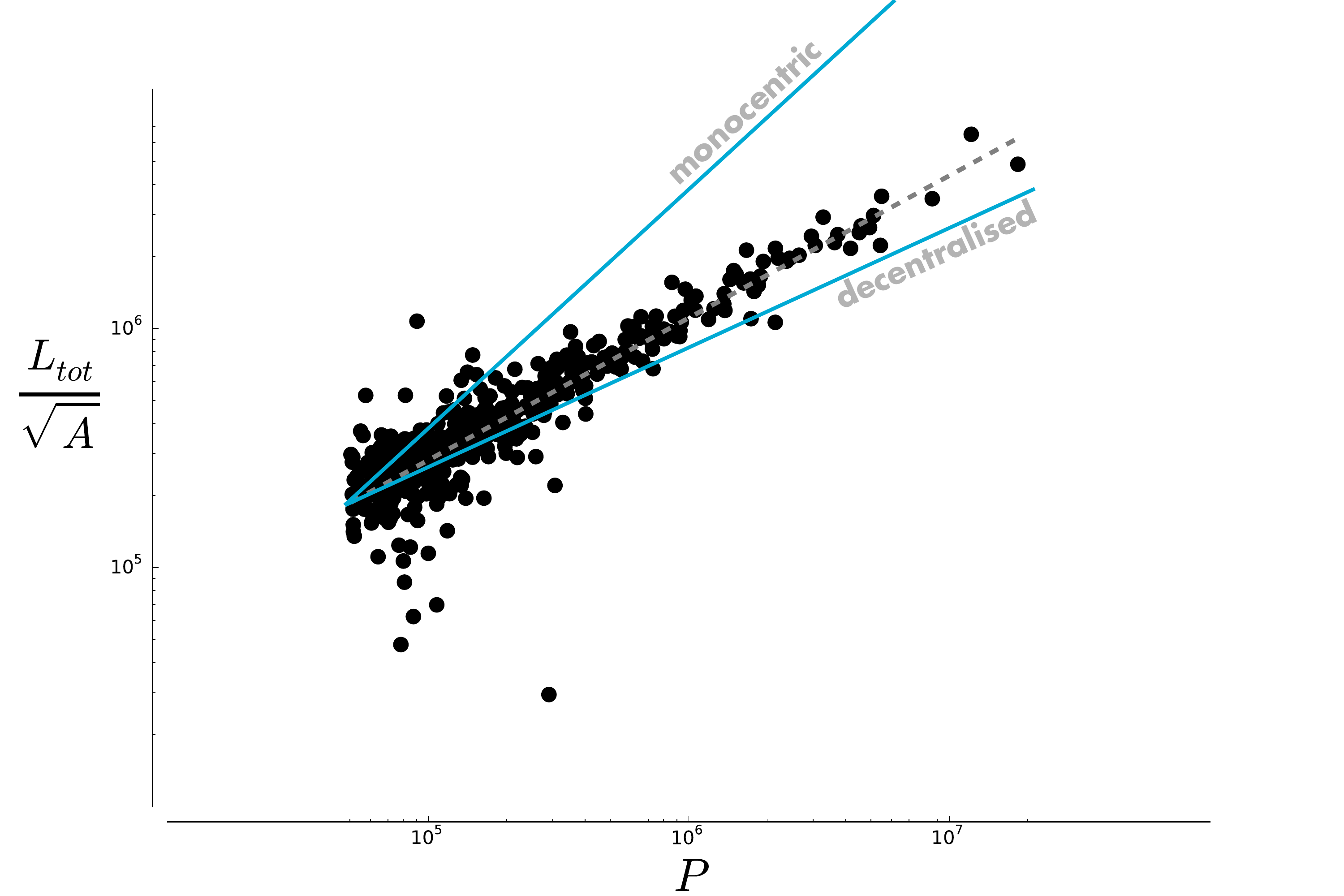}
    \caption{{\bf Commuting distance \& city structure.} Scaling of the total yearly commuted distance normalised by the
    city's surface area with population size for US Urban Areas. The blue lines
show the behaviours that would be expected for a monocentric and a totally
decentralised city. The dashed line represents the fit assuming a powerlaw
dependence, which yields an exponent $\beta =  0.595 \pm 0.026\, (r^2 =
0.90)$.\label{fig:scaling_Ltot_norm}}
    
\end{figure}

\paragraph{Reality. } The scaling of the total driven distance for Urban Areas
(morphological definition) is shown on
Figure~\ref{fig:scaling_Ltot_norm}, and the exponent sits between the ones of the
monocentric and decentralised cities

\begin{equation*}
    \boxed{\beta_L =  0.595 \pm 0.026\; (r^2 = 0.90)}
\end{equation*}

This comes as another evidence -- different from that presented in
Chapter~\ref{chap:monocentric_introduction} -- that cities do not have a
strictly monocentric structure. This result casts some further doubts about the
model by Bettencourt~\cite{Bettencourt:2013} which implicitely assumes that
cities are monocentric.\\

So far, so good. But how can we understand the non-trivial exponent that is observed? This is
where the limiting case are helpful: if the exponent sits between the ones that
would be obtained in a monocentric or decentralised city, surely, cities must
adopt an intermediate structure. 

\begin{figure}[!h]
    \centering
    \includegraphics[width=0.4\textwidth]{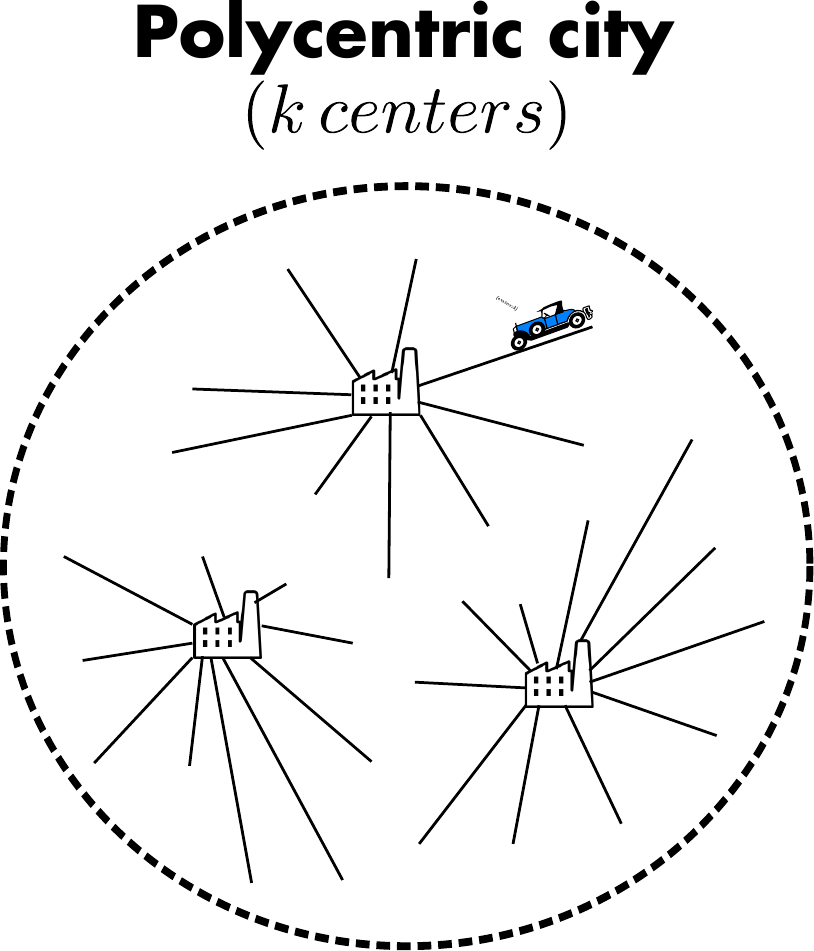}
    \caption{{\bf Polycentric structure.} City with a polycentric structure, intermediate between the
    monocentric and totally decentralised situations. \label{fig:polycentric}}
\end{figure}

One candidate stands out: the polycentric city (see
Figure~\ref{fig:polycentric}). Let us thus consider a polycentric city with $k$
employment centers. The typical distance commuted by individuals is then given by

\begin{equation}
    \ell_c \sim \sqrt{\frac{A}{k}}
\end{equation}

So that

\begin{equation}
    \frac{L_{tot}}{\sqrt{A}} = \frac{P}{\sqrt{k}}
\end{equation}

Therefore if, as we showed in the previous part,  the number of centers
increases sublinearly with population, we would have a scaling of the form
$L_{tot}/\sqrt{A}\sim P^{\,\beta_L}$ where $\beta_L \in [1/2,1]$. The previous
expression is consistent with that of $A/\lambda^2$ and $L_{tot}/P$ if

\begin{equation} 
    \beta_L = 1-\frac{\beta_A}{2} 
    \label{eq:consis} 
\end{equation}

which is indeed what we observe empirically (up to error bars). We conclude from
this preliminary empirical analysis that, in order to compute the various
exponents, we need to better describe the structure of commuting patterns. In
other words, we need to find a description of cities that goes beyond the naive
monocentric or totally decentralized views, and which accounts for the observed
sub-linear scaling of the surface area $A$.

\begin{table*}[!h]
    \centering
\begin{tabular}{|c|c|c|l|}
\hline
Quantity & Naive exponent &  Measured value\\
\hline
$A $ & $1$ & $0.85\; (r^2=0.93)$\\
\hline
$L_N / \sqrt{A}$ & $0.5$ & $0.42\; (r^2=0.83)$\\
$L_N $ & $1$ & $0.89\;(r^2=0.77)$\\ 
\hline
$L_{tot} / \sqrt{A}$ &  $\left\{0.5,1\right\}$ & $0.60\; (r^2=0.90)$\\
$L_{tot} /P$ &  $1$ & $0.03\; (r^2=0.04)$\\
\hline
\end{tabular}
\caption{{\bf Naive exponents and measured values.} This table displays the value of the exponent governing the behavior with the population $P$ obtained by naive arguments and the value obtained from empirical data. The discrepancies reveal the failure of the naive scaling arguments and the necessity to go further and model mobility patterns.}
\label{table:naive}
\end{table*}

\section{Beyond naive scalings: modeling the mobility patterns}

The previous results, in particular the behaviour of the total commuting length
with population, hint at the necessity to better describe the structure of the
mobility patterns (Table~\ref{table:naive}). This is exactly what the model presented in the previous
chapter does. 

Using the relation that we derived for the number of centers, we will see
how we can understand the values of the exponents presented earlier in this
chapter. We will also see how the model allows us to understand the scaling of
other quantities, namely the total time spent in traffic and the total $CO_2$
emissions due to transportation.

\subsection{Area}

According to the model introduced in Chapter~\ref{chap:monocentric_model}, the number of centers is a function of population and the area

\begin{equation}
    k = F\left(A,P\right)
\end{equation}

and we need an additional equation in order to get a closed system. Here we
focus on the area and its evolution with the population size, which reflects the
growth process of the city. 

In the following, we will investigate two different
approaches. It is worth noting that both approaches give results in qualitative
agreement, showing that some stylized facts ---such as super- or sublinearity---
are very robust.\\ 

\paragraph{Fitting procedure.}

In the absence of knowledge of the processes responsible for urban sprawl, we
can assume that the area behaves as 

\begin{equation}
    A \sim P^{\,a}
    \label{eq:fit}
\end{equation}

where $a$ is the exponent to be determined by fitting data. The empirical
value for the exponent for the US data is $a\simeq 0.85$. Once this exponent is
given we can then compute the various exponent for the quantities of interest.
We get for the number of centers $k$

\begin{equation}
    k \sim P^{\frac{\mu + a/2}{\mu+1}}
\end{equation}

which is sublinear as long as $a<2$, in agreement with the empirical results for
US cities. As we will see, this approach yields the same qualitative behaviours
as those predicted with the method of the next section. In other words, even if
the main mechanism behind urban sprawl is not congestion, the conclusions of
this paper are not affected as long as the area scales \emph{sublinearly} with
population.\\

\paragraph{Coherent growth.}

Let us now assume that the scaling of $A$ with population is determined by the
number of activity centers and the constant commuting length of individuals.
This means that the growth of the area is controlled by the appearance of new
activity centers. 

If we assume that a city is organized around $k$ activity
centers and that the attraction basin of each of these centers are spatially
separated~\cite{Louf:2013_polycentric} (See on Figure~\ref{fig:polycentric}), we then have  $A \sim k\, A_1$ where $A_1$ is the
area of each subcenter's attraction basin. This area $A_1$ is related to the
average individual commuting distance by $\sqrt{A_1} \sim L_{tot} / P$, and we
obtain

\begin{equation}
    A \sim k\,  \left( \frac{L_{tot}}{P} \right)^2 = k\, \ell_c^2
    \label{eq:area_poly}
\end{equation}

This leads to expression for the number of centers

\begin{equation}
    k \sim P^{\frac{2 \mu}{2\mu+1}}
\end{equation}

which is always smaller than $1$, also in agreement with the empirical results
for US cities. We can now also compute the scaling of the surface area

\begin{equation}
    \frac{A}{\ell_c^2} \sim \left( \frac{P}{c} \right)^{\frac{2 \mu}{2\mu+1}}
\end{equation}

We further assume that $L_{tot} / P$ is a fraction of the longest possible
journey $\ell$ individuals can afford, that is to say 

\begin{equation}
    \ell_c \sim \ell
\end{equation}

It is important to note that if $\ell_c$ is independent from $\ell$, the
quantitative predictions of our model would still hold. 

The final expression for the area is then here given by

\begin{equation}
    \frac{A}{\ell^2} \sim \left( \frac{P}{c} \right)^{\,2\,\delta}
    \label{eq:area}
\end{equation}

where $\delta=\frac{\mu}{2\mu+1}$. The exponent $\delta$ is smaller than $1/2$
whatever $\mu\geq 0$, which implies that the surface area of cities increases
\emph{sublinearly} with population. In other words, the density of cities
\emph{increases}  with population. This prediction is verified with data about
land area of urbanized areas in the US (Figure~\ref{fig:scaling_area}). We find
$\beta_A = 0.85 \pm 0.01\;$ which is not too far from the
theoretical value $2\delta_{th} = 0.64 \pm 0.12$, equal to
$\alpha$ in this case.\\

Because the area of a city results from centuries of evolution, we do
not a priori expect our model -- where individual vehicles are assumed to be the
only vector of mobility -- to give a prediction valid for all countries and all
times. Nevertheless, these results give us reasons to believe that the spatial
structure of the journey-to-work commuting might be the dominant factor
in the dependence of land area on population. In the following, we will use the
above numerical value to compute other scaling exponents.

\subsection{Total commuting distance}

Using Eq.~\ref{eq:assum} and Eq.~\ref{eq:area} we are now able to compute $L_{tot}/\sqrt{A}$

\begin{equation}
    \frac{L_{tot}}{\sqrt{A}} = P\; \left(\frac{P}{c}\right)^{-\delta}
    \label{eq:travelled_length}
\end{equation}

We plot $L_{tot} / \sqrt{A}$ for urbanized areas in the US on
Figure~\ref{fig:scaling_Ltot_norm}, and one
can verify in Table~\ref{table:results} that the exponent predicted from the previously measured
value of $\alpha$ agrees well with the exponent measured on the data.

\subsection{Total length of roads}

If we use the previously derived expression for the area $A$, we find

\begin{equation}
    L_N \sim \ell \; \sqrt{P}\; \left(\frac{P}{c}\right)^{\,\delta}
\end{equation}

The quantity $\delta$ is less than $1/2$, which implies that $L_N$ scales
\emph{sublinearly} with the city's population size. In other words, larger
cities need less roads per capita than smaller ones: we recover the fact that
 the agglomeration of people in urban centers involves economies of scale for
infrastructures.

\subsection{Total delay due to congestion}

\begin{figure}
    \centering
    \includegraphics[width=\textwidth]{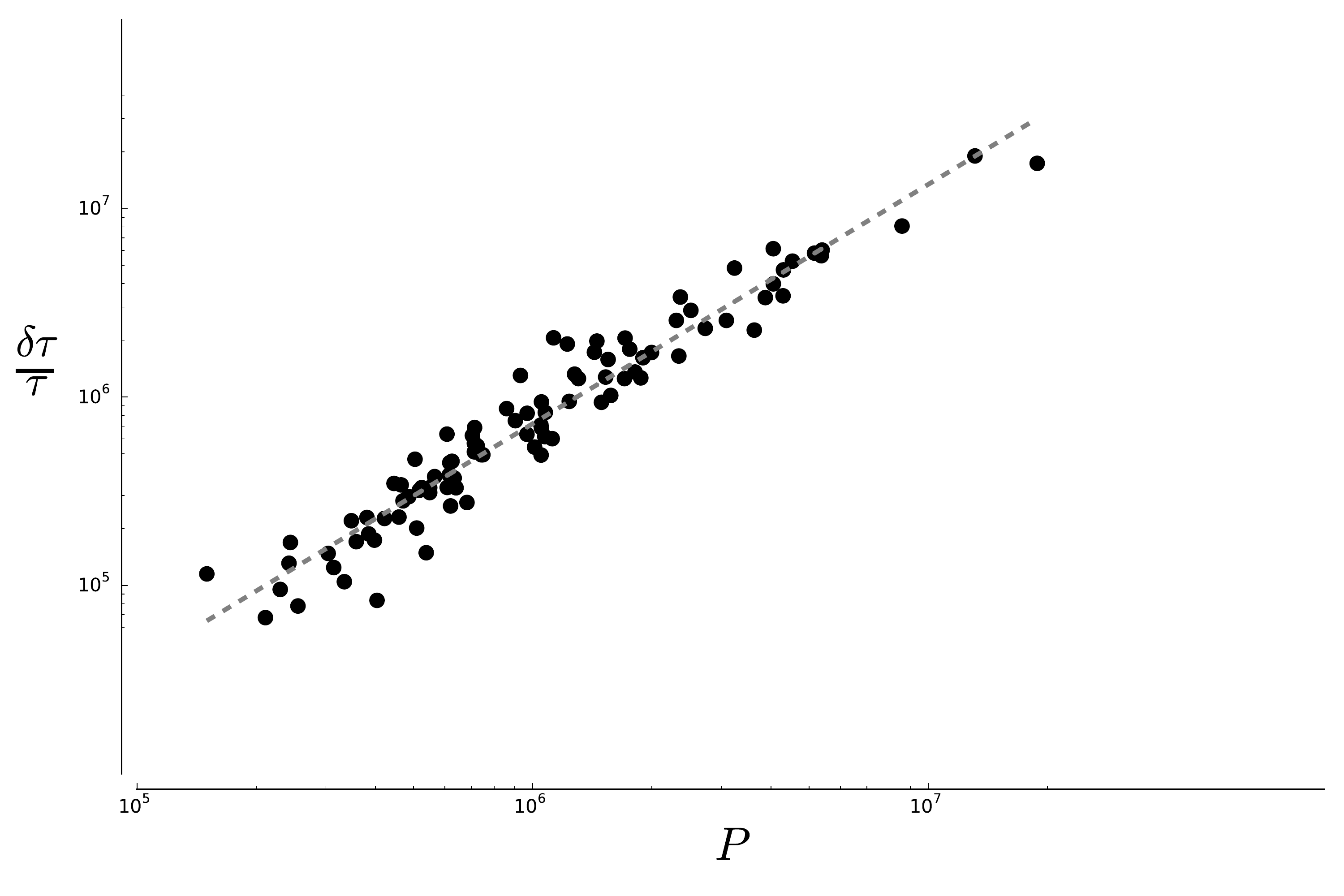}
    \caption{{\bf Congestion and delay.} Scaling of the total delay due to congestion of US urban areas with
    population size. A fit assuming a powerlaw dependence of the total delay on
population size yields an exponent $\beta_D = 1.270 \pm 0.067\;(r^2=0.97)$.\label{fig:scaling_delay}}
\end{figure}

Unfortunately, the agglomeration of activities in cities does not only generate economies.
Congestion, for instance, is a major diseconomy associated with the
concentration of people in a given area. A simple way to quantify the
impairement caused by traffic congestion is through the total delay it
generates. If we make the first order approximation that the average free-flow
speed $v$ is the same for everyone, the total delay due to congestion is given
--according to our model-- by

\begin{equation} 
    \delta \tau = \frac{1}{v} \sum_{i,j} d_{ij} \left(\frac{T_j}{c} \right)^\mu 
\end{equation}

If we assume that all the centers share the same number of commuters -- a
reasonable assumption within the model presented in
Chapter~\ref{chap:monocentric_model}~\cite{Louf:2013_polycentric} -- we obtain

\begin{equation} 
    \delta \tau \sim \frac{L_{tot}}{v} \left( \frac{P}{k}
\right)^{\mu} 
\end{equation} 

which, using the expressions for $L_{tot}$ and $A$
given in Eq.~\ref{eq:travelled_length} and Eq.~\ref{eq:area} respectively, gives

\begin{equation} 
    \delta \tau \sim \frac{\ell\; P}{v}\;\left(\frac{P}{c}\right)^{\delta} 
\end{equation}

The total commuting time corresponding to the same distance but without
congestion scales as $\tau_0\sim L_{tot}$ and thus less rapidly than the total
delay which scales \emph{super-linearly} with population (even when
polycentricity is taken into account). This means that, for the largest cities,
delays due to congestion actually dominate the time spent in traffic, and that
economical losses \emph{per capita} due to the time lost in congestion --and the
corresponding strain on people's life-- increase with the size of the city. 

The prediction $1+\delta = 1.32$ agrees well with the empirical measure (see
Table~\ref{table:results} and Figure~\ref{fig:scaling_delay})

\begin{equation}
    \boxed{\beta_D = 1.270 \pm 0.067\;(r^2 = 0.97)}
\end{equation}

\subsection{Transport related $CO_2$ emissions}

Another diseconomy associated with congestion is the quantity of $CO_2$ emitted
by cars and the gasoline consumed by motor vehicles. This amount not only
depends on the distance that has been driven, but also on the traffic during the
journey. It indeed turns out that for the same length driven, a car burns more
oil when the traffic is heavy than when the road is clear.  Within our model,
the presence of traffic is seen in the time spent to cover a given distance, and
we write that the quantity of $CO_2$ emitted by a vehicle is proportional to the
total time spent in traffic, leading to

\begin{equation}
    Q_{CO_2}  = q \sum_{i,j} d_{ij} \left[ 1+ \left( \frac{T_j}{c} \right)^\mu \right]
\end{equation}

where $q$ is the average quantity of $CO_2$ produced per unit time. In the
polycentric case with $k=k(P)$ subcenters, the typical trip length
$\overline{d_{ij}}$ is given by $\sqrt{A/k}$ and we obtain

\begin{equation}
    Q_{CO_2} = q\, \ell\, P \left[ 1 + \left(\frac{P}{c}\right)^{\delta} \right]
\end{equation}

The first term in brackets is a constant, and the quantity of $CO_2$ is thus
dominated by congestion effects at large populations

\begin{equation}
    Q_{CO_2} \sim q\; \ell\; P \left(\frac{P}{c}\right)^{\delta}
\end{equation}

and the total daily transport-related $CO_2$ emission per capita thus scales as 

\begin{equation}
    \frac{Q_{CO_2}}{P} \propto  q\ell \left(\frac{P}{c}\right)^{\delta}
\end{equation}

\begin{figure}
    \includegraphics[width=0.9\linewidth]{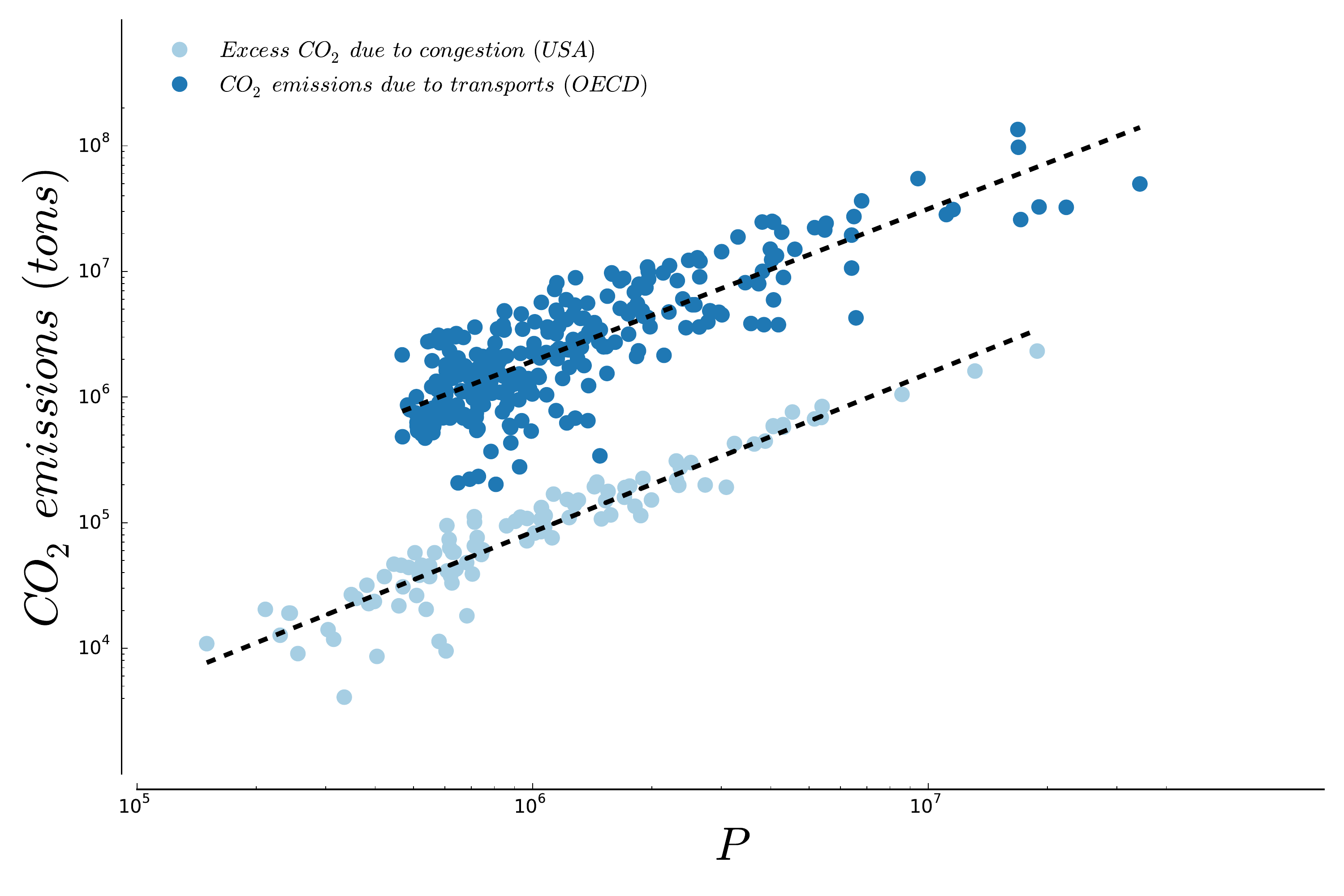}
    \caption{{\bf Congestion and $CO_2$ emissions.} Variation of $CO_2$
emissions due to transport with city size. In blue, excess $CO_2$ (in tons) due
to congestion, as given by the Urban Mobility Report (2010) for 101 metropolitan
areas in the US. In green, we show the estimated $CO_2$
emissions (in tons) due to transports, as given by the OECD for 268 metropolitan
areas in 28 different countries. The dashed yellow lines represent the least-square fit
assuming a power-law dependency with multiplicative noise, which gives
respectively $Q_{CO_2} \sim P^{1.262 \pm 0.089} (r^2=0.94)$ for the US data and
$Q_{CO_2} \sim P^{1.212 \pm 0.098} (r^2=0.83)$ for the OECD data.
\label{fig:scaling_co2}} 
\end{figure}

The quantity of $CO_2$ emitted per capita in cities thus increases with the size
of the city, a consequence of congestion. This prediction agrees with the
exponent we measure (Figurere~\ref{fig:scaling_co2}) on data gathered for US and OECD cities (see
Table~\ref{table:results})

\begin{equation}
    \boxed{\beta_C = 1.262 \pm 0.089\;(r^2=0.94)}
\end{equation}

\section{Discussion}

\subsection{Travel-time budget and congestion}

The total commuting time $T$ can be written as

\begin{equation}
    T = \tau_0 + \delta\,\tau
\end{equation}

where $\tau_0 = L_{tot} / v \sim P$ is the free-flow commuting time and $\delta
\tau \sim P^{1+\delta}$ the excess commuting time computed above. The first
thing we notice when looking at the respective population dependence of both
quantities, is that, in large cities, the total commuting time is dominated by
the time spent in congestion. Indeed, we have

\begin{equation}
    \displaystyle \frac{T}{\delta_\tau} \xrightarrow[\:P \gg 1\:]{} 1
\end{equation}

Which agrees with one's (at least our) experience of driving in large cities.

The second remark is linked to a long-standing belief in the study of urban
systems that individuals possess a constant travel-time
budget~\cite{Zahavi:1974}. We can easily see, however, that this hypothesis is
wrong. Indeed, in the limit of large cities, the individual commuting time is
given by

\begin{equation}
    \frac{\delta_\tau}{P} \sim P^{\,\delta}
\end{equation}

In other words, the \emph{individual commuting time increases with the size of the
city}. Note that not only is this a consequence of the model, but also of the
data analysis (see Figure~\ref{fig:scaling_delay}). The constant travel-time
budget hypothesis is thus refuted. The reason for the discrepancy between
previous measures and our results comes from the fact that these studies
considered averages over large regions, rather than averages at the city
level.

\subsection{Newman \& Kenworthy}

\begin{figure}
    \centering
    \includegraphics[width=\textwidth]{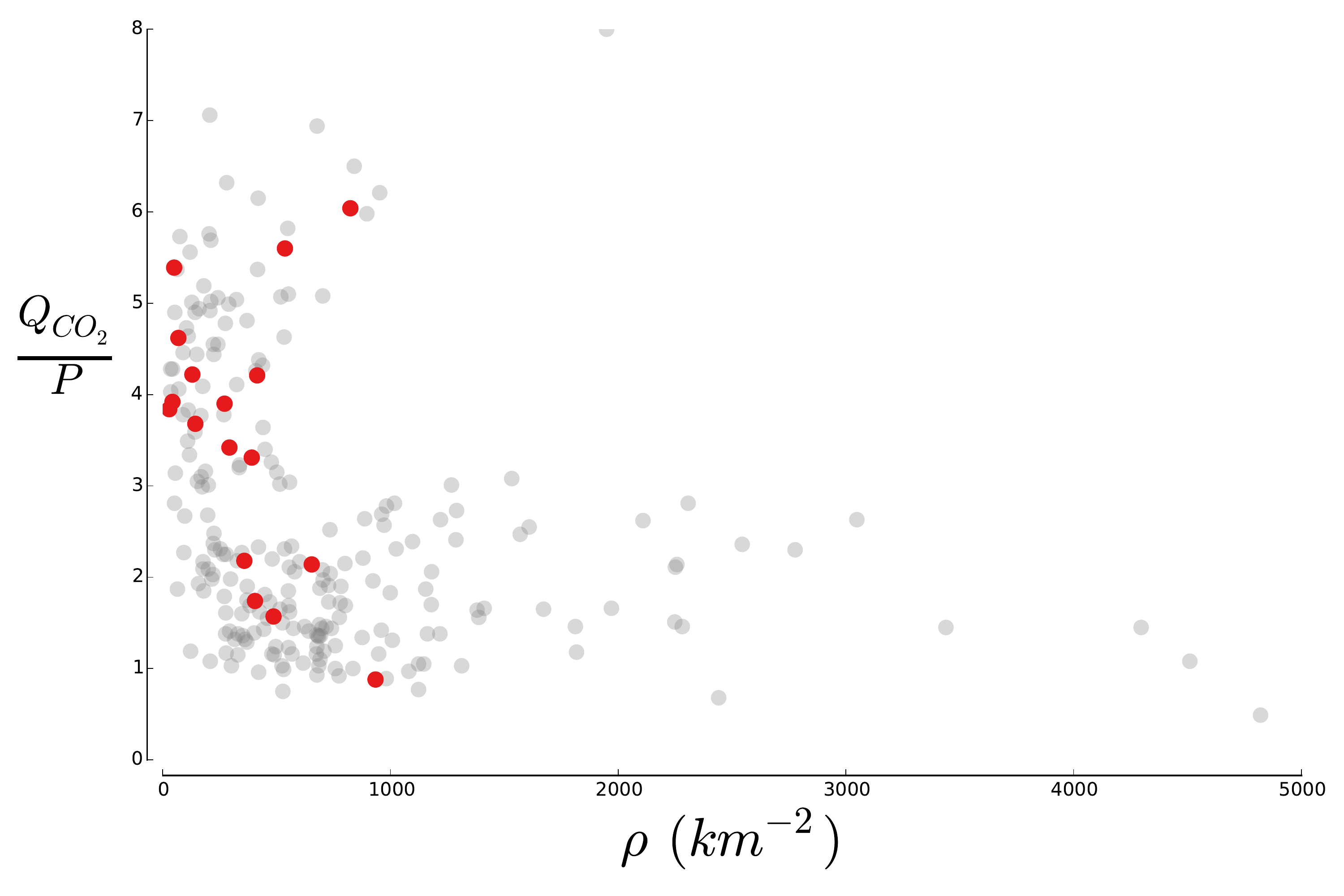}
    \caption{{\bf Newman \& Kenworthy.} Per capita $CO_2$ emissions versus the population density of cities
    belonging to OECD countries. The cities also present in the Newman \&
    Kenworthy dataset are represented in red. This curve casts serious doubt on
the fact that energy consumption is a simple funtion of
density.\label{fig:newman_kenworthy}}
\end{figure}

The consumption of gasoline is proportional to the emission of $CO_2$ and the time spent driving.
The total daily gasoline consumption is thus given by

\begin{equation} 
    Q_{gas} \sim q\; \ell\; P \left(\frac{P}{c}\right)^{\delta}
\end{equation}

where $q$ is the average quantity of gasoline needed per unit time. From this
expression, we see that the total daily gasoline consumption per capita scales
as

\begin{equation} 
    \frac{Q_{gas}}{P}\sim \ell\, \sqrt{\frac{P}{\rho}} = \ell
\sqrt{A} 
    \label{eq:nk_area}
\end{equation}

and is therefore not a simple function of population density, in contrast with
what was suggested by the seminal paper of Newman and
Kenworthy~\cite{Newman:1989}. We plot on Figure~\ref{fig:newman_kenworthy},  the average individual $CO_2$ emissions (used as a proxy for gasoline
consumption) as a function of the density for OECD cities. The points
corresponding to cities that were in the original study~\cite{Newman:1989} are
highlighted. The relation is a lot less clear than the one presented originally.  

\begin{figure}
    \centering
    \includegraphics[width=\textwidth]{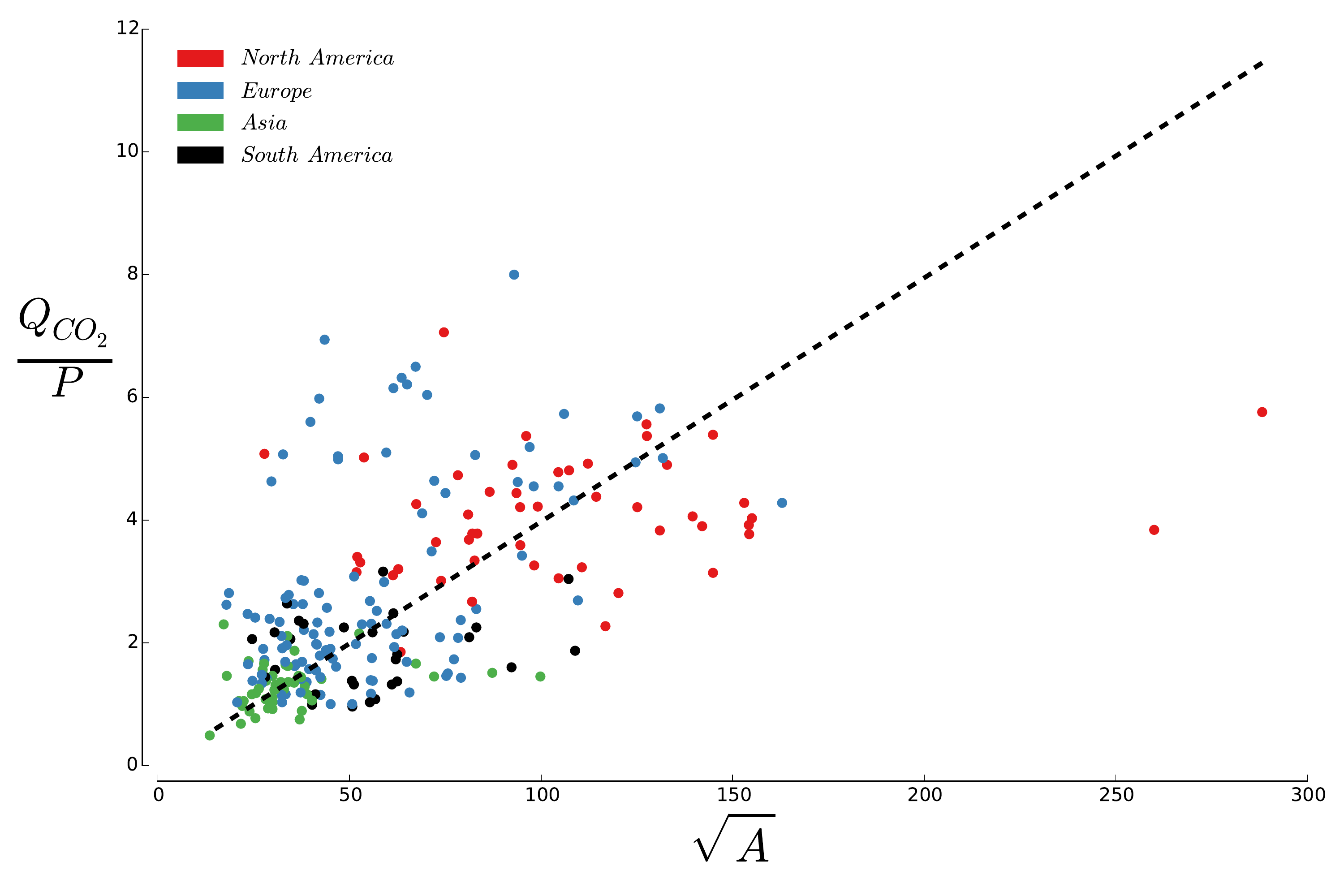}
    \caption{{\bf Newman \& Kenworthy revisited.} Per capita $CO_2$ emissions versus $\sqrt{A}$ for cities of
    countries that belong to the OECD. The dashed line represents the obtained
linear fit, as predicted by Eq.~\ref{eq:nk_area} ($r^2=0.55$). The agreement is
poor, which may be due to the fact that cities all belong to different systems
of cities (and thus have a different prefactor).\label{fig:nk_model}}
\end{figure}

We then plot the same quantity as a function of $\sqrt{A}$, the prediction given
by Eq.~\ref{eq:nk_area}, on Figure~\ref{fig:nk_model}. As one can see, the
prediction is far from perfectly followed. If anything, this figure, combined to
Figure~\ref{fig:newman_kenworthy} show that the debate, in the absence of a
clear-cut conclusion, is not over. At this stage, more data about gasoline
consumption -- preferably for cities belonging to the same system of cities --
is needed to explore this prediction.

\subsection{Monocentric versus polycentric}

Although polycentricity emerges naturally from our model as a result of
congestion, many circumstances can prevent or foster the appearance of new
activity centers in a city. There are many debates as to whether policies should
favour polycentric or monocentric developement of cities. Most of them are
based on ideologies and opinions about how cities should be, very few are based
on a quantitative understanding of the city as a complex system. Although this
only represents a small part of the debate, our model allows to quantify the
effect of polycentricity on the total delay due to congestion.

We can indeed compute the total delay due to congestion in the case of a
monocentric configuration. In this situation, all the population commutes to a
single destination $1$ and we have

\begin{equation}
    \delta \tau_{mono} = \frac{1}{v}\; \sum_i d_{i1} \left(\frac{P}{c} \right)^\mu = L_{tot} \left(\frac{P}{c} \right)^\mu
\end{equation}

It follows, using the expression given above for $L_{tot}$

\begin{equation}
    \delta\tau_{mono} = \frac{\ell}{v}\; P^{1+\mu}
\end{equation}

From the fact that $1+\mu > 1+\frac{\mu}{2\mu+1}$, we indeed find that the total
delay due to congestion is worse for monocentric cities than it is for
polycentric cities with the same population, which agrees with the usual
intuition. More precisely the ratio of delays is given by

\begin{equation}
    \frac{\delta\tau_{mono}}{\delta\tau_{poly}}\sim
    \left(\frac{P}{c}\right)^{\,\beta}
\end{equation}

where the exponent is of order $\beta \approx 0.57\;$. Therefore, even though
diseconomies associated with polycentric cities scale superlinearly with
population, it would be even worse if we did not let cities evolve from the
monocentric situation. The same reasoning applies to the consumption of gasoline and
the $CO_2$ emissions. 

This suggests that, everything else being equal,
polycentricity should be favoured for quality of life and environmental reasons.

\begin{table*}[!h]
\begin{tabular}{|c|c|c|c|}
\hline
Quantity & Theoretical expression & Predicted exponent & Measured value\\
  $\;\;$  & ($\delta=\alpha/\alpha+1$) & & \\ \hline
$L_{tot}$ & $P$ & $1$ & $1.03 \pm 0.03\;(r^2=0.95)$\\ 
$A / \ell^2$ & $\left( \frac{P}{c} \right)^{\,2\,\delta}$ & $2 \delta = 0.78 \pm 0.20$ & $0.853 \pm 0.011\; (r^2=0.93)$\\
$L_N / \ell$ & $\sqrt{P}\; \left(\frac{P}{c}\right)^{\,\delta}$ & $\frac{1}{2} + \delta = 0.89 \pm 0.10 $ & $0.765 \pm 0.033\; (r^2=0.92)$\\
$\delta \tau / \tau$ & $P\; \left( \frac{P}{c} \right)^{\, \delta}$ & $1 +
\delta = 1.39 \pm 0.10$ & $1.270 \pm 0.067\; (r^2=0.97)$\\
$Q_{gas,CO_2}/\ell$  & $P\left( \frac{P}{c}\right)^{\delta}$ & $1+\delta=1.39 \pm 0.10$ & $1.262 \pm 0.089\; (r^2=0.94)$\\
& & & $1.212 \pm 0.098\, (r^2=0.83)$\\
\hline \hline
$L_N / \sqrt{A}$ & $\sqrt{P}$ & $0.5$ & $0.42 \pm 0.02\;(r^2=0.83)$\\
$L_{tot} / \sqrt{A}$ &  $P\;
\left(\frac{P}{c}\right)^{-\delta}$ & $1 - \delta = 0.61 \pm 0.10$ & $0.595 \pm 0.026\; (r^2=0.90)$\\
\hline
\end{tabular}
\caption{{\bf Summary of the scaling exponents.} This table displays the predicted theoretical behavior and
  the empirical observations versus the population size $P$ for
  different quantities: $L_{tot}$ is the daily total driven distance,
  $A$ is the area of the city, $L_{N}$ is the total length of the road
  network, $\delta\tau$ is the daily total delay due to congestion,
  $Q_{gas}$ is the yearly total consumption of gasoline and $Q_{CO_2}$
  is the total $CO_2$ emissions emitted yearly due to
  transportation. In the third column, we show the predicted values of
  the exponent of $P$ using the value of $\alpha$ measured on US
  employment data, and in the fourth column, the value of the
  exponents directly measured on data about US and OECD cities. The
  measured values are in good agreement with the prediction. In
  particular, the exponents for $L_N$ and $\delta \tau$ are consistent
  with our prediction that their difference should be $1/2$.}
\label{table:results}
\end{table*}

\subsection{Outlook}

The superlinear increase of congestion delay with population, and thereby of
gasoline consumption and of $CO_2$ emissions, has terrible consequences on the
economy, the environment, health and well-being. The outlook is nothing short of
grim in our ever-urbanising world. As the proportion of human beings living in
cities dramatically increases -- the UN expects the world population to be $67\%$
urban in 2050 -- wages are likely to \graffito{Estimates are given in the United
Nations' 2011 World Urban Propects.}
increase~\cite{Bettencourt:2007} but not enough to compensate for the negative
effects of congestion. As a result, if the individual car stays the dominant
transportation mode, cities will put more strain on people's life, while acting
as catalysts for the production of $CO_2$ greenhouse gas, which is responsible for an
overall increase of the planet's temperature~\cite{Oreskes:2004}. 

It is currently believed that advantages associated with living in a large city
outweigh the costs. Our results reveal however the existence of very rapidly
growing problems such as congestion and $CO_2$ emissions, which inevitably begs
the question of the sustainability of large cities. It might be time to cut down
considerably the use of individual vehicles, or to consider the possibility of
living in smaller or medium sized cities: the infrastructure costs ($L_N$) may
be larger, but the impact on the environment ($CO_2$ emissions) and on the
well-being of people (delays in congestion) would be beneficial.\\

The most striking fact about the above results is that despite the apparence of
complexity that is conveyed by cities, most of their structure can be explained
by the very simple and universal desire for the best achievable balance between
income and commuting costs. Our model unifies mobility patterns, spatial
structure of cities and allometric scalings in a framework that can be built
upon. 

\chapter{Interpretations and implications of scaling laws}
\label{chap:scaling_implications}

\begin{flushright}{\slshape    
There are no facts, only interpretations.} \\ \medskip
--- Friedrich Nietzsche
\end{flushright}

\bigskip

Although allometric scaling relationships are a powerful tool to explore the
behaviour of cities, there are several continuing controversies in the
literature. First, about their interpretation: do these relationships say
something about cities and the processes they host, or cities as they relate to
one another in a system of cities? Second, recent
studies~\cite{Arcaute:2014,Louf:2014_mobility,Cottineau:2015} have shown that
the measured exponents are very sensitive to the way cities are defined. What
does it imply for the study of these scalings and, more generally, cities?

\section{What scaling laws tell us about cities}

Scaling laws are, in essence, cross-sectionnal
studies of cities. As opposed to dynamical studies where one would follow the
evolution of individual cities over time, scaling laws tells us about the behaviour
of an \emph{ensemble} of cities at a give point in time. Throughout
Chapters~\ref{chap:scaling_introduction} and~\ref{chap:scaling_model}, we have
implicitely assumed that scaling laws are the signature of phenomena occuring at
the intra-urban level. This assumption, we call \emph{evolution interpretation},
is however not completely obvious. 

Maybe the easiest way to understand the issues posed by this interpretation is
through the comparison with Biology, where allometric scaling laws are also
widely used. The interpretation of allometric scaling laws in Biology is
straightforward, because the compared organisms are independent. Consider, for
instance, the scaling of the metabolic rate of animals with their body
mass~\cite{West:1997,Banavar:1999}. The mass of a given elephant at a point in
time $t$ is not correlated to the mass of any other living creature in the
world. Therefore, the scaling relationship can only be understood as resulting
from the existence  of similar processes in the growth of these different
animals.  Cities are different. They are part of a bigger system -- the system
of cities -- and interact constantly with one another. People change residence,
companies relocate, goods are shipped and money is transfered. Therefore, as
argued by Denise Pumain~\cite{Pumain:2012}, scaling laws can also be construed
as reflecting the redistribution processes within this system of cities. We call
this the \emph{differentiation interpretation}.

\subsection{The evolution interpretation}
\label{sub:the_evolution_interpretation}

\begin{figure}[!h]
    \centering
    \includegraphics[width=\textwidth]{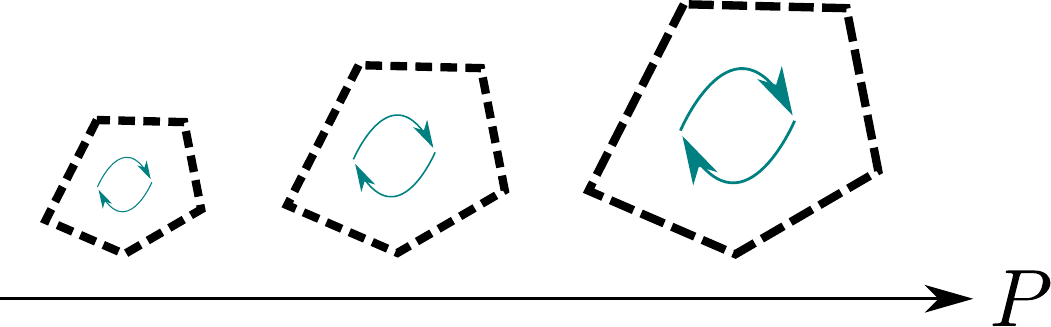}
    \caption{{\bf Evolution interpretation.} In this interpretation we consider
    that cities are different realisations of the same system. The intra-urban
processes -- and the way they respond to population changes -- are responsible for the non-linear scaling of the different
quantities.\label{fig:evolution_interpretation}}
\end{figure}

The evolution interpretation (Figure~\ref{fig:evolution_interpretation}) has
been widely adopted in the scaling literature~\cite{Bettencourt:2007,
Bettencourt:2013,Louf:2014_scaling} without ever being clearly stated, let alone
justified. It is based on two assumptions. The first assumption is that cities in the dataset
are different realisations of the same system. Thus, as stated in
Chapter~\ref{chap:scaling_introduction}, looking at the scaling of various
quantities with population size is a way to probe the system's internal
processes.

The second assumption has to do with the time scales over which the different
processes occur. Indeed, if the processes responsible for the change in the
value of the quantity $Y$ being studied occur on timescales significantly larger
than the timescale over which the population size changes, we cannot be sure the
exponent value actually reflects the internal processes at the time we measure
it.  For instance, an abrupt increase in population size is not likely to be
immediately reflected in the length of streets, while the evolution of the total
commuted length will be almost instantaneous. In practice, the rate of population change in cities is small enough for the
processes to follow, or the amplitude small enough for the induced error to be
insignificant. Hence the observed stability in the value of some exponents.

The previous discussion has several important consequences. First, it hints at
the difficulty to intepret the values of the observed deviations to scaling
laws~\cite{Bettencourt:2010a}. It is indeed difficult to assess to what extent
deviations account for a real over- or under-performance of the city compared to
the other cities, or for the time it takes for the studied quantity to react to
population changes.  Worse, the delayed adjustment to population changes
introduces an irreducible uncertainty in the numerical values of the exponents
themselves. Thus, the real error on the measured value of the exponent is very
likely larger than what is usually indicated by the statistical error bars.
Unfortunately, we cannot get a better estimate of the error until we understand
in details the mechanisms responsible for the time evolution of the corresponding
quantities. Until then, we should focus on (1) trying to understand the
qualitative behaviour, more than the exact numerical value of the exponents (2)
be wary of interpreting exponent values that are close to 1 (typically between $0.90$ and
$1.10$); in the absence of an alternative mechanistic explanation, the linear
relationship has to be favoured due to its simplicity.

\subsection{The differentiation interpretation}
\label{sub:the_differentiation_interpretation}

As Denise Pumain judiciously claims~\cite{Pumain:2006,Pumain:2012}, the evolution
interpretation is not the only possible interpretation for scaling laws. In some
cases and the mechanisms responsible for scaling relationships should be sought
after in the hierarchical organisation of cities and their interactions.\\

\begin{figure}[!h]
    \centering
    \includegraphics[width=0.5\textwidth]{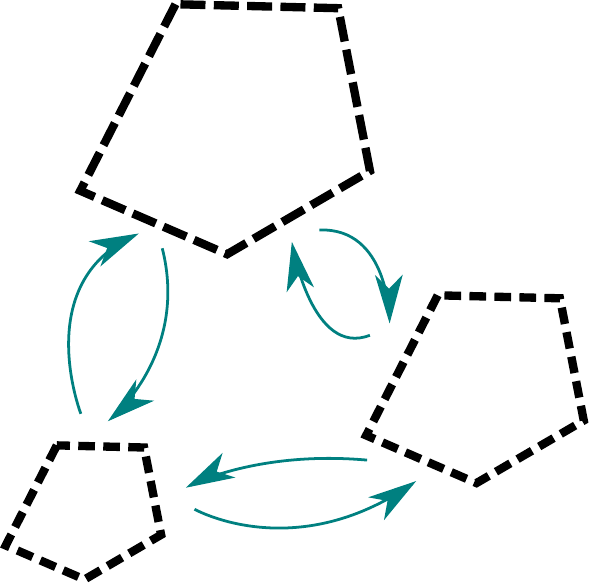}
    \caption{{\bf Differentiation interpretation.} In this interpretation, we
    consider that the redistribution processes occuring within systems of cities
are reponsible for the non-linear scaling of quantities with city size in this
system.\label{fig:label_fig}}
\end{figure}

We briefly mentioned in Chapter~\ref{chap:scaling_model} that allometric scaling relationships
could only be obtained when considering cities that belong to the same system of
cities. The fact that we observe scalings when taking a single country into
account, and a cloud of points when mixing two different countries, is a
signature of the integration of cities into systems of cities. It is
not clear at the moment what mechanisms are reponsible for the
coherence that permits the existence of scaling at the system level. But
clearly, the fact that cities are tightly connected through
the flow of commodities, populations, information and funds must be a key
factor.\\

Now, the same connections may be responsible for the scaling relationships
themselves, and the value of the exponent. As an example, Pumain et
al.~\cite{Pumain:2006} study the scaling of the number of employees from
different economic sectors in France with population size (see also
Chapter~\ref{chap:scaling_introduction}). They find that the number of employees
in innovative sectors (such as research and development) scales superlinearly
with population size, while the number of employees in mature economic sector
(such as the manufacture of food products) scales sublinearly with population
size. Using historical data, they further show that the scaling behaviour of
some activities has significantly changed over time: the exponent of
manufacturing activities has continuously decreased since $1960$, while that of
research and developement has continuously increased. This could be explained,
they claim, by the hierarchical diffusion of innovations in systems of cities.
Innovative activities first appear in large cities, entailing a larger
proportion of the active population working in these sectors than in smaller
cities, thus a superlinear scaling.  Over time, the innovations progressively
diffuse through the system of cities, the proportions are equilibrated and the
value of the scaling exponent decreases.

Although the mechanism is plausible, the current issue with this interpretation
is the lack of predictive model that explains the values of the various
exponents.

\subsection{Cities, or systems of cities?}
\label{sub:cities_or_systems_of_cities_}

So, are scaling relationships properties of cities, or of systems of cities?
Probably both. The above discussion is very general, and the origin of scalings
should be evaluated on a case-by-case basis. The scaling of some quantities,
such as the total quantity of CO\textsubscript{2} emitted or the total length of
roads are undoubtedly due to intra-urban processes (at least as long as the
explanation presented in Chapter~\ref{chap:scaling_model} holds). Indeed, the
total length of roads in Los Angeles only depends on what happens in Los
Angeles. Others, such as the linear
scaling of total income, are probably due to the interactions of cities within
the same system of cities. However, it is impossible to discriminate between
both interpretations on a purely empirical basis. Ultimately, we need models that are
able to reproduce at least the qualitative scaling behaviour. Plausible
narratives are not enough.

\section{What cities?}
\label{sec:what_cities_}

As we have argued up to this point, scaling relations are a signature
of various processes governing the phenomenon under study, especially when the exponent
$\beta$ is not what is naively expected~\cite{Barenblatt:1996}. However, as more and more scaling
relationships are being reported in the literature, it becomes less and less clear what we really
learn from these empirical findings. Mechanistic insights about these scalings are usually
nonexistent, often leading to misguided interpretations.

A striking example of the fallacies which hinder the interpretation and
application of scaling is given by different studies on $CO_2$ emissions due to
transportation~\cite{Fragkias:2013,Glaeser:2010,Oliveira:2014,Rybski:2013}. The
topic is particularly timely: pollution peaks occur in large cities worldwide
with a seemingly increasing frequency, and are suspected to be the source of
serious health problems~\cite{Bernstein:2004}. Glaeser and
Kahn~\cite{Glaeser:2010}, Rybski et al~\cite{Rybski:2013}, Fragkias et
al~\cite{Fragkias:2013}, and Oliveira et al~\cite{Oliveira:2014} are interested
in how $CO_2$ emissions scale with the population size of cities. The question
they ask is simple: Are larger cities greener---in the sense that there are
fewer emissions per capita for larger cities---or smoggier? Surprisingly, these
different studies reach contradictory conclusions. We identify here two main
sources of error which originate in the lack of understanding of the mechanisms
governing the phenomenon.

The first error concerns the estimation of the quantity $Q_{CO_2}$ of $CO_2$ emissions due to
transportation. In the absence of direct measures, Glaeser and Kahn~\cite{Glaeser:2010} have chosen
to use estimations of $Q_{CO_2}$ based on the total distance traveled by commuters. This is in fact
incorrect, and in heavily congested urban areas the relevant quantity is the total time spent
in traffic~\cite{Louf:2014_mobility}. Using distance leads to a serious underestimation of
$CO_2$ emissions: the effects of congestion are indeed strongly nonlinear, and the time spent
in traffic jams is not proportional to the traveled distance. As a matter of fact, commuting
distance and time scale differently with population size, and the time spent commuting and
$CO_2$ emissions scale with the same exponent~\cite{Louf:2014_mobility}.

\begin{figure}
    \centering
    \includegraphics[width=\textwidth]{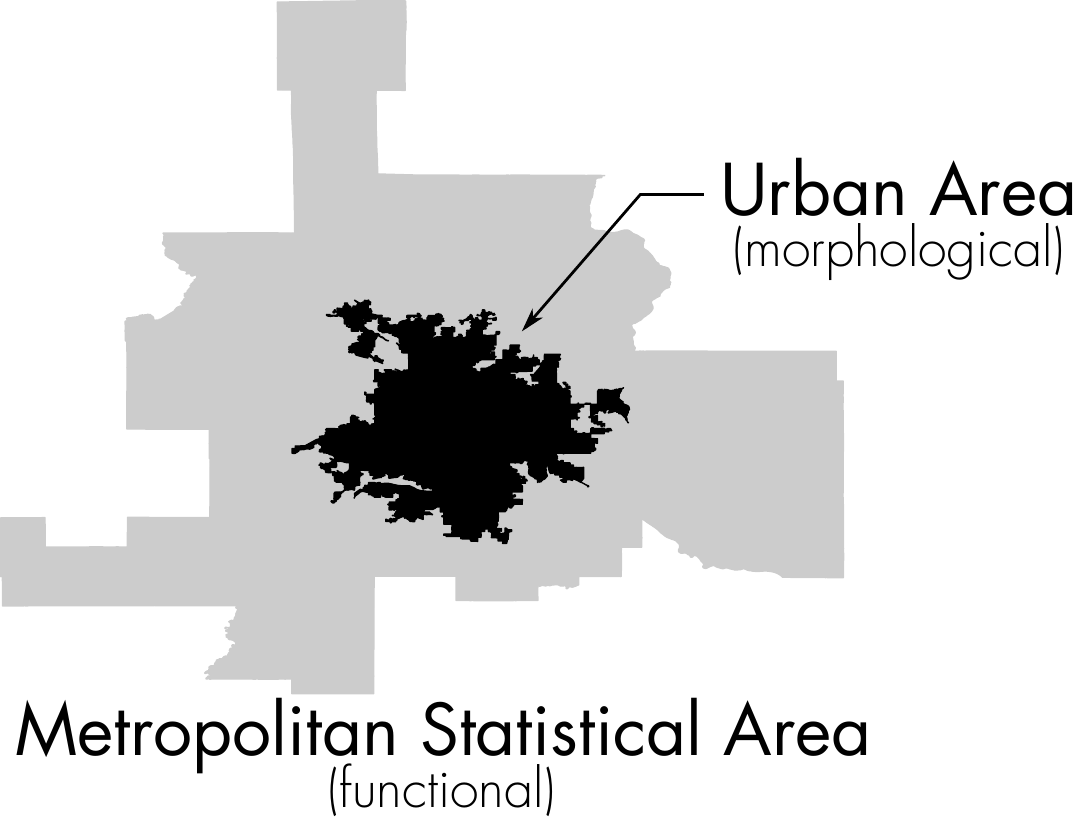}
    \caption{{\bf City definitions in the US.} The Minneapolis Urban Area (in
    black) is defined by the Census Bureau as contiguous block groups with at
least $1000$ inhabitants per square mile. The Minneapolis-St. Paul Metropolitan
Statistical Area (in grey) is defined as the counties containing the urban area
as well as any adjacent county that have a high degree of integration with the
core, as measured with commuting flows.\label{fig:two_definitions}}
\end{figure}

The second, subtler, issue lies in the definition of the city itself, and over
which geographical area the quantities $Q_{CO_2}$ and $P$ should be aggregated.
There is currently great confusion in the literature about how cities should be
defined, and scientists, let alone the various statistical agencies in the
world, have not yet reached a consensus. For instance, the US Census Bureau
defines two types of cities for statistical purposes
(see Figure~\ref{fig:two_definitions} for an illustration on the city of Minneapolis).
First, the Urban Areas are defined as a set of contiguous high-density areal
units with a threshold on the total population (morphological definition). The Metropolitan Statistical
Areas, on the other hand, include core Urban Areas, and the areal units that
sends more than a given percentage of its working population to work in the
core (functional definition).

\begin{figure}[!h]
	\centering
	\includegraphics[width=\textwidth]{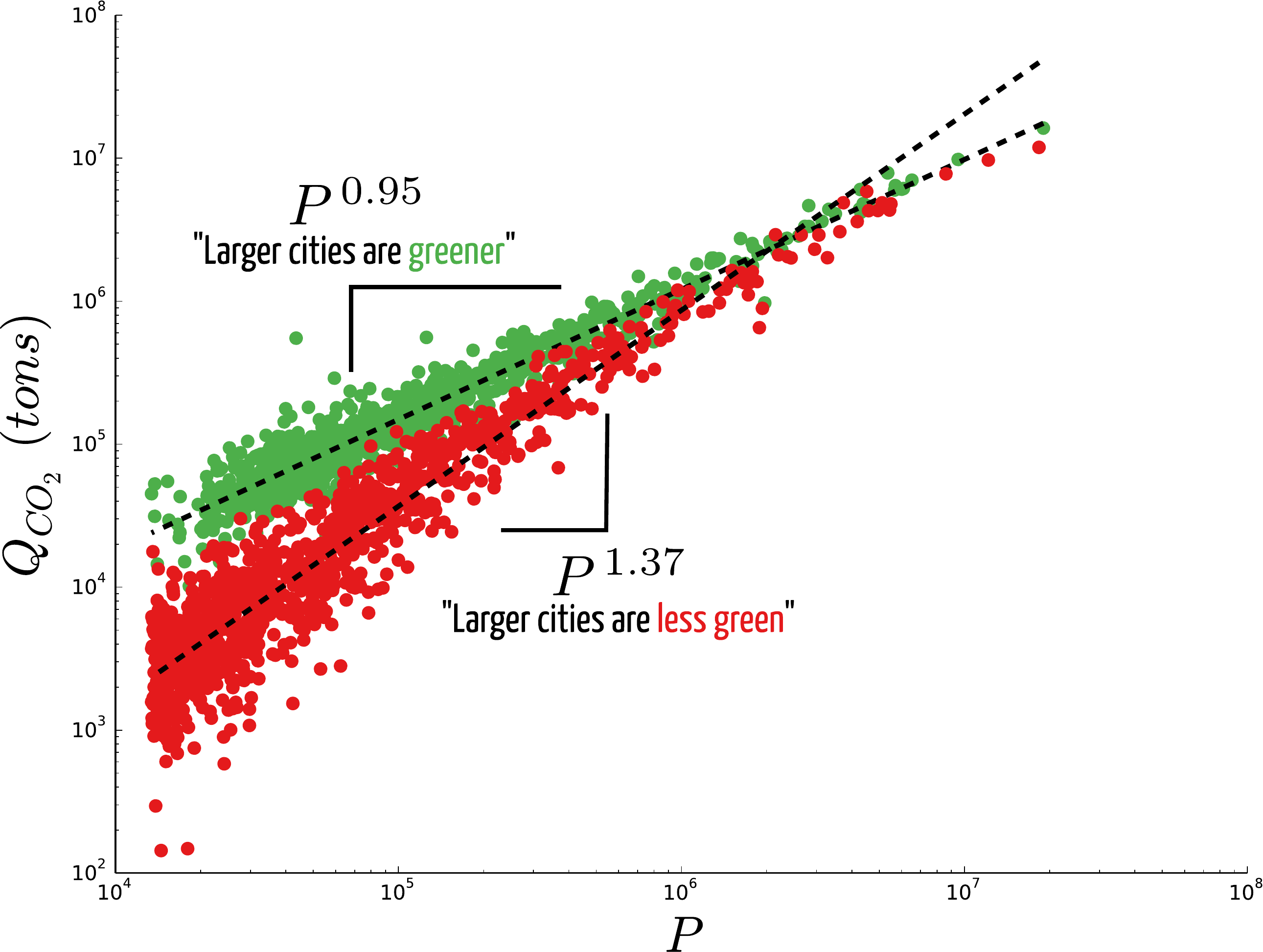}
	\caption{ {\bf Are larger cities greener or smoggier?} Scaling of transport-related $CO_2$
emissions with the population size for US cities from the same dataset but at different aggregation
levels. In red, the aggregation is done at the level of urban areas and in green for combined statistical
areas. Depending on the definition of the city, the scaling exponents are qualitatively different, leading
to two opposite conclusions. Data on $CO_2$ emissions were obtained from the Vulcan Project (\url{http://
vulcan.project.asu.ed}) (see~\cite{Fragkias:2013,Oliveira:2014}). Data on the population of urban areas and
metropolitan statistical areas were obtained from the Census Bureau
(\protect\url{http://www.census.org}). \label{fig:lost}}
\end{figure}

This is a crucial issue as scaling exponents are very sensitive to the
way city boundaries are delineated~\cite{Arcaute:2014}.  $CO_2$ emissions are no exception:
aggregating over Urban Areas or Metropolitan Statistical Areas entails radically
different behaviours (see Figure~\ref{fig:lost}). For the US, using the
definition of urban areas provided by the Census Bureau
(\url{http://www.census.org}), one finds that $CO_2$ emissions per capita
sharply increase with population size, implying that larger cities are less
green. Using the definition of metropolitan statistical areas, also provided by
the Census Bureau, one finds that $CO_2$ emissions per capita decrease slightly
with population size, implying that larger cities are greener.\\

Faced with these two opposite results, what should one conclude? Our point is that, in
the absence of a convincing model that accounts for these differences and how they arise,
nothing. Scaling relationships, and more generally data analysis, have an important role
to play in the rising new science of cities. But, as the previous discussion
illustrates (as well as the discussion in Chapter~\ref{chap:methodology}), it is
dangerous to interpret empirical results without any mechanistic insight. Conclusions cannot
safely be drawn from data analysis alone.\\

Does it mean that we should throw away scaling relationships altogether, as
suggested by Arcaute et al.~\cite{Arcaute:2014}? No, this would be tackling the
problem from the wrong end. Scaling relationships are the signature of processes
occuring at the system (city or system of cities) level. The issue encountered
here is that the system we study is not properly defined. We don't really know
what cities we are talking about! 

Cities are doubtlessly a real pattern. Yet, the way we unveil this pattern with
empirical data is, at best, imprecise. It is not based on a theoretical
understanding of what cities are. As a result, we cannot fully make sense of the
exponents found in empirical data. We therefore believe that future research in
this area should focus on

\begin{itemize}
    \item Understanding the basic object we are working on, cities. How they 
        should be defined, on what theoretical grounds.
    \item Accounting for the different qualitative behaviours of scaling
        exponents when different definitions are used.
\end{itemize}

Indeed, as long as we do not \emph{know} what system we should be probing, it is
not quite clear what our results mean. As long as we do not \emph{understand}
why values of exponents are different when the city definition changes, we
cannot draw reasonable conclusions.\\

The last years have seen many scholars coming forward with policy advice based
on empirical scaling relationships. It should now be clear that, given the
current state of knowledge, it is a risky game. Indeed, let us consider the
above CO\textsubscript{2} example: what should one do to curb CO2 emissions?
Favour the growth of large urban areas or the repartition of population in less
populated cities?  Both can be argued by considering data analysis alone. It
should therefore be obvious that, until they have a satisfactory understanding
of the mechanisms responsible for the observed behaviours, scientists should
refrain from giving policy advice that might have unforeseen, disastrous
consequences. If they choose to do so anyway, policy makers should be wary about
what is, at best, a shot in the dark

\section{Conclusion and perspective}
\label{sec:conclusion_and_perspective}

Scaling laws are useful tool to probe the internals of cities, but they are not
everything. They provide an extraordinarily easy way to explore the properties
of urban systems: the amount of data required is minimal, the statistical treatment
trivial. Allometric scaling is thus useful to declutter the field of investigation, help clear
a couple of paths, and establish a large-scale understanding of the system. But
this is done at the expense of an extensive coverage of the underlying
phenomena. Scalings can be seen as a gateway to the study of cities, but they
cannot be the study itself.

Furthermore, there are pressing issues that need to be solved if we want to make
sense of these empirical results. First, we need to question the definition of
cities, and understand what systems exactly we are studying. Second, measuring
exponents is not enough, and we need to understand the main processes that are
responsible for the measured values. This is what we have tried to do in the
previous chapter.


\ctparttext{Residential segregation is a reality. A reality so rife that it has
    pervaded even our everyday language though the expressions 'poor
    neighbourhood' or 'rich neighbouhood'. But despite its intuitive appeal,
    segregation is difficult to define.\\ 
    In this part, we propose to define segregation as a deviation to the
    unsegregated city, thereby providing a firm theoretical basis for any
    study of segregation patterns. We further propose
    a measure of attraction/repulsion of the different categories, which allows us to
    define unambiguously income classes from the original categories. We also study the
properties of neighbourhoods in which the different classes concentrate, and revisit the traditional poor
center/rich suburb dichotomy.} 

\part{Segregation}
\label{part:segregation}

\chapter{What segregation is not}
\label{chap:segregation_introduction}

\begin{flushright}{\slshape    
The limits of my language\\
Mean the limits of my world.} \\ \medskip
--- Ludwig Wittgenstein~\cite{Wittgenstein:1998}
\end{flushright}

\bigskip

\newcommand{\E}{\mathrm{E}}
\newcommand{\Var}{\mathrm{Var}}

\section{Studying segregation}
\label{sec:studying_segregation}

We cannot judge the spatial repartition of people. There is no criterion of
`good' or `bad' for the way people arrange themselves, no moral values attached
to any spatial pattern. It is the \emph{processes} that lead to such patterns,
the intentions behind people's decisions that make segregation condemnable. It
is the \emph{consequences} of segregation that may make it undesirable, something
worth fighting against.\\

As a matter of fact, social residential segregation has terrible consequences.
As shown in~\cite{Massey:1993}, residential segregation is the cause of major
economic disadvantages that affect the least affluent segments of the
population, through the isolation from social networks, or the presence of
deficient public service in the poorest areas. Worse, it has been shown that
increased levels of segregation in urban areas is associated with a higher
mortality burden~\cite{Lobmayer:2002}. For all these reasons, there is a
somewhat urgent need to measure the extent of segregation, especially its local
component, and understand the underlying mechanisms.\\

In the literature, authors systematically design a single index of segregation for
territories that can be very large, up to thousands of square
kilometers~\cite{Apparicio:2000}. In order to mitigate segregation, a more
local, spatial information is however needed: local authorities need to locate
where the poorest and richest concentrate if they want to design efficient
policies to curb, or compensate for, the existing segregation. In other words,
we need to provide a clear {\it spatial} information on the pattern of
segregation. We need to identify the areas where levels of segregation are high.

Besides, if we want to design policy or incentives to reduce socio-spatial
stratification and its consequences, we need to understand the processes at
play. We need to understand why segregation patterns exist, and why they
persist. Without mechanistic insights, attempts at regulating segregation may
have unforeseen, possibly damaging consequences.  The processes behind
segregation are however unclear. Schelling's cellular automata
model~\cite{Schelling:1971}, although intellectually stimulating, is very
limited in terms of predictions. More sophisticated models appeared
recently~\cite{Brueckner:1999,Glaeser:2008,Gauvin:2013}, yet the link with the
empirical reality is too thin, and processes are yet to be validated.  

In fact, we believe that the lack of an appropriate model is likely due to the lack
of identification of a clear structure, or clear behaviours in the data. In
order to identify the processes at play, we urgently 
need to properly describe the spatial patterns of segregation; the
dynamics of households (how they move, how their characteristics evolve over
time) and neighbourhoods (how their population changes).\\

In the following, we will therefore focus on the empirical characterisation of
the patterns of segregation. But first, we need to define what we mean when we
talk about residential segregation.

\section{Think first, measure later}
\label{sec:introduction}

As stated many times, and at different periods in the sociology
literature~\cite{Duncan:1955m,James:1982,Massey:1988,Reardon:2002}, the study of
segregation is cursed by its intuitive appeal. Pretty much everyone has heard of
segregation, and has an opinion about it. This familiarity with the concept
favours what Duncan and Duncan~\cite{Duncan:1955m} called `naive operationalism':
the tendency to force a sociological interpretation on measures that are at odds
with the conceptual understanding of segregation. In their own words

\begin{quote}
    [Segregation] is a concept rich in theoretical suggestiveness and of
    unquestionable heuristic value. Clearly we would not wish to sacrifice the
    capital of theoretization and observation already invested in the concept.
    Yet this is what is involved in the solution offered by naive
    operationalism, in more or less arbitrary matching some convenient numerical
    procedure with the verbal concept of segregation... (Duncan and Duncan,
    1955~\cite{Duncan:1955m})
\end{quote}

For all its intuitive appeal, segregation is however an intricate, compound
notion whose complexity only reveals itself through careful study. However
tempting it is to start writing measures of segregation that seem `reasonable',
it is necessary to stop and think about the meaning of the notion first. We need to
\emph{think} segregation to be able to provide \emph{useful} measures of
segregation.

\section{The dimensions of segregation}
\label{sec:the_dimensions_of_segregation}

Segregation has been extensively studied in the Sociology and Geography
literature. The most important conceptual heritage of this literature is the
distinction between residential segregation's different dimensions.
Massey~\cite{Massey:1988} first proposed a list of $5$ dimensions (and related
existing measures), which was recently reduced to $4$ by
Reardon~\cite{Reardon:2004}. 

\begin{description}
    \item[Evenness] (and clustering in the continuous
        limit, as shown by Reardon~\cite{Reardon:2004}) is the extent to which
        populations are evenly spread in the metropolitan area.
        Measures of evenness are affected by the fact that
        individuals are not spread uniformly across space in urban areas,
        disregarding of their respective category;

    \item[Exposure] is the extent to which different
        populations share the same residential area. This presupposes defining
        what is meant by `residential area';

    \item[Concentration] is the extent to which populations concentrate in their
        residential area;

    \item[Centralisation] is the extent to which populations concentrate in the
        center of the city. As we have seen in
        Chapter~\ref{chap:monocentric_introduction}, the notion of center
        is meaningless in large, polycentric urban areas;
\end{description}

We will discuss in details the shortcomings of the measures currently proposed
for each of these dimensions in Chapter~\ref{chap:patterns_segregation}.

\section{The unsegregated city}
\label{sec:null_model_the_unsegregated_city}

The fundamental issue with the picture given by these 4 dimensions lies in the
lack of a general theoretical framework in which all existing measures can be
interpreted.  Instead, we have a patchwork of seemingly unrelated measures that
are labelled with either of the aforementioned dimensions. Already in $1986$,
Michael White~\cite{White:1986} regretted the fact that segregation was never
defined in the literature, and always considered as a given. Each index implied a different definition of segregation, which lead to endless
debates about the virtues of such or such measure (dubbed the `index war').
Unknowingly, authors were trying to squeeze the social reality into existing
measures. When, in fact, one should start by defining the social reality, before
attempting to capture it with appropriate measures. As of today, no such
definition of segregation exists. We shall begin our study of segregation
patterns by an attempt at defining segregation. All the measure we propose then
naturally follow.\\

Segregation manifests itself in different ways, which makes it very difficult to
define. It is however easy to define what is \emph{not} segregation: a spatial
distribution of different categories that is undistinguishable from a uniform
random situation~\cite{Jahn:1947}. Therefore, we propose to define segregation
as the following

\begin{quote}
Segregation is any pattern in the spatial distribution of populations that
significantly deviates from a situation where individuals would have chosen
their residence at random (densities and overall category proportions being
equal).  
\end{quote} 

It is then easy to understand the different dimensions
of~\cite{Massey:1988,Reardon:2004}: each of the dimensions correspond to a
different ways in which a multi-dimensional pattern can deviate from its
randomized counterpart. Our definition is perfectly agnostic with regards to
the features of the population density pattern. It is also not concerned with
the overall inequality levels.\\

In the context of residential segregation in urban areas, a natural null model
is therefore the \emph{unsegregated city}. In the unsegregated city, all
households are distributed at random within the urban space with the further
constraints that

\begin{itemize}
    \item The total number $N_\alpha$ of people belonging to a category
	    $\alpha$ is fixed and equal to that found in the data;
    \item The total number $n(t)$ of households living in the areal unit $t$ is
	    fixed and equal to that found in the data.
\end{itemize}

which also fixes the total number of individuals $N$ in the city. The problem of
finding the numbers $\left( n_\alpha(1), \dots, n_\alpha(T) \right)$ of
individuals belonging to a certain category $\alpha$ in the $T$ areal units of
an unsegregated city is reminiscent of the traditionnal occupancy problem in
combinatorics~\cite{Feller:1950}. Their distribution is given by the multinomial
distribution $f \left( n_\alpha(1), \dots, n_\alpha(T) \right)$, and the number
of people of category $\alpha$ in the areal unit $t$ by a binomial distribution.
Therefore, in an unsegregated city, we have

\begin{align}
    \begin{split}
	\E \left[ n_\alpha(t) \right] &= N_\alpha\,\frac{n(t)}{N} \\
	\Var \left[ n_\alpha(t) \right] &= N_\alpha\,\frac{n(t)}{N} \left( 1 - \frac{n(t)}{N}  \right) 
    \end{split}
\end{align}

where $N$ is the total number of households in the city. In metropolitan areas
$N_\alpha$ is larged compared to $1$, and the distribution of the $n_\alpha(t)$
can be approximated by a Gaussian with the same mean and variance.

Most studies exploring the question of spatial segregation define measures
before comparing their value for different cities. Knowing that two quantities
are different is however not enough: we also have to know whether this
difference is {\em significant}. In order to assess the significance of a
result, we have to compare it to what is obtained for a reasonable null model.
As we will see in Chapter~\ref{chap:patterns_segregation}, the 
unsegregated city model allows us to assess whether a given pattern is
the result of a segregation process or not.

Any spatial distribution patterns could theoretically obtained via a random
repartition of households. They are however not equally likely. We propose to
measure the total segregation by the likelihood of obtaining a given pattern,
assuming a random distribution.

\bigskip

In this chapter, we have discussed some of the improvements that could be
brought to the existing measures in the literature. In particular, we have
emphasized the need for a \emph{local} knowledge of the patterns of segregation.
We have also laid the theoretical foundation upon which we are going to design
new measures.
In Chapter~\ref{chap:patterns_segregation}, we start from the above-defined null
model to propose a way to quantify the presence of various categories in parts
of the city. This allows us to identify and delineate neighbourhoods, measure the interactions
between the categories, and extract a class structure from the spatial pattern
alone.

%
\chapter{Patterns of segregation}
\label{chap:patterns_segregation}

\begin{flushright}{\slshape    
To understand is to perceive patterns.} \\ \medskip
--- Isaiah Berlin~\cite{Berlin:2013}
\end{flushright}

\bigskip

\section{Introduction}
\label{sec:introduction}

\subsection{Shortcomings in the current empirical picture}
\label{sub:shortcomings_in_the_current_empirical_picture}

There are many different ways in which a spatial pattern can deviate from its
randomised counterpart, and at least as many different measures one could
perform. In this chapter, we will try to quantify these patterns in a way that
may allow us to \emph{understand} the phenomenon of segregation.\\

Of course, segregation has been extensively studied in the literature. However,
we identify several difficulties in the current empirical picture.

First, some issues are tied to the existence of several categories in the
underlying data. Historically, measurements of racial segregation were limited
to measures between $2$ population groups. However, most measures generalise
poorly to a situation with many groups, and the others do not necessarily have a
clear interpretation~\cite{Reardon:2002}. Worse, in the case of groups based on
a continuum (such as income), the thresholds chosen to define classes are
usually arbitrary~\cite{Jargowsky:1996}. We propose to solve this issue by
defining classes in a unambiguous and non-arbitrary way through their pattern of
spatial interaction. Applied to the distribution of income categories in US
cities, we find $3$ emergent categories, which are naturally intepreted as the
lower-, middle- and higher-income classes.

Second, most authors systematically design a single index of segregation for
territories that can be very large, up to thousands of square
kilometers~\cite{Apparicio:2000}. In order to mitigate segregation, a more
local, spatial information is however needed: local authorities need to locate
where the poorest and richest concentrate if they want to design efficient
policies to curb, or compensate for, the existing segregation. Furthermore, 
a local description of the repartition of the different categories is the first
step towards the exploration of the mechanisms responsible for segregation: it
is necessary to gather hints (as well as empirical regularities) that are
essential to build a reasonable model. 
In other words, we need to provide a clear {\it spatial} information on the
pattern of segregation.\\

The lack of clear spatial characterization of the distribution of individuals is
not tied to the problem of segregation in particular, but pertains to the field
of spatial statistics~\cite{Ripley:1981}. Many studies avoided this spatial
problem by considering cities as monocentric and circular, and rely on either an
arbitrary definition of the city center boundaries, or on indices computed as a
function to the distance to the center (whatever this center may be, see
Part~\ref{part:polycentricity}). However, most if
not all cities are anistropic, and the large ones, polycentric
(see Chapter~\ref{chap:monocentric_introduction}), casting some doubt about
the application of the monocentric city picture. Many empirical studies and
models in economics aim to explain the difference between central cities and
suburbs \cite{Glaeser:2008, Brueckner:1999}. Yet, the sole stylized fact upon
which they rely -- city centers tend are allegedly poorer than suburbs (in the US) --
lacks a solid empirical basis.\\

In the following, we propose to answer the following questions

\begin{itemize}
    \item How can we quantify the presence of the different categories in areal
        units? Can we say whether they are overrepresented or normally
        represented? How can we define neighbourhoods?
    \item Can we quantify interactions between the different categories?
    \item Can we define meaningful classes from the original data?
    \item Do classes tend to leave in geographically coherent areas, or are they
        scattered across the city?
    \item Is there a difference between the city center and the suburbs? How
        can we quantify this adequately?
\end{itemize}

\subsection{Notations}
\label{sub:notations}

In the following, we will illustrate our measures using data from the $2000$ US
Census on the income of households per Census blockgroup. Data present
themselves as a number of households per blockgroup, sorted in different income
categories. There are $N$ individuals and  $T$ tracts in the considered
geographical area, and we note $N_\alpha$ the number of individuals belonging to
the category $\alpha$.  Finally, we write $n(t)$ the total number of individuals
living in the tract $t$, and $n_\alpha(t)$ the total number of individuals who
belong to category $\alpha$ living in the tract $t$.

\section{Presence of categories}
\label{sec:presence_of_categories}

In order to quantify segregation, we first need to measure the extent to which
categories are spread unevenly across space. Therefore, we start our analysis
with a discussion on how to quantify the presence of a category in areal units.
Several indicators exist, and one needs to be aware of their meaning, their
qualities and their shortcomings.

\subsection{Concentration index}
\label{sub:concentration}

The concentration index measures the proportion of individuals from category $\alpha$
in the areal unit $t$. 

\begin{equation}
    c(t) = \frac{n_\alpha(t)}{N_\alpha}
    \label{eq:concentration}
\end{equation}

The concentration is composition-invariant: it
does not depend on the relative proportion of category $\alpha$ in the
geographical zone as a whole. 

Nevertheless, its value strongly depends on the total population of the areal
unit we are studying: more populated areal units mechanically entail higher
values of concentration. Segregation measures based on the concentration
(such as the dissimilarity index) will therefore be dominated by the values in
highly populated areal units. This also makes values of concentration difficult
to intepret: we don't know whether large (repectively low) values of
concentration are the result of a large (respective low) population, or of a
local concentration of individuals in the area.

\subsection{Proportion index}
\label{sub:proportion}

Sometimes, we would prefer to know the proportion of individuals of a given
category in a unit. In our notations, the proportion index is simply defined as 

\begin{equation}
    p(t) = \frac{n_\alpha(t)}{n(t)}
\end{equation}

Although the values of the proportion index are easier to interpret (``x\% of the
individuals living in this areal unit belong to such category''), they are
not a good indicator of segregation. 

Indeed, they strongly depend on the relative proportion of individuals of the
category in the geographical area being studied. For instance, in a city where
$90\%$ of the individuals belong to category $A$, the proportion of people
belonging to category $A$ is very likely to be high in all areal units in the city.
The measure of proportion is therefore strongly tied to the overall inequality
levels.

\subsection{An unbiaised measure: representation}
\label{sub:an_unbiaised_measure_the_representation}

\subsubsection{Definition}
\label{ssub:definition}

The representation solves the problems linked to both measures of concentration
and proportion. The idea behind the measure of representation is that
segregation is, as we argued in Chapter~\ref{chap:segregation_introduction}, a
departure from the situation where households would be spatially distributed
at random. The properties of such a `random', unsegregated city are well known,
and the distribution of categories in each areal unit is given by a binomial
distribution. The representation is
thus defined as the number $n_\alpha(t)$ divided by its expected value in an
unsegregated city, $N_\alpha\,\frac{n(t)}{N}$

\begin{equation}
    r_\alpha(t) = \frac{n_\alpha(t)/n(t)}{N_\alpha/N}
    \label{eq:representation}
\end{equation}

Another way to understand the representation is to compare it to the
above-defined concentration and proportion. We can indeed write

\begin{equation}
    r_\alpha(t) = \frac{c(t)}{n(t) / N} = \frac{p(t)}{N_\alpha/N}
\end{equation}

The representation can thus be interpreted as the concentration
normalised by the local population concentration, or the proportion renormalised
by the proportion of the category at the city level, thereby addressing the
aforementioned shortcomings.\\

\subsubsection{Measuring significant deviations}
\label{ssub:measuring_significant_deviations}

The representation $r_\alpha(t)$ takes values between $0$ (when no
individuals from the category $\alpha$ are present in $t$) and
$\frac{N}{N_\alpha}$ (when all individuals in $t$ belong to the category
$\alpha$). In a city where individuals are distributed uniformly
(see Chapter~\ref{chap:segregation_introduction}), $r_\alpha(t) = 1$ in every
tract $t$.\\

In an unsegregated situation, the values of the representation are likely to be
close to $1$, but not necessarily strictly equal to $1$. There is indeed a
non-zero probability for any distribution to be obtained by chance. It is
therefore not obvious whether a given value of representation could have been
obtained in the unsegregated configuration. However, to quantify segregation, we
need to know how \emph{likely} it is that the present pattern is not the result
of a random repartition of individuals.  In other words, we need to know
whether areal units depart \emph{significantly} from the unsegregated situation. 

The distribution of individuals in a tract $t$ in the unsegregated city follows
a binomial distribution. We can therefore easily compute how likely it is that
the representation $r_\alpha(t)$ we measure has been obtained by chance. To do
that, we first compute the variance of the representation in the unsegregated
configuration:

\begin{equation}
    \mathrm{Var}\left[r_\alpha(t)\right] = \sigma_\alpha(t)^2 = \frac{1}{N_\alpha} \left[\frac{N}{n(t)} - 1\right]
\end{equation}

We say that the representation departs \emph{significantly} from the
unsegregated configuration if we can be sure with $99\%$ confidence that the
pattern has not been obtained at random. It follows that

\begin{itemize}
    \item $\alpha$ is {\bf overrepresented} in $t$ iff $r_\alpha(t) > 1 +
  2.57\;\sigma_\alpha(t)$
    \item $\alpha$ is {\bf underrepresented} in $t$ iff $r_\alpha(t) < 1 +
  2.57\;\sigma_\alpha(t)$
\end{itemize}

Note that the expression of the representation (Eq. \ref{eq:representation}) is very
similar to the formula used in economics to compute comparative
advantages~\cite{Balassa:1965}, or to the localisation quotient used in various
contexts~\cite{Apparicio:2000, Schwabe:2011}. To our knowledge, however, this
formula has never been justified by a null model in the context of residential
location. 

The representation allows to assess the significance of the deviation
of population distributions from the unsegregated city. As we will show below,
it is also the building block for measuring the level of repulsion or attraction
between populations -- allowing us to uncover the different classes -- and to
identify the neighbourhoods where the different categories concentrate. 

Last, but not least, the representation defined here does not depend on the
class structure at the city scale, but only on the spatial repartition of
individuals belonging to each class. This is essential to be able to compare
different cities where the group compositions -- or inequality -- might differ.
Inequality and segregation are indeed two separate concepts, and the way they
are measured should be distinct from one another. In that sense, the
representation is preferable to the measures of concentration or representation
as a basis to quantify segregation.

\section{Measuring the attraction and repulsion of categories}
\label{sec:measuring_the_attraction_and_repulsion_of_categories}

\subsection{Exposure}
\label{sub:exposure}

If we want to uncover the mechanisms underlying segregated patterns, it is
important to measure and understand the interactions between categories.
However, existing measures do not allow to quantify to which extent different populations
attract or repel one another. What we mean here by interaction is the
co-presence of the different categories in the same areal units, thus potential
interactions. This is the best one can do in the absence of data on the actual
interactions between individuals.

The measure we define is inspired by the M-value first introduced by Marcon \&
Puech in the economics literature~\cite{Marcon:2009} and used as a measure of
interaction in \cite{Jensen:2006}.  These authors were interested in measuring
the geographic concentration of different types of industries. While previous
measures (such as Ripley's K-value) allow to identify departures from a random
(Poisson) distribution, the M-value's interest resides in the possibility to
evaluate different industries' tendency to co-locate.\\

The idea, in the context of segregation, is simple. We consider two categories
$\alpha$ and $\beta$ and we would like to measure to which extent they are
co-located in the same areal unit. Essentially, we measure the representation of
the category $\beta$ as witnessed on average by the individuals in category
$\alpha$, and obtain the following quantity $E_{\alpha\beta}$

\begin{equation} 
    E_{\alpha \beta} = \frac{1}{N_\alpha} \sum_{t=1}^{T} n_\alpha(t)\,r_\beta(t) 
\end{equation}

Although it is not obvious with this formulation, this measure is symmetric:
$E_{\alpha \beta} = E_{\beta \alpha}$. Effectively, the E-value is a measure of exposure,
according to the typology of segregation measures found in~\cite{Massey:1988}.
It is however different from the traditional measure of exposure found in the
literature~\cite{Bell:1954}, as it allows to distinguish between the situations
where categories attract, or repel one another.\\

In the case of an unsegregated city, every household in $\alpha$ sees
on average $r_\beta = 1$ and we have $E_{\alpha \beta} = 1$. 
If populations $\alpha$ and $\beta$ {\em attract} one another, that is if they
tend to be overrepresented in the same areal units, every household $\alpha$
sees $r_\beta > 1$ and we have $E_{\alpha \beta} > 1$ at the city scale. 
On the other hand, if they
{\em repel} one another, every household $\alpha$ sees $r_\beta < 1$
and we have $E_{\alpha \beta} <1$ at the city scale.

\subsection{Extreme values}
\label{sub:extreme_values}

The minimum of the exposure for two classes $\alpha$ and $\beta$ is obtained
when these two categories are never present together in the same areal unit. Then

\begin{equation}
    E_{\alpha\,\beta}^{min} = 0 
\end{equation}

and the theoretical maximum is obtained when the two classes are alone in the system and
otherwise distributed at random

\begin{equation}
    E_{\alpha\,\beta}^{max} = \frac{N^2}{4\,N_\alpha\, N_\beta}
\end{equation}

These extrema are useful when comparing the exposure values for different
categories, and across different cities.

\subsection{Isolation}
\label{sub:isolation}

In the case $\alpha = \beta$, the previous measure represents the
`isolation' defined as

\begin{equation}
    I_\alpha = \frac{1}{N_\alpha}\sum_{t=1}^{t} n_\alpha(t)\,r_\alpha(t)
\end{equation}

and measures to which extent individuals from the same category are
exposed to their kins. In the unsegregated city, where individuals are
indifferent to others when choosing their residence, we have
$I_\alpha^{min}=1$. On the other hand, in the extreme situation where
individuals belonging to the class $\alpha$ live isolated from the
others, the isolation reaches its maximum value

\begin{equation}
    I_\alpha^{max} = \frac{N}{N_\alpha}
\end{equation}


\section{Emergent social classes}
\label{sec:the_emergent_social_classes}

\subsection{Defining classes}
\label{sub:defining_classes}

The study of income segregation must be rooted in a particular definition of categories
(or classes). There is however no consensus in the literature about how to
separate households in different classes according to their income, and studies
generally rely on more or less arbitrary divisions. While in some particular
cases grouping the original categories in pre-defined classes is justified,
most authors do so for mere convenience. However, as some sociologists
have already pointed out~\cite{Emirbayer:1997}, imposing the existence of
absolute, artificial entities is necessarily going to skew our reading of the
data. Entities such as social classes do not have an existence of their own.
Grouping the individuals into arbitrary classes when studying segregation is
thus problematic: it amounts to imposing a class structure on the society before
assessing the existence of this structure (which manifests itself by the
differentiated spatial repartition of individuals with different income,
segregation).
Furthermore, in the absence of recognized standards, different authors will
likely have different definitions of classes, making the comparisons between
different results in the literature difficult.\\

Here, instead of imposing an arbitrary class structure , we let the class
structure emerge from the data themselves. Our starting hypothesis is the
following: if there is such a thing as a social stratification based on income,
it should be reflected in the households' behaviours. The hypothesis is that households belonging to
the same class should tend to live together, while households belonging to
different classes should tend to avoid one another (It is worth noting that this
\emph{horizontal} definition of segregation is not relevant in every context; in
the 19th century Paris for instance, segregation was also vertical, with rich families
living in the lowest floors of buildings while poor individuals did tend to
live in the highest flats). The idea is thus to define
classes based on the way they manifest themselves through the spatial
repartition of the different categories. Of course, spatial proximity does not
necessarily imply social proximity. In particular, Chamboredon showed that in
some big French housing projects, households belonging to different social
classes were artificially brought in close proximity to one another but did not
necessarily interact with one another~\cite{Chamboredon:1970}\graffito{The work of
Chamboredon was kindly brought to my attention by Yann Renisio.}. We thus assume
here that the social class of housing tenants is not determined in a top-down
fashion, so that the spatial repartition of different income classes reflects
the nature of the interaction between these classes.

\subsection{Income classes in the US}
\label{sub:income_classes_in_the_u_s_}

\begin{figure}
    \includegraphics[width=\textwidth]{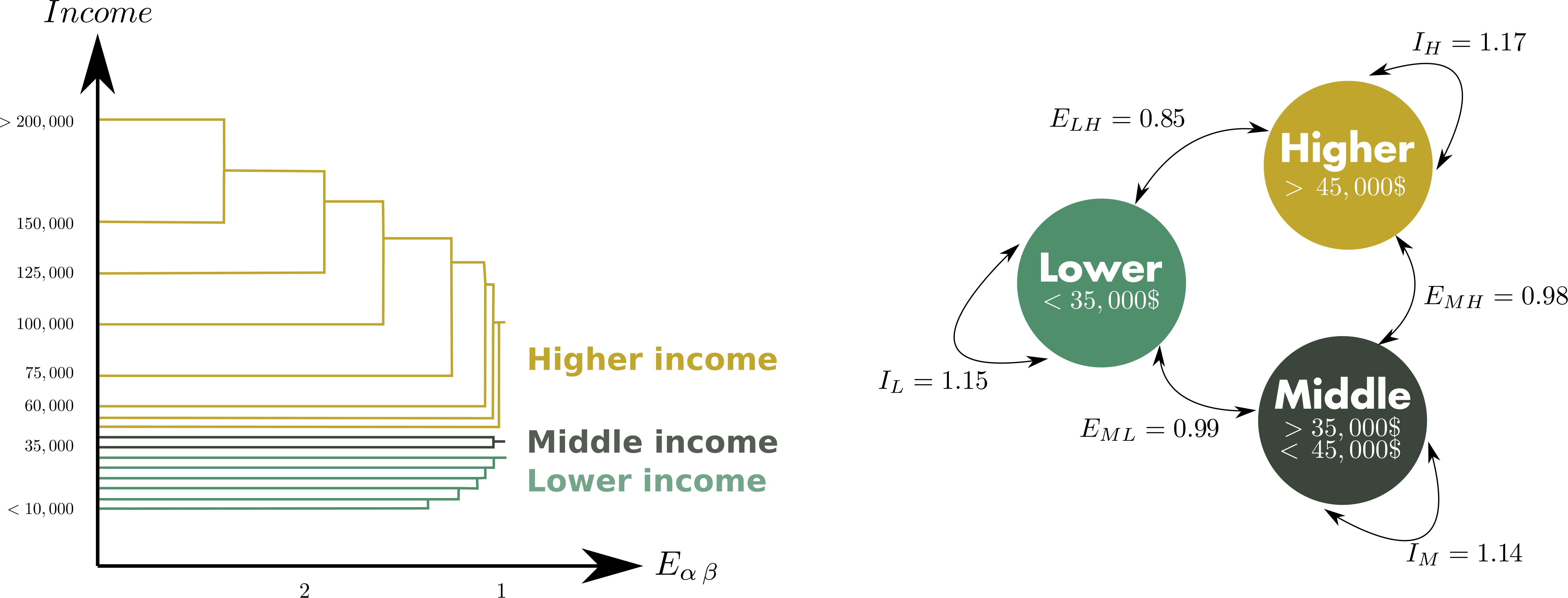}
    \caption{{\bf Emergent classes.} (Left) Alluvial diagram showing the successive aggregations
      of income categories in the clustering process, and
      the value of the exposure at which the aggregation took
      place. The aggregation stops when there is no pair of category
      for which $E>1$, that is when all classes are at best
      indifferent to one another. One can see on this diagram that
      the highest income categories attract one another more (higher values of
      $E_{\alpha \beta}$) than the lowest income categories. (Right) The classes that emerge from our
      analysis, and their respective exposure and isolation values. The lower
      and higher income classes repel one another, while the middle
      income class is indifferent to either other classes.  The
      higher-income class is slightly more coherent than the lower-income,
      which is more coherent than the middle-income class, as
      reflected by the isolation coefficient $I$.}
\label{fig:classes_alluvial}
\end{figure}

We choose as a starting point the finest income subdivision given by the Census
Bureau ($16$ subdivisions) and compute the $16 \times 16$ matrix of $E_{\alpha
\beta}$ values for all cities. We then perform a hierarchical clustering on this
matrix, successively aggregating the subdivisions with the highest $E_{\alpha
\beta}$ values. We stop the aggregation process when the only classes left are
indifferent ($E_{\alpha \beta} = 1$ with $99\%$ confidence) or
repel one another ($E_{\alpha \beta} < 1$ with $99\%$
confidence)~\cite{Louf:2015}. We obtain the dendrogram presented on
Figure~\ref{fig:classes_alluvial}.

Strikingly, the outcome of this method is the emergence of 3 distinct classes:
the higher-income ($47\%$ of the US population) and the lower-income ($42\%$ of the US population) classes
 -- which repel one another strongly while being
respectively very coherent -- and a somewhat meagre middle-income class ($11\%$
of the population) that is relatively indifferent to the other classes. This
result implies that there is some truth in the conventional way of dividing
populations into $3$ income classes, and that what we casually perceive as the
social stratification in our cities actually emerges from the spatial
interaction of people. Surprisingly, however, the middle-income class as
obtained here represents a significantly smaller part of the population than
other definitions.

Our method has several advantages over a casual, arbitrary definition: it only
depends on single tunable parameter, the size of the confidence interval.
Although, once an agreement has been reached, the class structure does not
depend on who is performing the analysis. Its origins are tractable, and can be argued on a
quantitative basis. Because it is quantitative, it allows comparison of the
stratification between different points in time, or between different countries. It
can also be compared to other class divisions that would be obtained using a
different medium for interaction, for instance mobile phone
communications~\cite{Eagle:2010}. 

In the following, we will systematically use the classes thus obtained.

\section{Larger cities are richer}
\label{sec:inter_urban}

\begin{figure}[!h]
    \centering
    \includegraphics[width=0.7\textwidth]{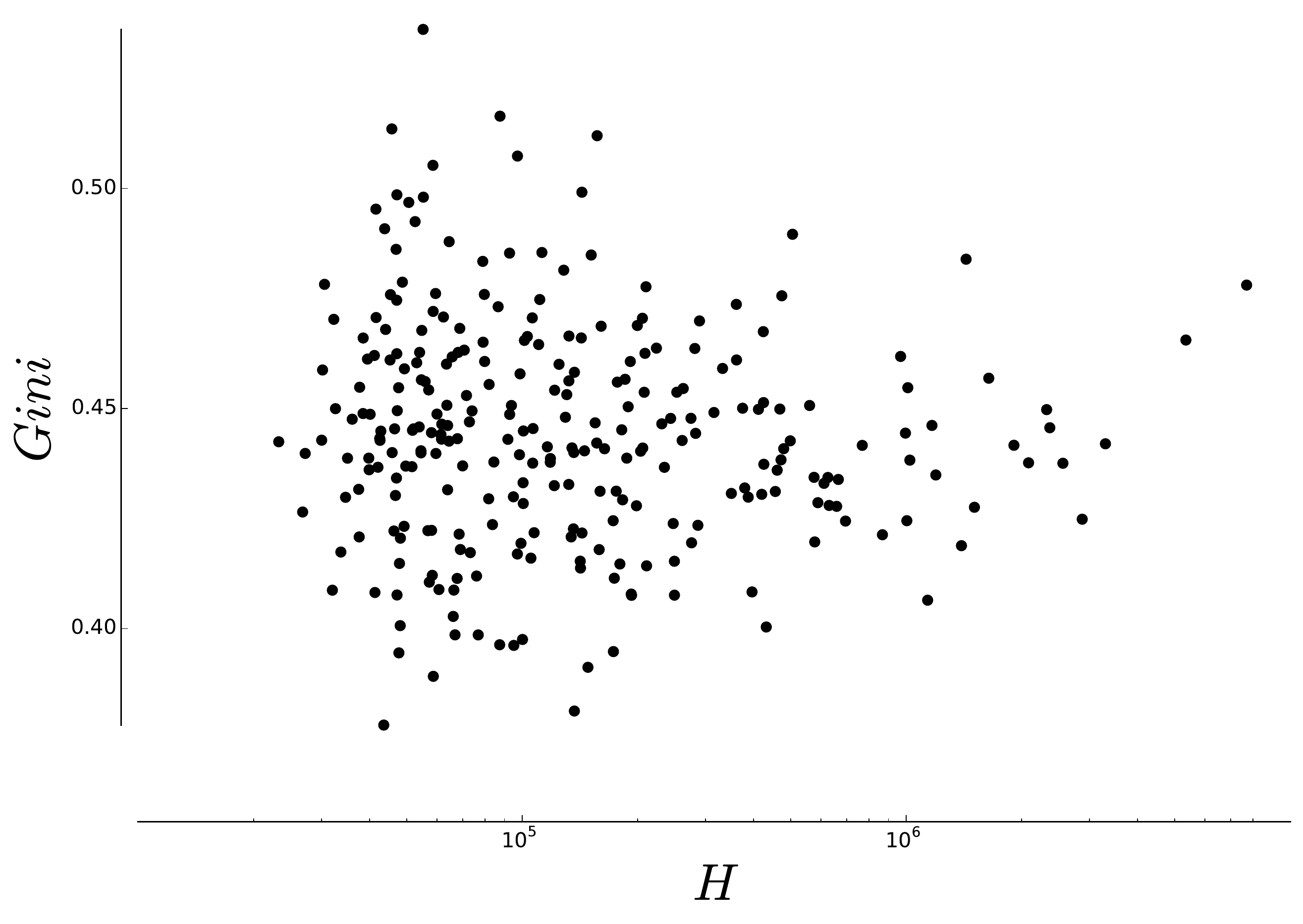}
    \includegraphics[width=0.7\textwidth]{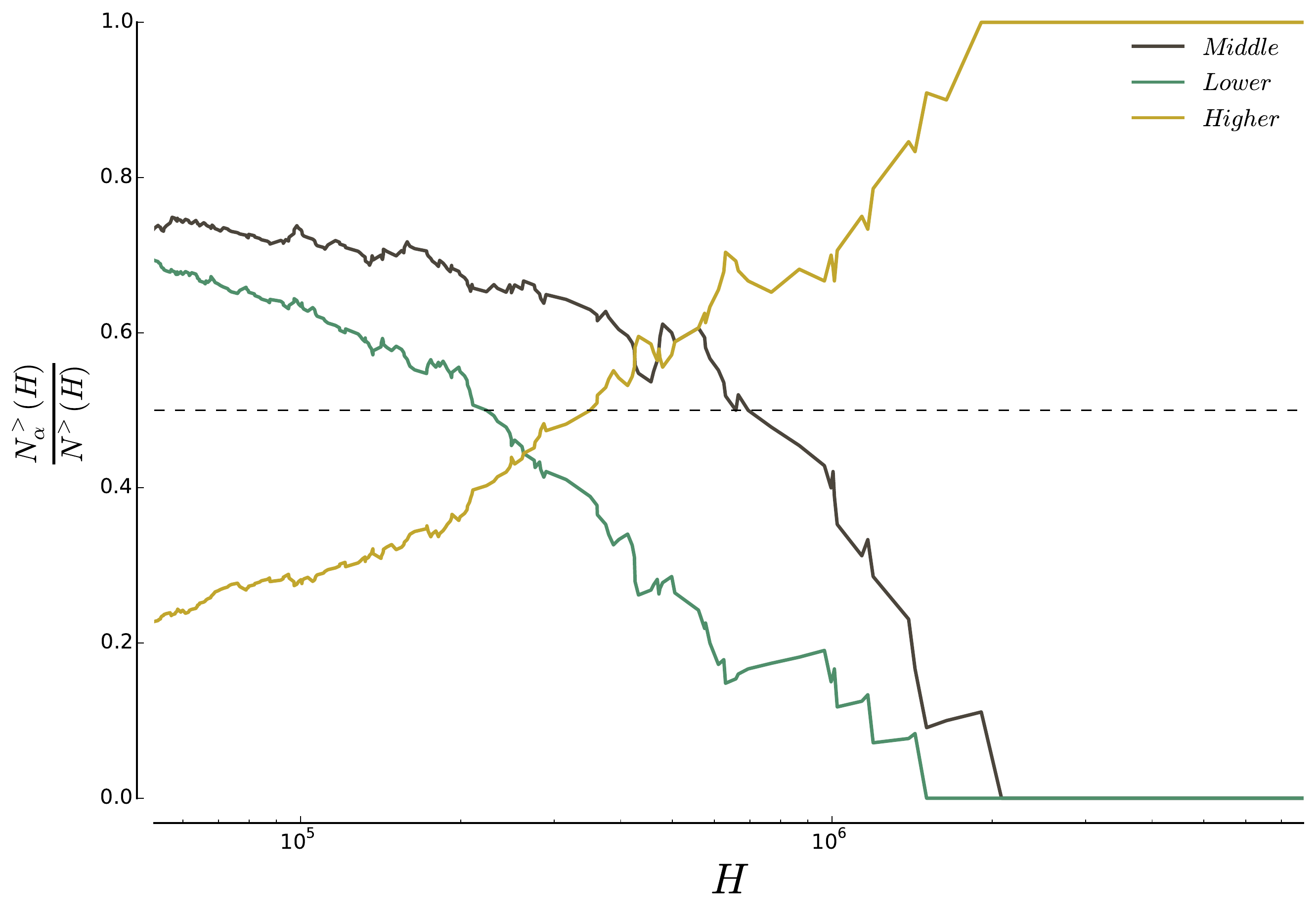}
    \caption{{\bf Larger cities are richer.} (Top) Gini coefficient of the income distribution of the $280$ MSA in
    $2000$ versus the number of households in the city. As one can see, there is
    no clear trend. (Bottom) Proportion of cities in which the different classes are
    overrepresented, as a function of the total population of the city. One can
    clearly see that as cities get larger rich people will
    be overrepresented and poor people underrepresented (compared to national
    levels). \label{fig:inter-urban}}
\end{figure}

At the scale of an entire country, segregation can manifest itself in the
unequal representation of the income classes in different urban areas.
We plot on Figure~\ref{fig:inter-urban} the ratio $
N_\alpha^{>}(H)/N^{>}(H)$ where $N^{>}(H)$ is the number of cities of
population greater than $H$, and $N_\alpha^{>}(H)$ the number of cities of
population greater than $H$ for which the class $\alpha$ is overrepresented.

A decreasing curve indicates that the category $\alpha$ tends to be
underrepresented in larger urban areas, while an increasing curve shows that the
category $\alpha$ tends to be overrepresented in larger urban areas.  The
representation is measured with respect to the total population at the US level.

There is a clear differentiation between cities: among the $276$ MSA in our
dataset, no city exhibits a number of households per class 
that is representative of the US as a whole. Furthermore, the number of cities
where higher-income households are overrepresented increases with the size of
the cities, while the inverse trend is true for lower-income
households. Therefore, larger cities are not richer in the sense that rich
households tend to be overrepresented in large cities, and underrepresented in
small ones.

Surprisingly, this effect is not visible using the Gini coefficient (see
Figure~\ref{fig:inter-urban}). This hints at the limitations of the Gini index
to compare income inequalities across an entire country.

\section{Delineating neighbourhoods}
\label{sec:neighbourhoods}

\subsection{Defining neighbourhoods}
\label{sub:defining_neighbourhoods}

\begin{figure}
    \centering
    \includegraphics[width=0.7\textwidth]{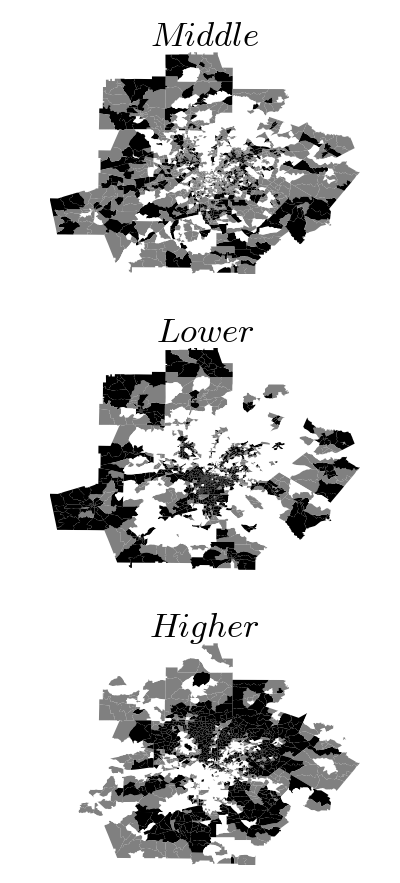}
    \caption{{\bf Neighbourhoods.} The neighbourhoods in Atlanta for the three different
      income category. In black, the tracts where the corresponding
      class is overrepresented, in white where it is
      underrepresented and in grey where its value is
      indistinguishable from the random distribution. All
      MSA defined for the $2000$ Census exhibit a total exclusion between
      lower-income and higher-income
      neighbourhoods: the pictures for lower- and higher-income classes are the
      perfect negative of one another. In contrast, middle-income households
      are scattered across the city.}
\label{fig:atlanta_neighbourhoods}
\end{figure}

Now that we can identify the areal units where classes are overrepresented, how
can we delineate neighbourhoods?

Considering a category $\alpha$,  we first look for the areal units where the
category is overrepresented. We then consider that two areal units in this set
are part of the same neighbourhood if they are contiguous. Of course, this
approach has limitations (some remarks that sprung in the discussion on the different methods to find
activity centers in Chapter~\ref{chap:monocentric_introduction} are relevant in
this context too), but it gives us a reasonable definition of neighbourhoods to
work with.
Let us now focus on the properties of these neighbourhoods.

\subsection{Clustering}
\label{sub:clustering}

\begin{figure}[!h]
    \centering
    \includegraphics[width=0.7\textwidth]{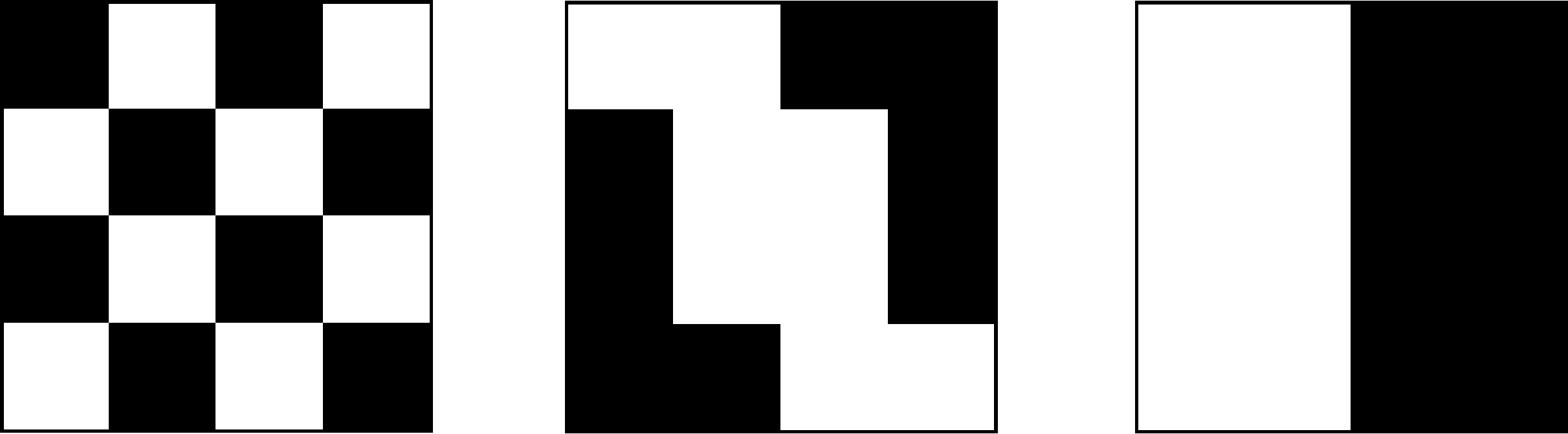}
    \caption{{\bf Spatial considerations.} Three situations that are identical for intra-areal unit measures,
        but that represent different segregation levels. (Left) The checkerboard
        city popularised by White~\cite{White:1983}, corresponding to a
        clustering value (defined in Eq.~\ref{eq:clustering}) of $C=0$ for the black squares. (Middle) An
        intermediate situation between the checkerboard and the divided city,
        corresponding to $C \approx 0.86$.(Right) The divided city, corresponding to
        $C=1$. \label{fig:checkerboard}} 
\end{figure}

Intra-tract measures such as the exposure are not enough to quantify
segregation. Indeed, areal units where a given class is overrepresented can
arrange themselves in different ways, without the intra-tract measures of
segregation being affected~\cite{White:1983}. In order to illustrate this, we
consider the schematic cases represented on Figure~\ref{fig:checkerboard}, and
assume that  they are obtained by reshuffling the various squares around.
Obviously, the checkerboard on the left depicts a very different segregation
situation from the divided situation on the right while intra-tract measures
would give identical results.\\

\begin{figure} 
    \centering
    \includegraphics[width=0.7\textwidth]{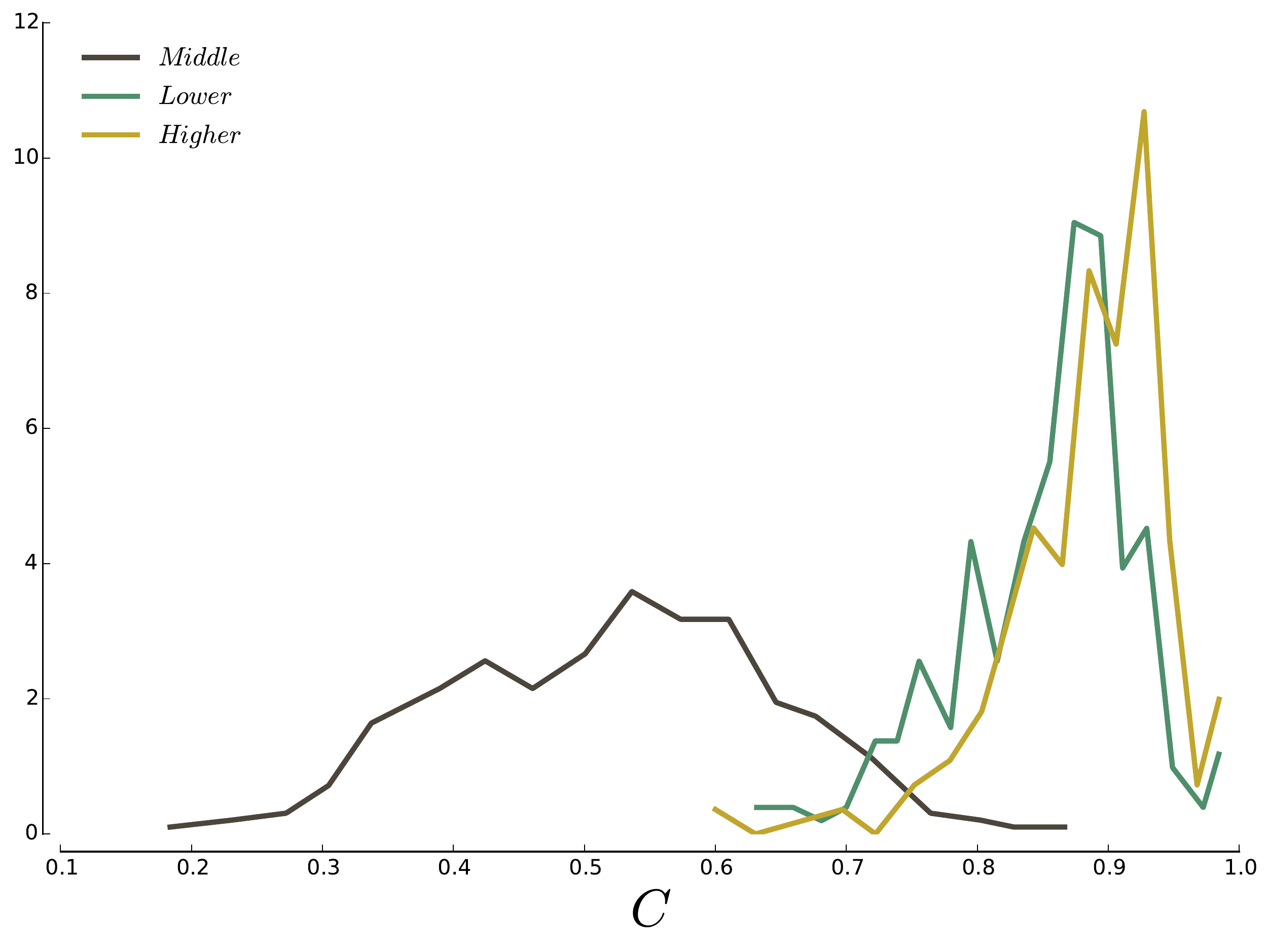}\\
    \caption{{\bf Clustering coefficient.} Distribution of the value of the clustering coefficient for
        all cities in our dataset, for the 3 classes. The higher income class
        exhibits the highest level of clustering, with an average of
        $\overline{C} = 0.90$, followed by the lower income class with on
        average $\overline{C} = 0.87$. The Middle income class households are
        significantly less clustered than the previous two, with $\overline{C} =
        0.56$ on average.} 
        \label{fig:clustering} 
\end{figure}

A way to distinguish between different spatial arrangements is then to measure
how clustered the overrepresented areal units are. We first aggregate adjacent
overrepresented areal units (for a given class) leading to consistent
neighbourhoods. The ratio of the number $N_n$ of neighborhoods (clusters) to the
total number $N_o$ of overrepresented areal units measures the level of
clustering and in 

\begin{equation} 
    C = \frac{N_{o}-N_{n}}{N_{o}-1}
    \label{eq:clustering}
\end{equation}

such that this quantity is $C = 0$ in a checkerboard-like situation, and $C = 1$
when all areal units form a unique neighbourhood. We show on
Figure~\ref{fig:clustering} the distibution of $C$ for the three classes over all
cities in our dataset. As one could infer from the maps on
Figure~\ref{fig:atlanta_neighbourhoods}, the rich and poor areal units are well
clustered, with a respective average clustering of $C = 0.80$ and $C = 0.74$.
The Middle class is on the other hand less coherent, with a average clustering
$C = 0.55$.\\

\subsection{Concentration in neighbourhoods}
\label{sub:concentration_in_neighbourhoods}

\begin{figure} 
    \centering
    \includegraphics[width=0.7\textwidth]{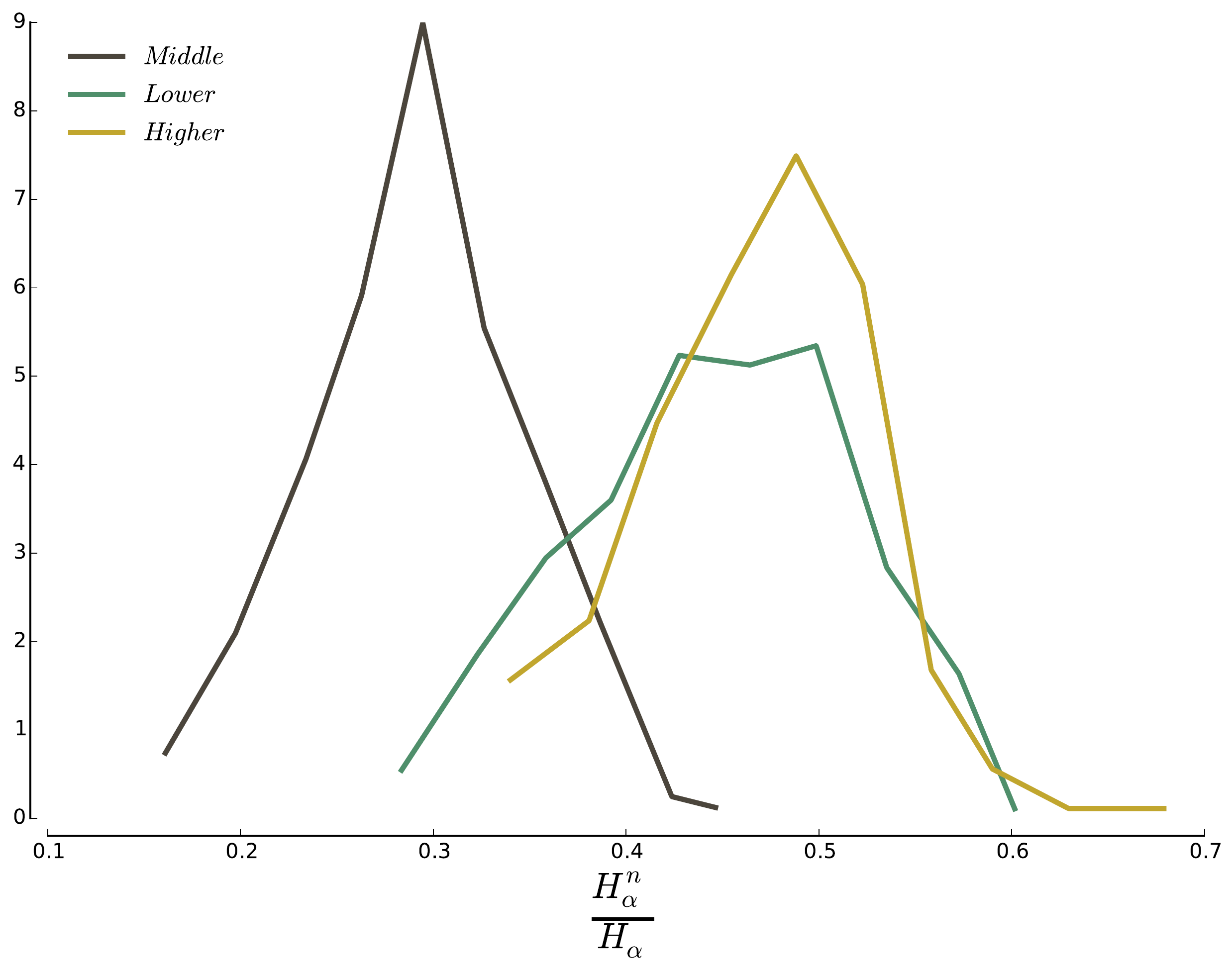}\\
    \caption{{\bf Concentration in neighbourhoods.} Distribution of the fraction of households belonging to a
      given class and that live in a neighbourhood where it is
      overrepresented (Middle, Lower, or Higher).} 
        \label{fig:content} 
\end{figure}

If a given class is overrepresented in a neighbourhood, it does not however mean
that most of the individuals belonging to this class live in this neighbourhood.
We compute the ratio of households of each income class that lives in a
neighbourhood over the total number of individuals in the income class (for
rich, poor, and middle class).  Results (Figure~\ref{fig:content}) indicate that
essentially less than $50\%$ of each class live in their respective neighbourhood, while
the rest is dispatched over the rest of the city. The average concentration
decreases from higher-income individuals ($50\%$), to lower-income ($48\%$) and
middle-income individuals ($32\%$).\\

\subsection{One large neighbourhood, or several small ones?}
\label{sub:one_large_neighbourhood_or_several_small_ones_}

\begin{figure} 
    \includegraphics[width=0.7\textwidth]{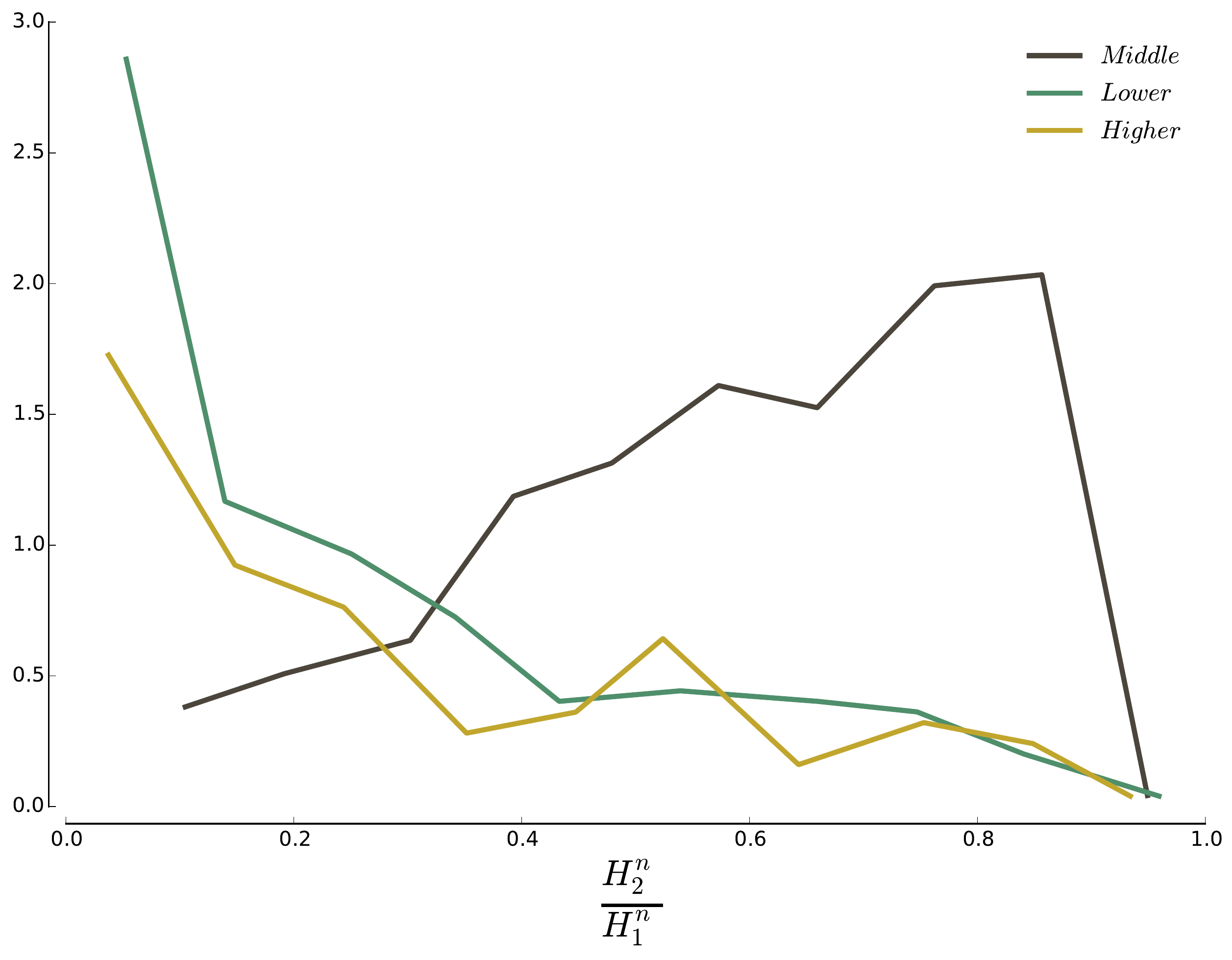}\\
    \caption{{\bf Poly-neighbourhoods. } Distribution of the ratio of the size of the
        largest and second largest neighbourhoods for each class for all MSA in
        the US. Higher- and lower-income househols tend to concentrate in single
        neighbourhood, with a secondary center that is on average $22\%$ and
        $26\%$ the size of the largest one, respectively. Middle-income
        households tend to be more dispersed, with a secondary neighbourhood that is on
        average $62\%$ of the size of the largest.} 
        \label{fig:polycentrism} 
\end{figure}

Finally, large values of clustering can hide different situations. We
could have on one hand a `giant' neighbourhood and several isolated areal units, which
would essentially mean that each class concentrates in a unique
neighbourhoods. Or on the other hand, several neighbourhoods of
similar sizes, meaning that the different classes concentrate in
several neighbourhoods across the city. In order to distinguish
between the two situations, we plot

\begin{equation} 
    P = H_{2}^N / H_{1}^N 
\end{equation}

where $H_{1}^N$ is the population of the largest neighbourhood, and $H_{2}^N$
the population of the second largest neighbourhood. The results are shown on
Figure~\ref{fig:polycentrism}, and again show a different behaviour for the
middle-income on one side, and higher-income and lower-income on the other side.
The size of the middle-income neighbourhoods are relatively balanced, with on average
$P=0.62$.  Higher- and lower-income neighbourhoods, on the other hand, are
dominated by one big neighbourhood, with respectively $P=0.22$ and
$P=0.26$ on average. 

 \subsection{Scaling of the number of neighbourhoods}
       \label{ssub:dependence_on_city_size}
       
\begin{figure}
    \centering
    \includegraphics[width=\textwidth]{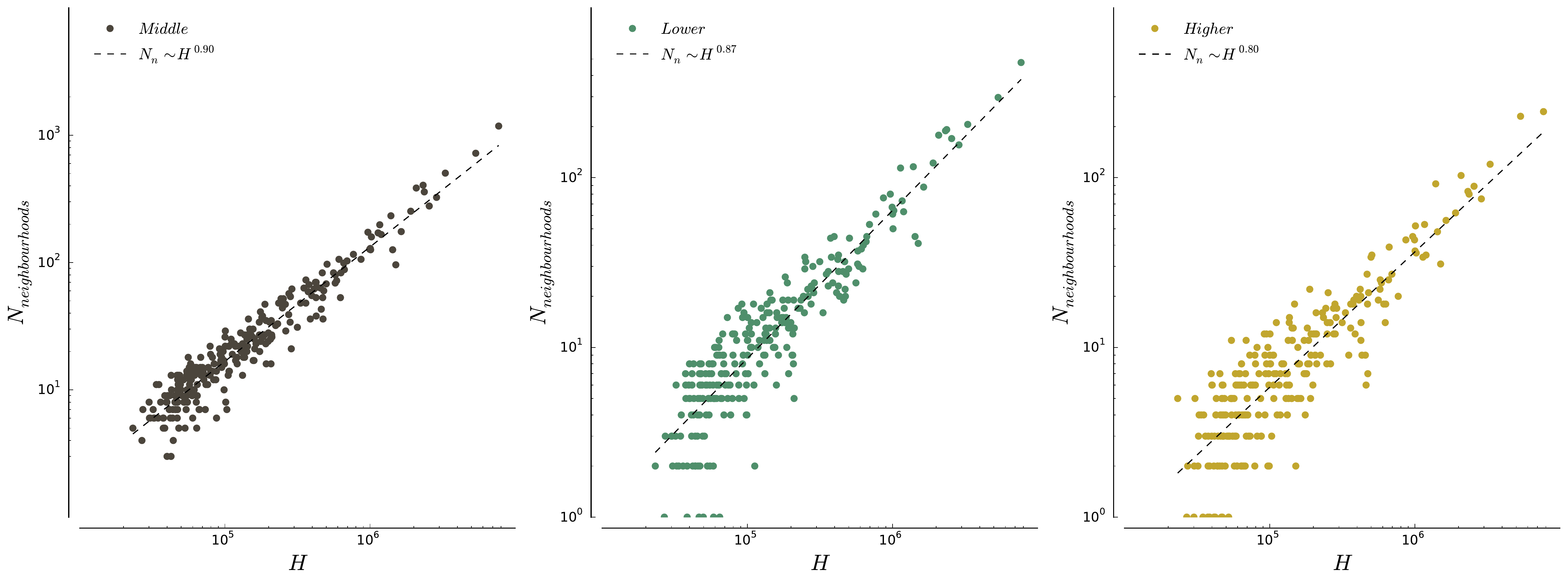}
    \caption{{\bf Number of neighbourhoods and city size.} Number of neighbourhoods for the three different classes as a
    function of the size of the city. These plots in loglog show that
    we have a behavior consistent with a power law with exponent less
    than one (and with different value for each class), with $r^2$ values that
    range between $0.88$ (higher-income) and $0.96$ (middle-income). Combined with the linear
    increase of the number of over-represented units with the number of
    households, this sub-linear increase in the number of neighbourhoods shows the tendency of
classes to cluster more as cities get larger.\label{fig:number_clusters_class}}
\end{figure}

The clustering values are high, indicating that the neighbourhoods occupied by
households of different classes are very coherent. We can now wonder whether
there is an effect of the city size on the number of neighbourhoods. We plot on
Figure~\ref{fig:number_clusters_class} the number of neighbourhoods found for all
three classes as a function of population. For each class, The curve is
well-fitted by a powerlaw function of the form

\begin{equation}
    N_n = b\,H^\beta
\end{equation}

where the exponent $\beta$ is less than one and depends on the class, 
indicating that there are proportionally less neighbourhoods
in larger cities (the number areal units scales proportionally with the
population size). The values of the exponents are

\begin{align*}
    \beta_{H} = 0.80\\
    \beta_{L} = 0.87\\
    \beta_{M} = 0.90\\
\end{align*}

\begin{figure}
    \centering
    \includegraphics[width=\textwidth]{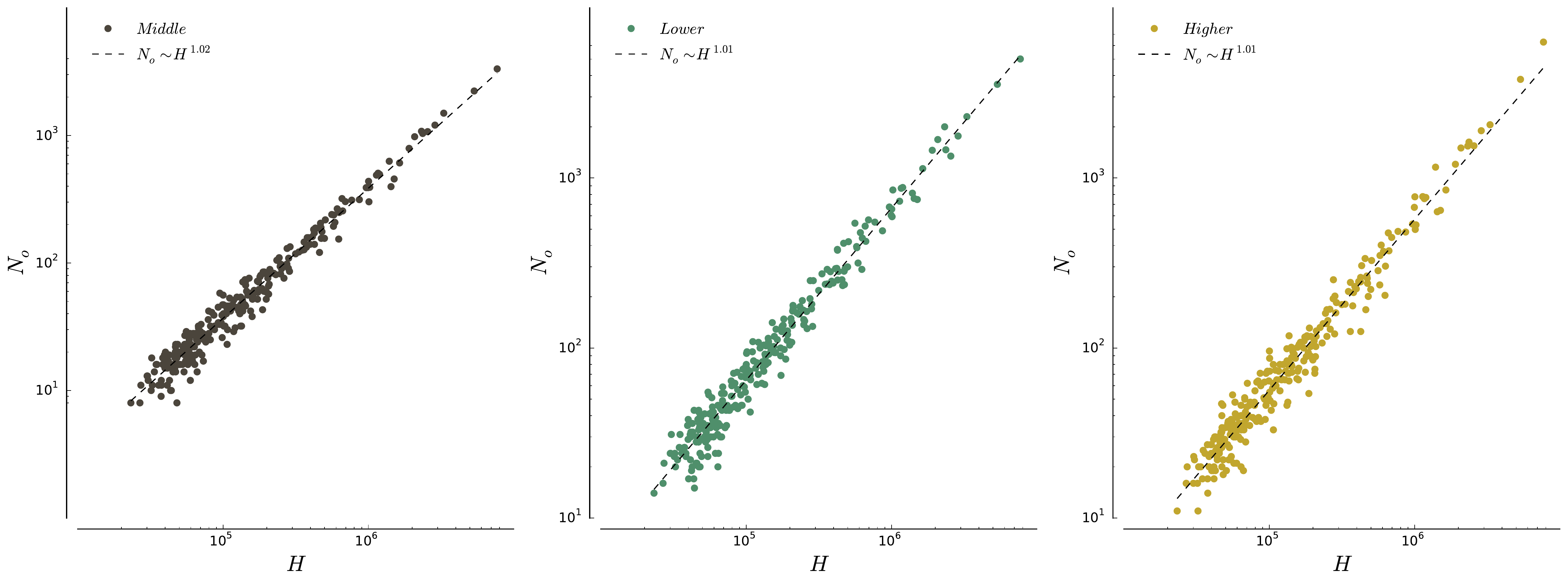}
    \caption{{\bf Number of overrepresented areal units and city size.} Number of areal units where each class is overrepresented as a
    function of the total number of households in the city. The behaviour is
consistent with a linear behaviour in the three cases.
\label{fig:overrepresented}}
\end{figure}

One is tempted to conclude from these numbers that the different classes
become more spatially coherent as the population increases. 
Yet, this conclusion only holds if the number of areal units in which each class
is overrepresented does not itself vary sublinearly with population size. We
plot on Figure~\ref{fig:overrepresented} these numbers as a
function of the size of the city. We find that the behaviour of the number of
overrepresented units is consistent with a linear behaviour for all three
classes. Together with the exponents above, this shows that the tendency
of the classes to cluster is greater as the city size increases.

In other words, the different classes are more spatially isolated as the city
size increases, implying higher levels of spatial segregation. We note that the
phenonemenon is more important for higher-income households than for lower- and
middle-income households, justifying to an extent the existence of the
expression `ghettos for the rich'.\\

\section{Poor centers, rich suburbs?}
\label{sec:poor_centers_rich_suburbs_}

In many studies, the question of the spatial pattern of segregation is limited
to the study of the center versus suburb and is usually adressed in two
different ways. First, a central area is defined by arbitrary boundaries and
measures are performed at the scale of the so-called center and at the scale of
the  rest, labelled as `suburbs'. The issue with this approach is that the
conclusions depend on the chosen boundaries and there is no unique unambiguous
definition of the city center: while some consider it to be the Central Business
District~\cite{Glaeser:2008}, others choose to define the center as the urban
core (urbanized area), where the population density is higher. The second
approach, in an attempt to get rid of arbitrary boundaries, consists in plotting
indicators of wealth as a function of distance to the
center~\cite{Glaeser:2008}. This approach, inspired by the monocentric and
isotropic city of many economic studies such as the Von Th\"unen or the
Alonso-Muth-Mills model~\cite{Brueckner:1987}, has however a serious flaw:
cities are not isotropic and are spread unevenly in space, leading to very
irregular shapes~\cite{Makse:1995}. Representing any quantity versus the
distance to a center thus amounts to average over very different areas and is necessarily misleading in
clear polycentric cases (as it is the case for large cities
~\cite{Louf:2013_polycentric}. See also
Chapter~\ref{chap:monocentric_introduction}). The notion of distance to the center is indeed
meaningless in polycentric situations.\\

We propose here a different approach that does not require the
definition of a distance to the center. Instead, we plot the average
representation computed over all areal units (Census blockgroups in this
dataset) with a given density population $\rho$, as a function of the density
$\rho$. Indeed, what is usually meant by `center' of a city are the areas with
the highest residential (or employment) densities. 

\begin{figure}
    \centering
    \includegraphics[width=\textwidth]{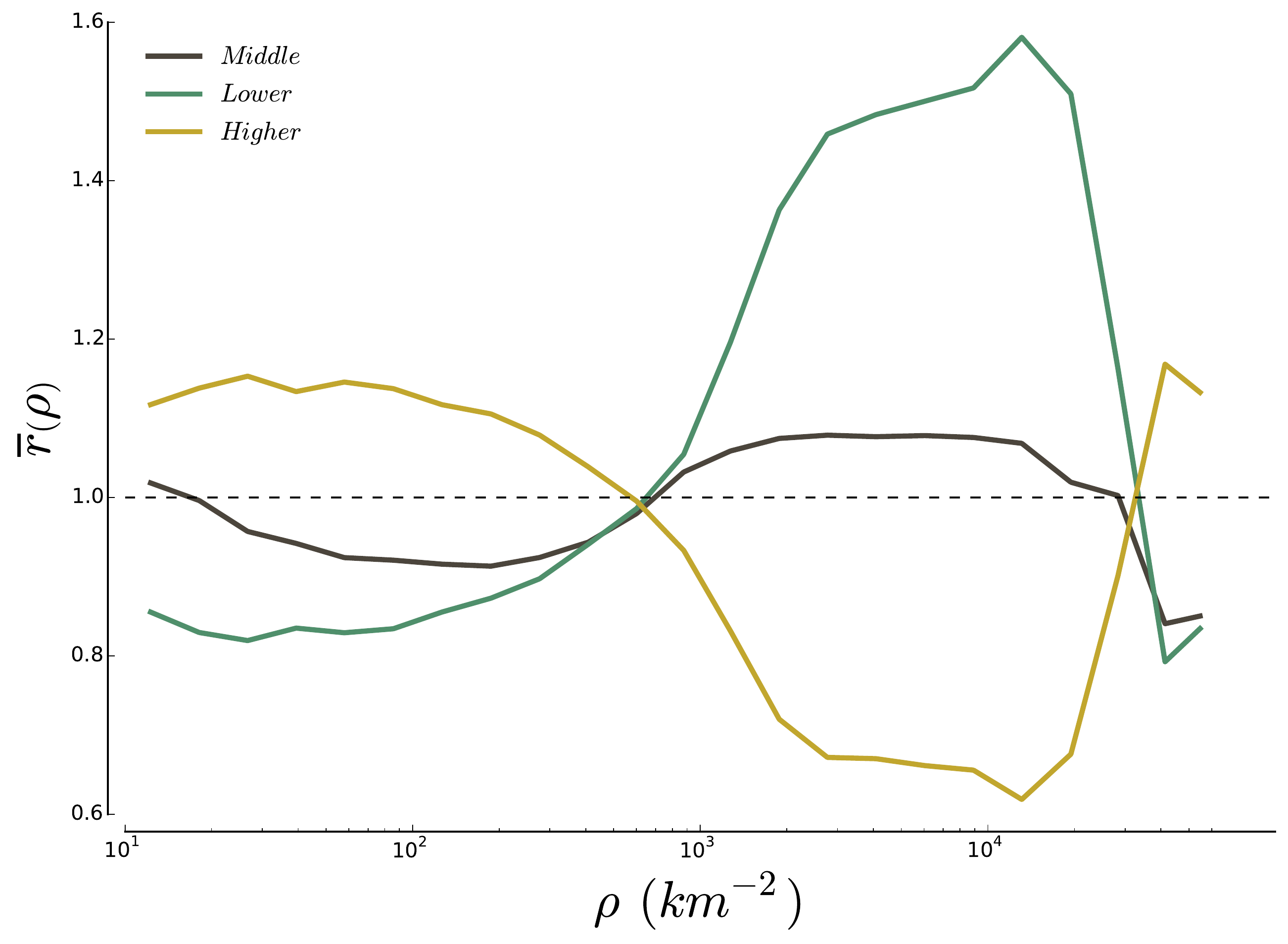}
    \caption{{\bf Representation and density.} Average representation of the higher-, middle- and
      lower-income classes over the $276$ MSA as a function of the
      local density of households. On average, we find that low-density regions (the
      suburbs) are rich, while high density regions (the center) are poor,
      confirming empirically on a large dataset a stylized facts that had previously
      emerged from local studies. Interestingly, we also
      find that  very large density areas ($\rho>20,000/km^2$) are rich on average,
      suggesting that density may be one relevant element in an eventual
      explanation of the differences between neighbourhoods~\cite{Jacobs:1961}.
      \label{fig:high_low_densities}
  } 
  \end{figure}

Our findings shed a new light on the difference of social composition
between the high-density and low-density areas in cities. As shown on
Figure~\ref{fig:high_low_densities}, we find that rich households are
overrepresented in low-density regions on average. While this agrees well with
the opinion people have of suburbian America, there is a more surprising result:
higher income households are also overrepresented in areas with very large densities
(typically above $20,000$ inhabitant$/\text{km}^2$). In between, neighborhoud
with intermediate values of density (between $1,000$ and $20,000$
inhabitants$/\text{km}^2$), are lower-income neighbourhoods. 

Only few cities in the US have neighbourhoods that reach the threshold of
$20,000$ inhabitants per km$^2$, which can explain why we observe in most cases
poor centers and rich suburbs. We can wonder whether the difference usually
discussed between North American and European cities does not come, in fact,
from differences in terms of densities. 

\section{Conclusion and perspective}
\label{sec:conclusion_and_perspective}

Instead of attempting to define segregation by enumerating its different
aspects, we took a radically different -- yet simpler -- approach. We chose to
define segregation through specifying what it is not.  This naturally lead to
defining the measure of representation, which is used in turn to delineate
neighbourhoods. We further defined the exposure (still based on the
representation), which measures the extent to which different categories
attract, repel or are indifferent to one another.

We then showed that we can define classes in a non-parametric way and 3 main
income classes emerged for the 2000 US Census data. The middle-income class
corresponds to a smaller income range than what is usually admitted, a curiosity
that certainly deserves further investigations. In terms of spatial arrangement,
although the fraction of the population that is contained in neighbourhoods does
not change with city size, the neighbourhoods are geographically more coherent
as cities get larger, which corresponds in effect to an increased level of
segregation as the size of the city increases. The behaviors of different
categories are very coherent and we showed that we could simplify the
description of these complex systems by reducing the sometimes large number of
categories to a small number of classes. This is an important point which will
simplify the description and modeling of stratification mechanisms.

Our results point to the intriguing fact that higher-income households are on
average overrepresented in very dense areas. Such high density areas are
relatively rare in the US, which might explain in part why authors have
traditionally simplified the picture, talking about poor
centers and rich suburbs. This result echoes Jane Jacobs' analysis~\cite{Jacobs:1961}
that neighbourhoods with the highest dwelling densities usually are the ones
exhibiting the most vitality, and therefore the most attractive. Of course, high
densities are not everything, and some high-density neighbourhoods also are
lower-income neighbourhoods. Further investigations along these lines may
provide quantitative insights into the mechanisms leading to urban decline or
urban regeneration.\\

In this Chapter, we have tried to highlight the \emph{spatial} pattern of
segregation.  We believe that the identification of neighbourhoods that our
method permits will allow a finer-scale investigation of these spatial patterns.
The fundamental issue that runs beneath, however,  is the need for a useful,
simplified description of spatial density. A problem yet to be solved, but that
has a huge potential of applications. We note that the problem is tightly
linked, if not identical, to the one we encountered while trying to describe the
spatial distribution of density in Chapter~\ref{chap:monocentric_discussion}.


\ctparttext{People, energy, information and goods are carried through cities
    (and across systems of cities) thanks to various networks. In this part, we
    succintly present our work on these---spatial---networks. 
    
    We first propose a quantative method to classify cities that is based on a new
    perspective on street patterns, and the use of the OpenStreetMaps database. In
    the second chapter, we propose a model for the growth of spatial networks
    based on cost-benefit analysis. The resulting networks exhibit a crossover
    between the star graph and the minimum spanning tree when the ratio of costs
    and benefits evolve. In the intermediate regime, the networks adopt a
    hub-and-spoke hierarchical structure that has many interesting properties.
    We conclude this part with a large-scale description of subway and
    railway networks. Using the model presented in the previous chapter, we are
    able to predict many of their properties based on the characteristics of the
underlying city or country.}

\part{Urban Networks} 
\label{part:networks}

\chapter{A typology of street patterns}
\label{chap:typology}

The following chapter is a reprint of an article, \emph{A typology of street
patterns}, that was previously published by the author of this
thesis with Marc Barthelemy~\cite{Louf:2014}.\\

Street networks of cities can be thought as a simplified schematic view of
cities, which however captures a large part of their structure and organization
\cite{Southworth:2003}.  Despite their apparent diversity, underlying universal
mechanisms are certainly at play in the formation and evolution of street
networks and extracting common patterns between cities is a way towards their
identification. This program is not new \cite{Haggett:1969}, but the
recent dramatic increase  of data availability such as digitized maps,
historical or
contemporary~\cite{Strano:2012,Barthelemy:2013,Porta:2014}\graffito{OpenStreetMap
data are freely available at \url{www.openstreetmap.org}} allows now to test
ideas and models on large scale cross-sectional and historical data.\\

Streets form a network which to a good approximation is planar (where nodes are
intersections and links are segment roads) and which is now fairly well
characterized
\cite{Jiang:2004,Rosvall:2005,Porta:2006,Porta:2006b,Lammer:2006,Crucitti:2006,Cardillo:2006,Xie:2007,Jiang:2007,Masucci:2009,Chan:2011,Courtat:2011}.
Due to spatial constraints, the degree distribution is peaked, the clustering
coefficient and assortativity are large, and most of the interesting information
lies in the spatial distribution of betweeenness centrality
\cite{Barthelemy:2011}. It is then tempting to use this information to compare
various cities with each other and to provide a classification. \\

The problem, from a fundamental point of view is however difficult: finding a
typology of street patterns amounts essentially to classify planar graphs, a non
trivial problem. For street networks, this problem has been addressed by the
space syntax community \cite{Hillier:1984,Penn:2003} and a good account can be found in
the book by Marshall~\cite{Marshall:2004}. These works, although based on
empirical observations, contain a large part of subjectivity and our goal is to
eliminate this subjective part to reach a non ambiguous, scientific
classification of these patterns. An interesting direction was provided in the
study of leaves and their classification according to their veination patterns
\cite{Katifori:2012,Mileyko:2012}, but with a notable difference which prevents us
from a direct application to streets and which is the existence of a hierarchy
of veins governed by their diameter. From a mathematical point of view there
exists an exact bijection between planar graphs and trees \cite{Bouttier:2004} which
provides an interesting direction. Using this bijection, classifying planar
graphs would amount to classify trees, which is a simpler problem. However, this
bijection does not take into account the geometrical shape of the planar graph:
indeed two street patterns can have the same topology but cells could be of very
different areas, leading to patterns visually different and to cities of
different structure. It is thus important to take into account not only the
topology of the planar graph --- as described by the adjacency matrix --- but
also the position of the nodes. In order to do that, we propose in this article,
a method to characterize this complex object by extracting the `fingerprint' of
a street pattern. These fingerprints allow us to define a measure of the
distance between two graphs and to construct a classification of cities.

\section{Streets versus blocks}

\begin{figure}
    \centering
    \includegraphics[width=0.7\textwidth]{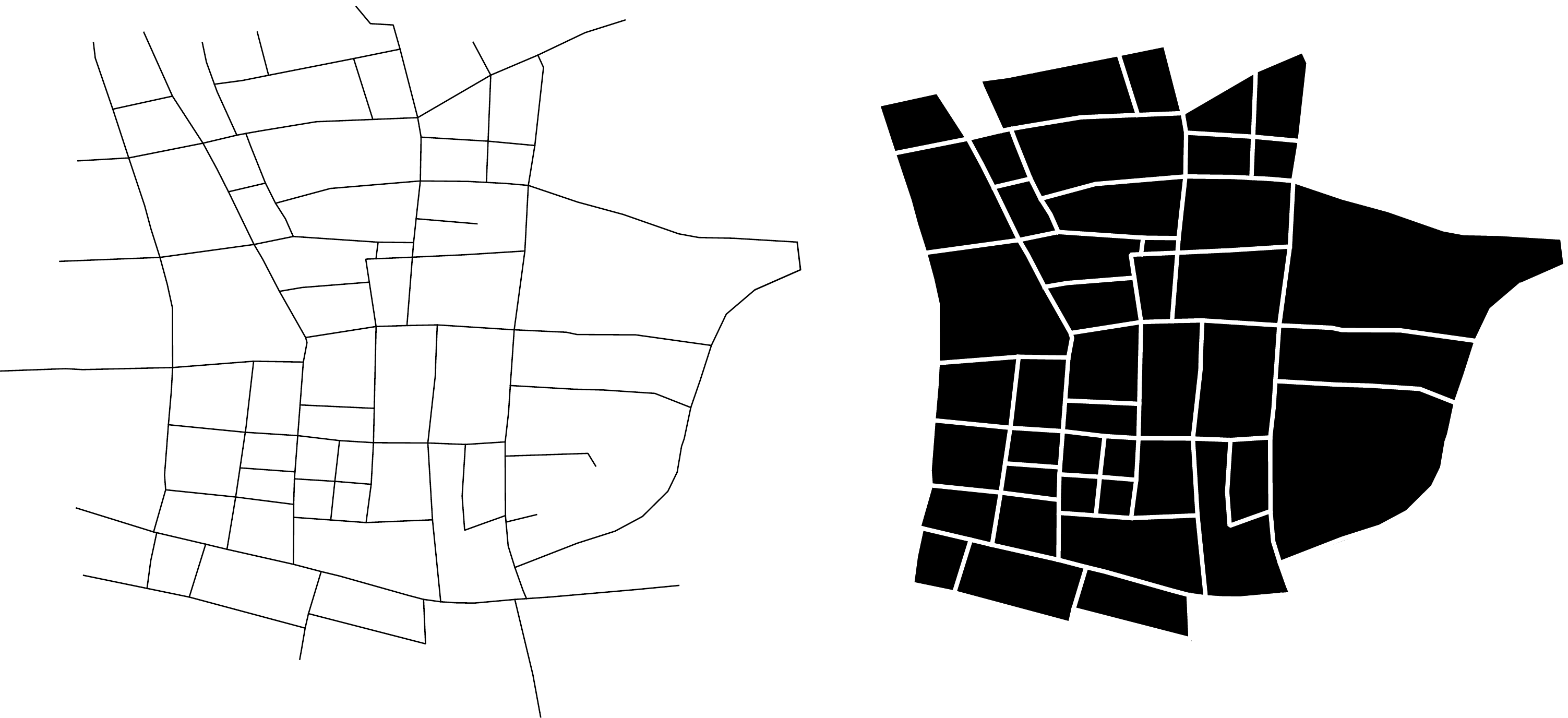}
    \caption{{\bf From the street network to blocks.} Example of a street pattern
    taken in the neighbourhood of Shibuya in Tokyo (Japan) and the corresponding set
    of blocks. Note that the block representation does not take into account
    dead-ends.  \label{fig:example}} 
\end{figure}

A major shortcoming of existing classifications  is that they are mostly based
on the street network. This is however problematic, for two different reasons.
First, there is no unambiguous, purely geometrical definition of what a street
is: we could define it as the road segment between two intersections, as an
almost straight line (up to a certain angular tolerance, see \cite{Porta:2006}), or we
could also follow the actual street names. There is a certain degree of
arbitrariness in each of these definitions, and it is not clear how robust a
classification based on streets would be. Second, it seems that what is
perceived by the human eye of a city map is not coming from streets but from the
distribution of the shape, area and disposition of blocks (see
Fig.~\ref{fig:example}). 

A natural idea when trying to classify cities is thus to focus on blocks (or
cells, or faces) rather than streets. A block can usually be defined without
ambiguity as being the smallest area delimited by roads (it has then to be
distinguished from a parcel which is a tax related definition). While the
information contained in the blocks and the streets are equivalent (up to
dead-ends), the information related to the visual aspect of the street network
seems to be easier to extract from blocks which are simple geometrical objects
--- polygons --- whose properties are easily measured. The block seems then to
be a good candidate for attempting a classification of city patterns.

\section{Characterizing blocks}

\begin{figure}
    \includegraphics[width=\textwidth]{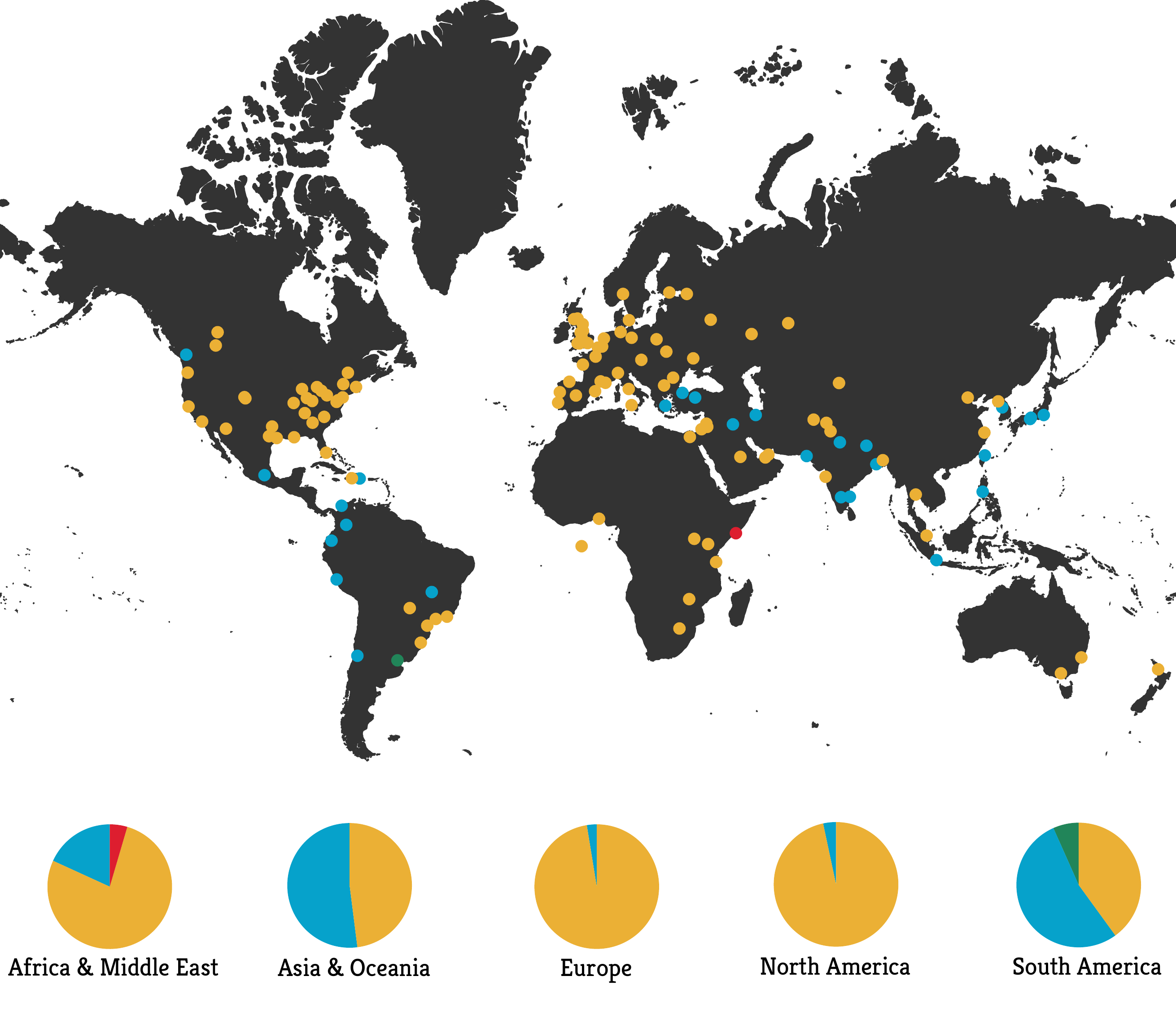}
    \caption{{\bf Location of the cities in our dataset and geographical repartition
    of the different groups.} The color of the dots indicates in which group the
    city falls, as defined on Fig~\ref{fig:groups}. On the bottom of the map, the
    pie charts display the relative importance of the different groups per continent
    for cities in our dataset (Group 1: 0.8\%, Group 2: 20.9\%, Group 3: 77.5\%,
    Group 4: 0.8\%). We see that the group $3$, composed of cities with blocks of
    various shapes and a slight predominance of larger areas is by far the most
    represented group in the world. \label{fig:world_map}} 
\end{figure}

Blocks are defined as the cells of the planar graph formed by streets, and it is
relatively easy to extract them from a map. We have gathered road networks for
$131$ major cities accross the world, spanning all continents (but Antartica),
and their locations are represented on the map Fig.~\ref{fig:world_map}. The
street networks have been obtained from the OpenStreetMap database,
and restricted to the city center using the Global Administrative Areas database
(or databases provided by the countries administration). We extracted the blocks
from the street network and cleaned undesired features. We end up with a set of
blocks, each with a geographical position corresponding to their centroids. 

\begin{figure}
    \center
    \includegraphics[height=2.1in]{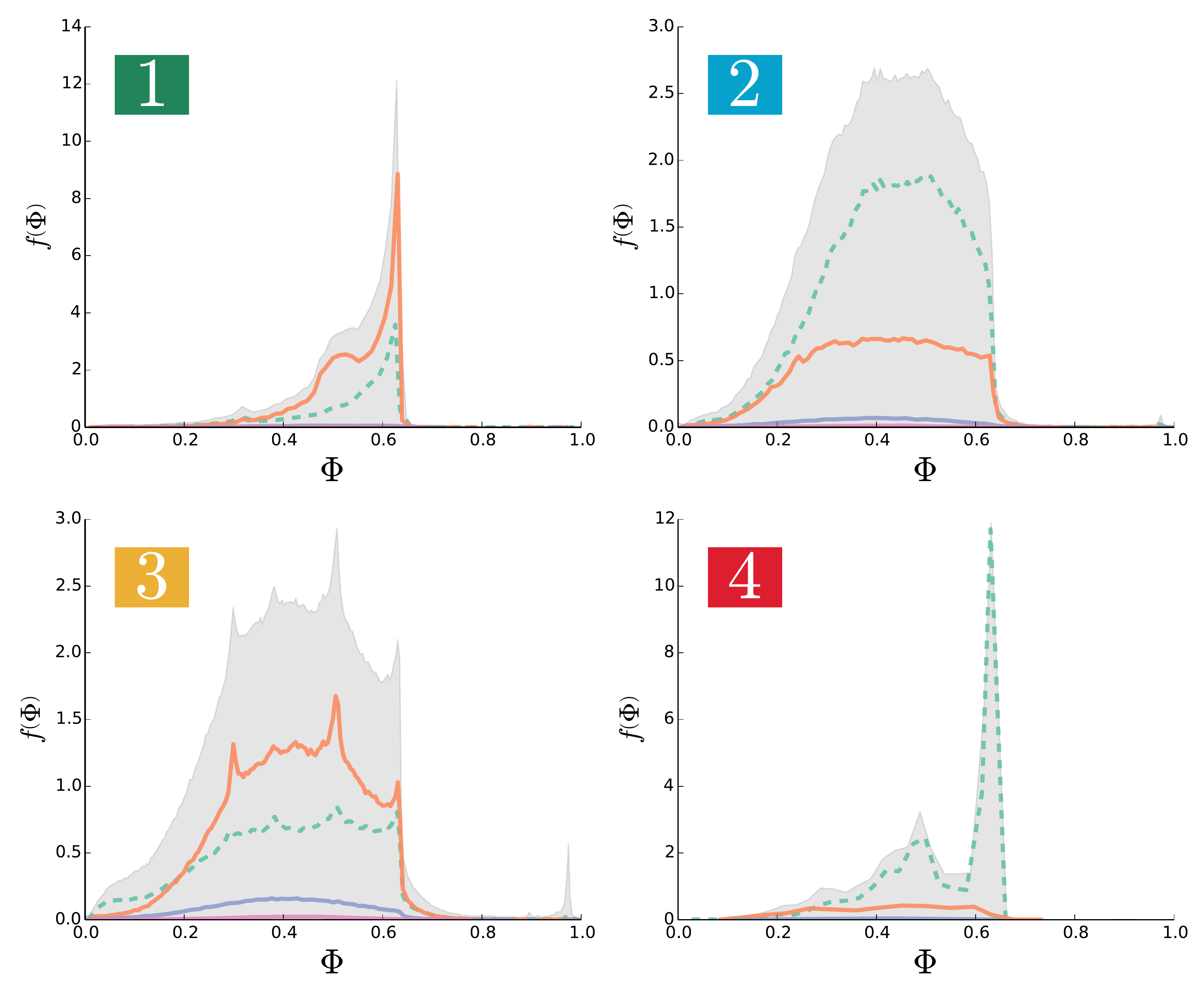}
    \includegraphics[height=2.1in]{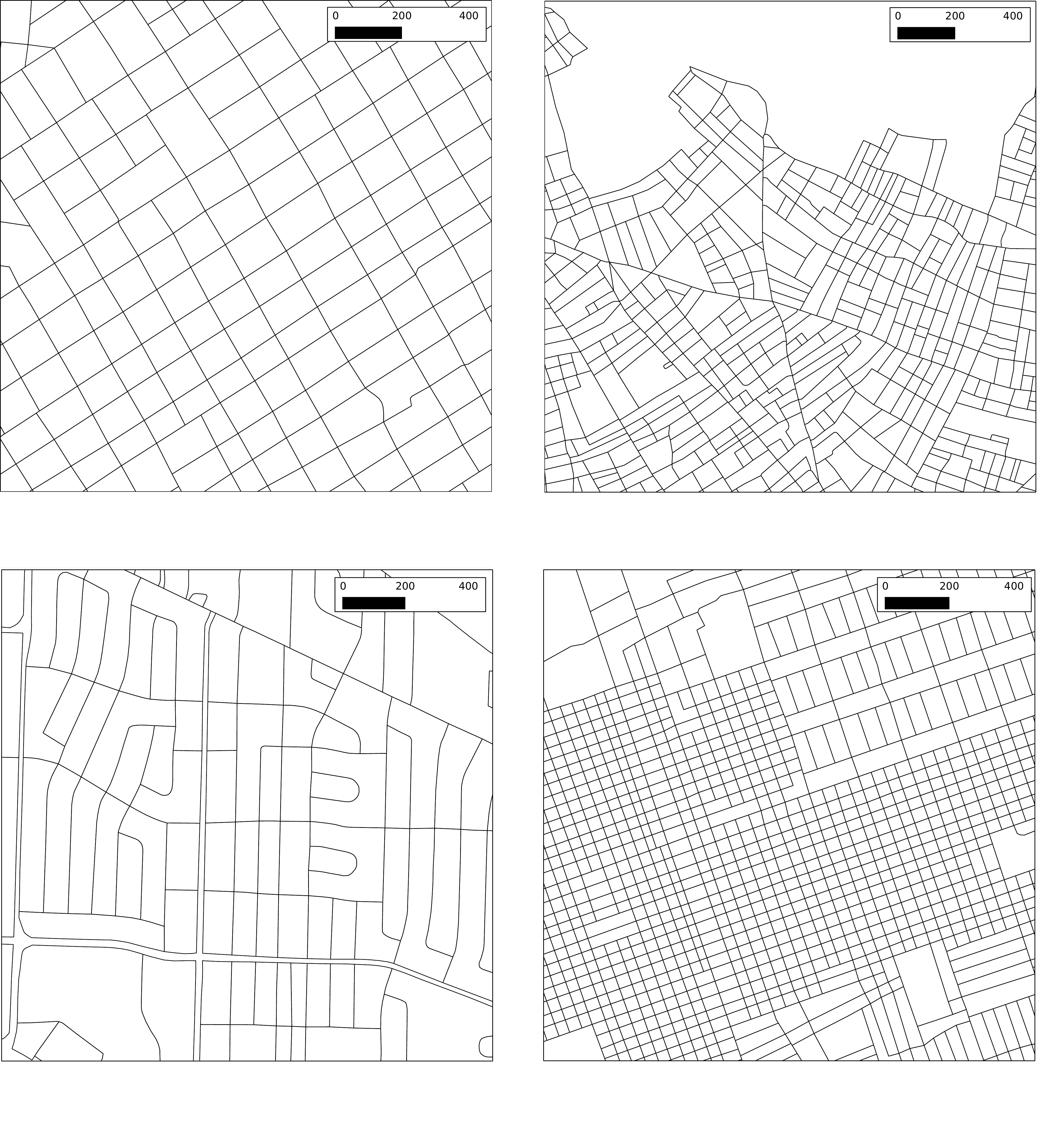}
    \caption{{\bf The four groups.} (Left) Average distribution of the shape
    factor $\Phi$ for each group found by the clustering algorithm (Right) Typical
    street pattern for each group (plotted at the same scale in order to observe
    differences both in shape and areas). Group 1 (top left): Buenos Aires | Group
    2: Athens | Group 3: New Orleans | Group 4: Mogadishu \label{fig:groups}}
\end{figure}

Blocks are polygons and as such can be characterized by simple measures. First,
the surface area $A$ of a block gives a useful indication, and its distribution
is an important information about the block pattern. As
in~\cite{Lammer:2006,Fialkowski:2008}, we find that for different cities the
distributions have different shapes for small areas, but display fat tails
decreasing as a power law \begin{equation} P(A)\sim \frac{1}{A^\tau}
\end{equation} with an exponent of order $\tau\approx 2$
\cite{Lammer:2006,Barthelemy:2011,Strano:2012,Barthelemy:2013}.
Although this seemingly universal behaviour gives a useful constraint on any
model that attempts at modeling the evolution of cities' road
networks, it does not allow to distinguish cities from each
other.

\begin{figure}
    \center
    \includegraphics[width=\textwidth]{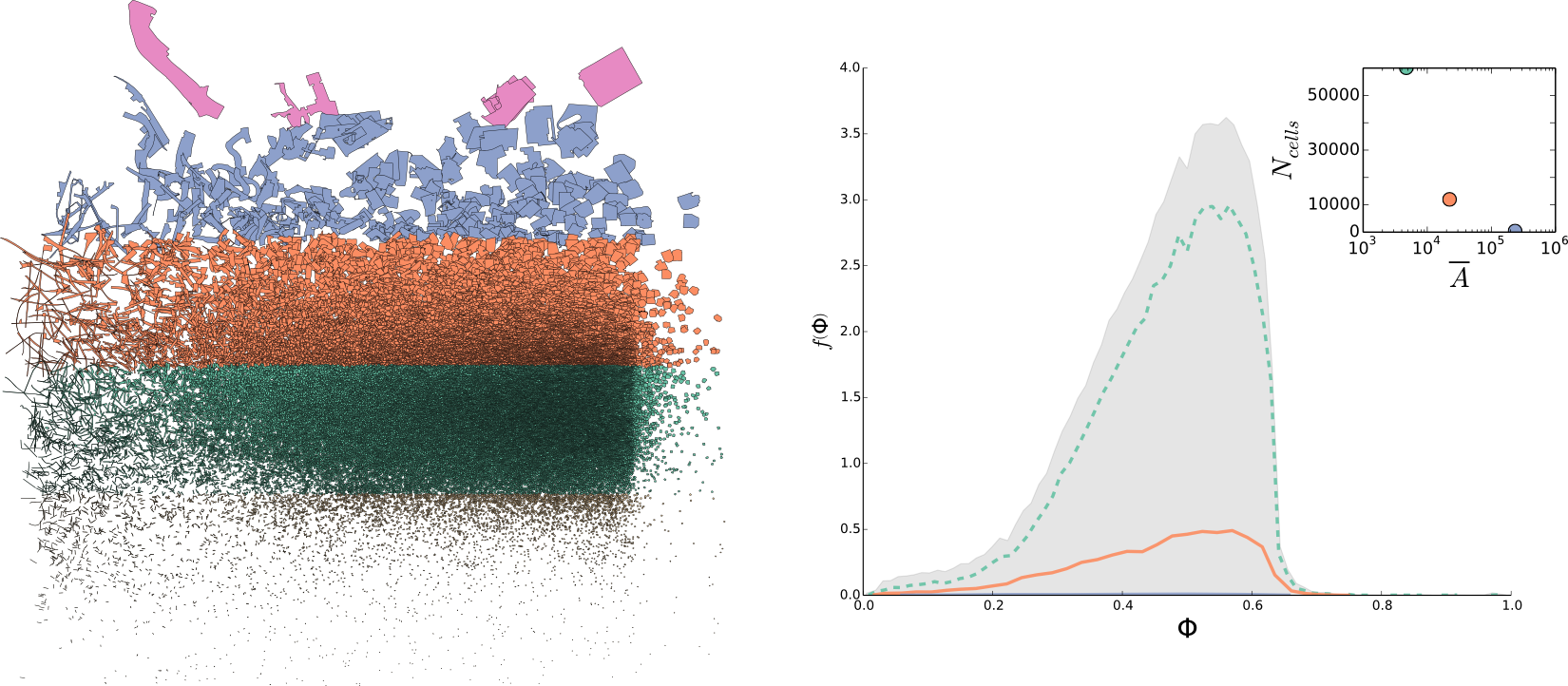}
    \includegraphics[width=\textwidth]{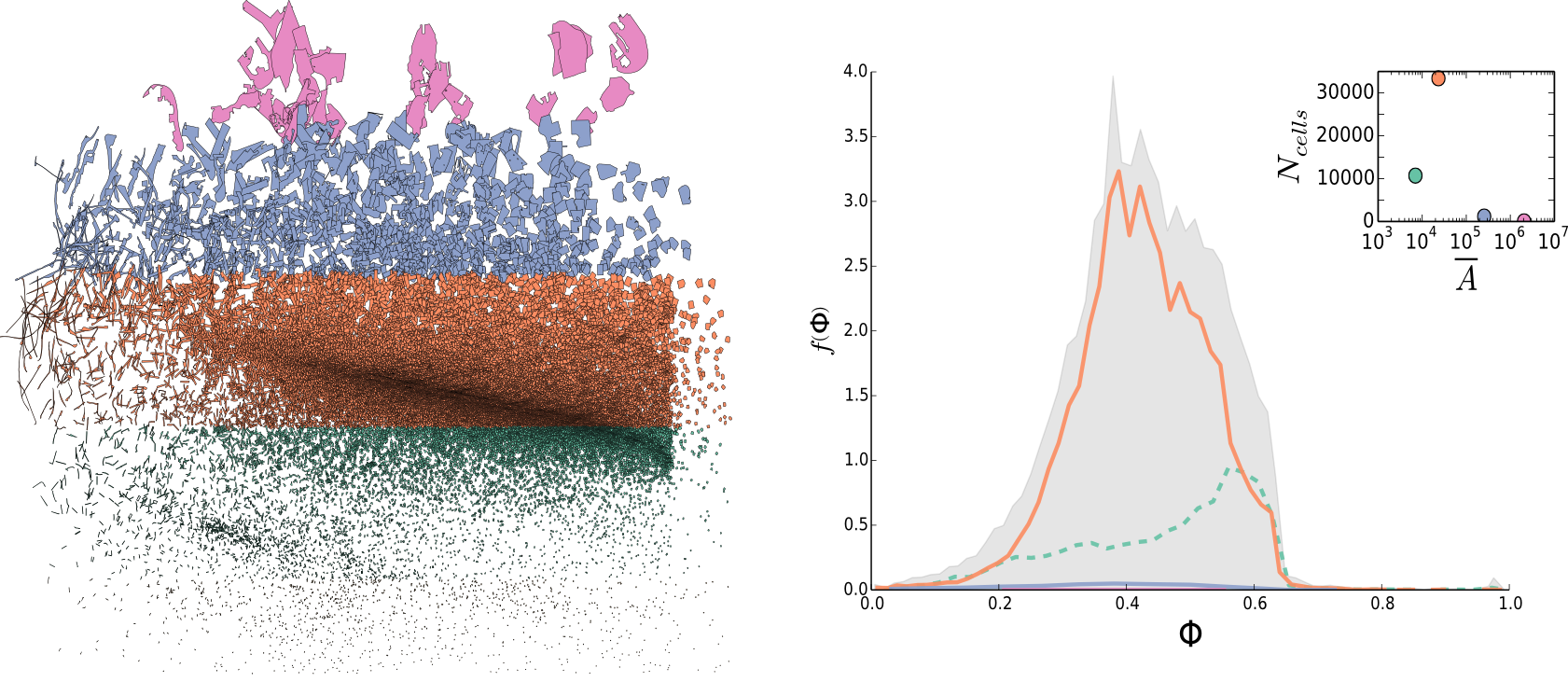}
    \caption{{\bf The fingerprints of Tokyo (top) and New-York, NY (bottom).}
    (Left) We rearrange the blocks of a city according to their area (y-axis), and
    their  $\Phi$ value (x-axis). The color of each block corresponds to the area
    category it falls into.  (Right) We quantify this pattern by plotting the
    distribution of shapes, as measured by $\Phi$ for each area category,
    represented by coloured curves. The gray curve is the sum of all the coloured
    curve and represents the distribution of $\Phi$ for all cells. As shown in the
    inset, we see that intermediate area categories dominate the total number of
    cells, and are thus enough for the clustering procedure.\label{fig:fingerprint}}
\end{figure}

A second characterization of a block is through its shape, with the form (or
shape) factor $\Phi$, defined in the Geography literature in~\cite{Haggett:1966}
as the ratio between the area of the block and the area of the circumscribed
circle $\mathcal{C}$
\begin{equation} \Phi = \frac{A}{A_{\mathcal{C}}} \end{equation}
The quantity $\Phi$ is always smaller than one, and the smaller its value, the
more anisotropic the block is. There is not a unique correspondence between a
particular shape and a value of $\Phi$, but this measure gives a good indication
about the block's shape in real-world data, where most blocks are relatively
simple polygons. The distributions of $\Phi$ displays important differences from
one city to another, and a first naive idea would be to classify cities
according to the distribution of block shapes given by $P(\Phi)$. The shape
itself is however not enough to account for visual similarities and
dissimilarities between street patterns. Indeed, we find for example that for
cities such as New-York and Tokyo, even if we observe similar distributions
$P(\Phi)$ (see Fig.~\ref{fig:fingerprint}), the visual similarity between both
cities's layout is not obvious at all. One reason for this is that blocks can
have a similar shape but very different areas: if two cities have blocks of the
same shape in the same proportion but with totally different areas, they will
look different.  We thus need to combine the information about both the shape
and the area.

In order to construct a simple representation of cities which integrates both
area and shape, we rearrange the blocks according to their area (on the y-axis)
and display their $\Phi$ value on the x-axis (Fig.~ \ref{fig:fingerprint}). We
divide the range of areas in (logarithmic) bins and the color of a block
represents the area category to which it belongs. We describe quantitatively
this pattern by plotting the conditional probability distribution $P(\Phi|A)$ of
shapes, given an area bin (Fig.~ \ref{fig:fingerprint}, right). The colored
curves represent the distribution of $\Phi$ in each area category, and the curve
delimited by the gray area is the sum of all the these curve and is the
distribution of $\Phi$ for all cells, which is simply the translation of the
well-known formula for probability conditional distribution

\begin{equation} 
    P(\Phi)=\sum_AP(\Phi|A)\,P(A) 
\end{equation}

These figures give a `fingerprint' of the city which encodes information about
both the shape and the area of the blocks. In order to quantify the distribution
of blocks inside a city, and thus the visual aspect of the latter, we will then
use $P(\Phi|A)$ for different area bins. The comparison between these quantities
will provide the basis for a classification of street patterns that we propose
here.

\section{A typology of cities across the world}

Two cities will display similar patterns if their blocks have both similar area
and shape. In other words, the shape distributions for each area bin should be
very close, and this simple idea allows us to propose a distance between street
patterns of different cities. More precisely, as one can see on
Fig.~\ref{fig:fingerprint}, the number of blocks of area in the range
$[10^3,10^5]$ (in square meters) dominate the total number of cells, and we will
neglect very small blocks (of area $<10^3\text{m}^2$) and very large ones (of
area $>10^5\text{m}^2$). We thus sort the blocks according to their area in two
distinct bins

\begin{align*} 
    \alpha_1 &= \left\{ \text{cells}\, |\, \mathcal{A} \in \left[10^3,
10^4\right]\right\}\\ 
    \alpha_2 &= \left\{ \text{cells}\, |\, \mathcal{A} \in
\left[10^4, 10^5\right]\right\}\\ 
\end{align*}

We denote by $f_\alpha(\Phi)$ the ratio of the number of cells with a form
factor $\Phi$ that lie in the bin $\alpha$ over the total number of cells for
that city. We then define a distance $d_\alpha$ between two cities $a$ and $b$
characterized by their respective $f^{a}_\alpha$ and $f^{b}_\alpha$

\begin{equation} 
    d_\alpha(a,b) = \int_0^1\: | f^{a}_\alpha(\Phi) -
f^{b}_\alpha(\Phi) |\: \mathrm{d}\Phi 
\end{equation} 

and we construct a global distance $D$ between two cities by combining all area
bins $\alpha$

\begin{equation} 
    D(a,b)= \sum_\alpha d_\alpha(a,b)^{\,2} 
\end{equation} 

\begin{figure}
    \includegraphics[width=0.75\textwidth]{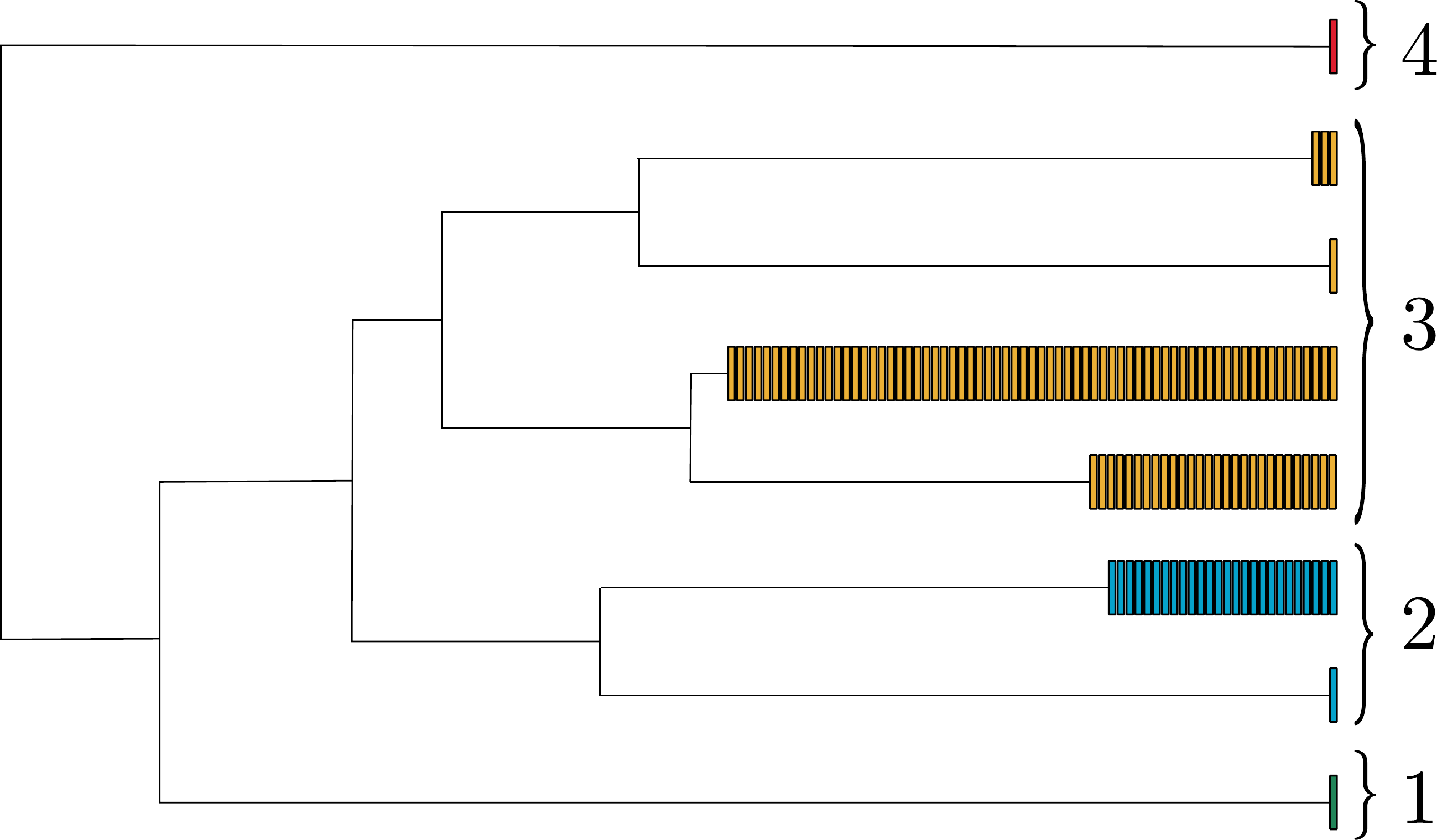}
    \caption{{\bf Dendrogram} We represent the structure of the hierarchical
    clustering at a given level. Interestingly, $68\%$ of american cities are
    present in the second largest sub-group of group $3$ (fourth from the top).
    Also, all european cities but Athens are in the largest subgroup of the group
    $3$ (third from top). This result gives a first quantitative grounding to the
    feeling that European and most American cities are laid out
    differently.\label{fig:dendrogram}} 
\end{figure}

At this
point, we have a distance between two cities' pattern and we measure the
distance matrix between all the $131$ cities in our dataset, and perform a
classical hierarchical clustering on this matrix \cite{Kaufman:2009}. We
obtain the dendrogram represented on Fig.~\ref{fig:dendrogram} and at an
intermediate level, we can identify $4$ distinct categories of cities, which
are easily interpretable in terms of the abundance of blocks with a given
shape and with small or large area. On Fig.~\ref{fig:groups} we show the
average distribution of $\Phi$ for each category and show typical street
patterns associated with each of these groups. The main features of each
group are the following.  

\begin{itemize} 
    \item{} In the group 1 (comprising
        Buenos Aires only) we essentially have blocks of medium size (in
        the bin $\alpha_2$) with shapes that are dominated by the square
        shape and regular rectangles. Small areas (in bin $\alpha_1$)
        are almost exclusively squares.  \item{} Athens is a
        representative element of group 2, which comprises cities with a
        dominant fraction of small blocks with shapes broadly distributed.
    \item{} The group 3 (illustrated here by New Orleans) is similar to the
        group 2 in terms of the diversity of shapes but is more balanced in
        terms of areas, with a slight predominance of medium size blocks.
    \item{} The group 4 which contains for this dataset the interesting
        example of Mogadishu (Somalia) displays essentially small,
        square-shaped blocks, together with a small fraction of small
        rectangles.  
\end{itemize}

The proportion and location of cities belonging to each group is shown on
Fig.~\ref{fig:world_map}. Although one should be wary of sampling bias here, it
seems that the type of pattern characteristic of the group $3$ (various shapes
with larger areas) largely dominates among cities in the world. Interestingly,
all North American cities (except Vancouver, Canada) are part of the group $3$,
as well as all European cities (except Athens, Greece). The composition of the
other continents is more balanced between the different groups. Strikingly, we
find that at a smaller scale within the group $3$ (Fig.~\ref{fig:dendrogram}),
all European cities (but Athens) in our sample belong to the same subgroup of
the group $3$ (the largest one, third from the top on
Fig.~\ref{fig:dendrogram}). Similarly, $15$ American cities out of the $22$ in
our dataset belong to the same subgroup of the group $3$ (the second largest
one, fourth from the top on Fig.~\ref{fig:dendrogram}. Exceptions are
Indianapolis (IN), Portland (OR), Pittsburgh (PA), Cincinnati (OH), Baltimore
(MD), Washington (DC), and Boston (MA), which are classified with European
cities, confirming the impression that these US cities have an european imprint.
These results point towards important differences between US and European
cities, and could constitute the starting point for the quantitative
characterization of these differences \cite{Bretagnolle:2010}.

\section{A local analysis}

\begin{figure}
    \includegraphics[height=4in]{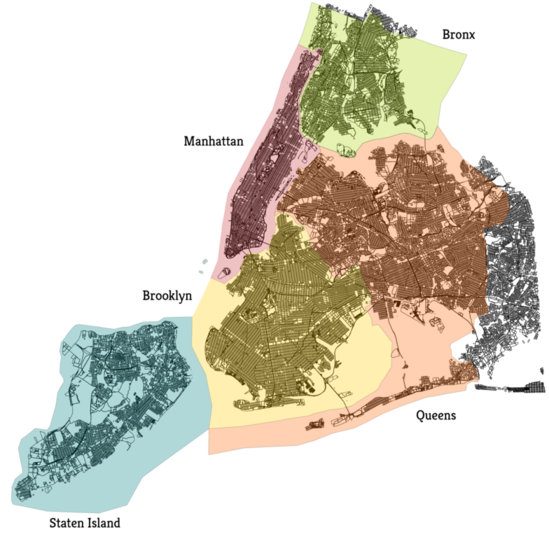}
    \includegraphics[width=\textwidth]{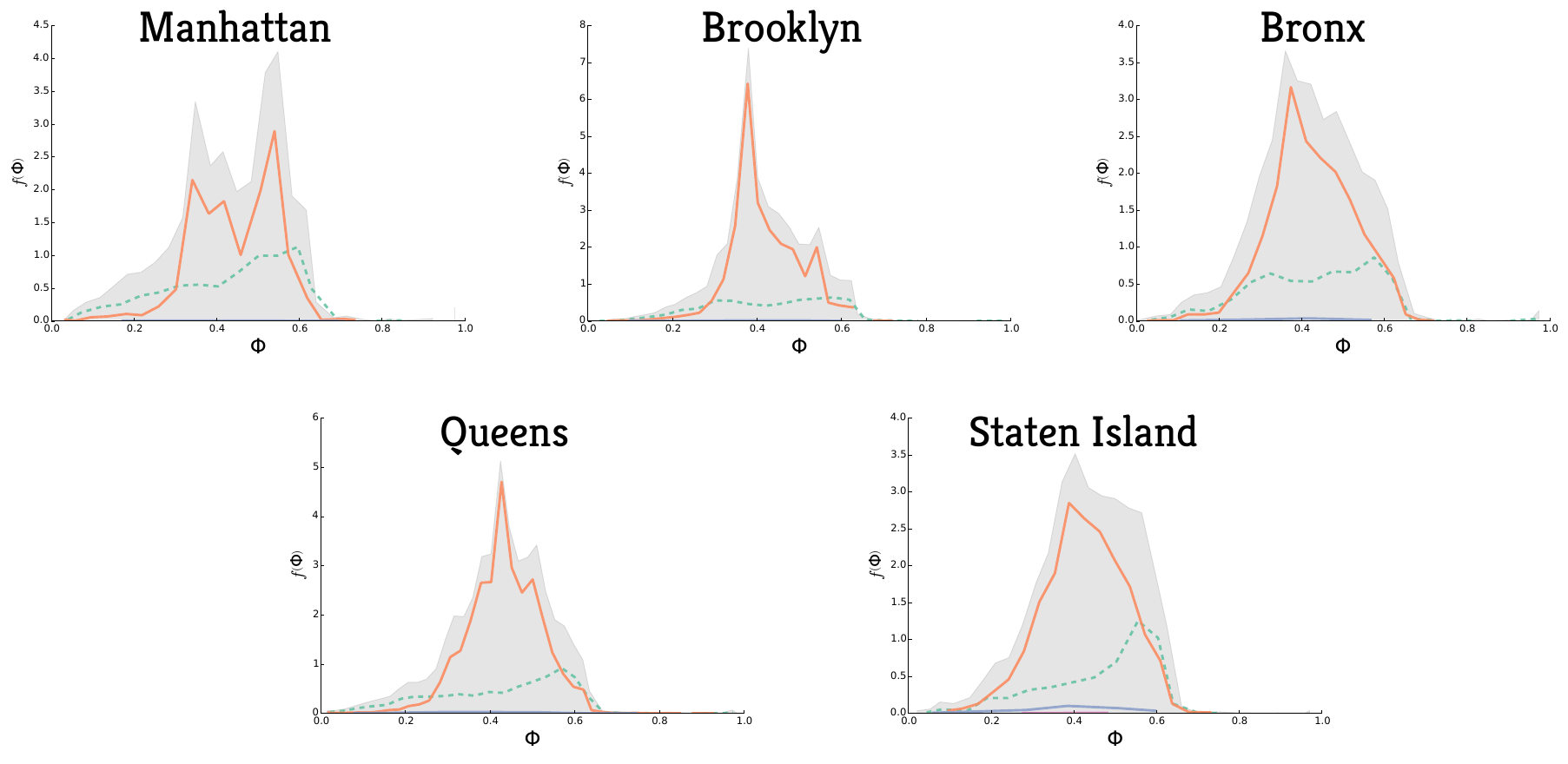}
    \caption{{\bf New-York, NY and its different boroughs} \label{fig:ny-boroughs}
    (Top) We represent New York City and its 5 boroughs: the Bronx, Brooklyn,
    Manhattan, Queens, and Staten Island. (Bottom) The corresponding fingerprints
    for each borough. Only Staten Island and the Bronx have similar fingerprints and
    the others are different. In particular,  Manhattan exhibits two sharp peaks at
    $\Phi \approx 0.3$ and $\Phi \approx 0.5$ which are the signature of a grid-like
    pattern with the predominance of two types of rectangles. Brooklyn and the
    Queens exhibit a sharp peak at different values of $\Phi$, signalling the
    presence of grid-like patterns made of different basic
    rectangles.\label{fig:boroughs}}
\end{figure}

Cities are complex objects, and it is unlikely that an object as simple as the
fingerprint can describe all its intricacies. Indeed, cities are usually made of
different neighbourhood which often exhibit different street patterns. In
Europe, the division is usually clear between the historical center and the more
recent surburbs. A striking example of such differences is the Eixample
neighbourhood in Barcelona, very distinct from other areas of the city. In order
to illustrate this difference, and to show that they also can be captured with
our method, we isolate the different Boroughs of New-York, NY: the Bronx,
Brooklyn, Manhattan, Queens and Staten Island. We extract the fingerprint of
each Borough, as represented on Fig.~\ref{fig:ny-boroughs}. The fingerprint of
New-York (bottom Fig.~\ref{fig:fingerprint}) is indeed the combination of
different fingerprints for each of the boroughs. While Staten Island and the
Bronx have very similar fingerprints, the others are different. Manhattan
exhibits two sharp peaks at $\Phi \approx 0.3$ and $\Phi \approx 0.5$ which are
the signature of a grid-like pattern with the predominance of two types of
rectangles. Brooklyn and the Queens exhibit a sharp peak at different values of
$\Phi$, also the signature of grid-like patterns with different rectangles for
basic shapes. 

\section{Discussion and perspectives}

We have introduced a new way of representing cities' road network that can be
seen as the equivalent of fingerprints for cities. It seems reasonable to think
that the possibility of a classification based on these fingerprints hints at
common causes behind the shape of the networks of cities in the same categories.
Of course, the present study has limitations: even if the shape of the blocks
alone is good enough for the purpose of giving a rough classification of cities,
we miss some aspects of the patterns. Indeed, the way the blocks are arranged
together locally should also give some information about the visual aspect of
the global pattern. Indeed, many cities are made of neighbourhoods, built at
different times, with different street patterns. What is lacking at this point
is a systematic, quantitative way to identify and distinguish different
neighbourhoods, and to describe their correlation. Indeed, the Boroughs taken as
examples in the last section are administrative, arbitrary definitions of a
neighbourhood. Reality is however more complex: similar patterns might span
several administrative regions, or a given administrative division might host
very distinct neighbourhoods. A further step in the classification would thus be
to find a method to extract these neighbourhoods, and integrate the spatial
correlations between different types of neighbourhoods.

Despite the simplifications that our method entail, we believe that the
classification we propose is an encouraging step towards a quantitative and
systematic comparison of the street patterns of different cities. This, together
with the specific knowledge of architects, urbanists, etc. should lead to a
better understanding of the shape of our cities. Further studies are indeed
needed in order to relate the various types that we observe to different urban
processes. For example, in some cases, small blocks are obtained through a
fragmentation process, and their abundance could be related to the age of the
city. A large regularity of cell shapes could be related to planning such as in
the case of Manhattan for example, but we also know with the example of Paris
\cite{Barthelemy:2013} that a large variety of shapes is also directly related
to the effect of a urban modification which does not respect the existing
geometry.\\

Finally, we believe that important empirical progress could be made. A first
limitation of the current study is the amount of data that we have. Although
$131$ cities is a larger number than what is used in most studies, the
OpenStreetMap database contains the street layout of many more cities. The more
cities we have, the better the classification. We should thus attempt to include
more cities.

The second limitation is the use of the administrative definition of cities to
delineate the boundaries of the street network. Although it is important to have
a large number of cities, it is at least as important to have a set of coherent,
similarly defined cities. Administrative definitions,
because they are based on political criteria, are completely arbitrary and do
not reflect any property of the contained networks. As a result, the chosen
boundaries are likely to vary from one country to another, from one city to
another. The measures we perform on each of the $131$ street patterns are thus, strictly
speaking, not comparable. A possible solution would be to use the delineation
method proposed by Masucci et al.~\cite{Masucci:2015}, which is parameter-free
and based only on the properties of the street network.
\chapter{Cost-benefit considerations in the growth of spatial networks}
\label{chap:cost-benefit}

The following chapter is a reprint of an article, \emph{Emergence of hierarchy
in the cost-driven growth of spatial networks}, that was previously published by the author of this
thesis with Pablo Jensen and Marc Barthelemy~\cite{Louf:2013_emergence}.\\

Our societies rely on various networks for the distribution of energy,
information and for transportation of individuals. These networks shape the
spatial organization of our societies and their understanding is a key step
towards the understanding of the characteristics and the evolution of our
cities~\cite{Batty:2013}. Despite their apparent diversity, these networks are
all particular examples of a broader class of networks --spatial networks--
which are characterised by the embedding of their nodes in space. As a
consequence, there is usually a cost associated with a link, leading to
particular structures which are now fairly well
understood~\cite{Barthelemy:2011}, thanks to the recent availability of large
sets of data. Nevertheless, the mechanisms underlying the formation and temporal
evolution of spatial networks have not been much studied. Different kinds of
models aiming at explaining the static characteristics of spatial networks have
been suggested previously in quantitative geography, transportation economics,
and physics (for a review, see ~\cite{Xie:2009}). Concerning the time
evolution of spatial networks, a few models only exist to describe in particular
the growth of road and rail networks~\cite{Levinson:2006, Gastner:2006,
Barthelemy:2008, Courtat:2011}, but a general framework is yet to be
discovered.\\

The earliest attempts can be traced back to the economic geography community in
the 60s and 70s (A fairly comprehensive review of these studies can be found
in~\cite{Haggett:1969}). However, due to the lack of available data and
computational power, most of the proposed models were based on intuitive,
heuristic rules and have not been studied thoroughly. Interestingly,
\cite{Black:1971} attempts to reproduce railway networks with the same
cost-benefits approach that will be adopted in the following.

A more recent trend is that of the optimization models. The common point between
all these models is that they try to reproduce the topological features of
existing networks, by considering the network as the realisation of the optimum
of given quantity (see section IV.E in~\cite{Barthelemy:2011} for an overview).
For instance, the hub-and-spoke models~\cite{OKelly:1998} reproduce correctly
with an optimization procedure the observed hierarchical organization of city
pair relations. However, the vast majority of the existing spatial networks do
not seem to result from a global optimization, but rather from the progressive
addition of nodes and segments resulting from a local optimization. By modeling
(spatial) networks as resulting from a global optimization, one overlooks the
usually limited time horizon of planners and the self-organization underlying
their formation.

Self-organization of transportation networks has already been studied in
transportation engineering~\cite{Levinson:2006, Xie:2009}. Using an agent-based
model including various economical ingredients, the authors
of~\cite{Levinson:2006} modeled the emergence of the networks properties as a
degeneration process. Starting from an initial grid, traffics are computed at
each time step and each edge computes its costs and benefits accordingly, using
any excess to improve their speed. After several iterations, a hierarchy of
roads emerges. Our approach is very different: we start from nodes and we do not
specify any initial network. Also, and most importantly, we deliberately do not
represent all the causal mechanisms at work in the system. Indeed, the aim of
our model is to understand the basic ingredients for emergence of patterns that
can be observed in various systems and we thus focus on a single, very general
economical mechanism and its consequence on the large-scale properties of the
networks.

Concerning spatial networks, as it is the case for many spatial structure, there
is a strong path dependency. In other words, the properties of a network at a
certain time can be explained by the particular historical path leading to it.
It thus seems reasonable to model spatial networks in an iterative way. Some
iterative models, following ideas for understanding power laws in the Internet
\cite{Fabrikant:2002} and describing the growth of transportation
networks~\cite{Gastner:2006} can be found in the literature. In these models,
the graphs are constructed via an iterative greedy optimization of geometrical
quantities. However, we believe that the topological and geometrical properties
of networks are consequences of the underlying processes at stake. At best,
geometrical and topological quantities can be a proxy for other --more
fundamental-- properties: for instance, it will be clear in what follows that
the length of an edge can be taken as a proxy for the cost associated with the
existence of that edge. Finding those underlying processes is a key step towards
a general framework within which the properties of networks can be understood
and, hopefully, predicted.\\

In this respect, cost-benefit analysis (CBA) provides a systematic method to
evaluate the economical soundness of a project. It allows one to appreciate
whether the costs of a decision will outweigh its benefits and therefore
evaluate quantitatively its feasibility and/or suitability. Cost-benefit
analysis has only been officially used to assess transport investments since
1960~\cite{Coburn:1960}. However, the concept comes accross as so intuitive in
our profit-driven economies that it seems reasonable to wonder whether CBA is at
the core of the emergent features of our societies such as distribution and
transportation systems. If the temporal evolution of spatial networks is rarely
studied, arguments mentioning the costs and benefits related to such networks
are almost absent from the physics litterature (\cite{Popovic:2012} is a notable
exception, although they do not consider the time evolution of the network.).
However, we find it intuitively appealing that in an iterative model, the
formation of a new link should --at least locally-- correspond to a cost-benefit
analysis. We therefore propose here a simple cost-benefit analysis framework for
the formation and evolution of spatial networks. Our main goal within this
approach is to understand the basic processes behind the self-organization of
spatial networks that lead to the emergence of their large scale properties.

\section{The model}

\subsection{Theoretical formulation}
\label{sub:theoretical_formulation}

We consider here the simple case where all the nodes are distributed uniformly
in the plane (see Methods for detailed description of the algorithm). For a rail
network, the nodes would correspond to cities and the network grows by adding
edges between cities iteratively; the edges are added sequentially to the graph
--as a result of a cost-benefit analysis-- until all the nodes are connected.
For the sake of simplicity, we limit ourselves to the growth of trees which
allows to focus on the emergence of large-scale structures due to the
cost-benefit ingredient alone.  Furthermore, we consider that all the actors
involved in the building process are perfectly rational and therefore that the
most profitable edge is built at each step. More precisely, at each time step we
build the link connecting a new node $i$ to a node $j$ which already belongs to
the network, such that the following quantity is maximum

\begin{equation}
    R_{ij} = B_{ij} - C_{ij}
    \label{eq:general_framework}
\end{equation}

The quantity $B_{ij}$ is the \textit{expected benefit} associated with the
construction of the edge between node $i$ and node $j$ and $C_{ij}$ is the
\textit{expected cost} associated with such a construction. Eq.
(\ref{eq:general_framework}) defines the general framework of our model and we
now discuss specific forms of $R_{ij}$. In the case of transportation networks,
the cost will essentially correspond to some maintenance cost and will typically
be proportional to the euclidean distance $d_{ij}$ between $i$ and $j$. We thus
write

\begin{equation}
    C_{ij} = \kappa d_{ij}
    \label{eq:cost}
\end{equation}

where $\kappa$ represents the cost of a line per unit of length per unit of
time. Benefits are more difficult to assess. For rail networks, a simple yet
reasonable assumption is to write the benefits in terms of distance and expected
traffic $T_{ij}$ between cities $i$ and $j$ 

\begin{equation} 
    B_{ij} = \eta T_{ij} d_{ij}
    \label{eq:benefits} 
\end{equation}

where $\eta$ represents the benefits per passenger per unit of length. We have
to estimate the expected traffic between two cities and for this we will follow
the common and simple assumption used in the transportation litterature, of
having the so-called gravity law ~\cite{Stewart:1948,Erlander:1990}

\begin{equation} 
    T_{ij} = k\frac{M_i \: M_j}{d_{ij}^a} 
    \label{eq:gravity} 
\end{equation}

where $M_{i(j)}$ is the population of city $i(j)$, and $k$ is the rate
associated with the process. We will choose here a value of the exponent $a>1$
($a<1$ would correspond to an unrealistic situation where the benefits
associated with passenger traffic would increase with the distance). This
parameter $a$ determines the range at which a given city attracts traffic,
regardless of the density of cities. The accuracy and relevance of this gravity
law is still controversial and improvements have been recently
proposed~\cite{Simini:2012,Lenormand:2012}. But it has the advantage of being
simple and to capture the essence of the traffic phenomenon: the decrease of the
traffic with distance and the increase with population. Within these
assumptions, the cost-benefit budget $R'_{ij}=R_{ij}/\eta$ now reads
\begin{equation} \label{eq:R0} R'_{ij} = k\frac{M_i M_j}{d_{ij}^{\;a-1}} - \:
\beta d_{ij} \end{equation} where $\beta = \frac{\kappa}{\eta}$ represents the
relative importance of the cost with regards to the benefits. We will assume
that populations are power-law distributed with exponent $\mu$ (which for cities
is approximatively $\mu\approx 1.1$, see Methods) and the model thus depends
essentially on the two parameters $a$, and $\beta$ (for a detailed description
of parameter used in this paper, see the next section). In the following we will
be working with fixed values of $\mu$ and $a$. The exact values we choose are
however not important as the obtained graphs would have the same qualitative
properties.

\subsection{Simulations} 

The simulation starts by distributing nodes uniformly in a square. We then
attribute to each node a random population distributed according to the power
law

\begin{equation}
    P_M(x) = \frac{\mu}{x^{\mu+1}}
\end{equation}

The choice of this distribution is motivated by Zipf's empirical results on city
populations~\cite{Zipf:1949} (which motivates the choice $\mu=1.1$ in our
simulations) but also because we can go from a peaked to a broad distribution by
tuning the value of $\mu$. Indeed, for $\mu>2$, both the first and the second
moment of this distribution exist and the distribution can be considered as
peaked. In contrast for $1<\mu<2$, only the first moment converges and the
distribution is broad.

Once the set of nodes is generated, we choose a random node as the root and add
nodes recursively until all the nodes belong to the graph. At each time step,
the nodes belonging to the graph constitute the set of `inactive nodes', and the
other -not yet connected - nodes the `active' nodes. At each time step we
connect an active node to an inactive node such that their value of $R$ defined
in Eq.~\ref{eq:R0} is maximum.

\section{Crossover between star-graph and Minimum Spanning Tree}

\subsection{Typical scale} 

The average population is $\overline{M}$ and the typical inter-city distance is
given by $\ell_1\sim 1/\sqrt{\rho}$ where $\rho=N/L^2$ denotes the city density
($L$ is the typical size of the whole system). The two terms of Eq.~\ref{eq:R0}
are thus of the same order for $\beta=\beta^*$ defined as

\begin{equation}
    \beta^* = k \overline{M}^2 \rho^{a/2}
    \label{eq:beta*}
\end{equation}

In the theoretical discussion that follows, we will take $k=1$ for simplicity
(but it should not be forgotten in empirical discussions). Another way of
interpreting $\beta^*$ which makes it more practical to estimate from empirical
data (see section Discussion), is to say that it is of the order of the average
traffic per unit time

\begin{equation} 
    \beta^* = < T > 
    \label{eq:beta*_traffic} 
\end{equation}

From Eq.~\ref{eq:beta*} we can guess the existence of two different regimes
depending on the value of $\beta$:

\begin{itemize}
    \item $\beta \ll \beta^*$ the cost term is negligible compared to the
    benefits term. Each connected city has its own influence zone depending on
its population and the new cities will tend to connect to the most influent
city. In the case where $a\approx 1$, every city connects to the most populated
cities and we obtain a star graph constituted of one single hub connected to all
other cities.  
    \item $\beta \gg \beta^*$ the benefits term is negligible
    compared to the cost term. All new cities will connect sequentially to their
    closest neighbour. Our algorithm is then equivalent to an implementation of
    Prim's algorithm~\cite{Prim:1957}, and the resulting graph is a minimum
    spanning tree (MST).  
\end{itemize}

The intermediate regime $\beta\simeq\beta^*$ however needs to be elucidated.  In
particular, we have to study if there is a transition or a crossover between the
two extreme network structures, and if we have a crossover what is the network
structure in the intermediate regime. In the following we answer these questions
by simulating the growth of these spatial networks.

\subsection{Evidence for the crossover}

\begin{figure}
    \centering
    \includegraphics[width=\textwidth]{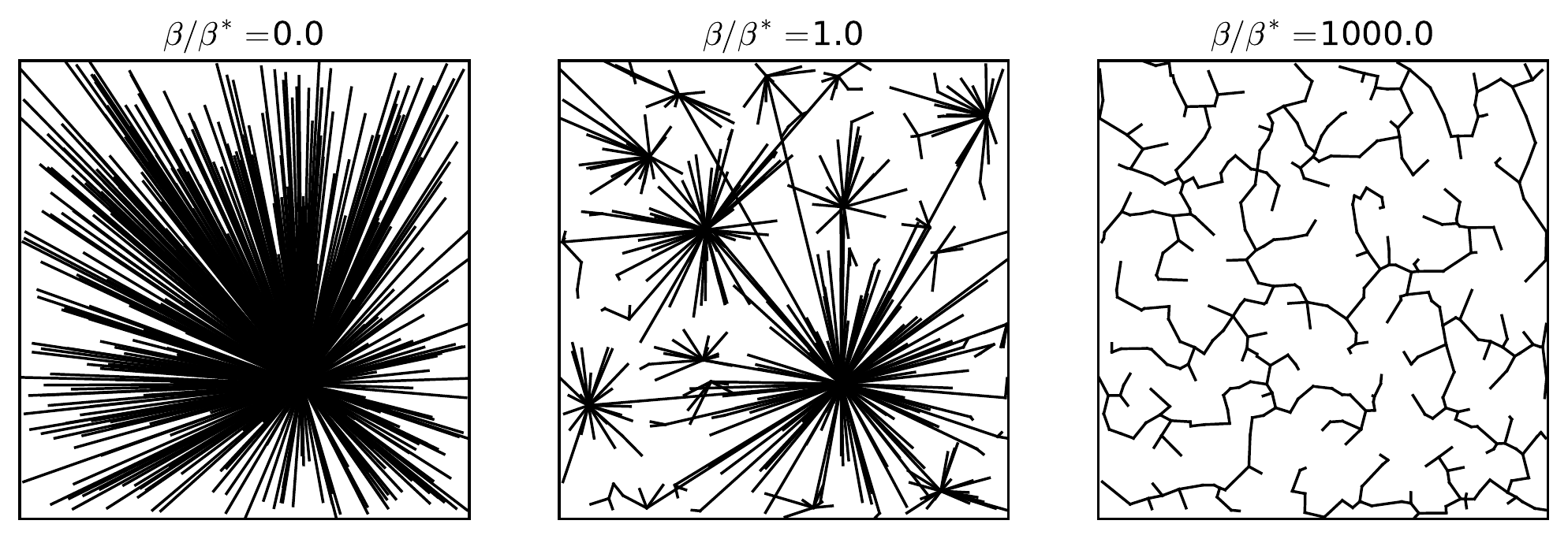}
    \caption{{\bf Simulated graphs.} Graphs obtained with our algorithm for the same set of cities
    (nodes) for three different values of $\beta^*$ ($a=1.1$, $\mu=1.1$, $400$
cities). On the left panel, we have a star graph where the most populated node
is the hub and on the right panel, we recover the minimum spanning tree.
\label{fig:plot_graphs}} 
\end{figure}

Fig.~\ref{fig:plot_graphs} shows three graphs obtained for the same set of
cities for three different values of $\beta/\beta^*$ ($a=1.1$, $\mu=1.1$)
confirming our discussion about the two extreme regimes in the previous section.
A visual inspection seems to show that for $\beta \sim \beta^*$ a different type
of graph appears, which suggests the existence of a crossover between the
star-graph and the MST. This graph is reminiscent of the hub-and-spoke structure
that has been used to describe the interactions between city
pairs~\cite{OKelly:1998,OKelly:1996}. However, in contrast with the rest of the
literature about hub-and-spoke models, we show that this structure is not
necessarily the result of a global optimization: indeed, it emerges here as the
result of the auto-organization of the system.

\begin{figure}
    \centering
    \includegraphics[width=0.8\textwidth]{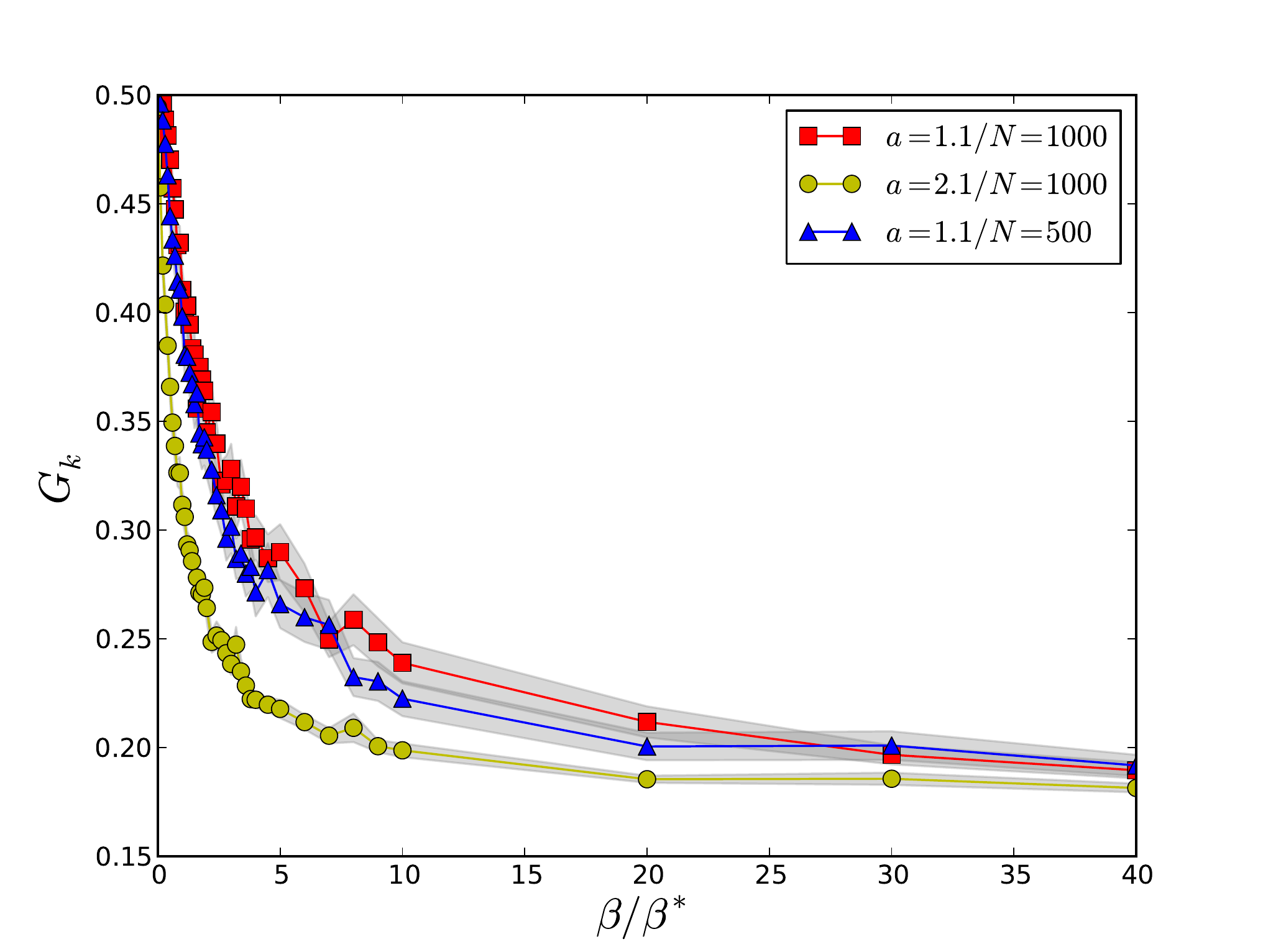}
    \caption{{\bf Gini on node degrees.} Evolution of the Gini coefficient with
$\beta/\beta^*$ for different values of $\beta^*$. The shaded area represents
the standard deviation of the Gini coefficient. Values decrease from $0.5$ in
the star-graph regime to below $0.20$ in the MST regime.\label{fig:gini}} \end{figure}

The MST is characterised by a peaked degree distribution while the star graph's
degree distribution is bimodal, and we therefore choose to monitor the crossover
with the Gini coefficient for the degrees defined as in~\cite{Dixon:1987}

\begin{equation}
    G_k = \frac{1}{2 N^2 \bar{k}} \sum_{i,j=1}^{N} | k_i - k_j |
    \label{eq:gini}
\end{equation}

where $\bar{k}$ is the average degree of the network. The Gini coefficient is in
$[0,1]$ and if all the degrees are equal, it is easy to see that $G=0$. On the
other hand, if all nodes but one are of degree 1 (as in the star-graph), a
simple calculation shows that $G=1/2$. Fig.~\ref{fig:gini} displays the
evolution of the Gini coefficient versus $\beta/\beta^*$ (for different values
of $\beta^*$ obtained by changing the value of $a$, $\mu$ and $N$). This plot
shows a smooth variation of the Gini coefficient pointing to a crossover between
a star graph and the MST,  as one could expect from the plots on
Fig.~\ref{fig:plot_graphs} (also, we note that for given values of $a$, $\mu$
all the plots collapse on the same curve, regardless of the number $N$ of nodes.
However for different values of $a$ or $\mu$ we obtain different curves).

Another important difference between the star-graph and the MST lies in how the
total length of the graph scales with its number of nodes. Indeed, in the case
of the star-graph, all the nodes are connected to the same node and the typical
edge length is $L$, the typical size of the system the nodes are enclosed in. We
thus obtain

\begin{equation}
    L_{tot} \sim L\; N
    \label{eq:Ltot_star}
\end{equation}

On the other hand, for the MST each node is connected roughly to its nearest
neighbour at distance typically given by $\ell_1\sim L/\sqrt{N}$, leading to

\begin{equation}
    L_{tot} \sim L\; \sqrt{N}
    \label{eq:Ltot_MST}
\end{equation}

\begin{figure}
    \centering
    \includegraphics[width=\textwidth]{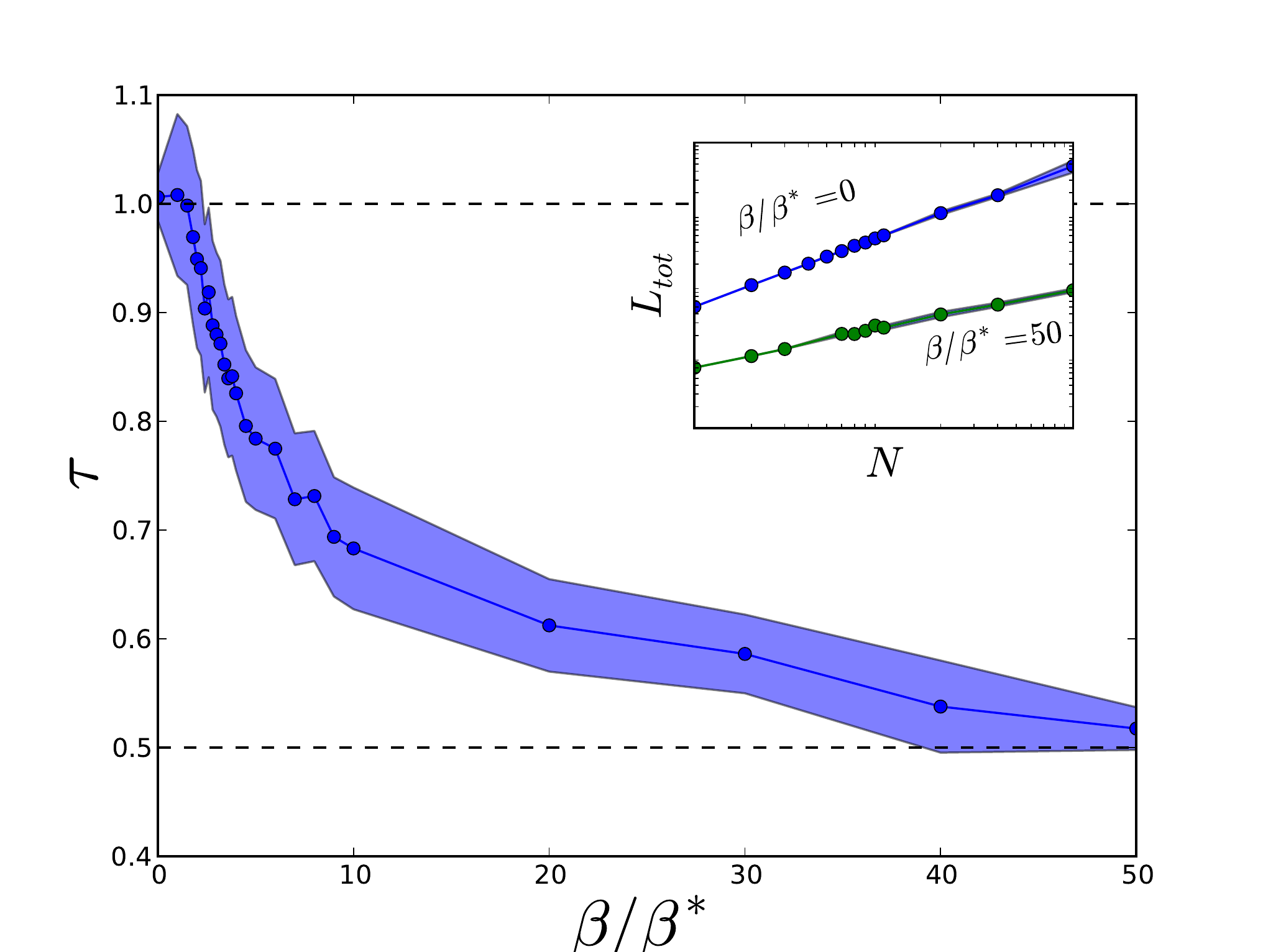} 
    \caption{{\bf Star graph to MST transition.} Exponent $\tau$ versus $\beta$.
For $\beta\ll \beta^*$ we recover the star-graph exponent $\tau=1$ and for the
other extreme $\beta\gg\beta^*$ we recover the MST exponent $\tau=1/2$. In the
intermediate range, we observe a continuously varying exponent suggesting a
non-trivial structure. The shaded area represents the standard deviation of
$\tau$.\label{fig:Ltot_vs_beta}. (Inset) In order to illustrate how we determined
the value of $\tau$, we represent $L_{tot}$ versus $N$ for two different values
of $\beta$. The power law fit of these curves gives $\tau$.} 
\end{figure}

More generally, we expect a scaling of the form $L_{tot}\sim N^\tau$ and on
Fig.~\ref{fig:Ltot_vs_beta} we show the variation of the exponent $\tau$ versus
$\beta$. For $\beta=0$ we have $\tau=1.0$ and we recover the behavior $L_{tot}
\propto N$ typical of a star graph. In the limit $\beta \gg \beta^*$ we also
recover the scaling $L_{tot} \propto \sqrt{N}$, typical of a MST. For
intermediate values, we observe an exponent which varies continuously in the
range $[0.5,1.0]$. This rather surprising behavior is rooted in the
heterogeneity of degrees and in the following, we will show that we can
understand this behaviour as resulting from the hierarchical structure of the
graphs in the intermediate regime. 

It is interesting to note that a scaling with an exponent $1/2<\tau < 1$ has
been observed~\cite{Samaniego:2008,Barthelemy:2011} for the total number
$\ell_T$ of miles driven by the population (of size $P$) of city scales as
$\ell_T \propto P^{\beta}$ with $\beta=0.66$. Understanding the origin of those
intermediate numbers might thus also give us insights into important features of
traffic in urban areas and the structure of cities.

It thus seems that from the point of view of interesting quantities such as the
Gini coefficient or the exponent $\tau$, there is no sign of a critical value
for $\beta$ and that we are in presence of a crossover and not a transition.

\section{Spatial Hierarchy}

The graph corresponding to the intermediate regime $\beta \approx \beta^*$
depicted on Fig.~\ref{fig:plot_graphs} exhibits a particular structure
corresponding to a hierarchical organization, observed in many complex
networks~\cite{Sales-Pardo:2007}. Inspired from the observation of networks in
the regime $\beta/\beta^* \sim 1$, we define a particular type of hierarchy
--that we call \emph{spatial hierarchy}-- as follows. A network will be said to
be spatially hierarchical if:

\begin{enumerate}
    \item  We have a hierarchical network of hubs that connect to nodes less and less far away as one goes down the hierarchy;
    \item Hubs belonging to the same hierarchy level have their own influence zone clearly separated from the others'. In addition, the influence zones of a given level are included in the influence zones of the previous level.
\end{enumerate} 

\begin{figure} 
    \centering
    \includegraphics[width=0.65\textwidth]{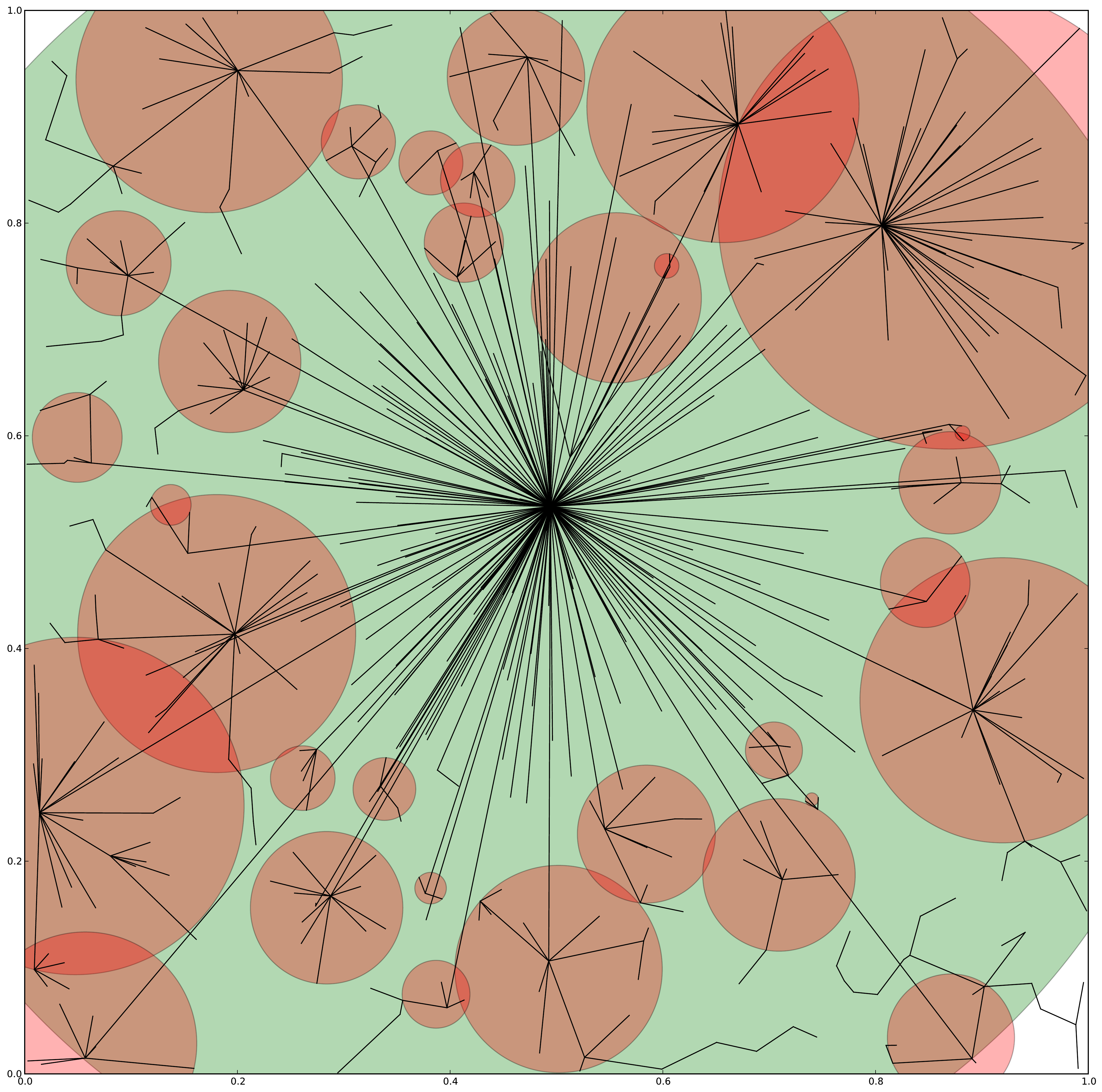} 
    \caption{{\bf Influence zones.} Example of a graph where we represent the influence zones for the
first two hierarchical levels.\label{fig:separation_example}} \end{figure}

The relevance of this new concept of hierarchy in the present context can be
qualitatively assessed on Fig.~\ref{fig:separation_example} where we represent
the influence zones by colored circles, the colors corresponding to different
hierarchical levels. In order to go beyond this simple, qualitative description
of the structure, we provide in the following a quantitative proof that networks
in the regime $\beta / \beta^*$ exhibit spatial hierarchy.

\subsection{Distance between hierarchical levels}
 
We propose here a quantitative characterisation of the part (1) in the
definition of spatial hierarchy. The first step is to identify the root of the
network which allows us to naturally characterize a hierarchical level by its
topological distance to the root. We choose the most populated node as the root
(which will be the largest hub for $\beta\ll\beta^*$) and we can now measure
various quantities as a function of the level in the hierarchy. In
Fig.~\ref{fig:distance_hierarchy}, we plot the average euclidean distance
$\overline{d}$ between the different hierarchical levels as a function of the
topological distance from the root node (for the sake of clarity, we also draw
next to these plots the corresponding graphs). For reasonably small  values of
$\beta/\beta^*$ (i.e. when the graph is not far from being a star-graph), the
average distance between levels decreases as we go further away from the root
node. This confirms the idea that the graphs for $\beta/\beta^*\simeq 1$ exhibit
a spatial hierarchy where nodes from different levels are getting closer and
closer to each other as we go down the hierachy. Eventually, as $\beta/\beta^*$
becomes larger than 1, the distance between consecutive levels just fluctuates
around $\ell_1\sim 1/\sqrt{\rho}$ the average distance between nearest
neighbours for a Poisson process, which indicates the absence of hierarchy in
the network.

\subsection{Geographical separation of hubs zones} 

We now discuss the part (2) of the definition of spatial hierarchy, that is to
say how the hubs are located in space. Indeed, another property that we can
expect from spatially hierarchical graph is that of \emph{geographical
separation}.

\subsubsection{Separation}
\label{ssub:separation}

We say that a graph is geographically separated if the influence zones of every
node of a given hierarchical level do not overlap and if they are included in
the influence zone of the nodes of the previous level in the hierarchy.
Formally, if we designate by $\mathcal{I}^i_l$ the influence zone of the node
$i$ located at level $l$ in the hierarchy, $\mathcal{I}_l = \cup_{i \in l}
\mathcal{I}^i_l$ the reunion of all the influence zones for nodes belonging to
the level $n$. We say that the graph is geographically separated if:

\begin{align}
    &\mathcal{I}_l \subset \mathcal{I}_{l+1} \; \forall l\\
    &\mathcal{I}^i_l \cap \mathcal{I}^j_l = \O \; \text{if} \; j \neq i, \forall l
\end{align}

The degree of geographical separability of a graph strongly depends on the
definition of the influence zone of a node. For instance, if we take the
influence zone of a node $i$ to be the surface of smallest area containing all
the nodes connected to $i$, it follows that all planar graph are totally
separated.  In the context of transportation networks, we expect hubs to radiate
up to a certain distance around them, that is to say connect to all the nodes
located in a convex shape. We simply define the influence zone of a node $i$ as
the circle centered on the barycenter of i's neighbours that belong to the next
level, of radius the maximum distance between the barycenter and those points. 

Figure~\ref{fig:separation_example} is intended to help the reader visualise
these influence zones on an example: The green circle represent the influence
zone of the root and the red circles the influence zones of the hubs connected
to it. One can see that the graph is geographically separated up to a good
approximation.

\begin{figure}
    \centering
    \includegraphics[width=0.45\textwidth]{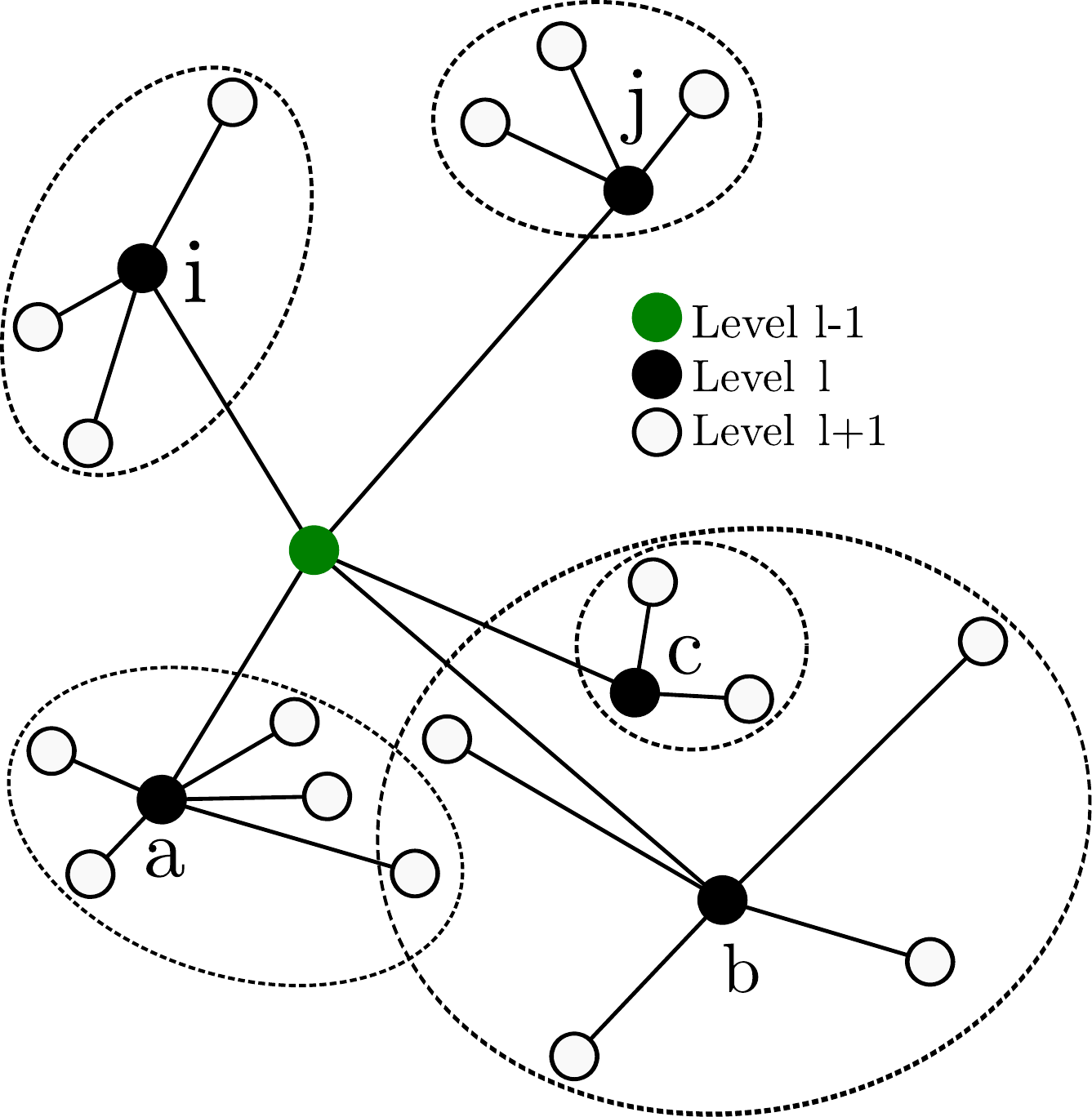}\\
    \caption{{\bf Influence zones.} Illustration of the influence zones (dotted
    lines) around several hubs. We have, according to the definition of the
separation index, $S(i,j)=0$, $0 < S(a,b)< 1$ and $S(b,c)=1$.
\label{fig:separation_illustration}} 
\end{figure}

In order to quantify this notion of geographical separability, we define the
separation index of the level $l$ as the average over all the nodes belonging to
$l$ of the separation function. The separation function is equal to $1$ if the
distance $d(i,j)$ between the centers of the influence zones of $i$ and $j$ is
larger than their respective radius (no overlap), and equal to 

\begin{equation}
    S(i,j) = 1-\frac{\text{Area of the overlap between} \; \mathcal{I}_l^i \; \text{and} \; \mathcal{I}_l^j}{\min \left(\text{Area of} \; \mathcal{I}_l^i\text{, Area of} \; \mathcal{I}_l^j\right)}
\end{equation}

One can see that the separation function is equal to 1 if the nodes' influence
zones do not overlap at all and 0 if they perfectly overlap (all the influence
zones overlapping, like Russian dolls). Therefore, the separation index is equal
to 1 if the level s is perfectly separated and 0 if the influence zones are
completely mixed. One can see on Fig.~\ref{fig:separation_illustration} an
illustration expliciting the value of the separation index for different
situations.

\subsubsection{Geographical separation in the intermediate regime}
\label{ssub:geographical_separation_of_simulated_graphs}

\begin{figure}
    \centering
    \includegraphics[width=0.80\textwidth]{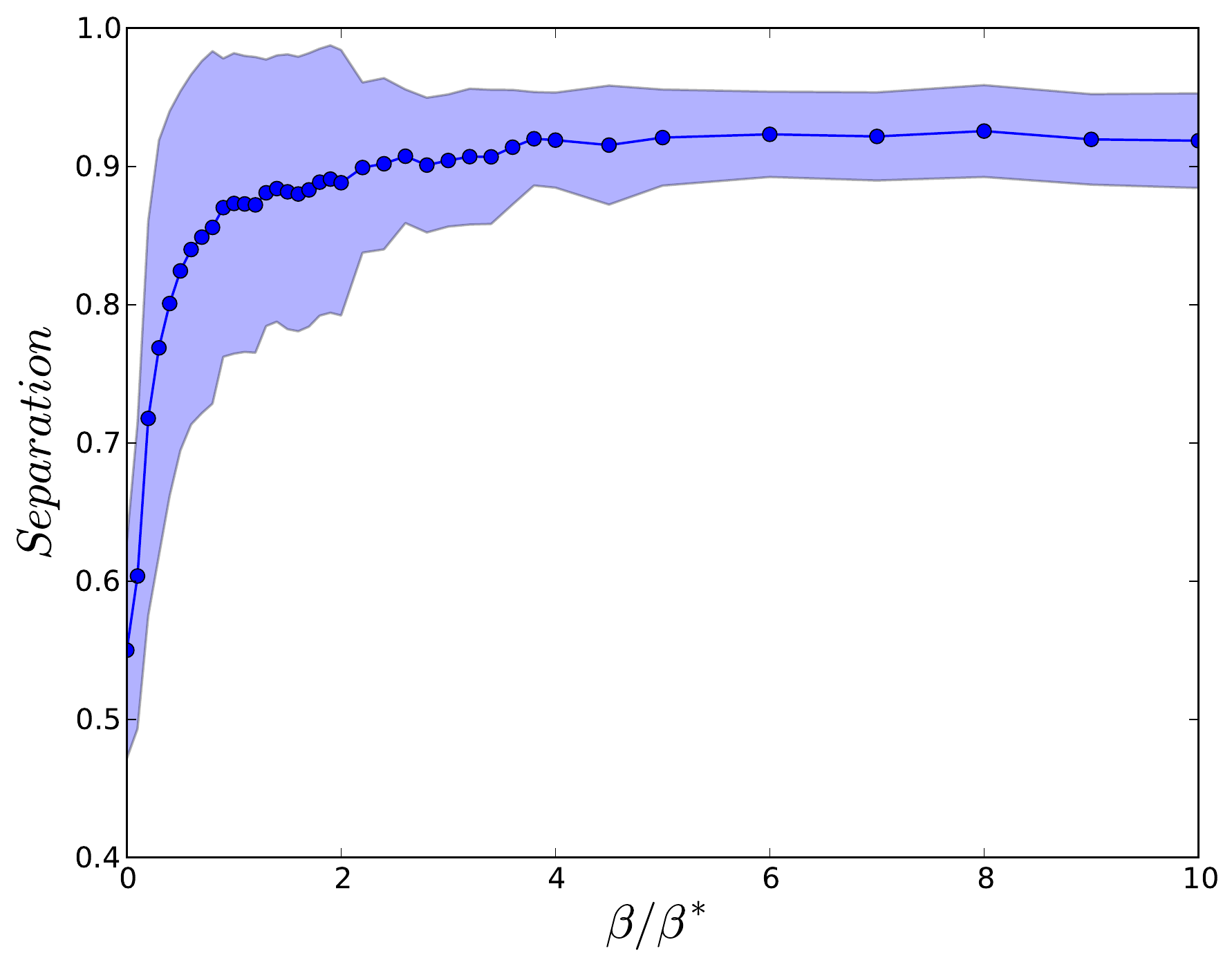}
    \caption{{\bf Separation index.} Separation index averaged over all the graph's level versus
$\beta/\beta^*$. The shaded area represents the standard deviation.
\label{fig:separation}} 
\end{figure}

We plot the separation index averaged over the all the graph's levels for
different values of $\beta/\beta^*$ on Fig.~\ref{fig:separation}. One can
observe on this graph that the separation index reaches values above $0.90$ when
$\beta/\beta^* \geq 1$, which means that the corresponding graphs indeed have a
structure with hubs controlling geographically well-separated regions.
Obviously, the choice of the shape of the influence zone (which is chosen here
to be a disk) strongly impacts the results but the
same qualitative behavior will be obtained for any type of convex shapes.

In conclusion, the graphs produced by our model in the regime $\beta / \beta^*$
satisfy the two points of the definition. They exhibit a spatially hierarchical
structure, characterised by a distance ordering and geographical separation of
hubs. We saw earlier that in this regime we have specific, non trivial
properties such as $L_{tot}$ scaling with an exponent depending continuously on
$\beta/\beta^*$. Using a simple toy model, we will now show that the spatial
hierarchy can explain this property.

\subsection{Understanding the scaling with a hierarchical model} 

\begin{figure}
    \centering
    \includegraphics[width=0.60\textwidth]{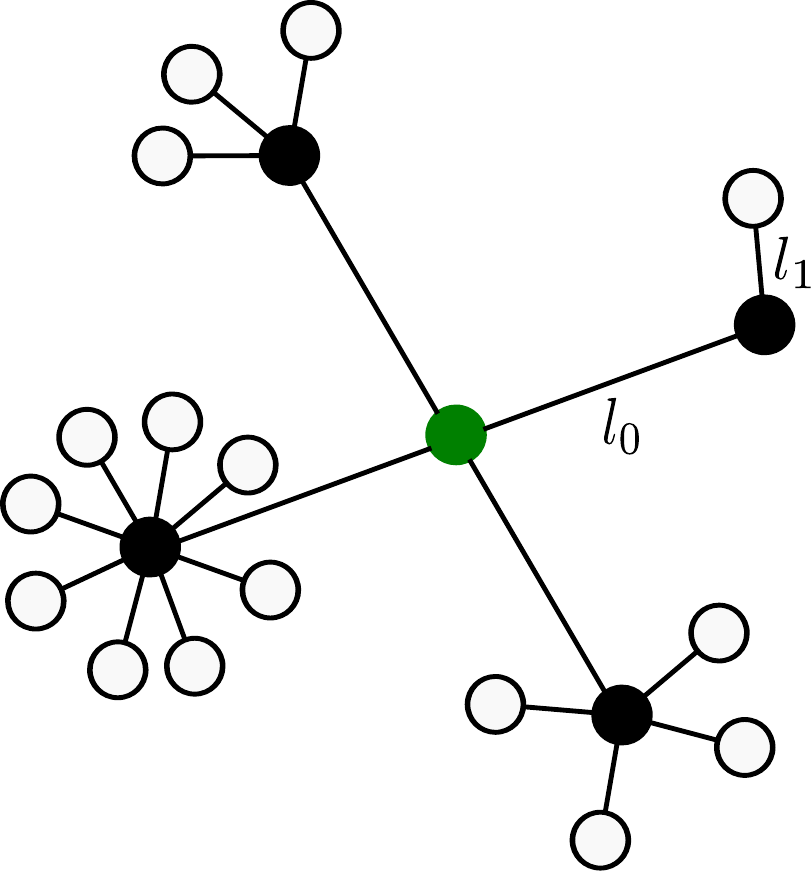}
    \caption{{\bf Fractal toy model.} A schematic representation of the hierarchical fractal network used as a toy model. 
    \label{fig:fractal_network}}
\end{figure}

The exponents $1$ and $0.5$ for the scaling for $L_{tot}$ with the total number
of nodes $N$ is well-understood. However, it is not clear how we can obtain
intermediate values. In the following we show with a simplified model that spatial
hierarchy can indeed lead to scaling exponents in the range $\left[0.5,
1\right]$. We consider the toy model defined by the fractal tree depicted on
Fig.~\ref{fig:fractal_network} for which the distance between the levels $n$ and
$n+1$ is given by

\begin{figure}
    \centering
    \includegraphics[width=0.80\textwidth]{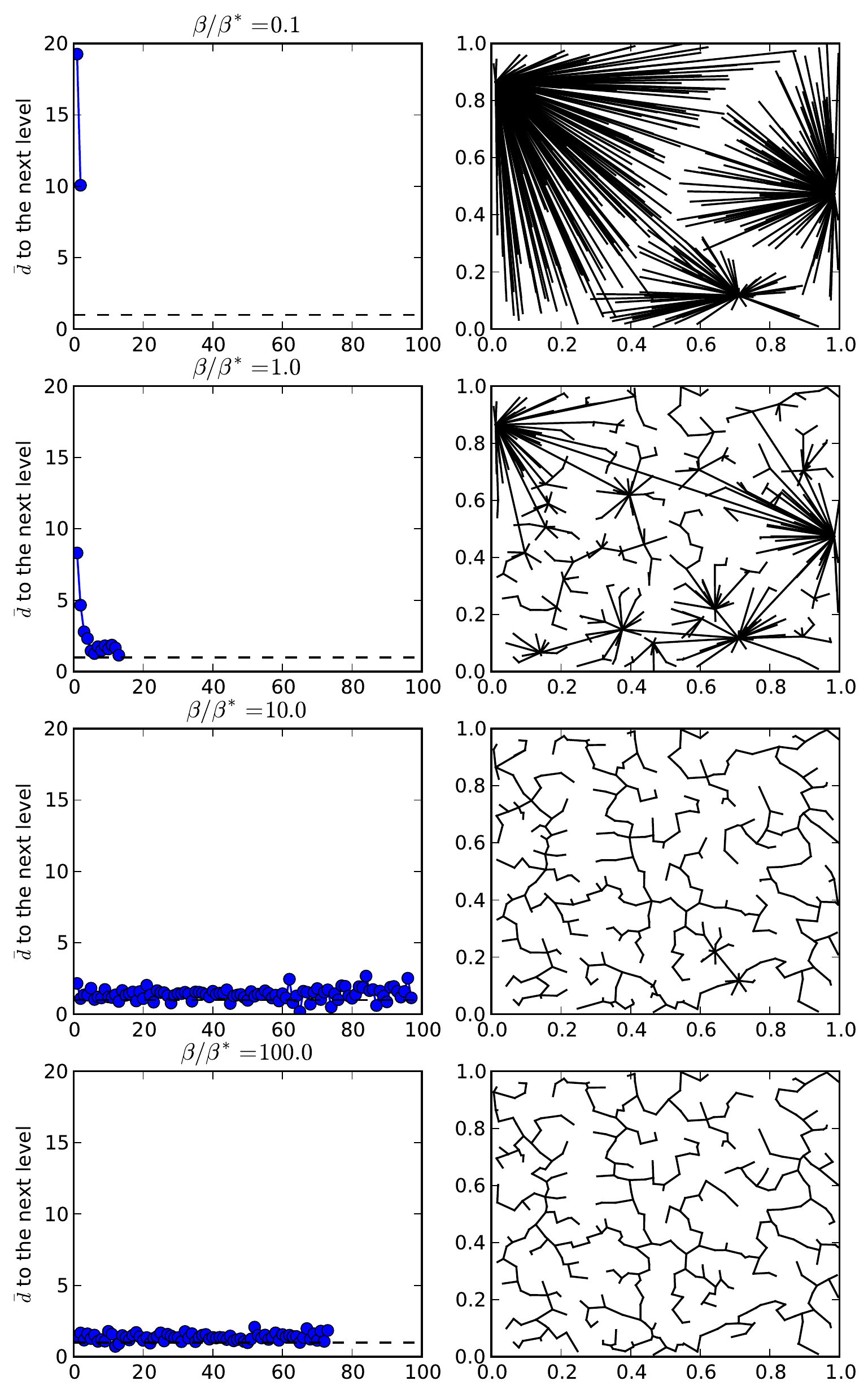}
    \caption{{\bf Distance between hierarchy levels.} Left column: Average
    distance between the successive hierarchy levels for different values of
    $\beta/\beta^*$, next to the corresponding graphs (on the right column). The
    most populated node is taken as the root node. \label{fig:distance_hierarchy}}
\end{figure}

\begin{equation}
    \ell_n=\ell_0 b^n
    \label{eq:fractal_distance}
\end{equation}

where $b\in [0,1]$ is the scaling factor. Each node at the level $n$ is
connected to $z$ nodes at the level $n+1$ which implies that

\begin{equation}
    N_n = z^n
    \label{eq:fractal_nodes}
\end{equation}

where $z>0$ is an integer. A simple calculation on this graph shows that in the
limit $z^g \gg 1$, the total length of the graph with g levels scales as

\begin{equation}
    L_{tot} \sim N^{\frac{\ln(b)}{\ln(z)}+1}
\end{equation}

where $\frac{\ln(b)}{\ln(z)} +1 \leq 1$ because $b \leq 1$ and $z>1$. This
simple model thus provides a simple mechanism accounting for continuous values
of $\tau$ whose value depends on the scaling factor $b$. It provides a
simplified picture of the graphs in the intermediate regime $\beta \simeq
\beta^*$ and exhibits the key features of the graphs in this regime: the hub
structure reminiscent of the star graph and where the nodes connected to each
hub form geographically distinct regions, organized in a hierarchical fashion.
It is also interesting to note that the parameter $z$ can be easily determined
from the average degree of the network, and that the parameter $b$ of the toy
model can be related to our model by measuring the decrease of the mean distance
between different levels of the hierarchy, as in
Fig.~\ref{fig:distance_hierarchy}. By plotting these curves for different values
of $\beta/\beta^*$, we find that the coefficient of the exponential decays
decreases linearly with $\beta / \beta^*$ and therefore that $b \sim
e^{\beta/\beta*}$ (However, the comparison only makes sense in the regime $\beta
\sim \beta^*$, as otherwise the graphs do not exhibit spatial hierarchy).

\section{Efficiency}

Most transportation networks are not obtained by a global optimization but
result from the addition of various, successive layers. The question of the
efficiency of these self-organized systems is therefore not trivial and deserves
some investigation. The model considered here allows us to test the effect of
various parameters and how efficient a self-organized system can be. In
particular, we would like to characterize the efficiency of the system for
various values of $\beta$. For this, we can assume that the construction cost
per unit length is fixed (ie. the factor $\eta$ in Eq.~\ref{eq:cost} is
constant), and since $\beta = \frac{\eta}{\kappa}$ a change of value for $\beta$
is equivalent to a change in the benefits per passenger per unit of length. 

A first natural measure of how optimal the network is, is given by its total
cost proportional to the total length $L_{tot}$: the shorter a network is, the
better for the company in terms of building and maintenance costs. In our model,
the behaviour of the total cost is simple and expected: for small values of
$\beta/\beta^*$, the obtained networks correspond to a situation where the users
are charged a lot compared to the maintenance cost, and the network is very long
($L_{tot}\propto N$). In the opposite case, when $\beta/\beta^* \gg 1$ the main
concern in building this network is concentrated on construction cost and the
network has the smallest total length possible (for a given set of nodes). 

\begin{figure}
    \centering
    \includegraphics[width=0.49\textwidth]{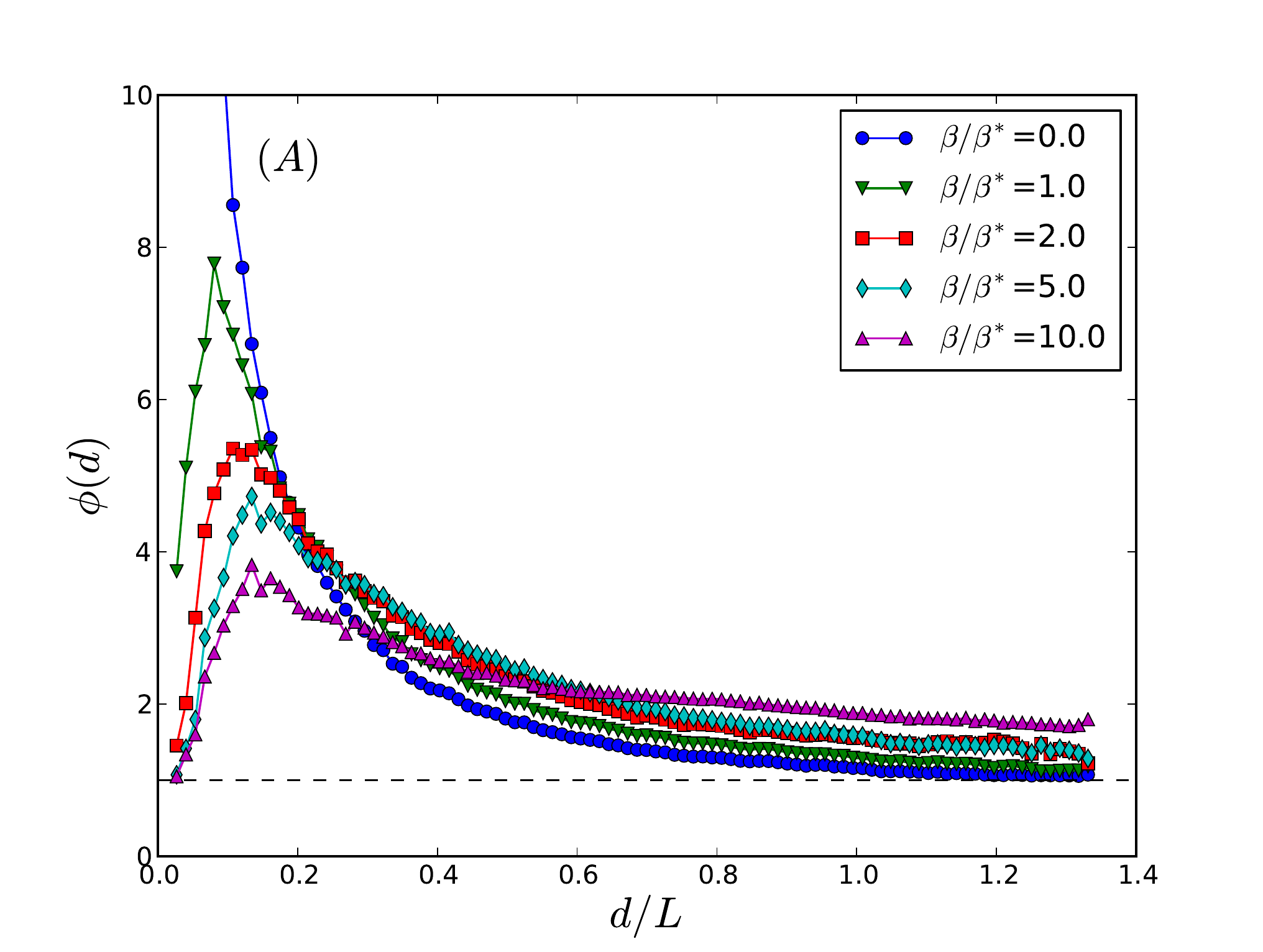}
    \includegraphics[width=0.49\textwidth]{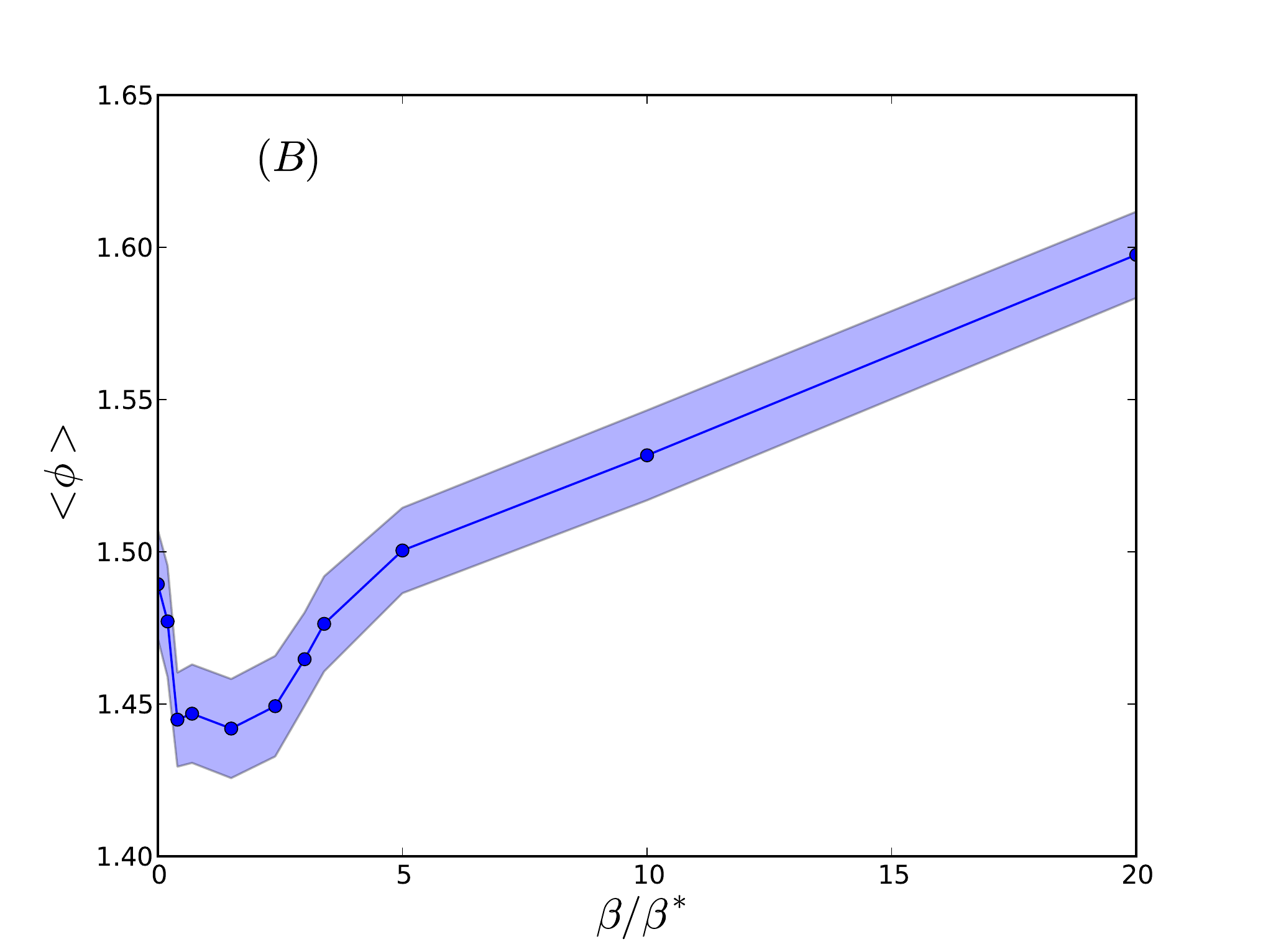}
    \caption{{\bf Detour function.} (Left) Detour function $\phi (d)$ versus the
        relative distance between nodes for different values of $\beta/\beta^*$.
        (Right) Average detour index $< \phi >$  for several realisations of the
        graphs as a function of $\beta/\beta^*$. The shaded area represents the
        standard deviation of $< \phi >$. This plot shows that there is a
        minimum for this quantity in the intermediate regime $\beta\sim\beta^*$.
\label{fig:RLE}} 
\end{figure}

The cost is however not enough to determine how efficient the network is from
the users' point of view: a very low-cost network might indeed be very
inefficient. A simple measure of efficiency is then given by the amount of
detour needed to go from one point to another. In other words, a network is
efficient if the shortest path on the network for most pairs of nodes is very
close to a straight line. The detour index for a pair of nodes $(i,j)$ is
conveniently measured by $D(i,j)/d(i,j)$ where $D(i,j)$ is the length of the
shortest path between $i$ and $j$, and $d(i,j)$ is the euclidean distance
between $i$ and $j$. In order to have a detailed information about the network,
we use the quantity introduced in~\cite{Aldous:2010}

\begin{equation}
    \phi(d) = \frac{1}{\mathcal{N}(d)} \sum_{\substack{i,j\\d(i,j) =d}} \frac{D(i,j)}{d(i,j)}
\end{equation}

where the normalisation $\mathcal{N}(d)$ is the number of pairs with $d(i,j)=d$.
We plot this `detour function' for several values of $\beta/\beta^*$ on
Fig.~\ref{fig:RLE}(A). For $\beta/\beta^* \ll 1$, the function $\phi(d)$ takes
high values for $d$ small and low values for large $d$, meaning that the
corresponding networks are very inefficient for relatively close nodes while
being very efficient for distant nodes. On the other hand, for $\beta/\beta^*
\gg 1$ we see that the MST is very efficient for neighboring nodes but less
efficient than the star-graph for long distances. Surprisingly, the graphs for
$\beta/\beta^* \sim 1$ exhibit a non trivial behaviour: for small distances, the
detour is not as good as for the MST, but not as bad as for the star graph and
for long distances it is the opposite. In order to make this statement more
precise we compute the average of $\phi(d)$ over $d$ (a quantity which has a
clear meaning for trees, see~\cite{Aldous:2010} for objections to the use of $<
\phi(d) >$ as a good efficiency measure in general), and plot it as a function
of $\beta/\beta^*$. The results are shown in Fig.~\ref{fig:RLE}(B) and confirm
this surprising behavior in the intermediate regime: we observe a minimum for
$\beta/\beta^* \sim 1$. In other words, there exists a non trivial value of
$\beta$, i.e. a value of the benefits per passenger per unit of length, for
which the network is optimal from the point of view of the users. 

\begin{figure}
    \centering
    \includegraphics[width=0.8\textwidth]{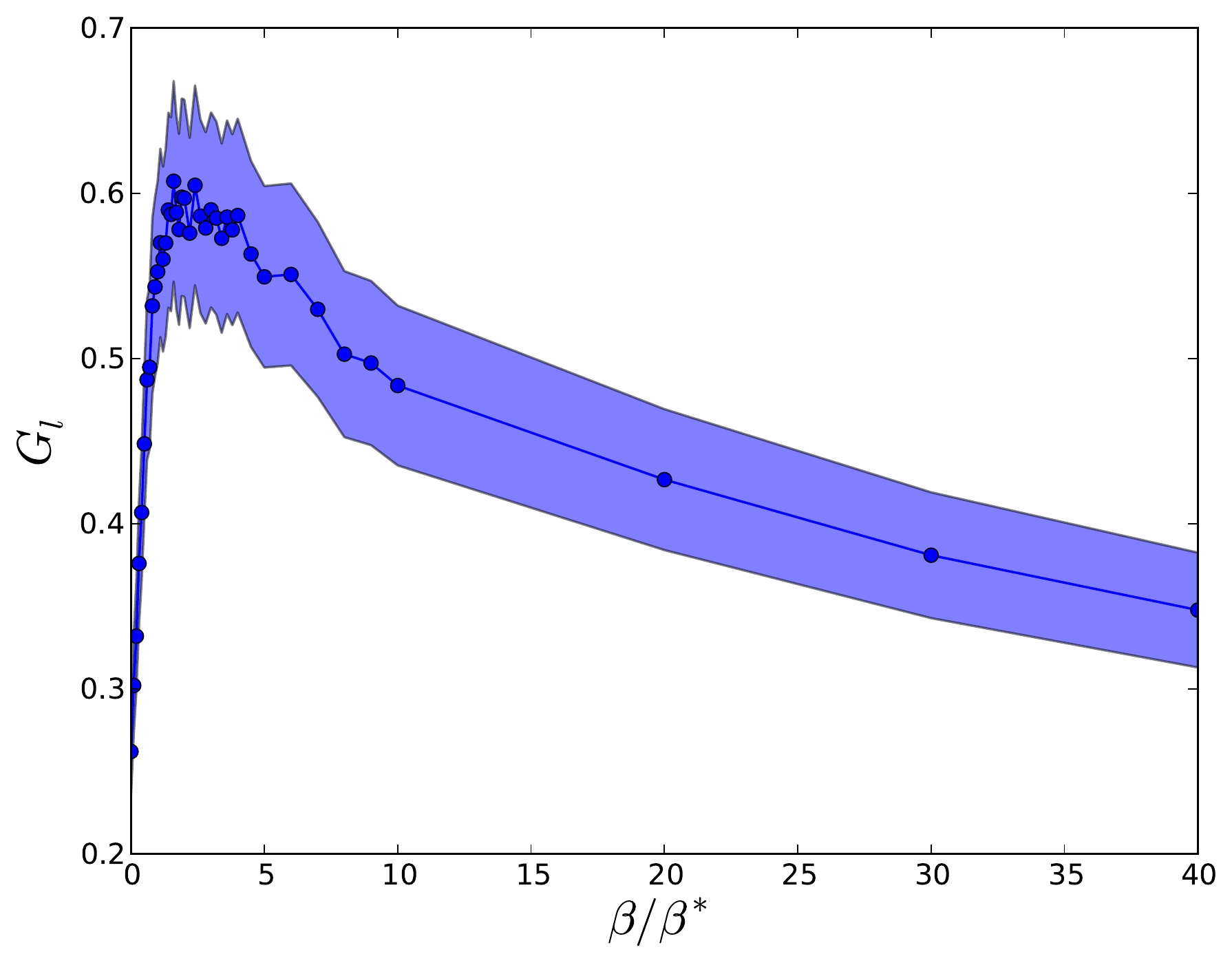}
    \caption{{\bf Gini on the length.} Evolution of the Gini coefficient for the length versus $\beta/\beta^*$ (for different values of $\beta^*$). The shaded area represents the standard deviation. \label{fig:gini_length}}
\end{figure}

The existence of such an optimum is far from obvious and in order to gain more
understanding about this phenomenon, we plot the Gini coefficient $G_l$ relative
to the length of the edges between nodes in Fig.~\ref{fig:gini_length}. We
observe that the Gini coefficient peaks around $\beta/\beta^* = 1$, which means
that in this regime, the diversity in terms of edge length is the highest. The
large diversity of lengths explains why the network is the most efficient in
this regime: indeed long links are needed to cover large distances, while
smaller links are needed to reach efficiently all the nodes. It is interesting
to note that this argument is similar to the one proposed by Kleinberg
\cite{Kleinberg:2000} in order to explain the existence of an optimal delivery
time in small-world networks.

\section{Discussion}

We have presented a model of a growing spatial network based on a cost-benefit
analysis. This model allows us to discuss the effect of a local optimization on
the large-scale properties of these networks. First, we showed that the graphs
exhibit a crossover between the star-graph and the minimum spanning tree when
the relative importance of the cost increases. This crossover is characterized
by a continuously varying exponent which could give some hints about other
quantities observed in cities such as the total length travelled by the
population. Secondly, we showed that the model predicts the emergence of a
spatial hierarchical structure in the intermediate regime where costs and
benefits are of the same order of magnitude. We showed that this spatial
hierarchy can explain the non trivial behaviour of the total length versus the
number of nodes. Finally, this model shows that in the intermediate regime the
vast diversity of links lengths entails a large efficiency, an aspect which
could of primary importance for practical applications.

An interesting playground for this model is given by railways and we can
estimate the value of $\beta/\beta^*$ for these systems. In some cases, we were
able to extract the data from various sources (in particular financial reports
of railway companies) and the results are shown in Table~1. We estimate for
different real-world networks, including some of the oldest railway systems,
$\beta$ using its definition (total maintenance costs per year divided by the
total length and by the average ticket price per km). In order to estimate
$\beta^*$ we use Eq.\ref{eq:beta*_traffic} in the following way

\begin{equation}
    \beta^*\simeq\frac{T_{tot}}{L_{tot}}
\end{equation}

where $T_{tot}$ is the total travelled length (in passengers$\cdot$kms$/$year)
and $L_{tot}$ is the total length of the network under consideration.
Remarquably, the computed values for the ratio $\beta/\beta^*$ shown in
Table~\ref{table:b_b*}.
are all of the order of $1$ (ranging from $0.20$ to $1.56$). In the framework of
this model, this result shows that all these systems are in the regime where the
networks possess the property of spatial hierarchy, suggesting it is a crucial
feature for real-world networks. We note that in our model, the value of $\beta
/ \beta^*$ is given exogeneously, and it would be extremely interesting to
understand how we could construct a model leading to this value in an
endogeneous way.

\begin{table*}
\scalebox{0.8}{
\begin{tabular}{lcccccc}
\hline
Country & $T_{tot}$ & $L_{tot}$ & Maintenance & Ticket price & $\beta/\beta^*$ \\
 & \footnotesize(kms$/$year) & \footnotesize(kms) & \footnotesize(euros$/$year)
& \footnotesize{euros} &  \\
\hline
France & $88.1 \, 10^9$ & $29,901$	&	$2.10 \, 10^9 $ & $0.12$ & $\mathbf{0.20}$ \\
Germany & $79.2 \, 10^9$ & $37,679$	&	$7.50 \, 10^9 $ & $0.30$ & $\mathbf{0.32}$ \\
India & $978.5 \, 10^9$ & $65,000$	& 	$3.00 \, 10^9 $ & $0.01$ & $\mathbf{0.31}$ \\
Italy & $40.6 \, 10^9$ & $24,179$	& 	$4.30 \, 10^9 $ & $0.20$ & $\mathbf{0.53}$ \\
Spain & $22.7 \, 10^9$ & $15,064$	& 	$3.16 \, 10^9 $ & $0.11$ & $\mathbf{1.26}$ \\
Switzerland & $18.0 \, 10^9$ & $5,063$	& 	$2.03 \, 10^9 $ & $0.17$ & $\mathbf{0.66}$ \\
United Kingdom & $62.7 \, 10^9$ & $16,321$	&	$12 \, 10^9 $ & $0.16$ & $\mathbf{1.19}$ \\
United States & $17.2 \, 10^9$ & $226,427$	& 	$2.96 \, 10^9 $ & $0.11$ & $\mathbf{1.56}$\\
\hline
\end{tabular}
}
\caption{{\bf Empirical estimates for $\beta / \beta^*$}. Table giving the total
ride distance (in km), the total network length (in km), the total annual
maintenance expenditure (in euros per year)  and the average ticket price (in
euros per km). All the given values correspond to the year 2011. From these data
we compute the experimental values of $\beta$, $\beta^*$ and their ratio (data
obtained from various sources such as financial reports of railway companies)
\label{table:b_b*}} \end{table*}

There are also several directions that seem interesting. First, various forms of
cost and benefits functions could be investigated in order to model specific
networks. In particular, there are several choices that can be taken for the
expected traffic. In this paper we limited ourselves to estimate the traffic as
a direct traffic from a node $i$ to a node $j$, but it is likely that part of
the traffic will come from other nodes. In order to take this into account, we
think that the following extensions are probably interesting:

\begin{enumerate}
    \item A given city (denoted by $0$ with population $M_0$) plays a particular role in the network (the capital city in a relatively small country, for example). In that case it is beneficial to be close to that city through the network and we write
    \begin{equation}
        \label{eq:R1}
        R^{(1)}_{ij} = (1-\lambda)\frac{M_i M_j}{d_{ij}^{a-1}}  + \lambda \: \frac{M_i M_0}{\left(D_{0j} + d_{ij}\right)^{a-1}}- \beta \: d_{ij}
    \end{equation}
    where $\lambda \in \left[ 0,1 \right]$ is a coefficient weighing the relative importance of the traffic coming from the particular city.

    \item The most general case where all the network-induced traffic are taken into account. We then consider
    \begin{equation}
        \label{eq:R2}
        R^{(2)}_{ij} = \sum_{k \neq i}\frac{M_i M_k}{\left(D_{kj} + d_{ij}\right)^{a-1}}- \beta \: d_{ij}
    \end{equation}
\end{enumerate} 

Other ingredients such as the presence of different rail companies, or the
difference between a state-planned network and a network built by private
actors, etc, could easily be implemented and the corresponding models could
possibly lead to interesting results.

More importantly, we limited ourselves here to trees in order to focus on the
large-scale consequences of the cost-benefit mechanism. Further studies are
needed in order to uncover the mechanisms of formation of loops in growing
spatial networks and we believe that the model presented here might represent a
suitable modeling framework.

Finally, it seems plausible that the general cost-benefit framework introduced
at the beginning of the article could be applied to the modelling of systems
besides transportation networks. We believe it captures the fundamental features
of spatial network while being versatile enough to model the growth of a great
diversity of systems shaped by space.

\chapter{Scaling in transportation networks}
\label{chap:scaling_networks}

The following chapter is a reprint of an article, \emph{Scaling in
transportation networks}, that was previously published by the author of this
thesis with Camille Roth and Marc Barthelemy~\cite{Louf:2014_scaling}.\\

Almost $200$ subway systems run through the largest agglomerations in the world
and offer an efficient alternative to congested road networks in urban areas.
Previous studies have explored the topological and geometrical static properties
of these transit systems~\cite{Derrible:2009,Levinson:2012}, as well as their
evolution in time~\cite{Roth:2012}. However, subways are not mere geometrical
structures growing in empty space: they are usually embedded in large, highly
congested urban areas and it seems plausible that some properties of these
systems find their origin in the interaction with the city they are in. Previous
studies~\cite{Levinson:2008,Xie:2009} have shown that the growth and properties
of transportation networks are tightly linked to the characteristics of urban
environment. Levinson \cite{Levinson:2008} for instance, showed that rail
development in London followed a logic of both `induced supply' and `induced
demand'. In other words, while the development of rail systems within cities
answers a need for transportation between different areas, this development also
has an impact on the organisation of the city. Therefore, while the growth of
transportation cannot be understood without considering the underlying city, the
development of the city cannot be understood without considering the
transportation networks that run through it. As a result, the subway system and
the city can be thought as two systems exhibiting a symbiotic behaviour.
Understanding this behaviour is crucial if we want to get a deeper understanding
of how the city grows and how the mobility patterns organise themselves in urban
environments.

At a different scale, railway networks answer a need for fast transportation
between different urban centers. We therefore expect their properties to be
linked to the characteristics of the underlying country. The model of growth
presented in Chapter~\ref{chap:cost-benefit} relates the existence of a given
line to the economical and geographical features of the environment. An
interesting question is thus to know whether subways and railway networks behave
in the same way, but at different scales. In other words, we are interested to
know whether subways are merely scaled down railway networks, or whether they
are fundamentally different objects, following different growth mechanisms. 

In the spirit of the model proposed in the previous Chapter, we propose here a
large-scale framework which relates structural and economical properties of
subway and railway networks. Although many
studies~\cite{Kansky:1963,Derrible:2009,Levinson:2012} explore the interplay
between regional characteristics and the structure of transportation networks, a
simple picture relating the network's most basic quantities and the region's
properties is still lacking. It has been found that several biological and
man-made systems exhibit allometric scaling relationships between the output of
processes and size. These relationships are hints that very general processes
are at stake in the growth of these systems, and a first step towards their
understanding is to uncover these
processes~\cite{Banavar:1999,Louf:2014_mobility}. In
the spirit of what has recently been done for cities~\cite{Louf:2014_mobility}, we try in
the following to understand the way subways and railway networks scale with some
of the substrates' most basic attributes: population, surface area and wealth. 

We believe this should lay the foundations for more specific and involved discussions.

As a result, we are able to relate the total ridership, the number of stations,
the length of the network to socio-economical features of the environment. We
find that these relations are in good agreement with the data gathered for $138$
subway systems and $58$ railway networks accross the world. \graffito{Data for
$138$ subways accross the world were collected on Wikipedia,
and cross-referenced with the operators' data when possible}In particular, we
show that even if the main mechanisms are the same, the difference of scale at
which both systems operate is responsible for their different behavior.

\section{Framework}

A transportation network is at least characterized by its total number of nodes
(which are here train or subway stations), its total length, and the total
(yearly) ridership. On the other hand, a city (or a country in the railway case)
is characterized by its area, its population and its GDP. Because transportation
systems do not grow in empty space, but result from multiple interactions with
the substrate, an important question is how network characteristics and
socio-economical indicators relate to each other. Naturally, cost-benefit
analysis seems to be the appropriate theoretical framework. While this approach
has already been developed in the context of the growth of railway
networks~\cite{Black:1971,Louf:2013_emergence}, these studies considered an iterative
growth: at each step an edge $e$ is built such that the cost function

\begin{equation} 
    Z_e = B_e - C_e 
\end{equation}

is maximum. The quantity $B_e$ is the expected benefit and $C_e$ the expected
cost of $e$. In the following, we consider networks after they have been built,
and we assume that they are in a `steady-state' for which we can write a cost
function of the form

\begin{equation} 
    Z =\sum_eZ_e= B - C 
\end{equation}

where $B$ is the total expected benefits and $C$ the total expected costs, now
operating costs (mainly maintenance costs). We further assume that, during this
steady-state, operating costs are balanced by benefits. In other words

\begin{equation} 
    Z \approx 0 
\end{equation} 

Indeed, because lines and stations cost money to be maintained, we expect the
network to adapt to the way it is being used. Therefore we can reasonably expect
that at first order the cost of operating the system is compensated by the
benefits gained from its use.  In the following we will apply this general
framework to subway and railway networks in order to determine the behavior of
various quantities with respect to population and GDP.

\section{Subways}

In the case of subways, the total benefits in the steady-state are simply
connected to the total ridership $R$ and the ticket price $f$ over a given
period of time. The costs, on the other hand, are due to the maintenance costs
of the lines and stations, so that we can write (for a given period of time)

\begin{equation}
    Z_{sub} = R\,f - \epsilon_L L - \epsilon_S N_s
    \label{eq:cost-benefit}
\end{equation}

where $L$ is the total length of the network, $\epsilon_L$ the maintenance cost
of a line per unit of length, $N_S$ the total number of stations and
$\epsilon_S$ the maintenance cost of a station (for a given period time).

It is usually difficult to estimate the ridership of a system given its
characteristics and those of the underlying city. Due to the importance of such
estimates for planning purposes, the problem of estimating the number of
boardings per station given the properties of the area surrounding the stations
has been the subject of numerous studies~\cite{Matsunaka:2013,Kuby:2004}. Here
we are interested in the dependence of global, average behaviours of the
ridership on the network and the underlying city. Very generally, we write that
the number $R_i$ of people using the station $i$ will be a function of the area
$C_i$ serviced by this station --- the `coverage'~\cite{Derrible:2009} --- and
of the population density $\rho = \frac{P}{A}$ in the city

\begin{equation}
    R_i = \xi_i\, C_i\, \rho
\end{equation}

where $\xi_i$ is a random number of order one representing the ratio of people
covered who use the subway. The main difficulty is in finding the expression of
the coverage. It depends, a priori, on local particularities such as the
accessibility of the station, and should thus vary from one station to another.
We take here a simple approach and assume that on average

\begin{equation}
    C_i \sim \pi\, d_0^{\,2}
\end{equation}

where  $d_0$ is the typical size of the attraction basin of a given station. If
we assume that it is constant, the total ridership can be written as

\begin{equation}
    R = \sum_i R_i \sim \overline{\xi} \pi d_0^2 \rho \: N_s
    \label{eq:ridership}
\end{equation}

where $\overline{\xi} = \frac{1}{N_s}\,\sum_i \xi_i$ is of the order of 1.\\

We gathered the relevant data for $138$ metro systems across the world, which we
cross-verified when possible with the data given by network operators. While the
number of stations, the number of lines, total length of the networks and
ridership are relatively straightforward to define, the choice of population and
city area is more subtle. Indeed, most subway systems span an area greater than
the city core, and the relevant area therefore lies somewhere between the city
core's area and the total urbanized area. We chose to use the population and
surface area data for urbanized areas provided by Demographia.

\begin{figure}
\centering
    \includegraphics[width=.49\textwidth]{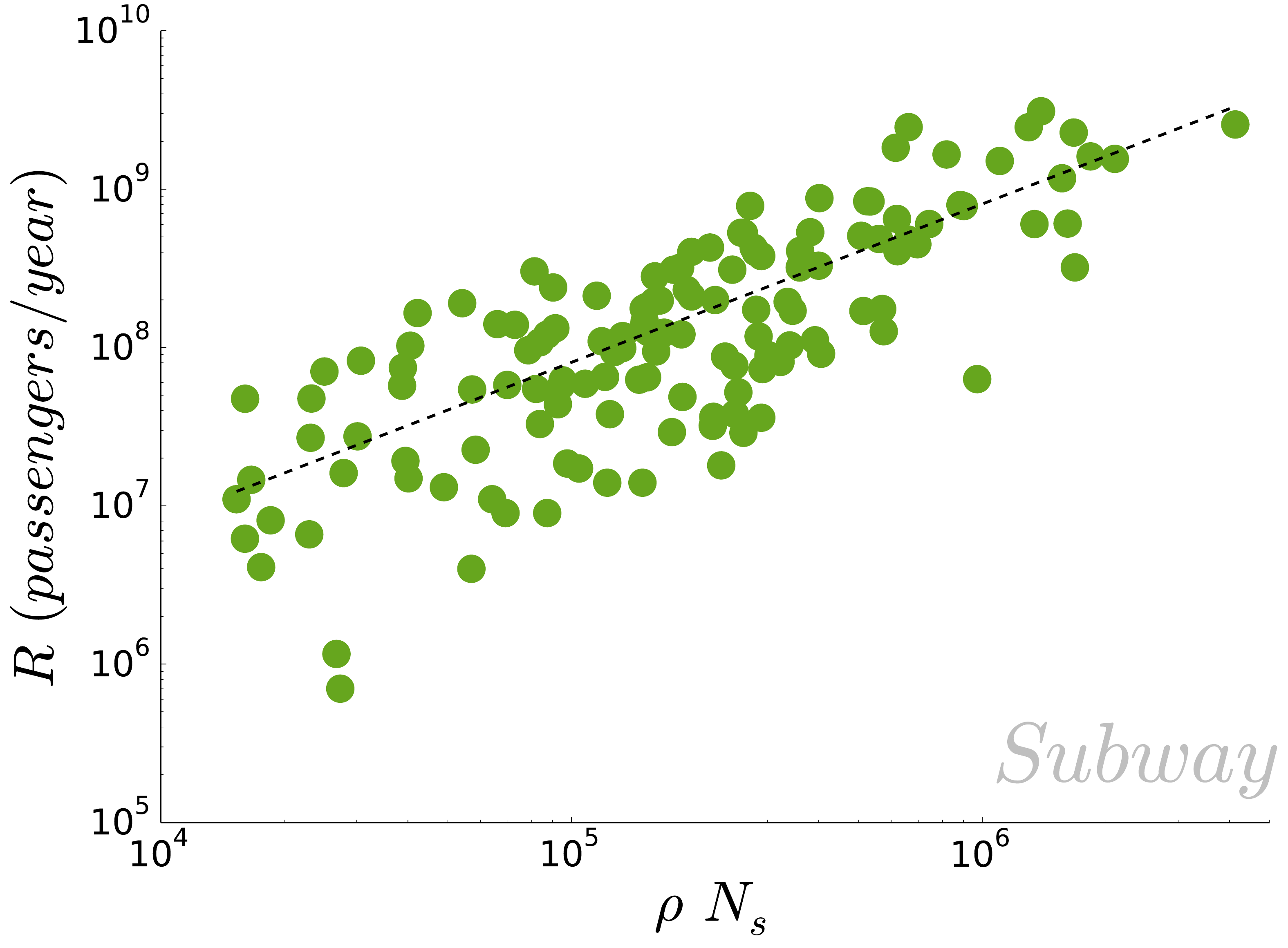}
    \includegraphics[width=.49\textwidth]{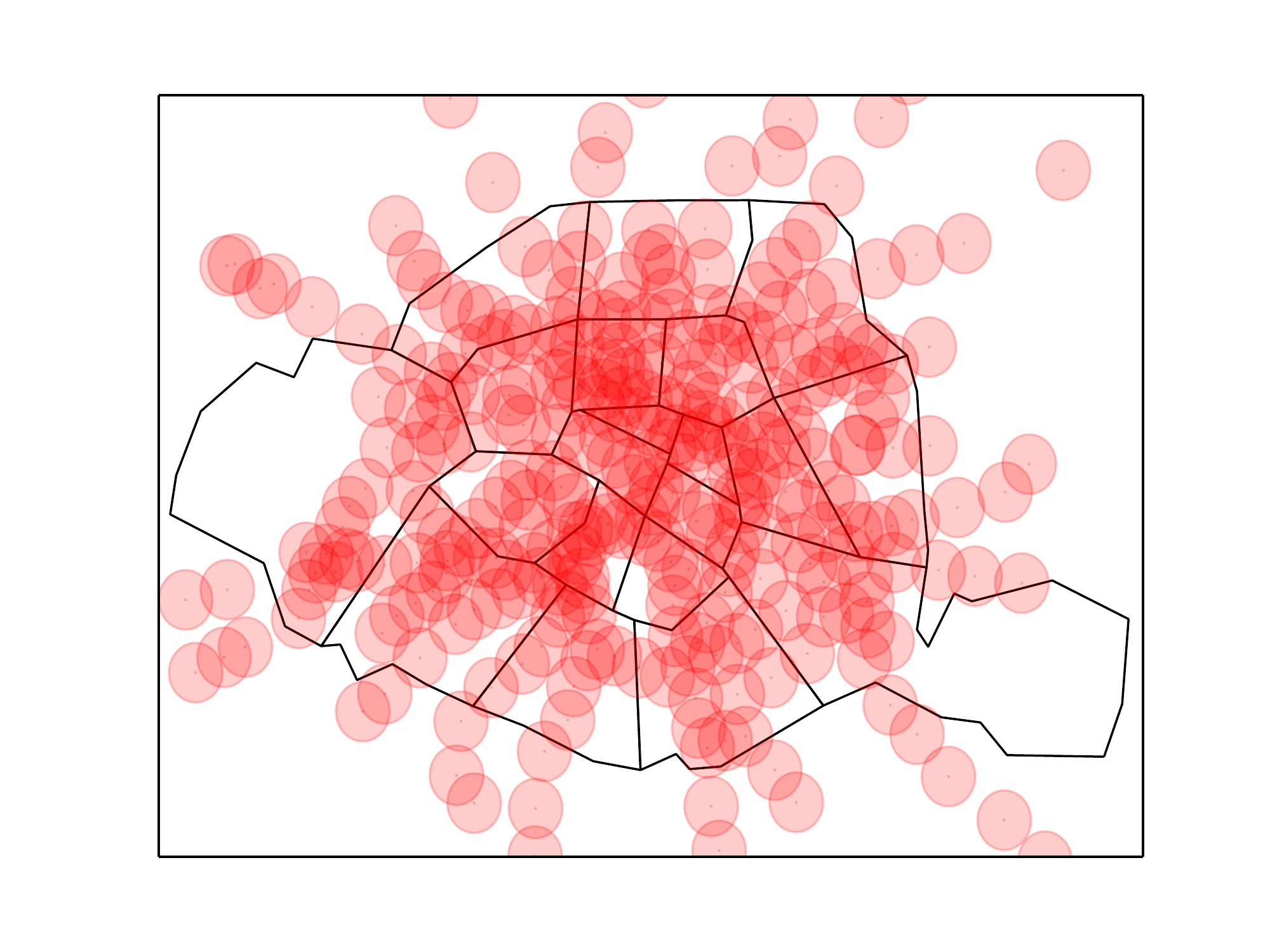}
    \caption{{\bf (Subway) The relationship between ridership and coverage} (Left)
    We plot the total yearly ridership $R$ as a function of $\rho\,N_s$. A linear
    fit on the $138$ data points gives $R \approx 800\:\rho N_s$ ($R^2=0.76$) which
    leads to a typical effective length of attraction $d_0 \approx 500\,\text{m}$
    per station. (Right) Map of Paris, France with each subway station represented
    by a red circle of radius $500\,\text{m}$.\label{fig:metro_ridership}}
\end{figure}

 We plot the ridership $R$ as function of $N_s\,\rho$ on
 Fig.~\ref{fig:metro_ridership} and observe that the data is consistent with a
 linear behavior. We measure a slope of $800\, \text{km}^2/\text{year}$ which
 gives an estimate for $d_0$

\begin{equation}
    d_0 \approx 500\,\text{m}
\end{equation}

We illustrate this result on Fig.~\ref{fig:metro_ridership} by representing the
subway stations of Paris each with a circle of radius $500\,\text{m}$. 

So far, the distance $d_0$ appears here an intrinsic feature of user's
behaviors: it is the maximal distance that an individual would walk to go to a
subway station.

The average interstation distance $\ell_1$ is another distance characteristic of
the subway system. Rigorously, this distance depends on the average degree $<k>$
of the network so that $\ell_1 = \frac{2\,L}{N_s <k>}$. It has however been
found that for the $13$ largest subway systems in the world, $<k> \in
\left[2.1,2.4\right]$, so that we can reasonably take $<k> / 2 \approx 1$ and
thus

\begin{equation} 
    \ell_1 \simeq \frac{L}{N_s} 
\end{equation}

The interstation distance depends in general on many technological and
economical parameters, but we expect that for a properly designed system it will
match human constraints. Indeed, if $d_0\ll\ell_1$, the network is not dense
enough and in the opposite case $d_0\gg\ell_1$, the system is not economically
interesting. We can thus reasonably expect that the interstation distance
fluctuates slightly around an average value given by twice the typical station
attraction distance $d_0$

\begin{equation} 
    d_0 = \frac{\ell_1}{2} = \frac{L}{2\,N_s} 
\end{equation}

\begin{figure} \centering
    \includegraphics[width=0.49\textwidth]{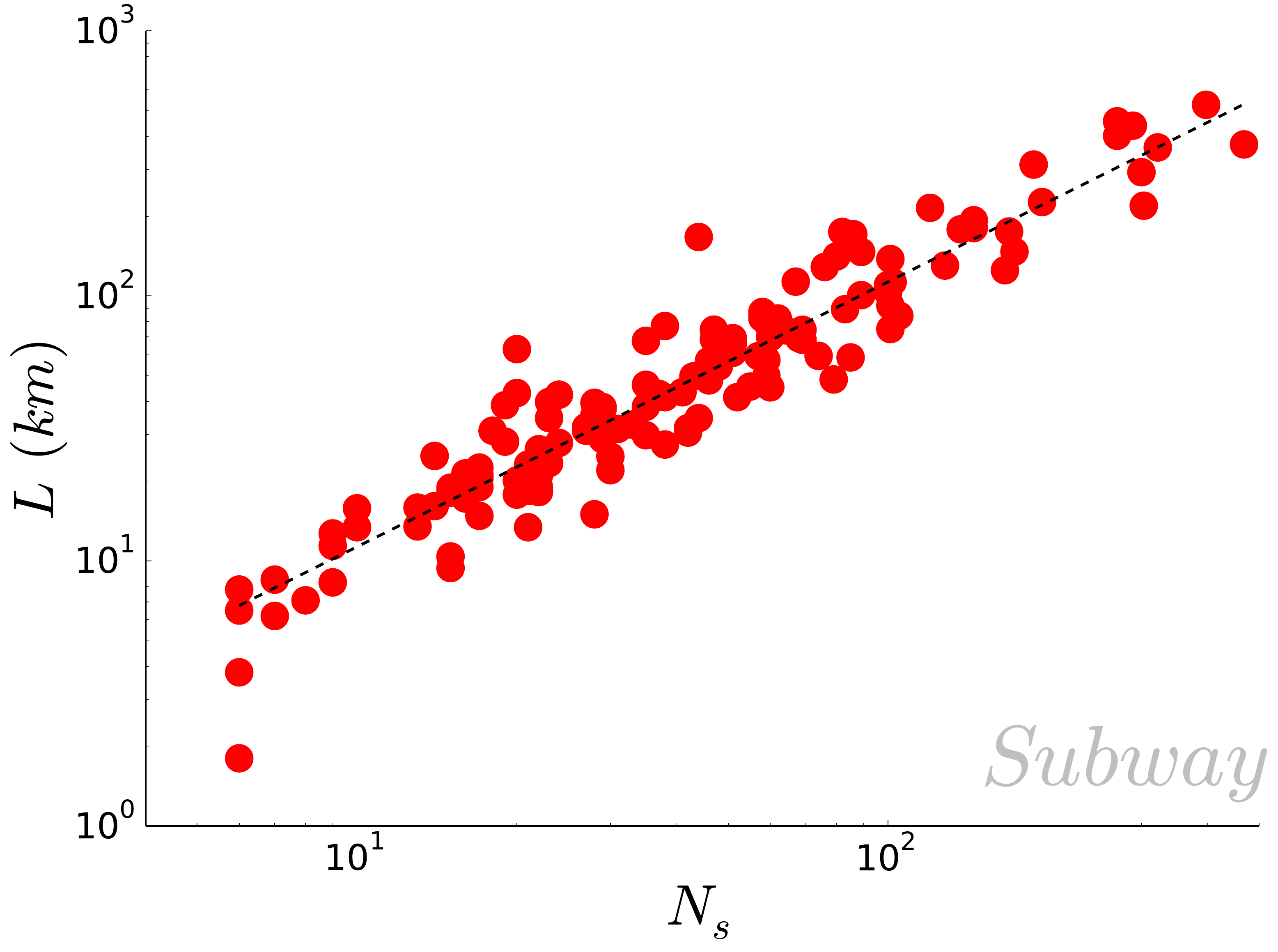}
    \includegraphics[width=0.49\textwidth]{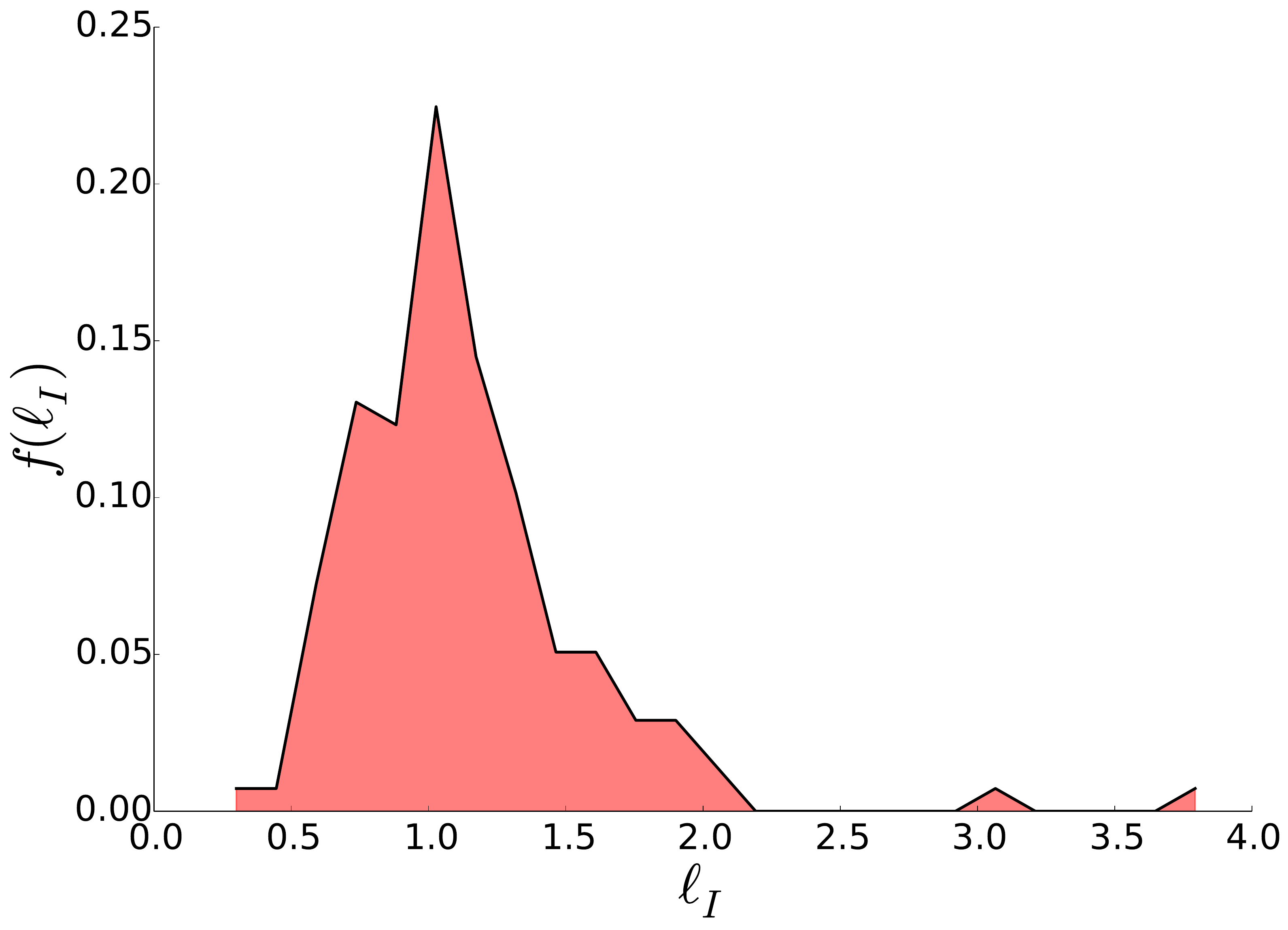}
    \caption{{\bf (Subway) Relation between the length and the number of
    stations} (Left) Length of $138$ subway networks in the world as a function
of the number of stations. A linear fit gives $L \sim 1.13\,N_S\,(R^2=0.93)$
(Right) Empirical distribution of the inter-station length. The average
interstation distance is found to be $\overline{\ell_1} \approx 1.2\, \text{km}$
and the relative standard deviation is approximately $440\,\text{m}$
\label{fig:metro_length_stations}}
\end{figure}

It follows from this assumption that the interstation distance is constant and
independent from  the population size. We plot on
Fig.~\ref{fig:metro_length_stations} the total length of subway networks as a
function of the number of stations. The data agrees well with a linear fit $L
\sim 1.13\,N_S\,(r^2=0.93)$. We also plot on
Fig.~\ref{fig:metro_length_stations} the histogram of the inter-station length,
showing that the interstation distance is indeed narrowly distributed around an
average value $\overline{\ell_1} \approx 1.2\,\text{km}$ with a variance $\sigma
\approx 400\,\text{m}$, consistently with the value found above for $d_0\approx
500\,\text{m}$. The outliers are San Francisco, whose subway system is more of a
suburban rail service and Dalian, a very large city whose metro system is very
young and still under development.

As a result of the previous argument, we can express $\ell_1$ in terms of the
systems characteristics. Indeed, the total ridership now reads

\begin{equation}
    R \sim \overline{\xi}\pi\rho\frac{L^2}{N_s}
    \label{eq:ridership-other}
\end{equation}

If we assume to be in the steady-state $Z_{sub} \approx 0$, using the results
from Eqs.~(\ref{eq:cost-benefit},\ref{eq:ridership-other}), we find that the
total length of the network and the number of stations are linked at first order
in $\epsilon_s/\epsilon_L$ by

\begin{equation}
    L \sim \left( \frac{4 \epsilon_L}{\pi\,\xi\,f\,\rho} + \frac{\epsilon_s}{\epsilon_L}\right) N_s
    \label{eq:length-stations}
\end{equation}

and that the interstation distance reads

\begin{equation}
    \ell_1 = \frac{4 \epsilon_L}{\pi\,\xi\,f\,\rho} + \frac{\epsilon_s}{\epsilon_L}
\end{equation}

This relation implies that the interstation distance increases with an increased
station maintenance cost, and decreases with increased line maintenance costs,
density and fare. We thus see that the adjustment of $\ell_1$ to match $2\,d_0$
can be made through the fare price (or subsidies by the local authorities or
national government). At this point, it would be interesting to get reliable
data about the maintenance costs and fare for subway systems in order to pursue
in this direction and test the accuracy of this prediction.\\

\begin{figure}
\centering
    \includegraphics[width=0.49\textwidth]{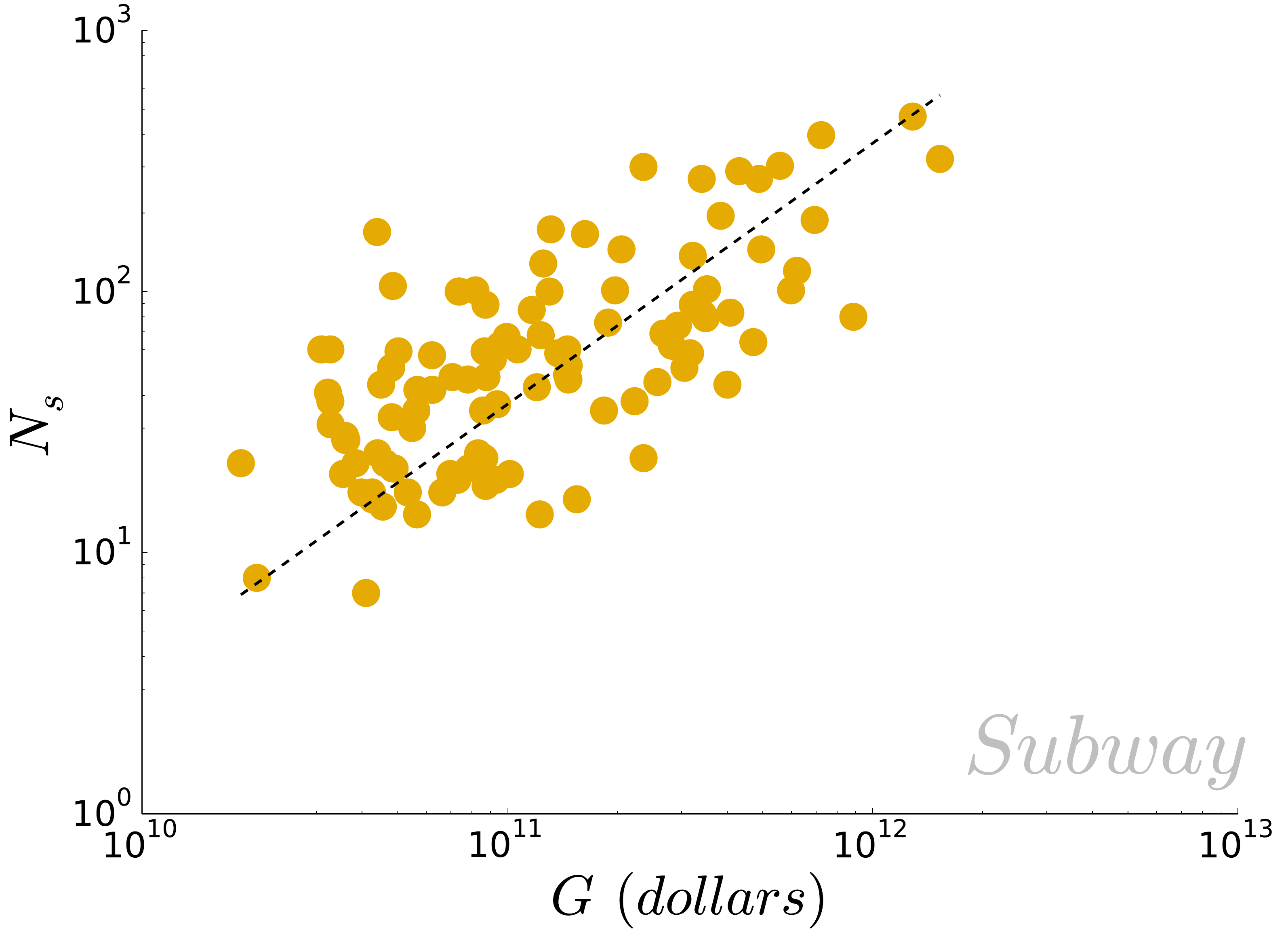}
    \includegraphics[width=0.49\textwidth]{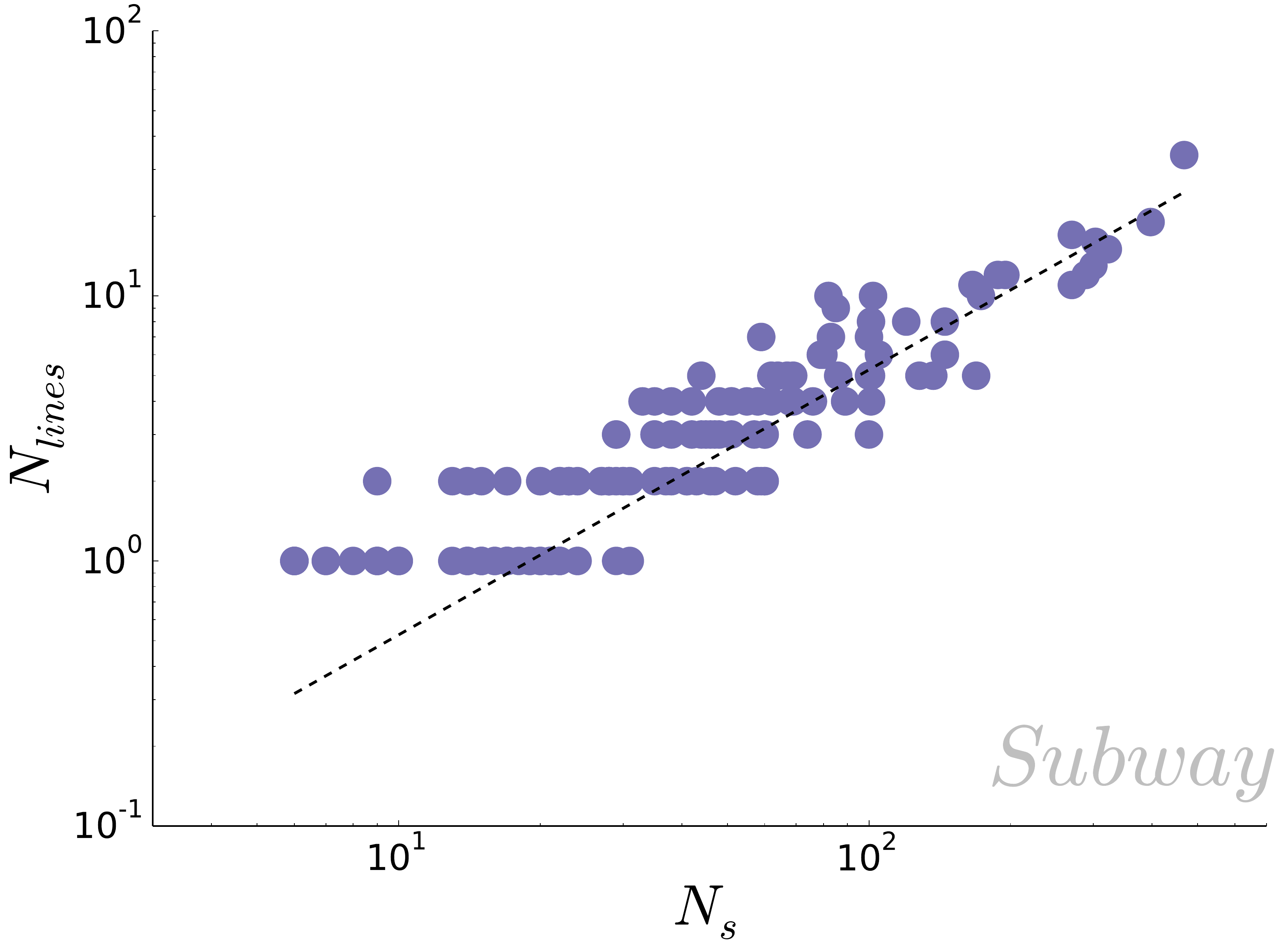}
    \caption{{\bf (Subway) Size of the subway system and city's wealth} We plot
    the number of stations for the different subway systems in the dataset as a
function of the Gross Metropolitan Product of the corresponding cities (obtained
for $106$ subway systems). A linear fit (dashed line) gives $N_s = 2.51\,
10^{-10}\,G$ ($R^2=0.73$). {\bf (Subway) Number of lines and number of stations}
We plot the number of metro lines $N_{lines}$ as a function of the number of
stations $N_s$. A linear fit on the $138$ data points gives $N_{lines} \approx
0.053\,N_s\,(R^2=0.94)$, or, in other words, metro lines contain on average $19$
stations.} 
\label{fig:metro_stations_gdp} 
\end{figure}

So far, we have a relation between the total length and the number of stations,
but we need another equation in order to compute their value. Intuitively, it is
clear that the number of stations --- or equivalently the total length --- of a
subway system is an increasing function of the wealth of the city. We assume a
simple, linear relation of the form

\begin{equation}
    N_s = \beta \frac{G}{\epsilon_s}
\end{equation}

where $G$ is the city's Gross Metropolitan Product, and $\beta$ the fraction of
the city's wealth invested in public transportation. \graffito{The cities' GDP
per capita was retrieved for $114$ cities from Brooking's Global
MetroMonitor.} On Fig.~\ref{fig:metro_stations_gdp} (left) we plot the number of
stations of different metro systems around the world as a function of the Gross
Metropolitan Product of the city. A linear fit agrees relatively well with the
data ($R^2=0.73$, dashed line), and gives $\frac{\epsilon_s}{\beta} \approx
10^{10}\,\text{dollars/station}$. However, the dispersion around the linear
average behaviour is important: more specific data is needed in order to
investigate whether differences in the construction costs and investments (or
the age of the system) can, alone, explain the dispersion.

Finally, we now consider the number of different lines with distinct tracks. A
natural question is how the number of lines $N_{lines}$ scales with the number
stations $N_s$, that is to say whether lines get propotionally smaller, larger
or the same with the size of the whole system. We plot the number of lines as a
number of stations on Fig.~\ref{fig:metro_stations_gdp} and find that the data
agree with a linear relationship between both quantities ($R^2=0.93$, see the
dashed black line). In other words, the number of stations per line is
distributed around a typical value of $19$, whatever the size of the system.

\section{Railway networks}

\graffito{Data about ridership, network length were easily retrievable for more
 than $100$ countries from the UIC Railisa 2011 database.}
We start by discussing an important difference between railway and subway
networks. In the subway case, the interstation distance is such that it matches
human constraints: $\ell_1\sim 2\,d_0$ where $d_0$ is the typical distance that
one would walk to reach a subway station. For the railway network, the logic is
however different: while subways are built to allow people to move within a
dense urban environment, the purpose of building a railway is to connect
different cities in a country. In addition, due to the long distance and hence
high costs, it seems reasonable to assume that each station is connected to its
closest neighbour. In this respect, the railway network appears as a planar
graph connecting randomly distributed nodes in the plane in an economical way.
If we assume that a country has an area $A$ and $N_s$ train stations, the
typical distance between nearest stations will be

\begin{equation} 
    \ell_N = \sqrt{\frac{A}{N_s}} 
\end{equation}

The total length $L \sim N_s\,\ell_N$ is then given by

\begin{equation} 
    L \sim \sqrt{A\, N_s} 
\end{equation} 

\begin{figure}
    \centering
    \includegraphics[width=0.5\textwidth]{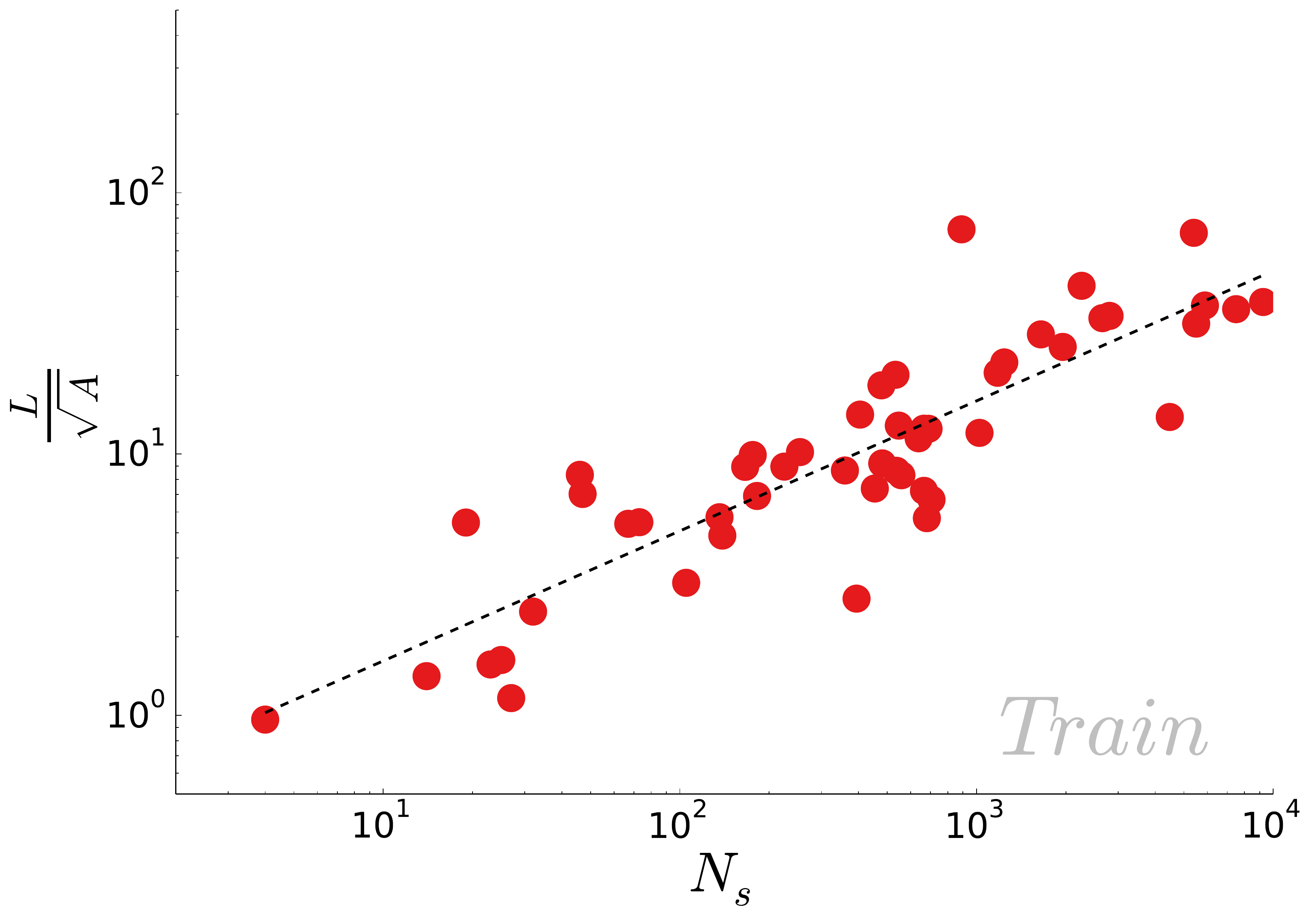}
    \caption{{\bf (Train) Total length and number of stations} Total length of
the railway network $L$ rescaled by the typical size of the country $\sqrt{A}$
as a function of the number of stations $N_s$. The dashed line shows the best
power-law fit on the $50$ data points with an exponent $0.50 \pm 0.08\,(R^2 =
0.87)$.\label{fig:length-stations}} 
\end{figure}

In order to test this relation for different countries, we plot the adimensional
quantity $\frac{L}{\sqrt{A}}$ as a function of the number of stations $N_s$ on
Fig.~\ref{fig:length-stations}. \graffito{The number of stations was more difficult to find. We had to use various
 data sources, mainly scrapping the operators' ticket booking websites.}
 A power law fit gives an exponent $0.50 \pm 0.08\,(R^2 = 0.87)$, which is
 consistent with the previous argument.

At this point, we have a relation between $L$ and $N_s$, but we need to find the
expressions for the other quantities. There are other differences with the
subway system. First, due to the distances involved, the ticket price usually
depends on the distance travelled and we will denote by $f_L$ the ticket price
per unit distance. The relevant quantity for benefits is therefore not the raw
number of passengers--as in subways--, but rather the total distance travelled
on the network $T$. Also, again due to the long distances spanned by the
network, the costs of stations can be neglected as a first approximation, and we
get for the budget the following expression

\begin{equation}
    Z_{train} \simeq T\, f_L - \epsilon_L\, L
\end{equation}

In the steady-state regime $Z_{train}\approx 0$ --- or in other words, the
revenue generated by the network use must be of the order of the total
maintenance costs~\cite{Louf:2013_emergence} (see
Chapter~\ref{chap:cost-benefit} --- we find that

\begin{equation}
    T \sim \frac{\epsilon_L}{f_L} L
\end{equation}

In addition, if we assume that the order of magnitude of a trip is given by
$\ell_N$, the total travelled length is simply proportional to the ridership
$T\sim \ell_N R$ leading to 

\begin{equation}
    R \sim \frac{\epsilon_LN_s}{f_L}
\end{equation}

\begin{figure}
    \centering
    \includegraphics[width=0.5\textwidth]{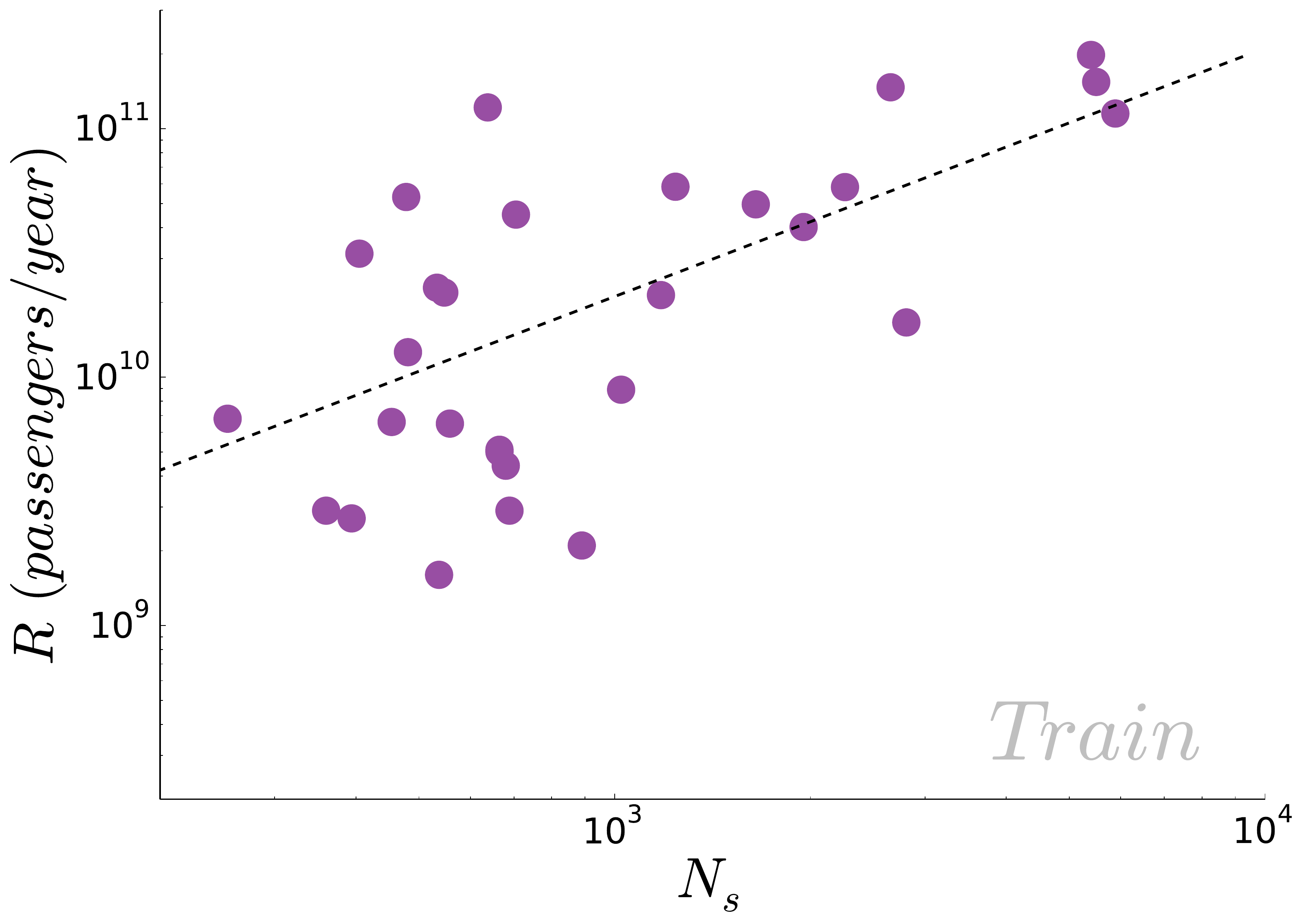}
    \caption{{\bf(Train) Ridership and number of stations} The total yearly
    ridership $R$ of the railway networks as a function of the number of
stations. A linear fit on the $47$ data points gives $R \sim 7.0\,10^8\:N_s$
($R^2 = 0.86$)} \label{fig:train_rider} 
\end{figure}

We thus plot the total daily ridership $R$ as a function of the total number of
stations $N_s$ (figure \ref{fig:train_rider}), and despite the small number of
available data points, a linear relationship between these both quantities seems
to agree with empirical data on average ($R^2 = 0.86$). This result should be
taken with caution, however, due to the important dispersion that is observed
around the average behaviour, and the small number of observations.

\begin{figure}
    \centering
    \includegraphics[width=0.5\textwidth]{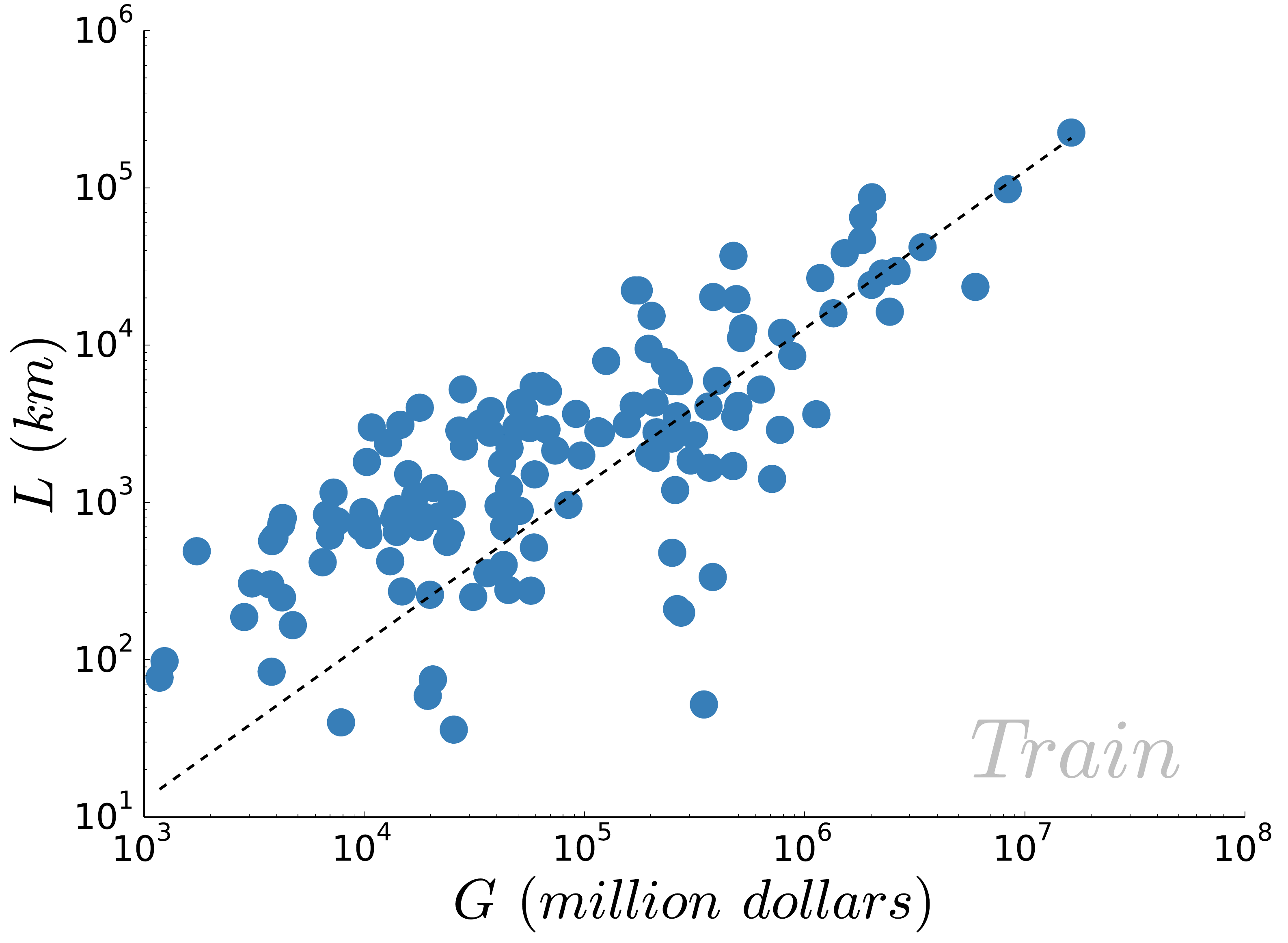}
    \caption{{\bf(Train) Total length of the network and wealth} Total length of
the railway network $L$ as a function of the country GDP $G$. The dashed line
shows the best linear fit on the $138$ data points which gives $\epsilon_L /
\alpha \approx 10^4\, \text{dollars.km}^{-1}\,(R^2 =
0.91)$.\label{fig:length-gdp}} 
\end{figure}

According to the previous result, the total length and the number of stations
are related to each other. We now would like to understand what property of the
underlying country determines the total length of the network. That is to say,
why networks are longer in some countries than in others. As in subway systems,
economical reasons seem appealing. Indeed, the railway networks of some large
african countries such as Nigeria are way smaller than that of countries such as
France or the UK of similar surface areas. A priori, when estimating the cost of
a railway network, one should take into account both the costs of building lines
and the stations. However, as stated above, considering the distances involved,
the cost of building a station is negligible compared to that of building the
actual lines. We thus can reasonably expect to have

\begin{equation} 
    L \sim \frac{\alpha\,G}{\epsilon_L} 
\end{equation}

where $G$ is here the country's Gross Domestic Product (GDP) used as an
indicator of the country's wealth, and $\alpha < 1$ the ratio of the GDP
invested in railway transportation. We plot $L$ as a function of $G$ on
Fig.~\ref{fig:length-gdp} and the data agree well ($R^2 = 0.91$) with a linear
dependence between $L$ and $G$. \graffito{Data about the GDP of different
countries were obtained from the World Bank.} Again, the dispersion indicates
that the linear trend should only be understood as an average behaviour and that
local particularities can have a strong impact on the important deviations
observed.  For instance, the United Arab Emirates are far from the average
behaviour, with a $52\,\text{km}$ network and a GDP of roughly $3\,10^5$ million
dollars. Yet, the construction of a $1,200\,\text{km}$ railway network has been
decided in 2010, which would bring the country closer to the average behaviour. 

\section{Summary}

We have proposed a general framework to connect the properties of railway and
subway systems (ridership, total length and number of stations) to the
socio-economic and spatial characteristics of the country or city they are built
in (population, area, GDP). Despite their simplicity, our arguments agree
satisfactorily with the data we gathered for more than $100$ subway systems and
$50$ railway networks accross the world. It should be noted that the noise
associated with these data (and sometimes their definition, see Material and
Methods) makes it difficult to infer behaviours from the empirical analysis
alone. Therefore, the most appropriate way to proceed, we believe, is to make
assumptions about the systems and build a model whose predictions can then be
tested against data.

This study suggests that the fundamental difference between railways and subways
comes from the determination of the interstation distance. While it is imposed
by human constraints in the subway case, the railway network has to adapt to the
spatial distribution of cities in a country. This remark is at the heart of the
different behaviors observed for railways and subways (see
Table~\ref{table:summary} for a summary of these differences). 

\begin{table}[!ht]
\centering
\begin{tabular}{|c|c|c|}
\hline
 & {\bf Subway} & {\bf Train} \\
 \hline
$L / N_s$ & cste. & $\sqrt{\frac{A}{N_s}}$\\
$R$ & $\frac{P}{A}\,N_s$ & $N_s$ \\
$G$ & $N_s$ & $L$ \\
\hline
\end{tabular}
\caption{{\bf Summary of the differences between subways and railways}
We summarize the difference of behaviour between subways and railways. The
scaling of the length $L$ of the network with the number of stations $N_s$
reveals the different logics behind the growth of these systems. Another
difference lies in the total ridership $R$: while it depends on the population
density $P/A$ for subways, it only depends on the number of stations $N_s$ for
train networks. Finally, the size of both types of network can be expressed as a
function of the wealth of the region, represented here by the GDP $G$. However,
because the interstation length is constant for subways, the size is better
expressed in terms of the number of stations $N_s$; in the case of railway
networks, the cost of stations are negligible compared to the building cost of
lines, and the size is better expressed in terms of the total length $L$.
\label{table:summary}} 
\label{tab:label} 
\end{table}

The previous arguments are able to explain the average behaviour of various
quantities. Nevertheless, it would be interesting to identify deviations from
these behaviours, and see whether they correlate --for instance-- with
topological properties of the system, as suggested in~\cite{Derrible:2009} or
other properties of the network and the region. We think that the relations
presented here provide  nevertheless a simple framework within which local
particularities can be discussed and understood. We also think that this
framework could be used as a useful null-model to quantify the efficiency of
individual transportation networks, and compare them to each other. This would
however require more specific data than those that were available to us. 

While we have focused on an average, static description of metro systems, we
believe that our study provides a better understanding of how these systems
interact with the region they serve. This new insight is a necessary step
towards a model for the growth of subway systems that takes the characteristics
of the city into account. Indeed, although models of network growth exist, the
length of networks and nodes at a given time is usually imposed exogeneously,
instead of being linked to the socio-economic properties of the substrate. This
study provides a simple approach to these complex problems and could help in
building more realistic models, with less exogeneous parameters.

It would be interesting to gather data about the exact structure of all the
studied network, so as to study whether there is a relationship between the
topology (degree distribution, detour index, etc.) of these networks and
properties of the substrate, as was done for the road network
in~\cite{Levinson:2012}.

Finally, gathering historical data should allow to address the problem of the
conditions for the appearance of a subway in a city. In particular, we observe
empirically that the GDP of the cities that have a subway system is always
larger than about $10^{10}$ dollars, a fact that calls for a theoretical
explanation.





\ctparttext{\centering Self-explanatory title.} 

\part{Conclusion} 
\label{part:conclusion}

%
\chapter{Conclusion}
\label{sec:conclusion}

\begin{flushright}{\slshape    
If people never did silly things\\
nothing intelligent would ever get done.} \\ \medskip
--- Ludwig Wittgenstein~\cite{Luckhardt:1979}
\end{flushright}

\bigskip

In this thesis, we have adopted a `physicist' approach to the study of a system
that traditionally belonged to the realm of social sciences: the city. We have
tried to show that simple approaches allow to better understand
these complex systems. Although simple models with a few variables cannot
reproduce all the properties and behaviours of the observed phenomena, they
allow us to uncover the dominant mechanisms that are responsible for their most
salient features. Does it mean that our approach is the only valid approach?
Probably not. Is it useful? Certainly, as it structures
our knowledge and sets a solid basis for future investigations.\\

In the first part, we have reviewed the evolution of the concept of
polycentricity in the literature, and the methods used to identify and count the
number of centers. Doing so, we provided evidence for the increasing number of
activity centers with population size, a phenomenon we called `polycentric
transition'. We then proposed an out-of-equibrium, stochastic model of
city growth that reproduces the empirical regularity, and explains the
transition with the increasing levels of congestion as cities get larger. This
model is a substantial improvement over the models presented in the Economics
literature: it makes predictions that are supported by data, and allows to
identify the mechanisms responsible for the observed phenomena. 

In the second part, we further use the model to give a prediction for the
scaling exponent of the total distance commuted daily, the total length of the
road network, the total delay due to congestion, the quantity of
CO\textsubscript{2} emitted, and the surface area with the population size of
cities. We successfully test these predictions with data gathered for US urban
areas.

In a third part, we focus on the quantitative description of the patterns of
residential segregation. For the first time in the quantitative literature, we
propose an explicit definition of segregation as a deviation from a random
distribution of individuals across the urban space. This definition provides a
unifying theoretical framework in which segregation can be empirically
characterised. We propose a measure of interaction between the different
categories. Building on the information about the attraction and repulsion
between categories, we are further able to propose a definition of classes that
is quantitative and unambiguous. The framework also allows us to identify the
neighbourhoods where the different classes concentrate, and characterise their
properties and spatial arrangement. Finally, we revisit the traditional
dichotomy between poor city centers and rich suburbs and provide a measure that
is adapted to anisotropic, polycentric cities. 

In the fourth and last part, we briefly reviewed the results we have obtained in
the study of spatial networks. We first presented a quantitative method to
classify cities based on their street patterns, which we applied to a set of
$131$ cities across the world. Then, we introduced an iterative model for the
growth of spatial networks that is based on cost-benefit considerations. The
model exhibits interesting features: a crossover between the Minimum Spanning
Tree and the star graph, with an intermediate regime characterised by the
emergence of spatial hierarchy. Finally, we proposed a general coarse-grained
approach -- based on a
cost-benefit analysis -- that accounts for the scaling properties of the main
quantities characterizing railway and subway networks (the number of stations, the total
length, and the ridership) with the substrate's population, area and wealth.
We showed that the length, number of stations and ridership of
subways and rail networks can be estimated knowing the area, population and
wealth of the underlying region. These predictions are in good agreement with
data gathered for about $140$ subway systems and more than $50$ railway networks in the
world.\\ 

The field is still in its infancy compared to more mature sciences, but there
are very good reasons to hope for the convergence of knowledge and methods into
a new discipline. Into what we may call -- following Michael Batty -- a Science of
Cities~\cite{Batty:2012,Batty:2013}. It is difficult at this stage to say what this Science will look like,
and what kind of results it can pretend to achieve. Nevertheless, it is tempting to
compare the current state of the field to the study of planetary motions
before Isaac Newton's \emph{Philosophi\ae  Naturalis Principia Mathematica}, or
the study of electromagnetism before James Clerk Maxwell's \emph{A Dynamical
Theory of the Electromagnetic Field}; a set of stylized facts and empirical laws
that are yet to be unified in a coherent theory. 

This is not to say that one should look for a unifying set of equations, or that
laws about urban system will have the same permanence as those describing
natural phenomena. No two theories are alike -- even in Physics. But we believe
that the underlying methodological principles have a universal character.
Nothing can go fundamentally wrong if data are the ultimate judge of the
validity of our theoretical endeavours.

\section{Lessons learned}
\label{sec:what_the_past_3_years_have_brought}

The last $3$ years have taught me lessons that go beyond simple scientific
knowledge.

\subsection{Thinking the city}
\label{sub:thinking_the_city}

A first lesson, painstakingly learned during this thesis is that \emph{thinking} the
city is as important as \emph{measuring} the city, or \emph{modeling} the city.
Concepts guide us and tell us what to measure, what to model. In the same way
measures and model can tell us what to think. It would be very naive to
believe that scientific enquiries are fueled by the sole discussion between
measures and models. In fact, many studies are based upon an hypothesis, a
pattern that the author has seen and whose existence she is trying to prove on a
quantitative basis. 

It is also certainly true that the most difficult and important problems are
conceptual in nature.  It is impossible to define a city quantitatively 
before you have formed---with words, possibly drawings---a conceptual picture of
what a city is.  It is impossible to study segregation before you have logically
clarified what one means by segregation. However quantitative, an investigation
built upon weak conceptual foundations is unlikely to go anywhere, or to say
anything substantial. On the other hand, when the thoughts have settled and the
question is clear, one can quickly make a substantial contribution. In this
sense, qualitative and quantitative investigations are not incompatible: they
are really two sides of the same coin.\\

\subsection{Disciplinary borders}
\label{sub:disciplinary_borders}

The topics I had the chance to tackle during these 3 years of PhD were very
diverse. In retrospect, this was a real chance. This pushed me to browse a wide
literature that encompassed many different disciplines. What I found striking
while perusing articles and books is the tendency of the different communities to
ignore one another. 

The problem, however, is not to blame on individuals. While there may be
deliberate omissions here and there, authors are generally willing to cite the
appropriate literature when they are aware of its existence. The issue, I
believe, is institutional. It stems from the academic organisation of Science,
and the existence of disciplinary borders.

But do disciplinary borders still mean anything? While there is an undeniable
historical justification to the existence of disciplines, do they still make
sense, scientifically speaking? Should the path-dependency in the evolution of
the man-made, academic classification of sciences dictate what research avenues
are worth being pursued today? At a time when some topics -- including cities
-- get an increasingly multi-disciplinary attention, these questions are worth
asking. Science is fueled by ignorance and questions, not knowledge. It may therefore be
time to organise communities around common questions, rather than (overlapping)
corpora of knowledge.

\section{If I had to write a second thesis (Future directions)}
\label{sec:limitations}

What would I write about -- or at least try to -- if I had to start my
thesis all over again? This is another way of saying: what are the next steps?
Many clues can be found in the various parts of this manuscript. Indeed, I have tried
to explicit the limitations of the empirical methods and models presented. In
these remarks lie many potential avenues for future research. In the following,
I will present some other ideas that sprung over the last $3$ years.\\

I would probably start with the basics, with the single noun that was most often
printed in these pages: Cities~\footnote{\emph{Not} verified on data.}. It is
indeed uncomfortable -- to say the least -- that our most fundamental object,
the city, is ill-defined, and that most empirical studies possibly rely on a definition
that is not suited to the investigation they undertake.  This lack of serious
definition compromises the comparison between cities of different countries, or
at different points in time. I am, of course,
not the first person to acknowledge this empirical shortcoming. In fact, it is a
long-lasting worry of geographers who have been trying to produce harmonised
database for many years~\cite{Pumain:2015}. Yet, we still lack of an
unambiguous, theoretically grounded definition of what a city is. And this is
problematic, since statistical institutes' results are based on what is believed
to be the best definition of the city at a time. Which in turn influences the
research on cities. If we want to exhibit robust empirical results, compare the
results obtained in different countries, we therefore need to start worrying
about the definition of the system we are studying. We need to know \emph{what}
cities we are talking about.\\

\begin{figure}
    \centering
    \includegraphics[width=1\textwidth]{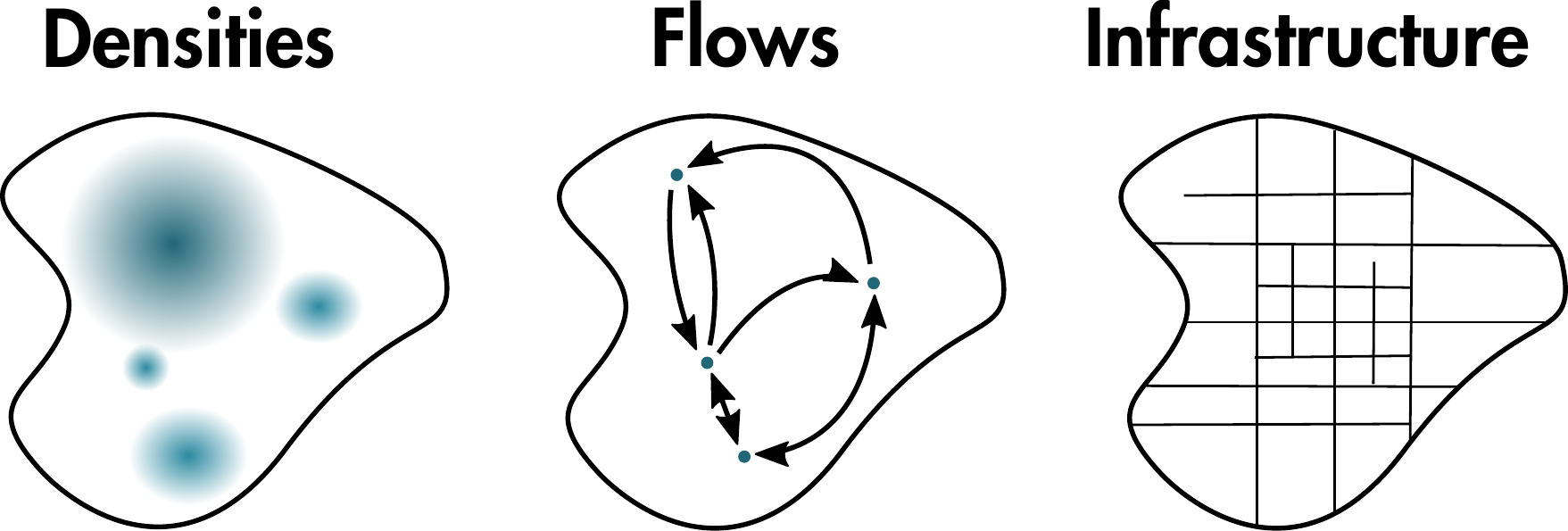}
    \caption{{\bf Intra-urban organisation.} Cities are first and foremost defined by
    the concentration of populations and various activities. The fact that
residences and activities have different locations is responsible for the
existence of flows of people, goods, etc. across the urban space. These flows
occur on appropriate infrastructure.\label{fig:intra_urban}}
\end{figure}

Once the boundaries are defined, we can start studying the way objects are
scattered within them. By objects, I mean buildings, roads, and first and
foremost people. The way we traditionally study the repartition of objects in
space is through the study of densities. But density profiles are too
complicated to comprehend for our brains, especially when cities get large. So
complicated, that an entire sub-field is dedicated to their study: urban
form~\cite{Tsai:2005,Schwarz:2010,LeNechet:2015}.
Authors attempt to solve this problem by providing simple measures 
that extract a single number from the profile. A single number is however too
simple to be able to describe accurately complex spatial distributions. What we need
is a meso-scale representation, somewhere between the micro-scale picture (the
density profile itself) and the macro-scale picture (a single number to
summarize the density profile). Hopefully, because `centers' are themselves a
mesoscopic structure, their definition should emerge naturally from such a
representation.\\

Once one is able to provide an accurate description of density profiles, the
possibilities start to diverge. An obvious worry, when one has a picture of the
city's population at different times of the day, is the way these profile
transform one into another. This is linked to commuting --but not only,
commuting representing only $20\%$ of total travels in the
US~\cite{FHWA-PL-11-022}-- and the study of congestion of networks. 

We could first try to explicit the link between the urban form (typically the
residential and employment densities) and mobility
patterns~\cite{Ma:2006,Chowdhury:2013}.  For instance, we could wonder: what
proportion of commuting flows is due to the spatial mismatch between jobs and
residences? 

A futher worry linked to commuting is that of congestion:
understanding how traffic jams are formed, how they propagate and devise
strategies to mitigate them, either by influencing the transportation
infrastructure, the spatial repartition of residences and employment, or the
behaviour of people themselves.  This is far from being a recent worry, but
there is room for new approaches that leverage the knowledge we have about
network and phase transition in physics. A first step in this direction has been
made by the authors of~\cite{Li:2015}, but there is surely more to be understood
and discovered.  

Modeling congestion also implies understanding the individual behaviour of
people when they are moving from a point to another in cities. Altough most
research nowadays assume that people choose the shortest (time or distance)
path, GPS data now provide overwhelming evidence that this is not the
case~\cite{Manley:2015}. So, while there is a clear need to understand the mesoscopic
picture (how congestion spread), there is also is a critical need to understand
the microscopic picture (how people behave).\\

So far we have talked about the movement induced by the spatial mismatch between
residential areas and activity areas. One might also want to study the
characteristics of the spatial repartition of people. Inhabitants of cities are
not just a combination of a latitude and a longitude, a point on a map. Like you
and me, they are characterised by different qualities, some of which are
measurable: their income, their education level, their ethnicity, etc. A
natural question, that has interested sociologist and geographers, is to wonder
whether people's residence is independent of these characteristics, or whether
these characteristics have an influence on the spatial repartition of
individuals.  

In this thesis, we provided a rigorous method to study the patterns of
segregation in the presence of multiple income categories. The method is far
more general, however. It could be used to study the concentration of any
category (be it ethnic categories, or certain business types, etc.)
in certain regions of the urban space, and quantify the resulting spatial pattern. As a
matter of fact, more work is needed to be able to identify the topology and
geometry of these distributions. The problem is very close to the description of
density pattern described above.

The definition of neighbourhoods (again, a mesoscopic structure) is also not
completely satisfactory. Often, it relies on non-overlapping census boundaries
that were drawn to maximise the intra-neighbourhood homogeneity and maximise the
inter-neighbourhood heterogeneity. Although this may be useful for political
institutions to target the most segregated regions of the city, this does not
account for how segregation is witnessed by individuals, at an individual level.
This has recently been questioned in the Sociology literature, and there has
recently been new attempts to define neighbourhoods based on social ties~\cite{Hipp:2012}.\\

There are many more ideas that would deserve to be explored, many more topics
that are worthy of attention. I hope the years to come will give me the opportunity
to address some of them. But not now; this thesis has to stop somewhere.  





\cleardoublepage

\label{app:bibliography} 

\manualmark
\markboth{\spacedlowsmallcaps{\bibname}}{\spacedlowsmallcaps{\bibname}} 
\refstepcounter{dummy}

\addtocontents{toc}{\protect\vspace{\beforebibskip}} 
\addcontentsline{toc}{chapter}{\tocEntry{\bibname}}

\bibliographystyle{plainnat}
\begin{btSect}{biblio/bibliography}
\btPrintCited
\end{btSect}



\end{document}